\documentclass[12pt]{article}
\pdfoutput=1
\usepackage{lmodern}
\usepackage{amsfonts,amsmath,amssymb,amstext,mathrsfs,epsfig,cite,setspace,bigstrut,framed}
\usepackage[scr=boondox]{mathalfa}
\usepackage{euscript}
\usepackage{bbm}
\usepackage{dsfont}
\usepackage{pifont}
\usepackage[all]{xy}
\usepackage{color,xcolor,soul}
\usepackage{latexsym}
\usepackage{graphics,graphicx} 
\usepackage[colorlinks = true, allcolors = blue!30!black!80!, linktocpage=true]{hyperref}
\usepackage[bbgreekl]{mathbbol}
\usepackage{fancyhdr}
\usepackage{lastpage}
\usepackage{etoolbox}
\usepackage{setspace}
\usepackage{titlesec}
\usepackage{cancel}
\usepackage{slashed}
\usepackage{comment}
\usepackage[framemethod=tikz]{mdframed}
\usepackage{empheq}
\usepackage{tikz}
\usepackage{tcolorbox}
\tcbuselibrary{theorems}
\usepackage[font=footnotesize,labelfont=bf,singlelinecheck=false,width=1.05\textwidth]{caption}
\usepackage[utf8x]{inputenc}
\usepackage[T1]{fontenc}

\DeclareMathAlphabet{\dutchcal}{U}{dutchcal}{m}{n}

\makeatletter \@addtoreset{equation}{section} \makeatother
\renewcommand{\theequation}{\thesection.\arabic{equation}}

\def\k{\kappa}
\def\r{\rho}
\def\a{\alpha}

\def\b{\beta}

\def\g{\gamma}
\def\G{\Gamma}
\def\d{\delta}
\def\D{\Delta}
\def\e{\epsilon}

\def\p{\pi}

\def\c{\chi}

\def\th{\theta}

\def\m{\mu}
\def\n{\nu}

\def\l{\lambda}
\def\L{\Lambda}
\def\s{\sigma}

\def\vf{{\varphi}}
\def\bvf{{\bar\varphi}}

\def\cg{{\cal{G} }}

\def\IR{\relax{\rm I\kern-.18em R}}

\def \bphi { {\bar \phi} }
\def \z { {\bar z} }

\def\diag{{\rm diag}}

\def\nb{{\nabla}}
\def\hve{{\hat\ve}}

\def\IR{\relax{\rm I\kern-.18em R}}
\def\IL{\relax{\rm I\kern-.18em L}}

\def\inv{^{\raise.15ex\hbox{${\scriptscriptstyle -}$}\kern-.05em 1}}

\def\tr{{\rm tr}}

\def\bes{\begin{split}}
\def\ens{\end{split}}

\def\del{\partial}

\setul{0.5ex}{0.3ex}
    \definecolor{Red}{rgb}{1,0.0,0.0}
    \setulcolor{Red}
\definecolor{Blue}{rgb}{0.0,1,0.0}
    \setulcolor{Blue}
    
\newmdenv[%
middlelinecolor=gray!30!,
middlelinewidth=1pt,
backgroundcolor=gray!10!,
roundcorner=3pt
]{myidentity}

\mdfdefinestyle{mystyle}{%
middlelinecolor=gray!80!white,
middlelinewidth=1pt,
backgroundcolor=blue!5!white,
roundcorner=10pt,
subtitlebelowline=true,
frametitlefont={\normalfont\bfseries\sffamily\color{gray!90!white}},
innertopmargin=.1cm,
innerbottommargin=.3cm,
}

\def\bebx #1 \eebx{\begin{empheq}[box={\tcbhighmath[colframe=gray!90!white,colback=white]}]{align} #1 \end{empheq}}

\def\bbxd#1\ebxd{\begin{myidentity} \vskip -.4cm #1 \end{myidentity}\vskip-.2cm}

\def\bframe#1{\begin{mdframed}[style=mystyle,frametitle={#1}]}
\def\eframe{\end{mdframed}}

\definecolor{origblue}{RGB}{0,118,186}
\definecolor{hatteal}{RGB}{0,171,142}
\definecolor{arrowteal}{RGB}{0,108,101}
\definecolor{redbranch}{RGB}{238,34,12}
\definecolor{redtick}{RGB}{181,23,0}

\newcommand{\red}[1]{\textcolor{red}{#1}}
\newcommand{\blue}[1]{\textcolor{blue}{#1}}

\def\){\right)}
\def\({\left( }
\def\]{\right] }
\def\[{\left[ }

\def\NO{\nonumber}

\newcommand{\be}{\begin{equation}}
\newcommand{\ee}{\end{equation}}

\def\bea{\begin{eqnarray}}
\def\eea{\end{eqnarray}}

\def\bal#1\eal{\begin{align}#1\end{align}}

\def\bald{\begin{aligned}}
\def\eald{\end{aligned}}

\def\bsub{\begin{subequations}}
\def\esub{\end{subequations}}

\def\beqx{\begin{displaymath}}
\def\eeqx{\end{displaymath}}

\newcommand{\bmat}{\left(\begin{array}}
\newcommand{\emat}{\end{array}\right)}

\def\a{\alpha}
\def\b{\beta}
\def\c{\chi}
\def\d{\delta}
\def\e{\epsilon}
\def\f{\phi}
\def\g{\gamma}
\def\h{\eta}
\def\j{\psi}
\def\k{\kappa}
\def\l{\lambda}
\def\m{\mu}
\def\n{\nu}
\def\o{\omega}
    
\def\p{\pi}

    \def\th{\theta}
\def\r{\rho}
\def\s{\sigma}
\def\t{\tau}
\def\x{\xi}
\def\z{\zeta}
\def\D{\Delta}
\def\F{\Phi}
\def\G{\Gamma}

\def\L{\Lambda}

\def\ve{\varepsilon}

    \def\vth{\vartheta}

\def\vf{\varphi}

\def\ca{{\cal A}}

\def\cd{{\cal D}}

\def\cf{{\cal F}}
\def\cg{{\cal G}}

\def\cj{{\cal J}}
\def\ck{{\cal K}}
\def\cl{{\cal L}}
\def\cm{{\cal M}}
\def\cn{{\cal N}}
\def\co{{\cal O}}
\def\cp{{\cal P}}

\def\cs{{\cal S}}
\def\ct{{\cal T}}

\def\cw{{\cal W}}

\def\cz{{\cal Z}}

\def\bb#1{\ensuremath{\mathbb{#1}}} 
\def\dc#1{\ensuremath{\dutchcal{#1}}} 

\def\bo{{\raise-.3ex\hbox{\large$\Box$}}}               
\def\pa{\partial}                                       
\def\face{{\raise.2ex\hbox{$\displaystyle \bigodot$}\mskip-2.2mu \llap {$\ddot
        \smile$}}}                                   
\def\>{\rangle}                                      
\def\<{\langle}                                      

\def\wt#1{\widetilde{#1}}                            
\def\Hat#1{\widehat{#1}}                             
\def\lbar#1{\ensuremath{\overline{#1}}}              
\def\leftrightarrowfill{$\mathsurround=0pt \mathord\leftarrow \mkern-6mu
        \cleaders\hbox{$\mkern-2mu \mathord- \mkern-2mu$}\hfill
        \mkern-6mu \mathord\rightarrow$}        
\def\dvec#1{\vbox{\ialign{##\crcr
        \leftrightarrowfill\crcr\noalign{\kern-1pt\nointerlineskip}
        $\hfil\displaystyle{#1}\hfil$\crcr}}}           
\def\tr{{\rm tr \,}}                                    
\def\diag{{\rm diag \,}}                                

\def\-{\hphantom{-}}

\begin{document}
\begin{titlepage}
\vfill
\begin{flushright}
{\normalsize KIAS-P26082}\\
\end{flushright}
\vfill

\begin{center}
\vbox{\center\LARGE\bf  Anomaly effective actions for 4d $\cn=1$ 
superconformal field theories}
\vbox{\center \large Pantelis Panopoulos$^{a}$ and  
Ioannis Papadimitriou$^{b,\,c}$}

\vbox{\center\em $^{a}$  School of Physics, Korea Institute for Advanced Study, Seoul 02455, Korea}
\vbox{\center\em $^{b}$Division of Nuclear and Particle Physics, Department of Physics,
National and Kapodistrian University of Athens, 157 84 Athens, Greece}
\vbox{\center\em $^{c}$Department of Physics, College of Science, Kyung Hee University, Seoul 02447, Korea}

\vbox{\center \href{mailto: ppanopoulos@kias.re.kr}{\tt ppanopoulos@kias.re.kr},\,\,\, 
\href{mailto: ioannis.papadimitriou@phys.uoa.gr}{\tt ioannis.papadimitriou@phys.uoa.gr}}
\end{center}
\vfill

\begin{abstract}

We present a comprehensive analysis of the Wess-Zumino anomaly effective action of four-dimensional $\cn=1$ superconformal field theories in components. After reviewing the coset construction of the Wess-Zumino action in the presence of an unbroken subgroup, the gauging of the conformal algebra and the Wess-Zumino effective action for non-supersymmetric conformal field theories with a U(1) symmetry in two and four dimensions, we revisit the superconformal anomalies of $\cn=1$ conformal supergravity. Solving the Wess-Zumino consistency conditions we determine both the bosonic and fermionic superconformal anomalies in a two-parameter family of schemes, to quadratic order in the gravitino. Turning on a chiral compensator multiplet and identifying a local super-Poincar\'e subalgebra of the conformal supergravity algebra without gauge-fixing the coset symmetries, we employ the coset construction and the superconformal anomalies in order to evaluate the Wess-Zumino action for generic $\cn=1$ superconformal field theories coupled to conformal supergravity. By construction, this action reproduces the Weyl, R-symmetry, and S-supersymmetry coset anomalies, but its transformation under the unbroken Poincar\'e supersymmetry differs from the corresponding anomaly of the conformal supergravity functional. We show that there exists a local functional of the coset-invariant old-minimal fields that, when added to the Wess-Zumino action, results in a generalized Wess-Zumino action that reproduces all anomalies of the conformal supergravity functional. The same local functional enables us to determine the super-Weyl anomalies and the Wess-Zumino action of old-minimal supergravity, providing a component version of the known superspace Wess-Zumino action.    

\end{abstract}

\vfill
\end{titlepage}

\tableofcontents
\addtocontents{toc}{\protect\setcounter{tocdepth}{2}}
\renewcommand{\thefootnote}{\arabic{footnote}}
\setcounter{footnote}{0}

\section{Introduction and summary}

Global symmetry anomalies impose strong constraints on the dynamics of quantum field theories (QFTs) and have important physical consequences, even if the underlying symmetry is spontaneously broken. 't Hooft anomaly matching \cite{tHooft:1979rat} and the $a$-theorem \cite{Komargodski:2011vj} are prominent examples. Much of the information described by anomalies is encoded in the Wess-Zumino (WZ) effective action \cite{Wess:1971yu}. This is a local functional of the background gauge fields that couple to the classically conserved Noether currents, as well as of the Goldstone fields that arise when the symmetry is spontaneously broken. Besides locality, the defining property of the WZ action is that its transformation under the spontaneously broken symmetries reproduces the corresponding anomalies, expressed as local functionals of the background gauge fields. 

This property can be used to define the WZ action even if the global symmetry is not spontaneously broken. The would-be Goldstone fields in this case can be understood as `compensators', redundant auxiliary background fields corresponding to the parameters of a finite gauge transformation of the background gauge fields. The WZ action, therefore, encodes the effects of anomalies even when the corresponding symmetries are not spontaneously broken. It determines, for example, the Casimir charges of a theory, and more generally the anomalous transformation of the current operators under finite symmetry transformations. A familiar example is the Schwarzian derivative that appears in the finite conformal transformation of the energy-momentum tensor of a two-dimensional conformal field theory (CFT). 

Here, we consider the WZ action of generic two- and four-dimensional CFTs, and focus in particular on $\cn=1$ supersymmetric conformal field theories (SCFTs) in four dimensions. Supersymmetry introduces both technical and conceptual complications, as well as stronger constraints on the dynamics. Besides the inherent difficulty of fermionic background fields, there often exists a choice between different background multiplets a supersymmetric QFT or CFT can be coupled to \cite{Komargodski:2010rb}. Each of these background multiplets defines a distinct current multiplet, set of superconformal anomalies, and WZ action. In this work, and in the companion article \cite{omct}, we consider $\cn=1$ off-shell conformal supergravity \cite{Kaku:1977pa,Kaku:1977rk,Kaku:1978nz,Townsend:1979ki} and old-minimal supergravity \cite{Stelle:1978ye,Ferrara:1978em}, both of which are consistent backgrounds for renormalized $\cn=1$ SCFTs. They define respectively the conformal and Ferrara-Zumino current multiplets. 

One of the main results of the present work is the component form of the superconformal anomalies of both the conformal and old-minimal multiplets for arbitrary values of the $a$ and $c$ anomaly coefficients, determined to quadratic order in the gravitino. For the conformal supergravity anomalies, we revisit the solution of the WZ consistency conditions in \cite{Papadimitriou:2019gel}, generalizing it to a two-parameter family of schemes. We also correct the form of the Q-supersymmetry anomaly given in \cite{Papadimitriou:2019gel} for the case $a\neq c$, and we provide the explicit form of the quadratic in the gravitino terms of the Weyl and R-symmetry anomalies, which were not given in \cite{Papadimitriou:2019gel}. The anomalies of conformal supergravity have also been determined to all orders in the gravitino using BRST techniques in \cite{Imbimbo:2023sph,Imbimbo:2025ffw}. 

The anomalies of old-minimal supergravity are determined in two different schemes, without explicitly solving the WZ conditions of the local old-minimal algebra. The first scheme is directly related with conformal supergravity and the corresponding anomalies can be simply deduced from those of conformal supergravity. This scheme possesses an anomalous Poincar\'e supersymmetry. However, we systematically construct a local counterterm that eliminates the Poincar\'e supersymmetry at the expense of introducing an anomaly in another local symmetry of old-minimal supergravity. The entire super-Poincar\'e algebra is anomaly free in this scheme, while all anomalies of old-minimal supergravity are placed in the super-Weyl symmetries, which are present when an SCFT, rather than an SQFT, is coupled to background old-minimal supergravity. This is the scheme implicitly selected by superspace. However, as we show in detail here and further discuss in \cite{omct}, the superspace counterterm constructed in \cite{Kuzenko:2019vvi} corresponds only to part of the complete counterterm that connects the two schemes.

The superconformal anomalies are a prerequisite for the other main result of this work, namely the construction of the WZ action in components for both the conformal and Ferrara-Zumino current multiplets of generic four-dimensional $\cn=1$ SCFTs. A major obstacle in the application of the coset construction to the computation of the WZ action of conformal supergravity is the identification of a super-Poincar\'e subalgebra of the conformal supergravity algebra. The usual construction of old-minimal supergravity via conformal calculus involves setting the compensators to zero, a configuration that is preserved by a super-Poincar\'e subalgebra of the conformal supergravity algebra. However, we cannot set the compensators to zero if our goal is to determine the WZ effective action. Instead, by accompanying diffeomorphisms, Lorentz and Q-supersymmetry transformations with suitable Weyl, R-symmetry and S-supersymmetry transformations with composite compensator-dependent parameters, we construct super-Poincar\'e generators that annihilate the compensators. This enables us to identify a local super-Poincar\'e subalgebra of the conformal supergravity algebra without any constraint on the compensators. The resulting action obtained by evaluating the WZ transgression formula reproduces the coset anomalies, but its Poincar\'e supersymmetry transformation differs from the corresponding anomaly by coset-invariant terms. Remarkably, the same local counterterm that interpolates between the two old-minimal schemes cancels this coset-invariant terms, leading to a generalized WZ action that reproduces both the coset and subgroup anomalies of conformal supergravity. The old-minimal WZ action can be obtained directly from this generalized WZ action of conformal supergravity, without explicitly applying the transgression formula to the old-minimal anomalies. The result is the component form of the superspace WZ action obtained in \cite{Schwimmer:2010za}.

The paper is organized as follows. In section \ref{sec:WZactions} we review the coset construction of the WZ effective action for a general non-Abelian gauge group $G$ with a generally anomalous subgroup $H$. By construction, the WZ action reproduces the coset anomalies, but we show that its transformation under the anomalous subgroup $H$ differs from the original subgroup anomaly by a coset-invariant term. This observation plays an important role in the construction of a generalized WZ action that reproduces all anomalies of four-dimensional $\cn=1$ SCFTs in the subsequent sections. In section \ref{sec:gengauging} we review the gauging of spacetime symmetries, in particular of the conformal algebra, in order to illustrate the necessity of curvature constraints and the appearance of diffeomorphisms and local Weyl transformations. The formalism developed in the previous two sections is utilized in section \ref{sec:nonSUSYCFTs} in order to compute the WZ action for generic non-supersymmetric CFTs with a U(1) symmetry in two and four dimensions. The corresponding WZ currents are computed explicitly and we show that they satisfy the anomalous Ward identities when the compensator fields are integrated out. 

In section \ref{sec:conf_sugra} we turn to four-dimensional $\cn=1$ SCFTs coupled to background $\cn=1$ conformal supergravity. Revisiting the solution of the WZ consistency conditions, we generalize and partly correct the expressions for both the bosonic and fermionic superconformal anomalies originally obtained in \cite{Papadimitriou:2019gel}, determining these in a two-parameter family of schemes and to quadratic order in the gravitino. The Bardeen-Zumino terms for all current operators in the conformal multiplet are also constructed, allowing us to obtain the corresponding covariant currents and anomalies. In section \ref{sec:compensator_multiplet} we introduce the chiral compensator multiplet that enables us to apply the coset construction to the conformal supergravity algebra. Accompanying diffeomorphisms, Lorentz and Q-supersymmetry transformations with suitable coset transformations with compensator-dependent composite parameters, we are able to identify a super-Poincar\'e subalgebra of the conformal supergravity algebra acting trivially on the compensator multiplet, without gauge-fixing the coset symmetries. This is the unbroken subalgebra in the coset construction of the WZ action of $\cn=1$ conformal supergravity. 

The WZ action of conformal supergravity is evaluated explicitly in section \ref{sec:WZaction}, using the coset construction and the superconformal anomalies determined in section \ref{sec:conf_sugra}. Its transformation under the coset symmetries reproduces the Weyl, R-symmetry and S-supersymmetry anomalies. Under the unbroken super-Poincar\'e subalgebra, however, it reproduces the corresponding anomalies of the conformal supergravity partition function, up to terms that involve the coset-invariant fields of old-minimal supergravity. We explicitly construct a local functional of the old-minimal fields whose super-Poincar\'e transformations precisely cancel the extra terms in the corresponding transformations of the conformal supergravity WZ action. This leads to a generalized WZ action that reproduces all anomalies of conformal supergravity, including those under the anomalous super-Poincar\'e subalgebra. The same local functional of the old-minimal fields enables us to explicitly construct a second WZ action, that of old-minimal supergravity. The old-minimal supergravity WZ action is invariant under the super-Poincar\'e subalgebra and can be identified with the component form of the superspace WZ action obtained in \cite{Schwimmer:2010za}.

Our conventions, together with a number of useful identities involving spinor manipulations and torsionful geometry are collected respectively in appendix \ref{appendix:spinors} and \ref{appendix:torsion}.

\section{Anomalies and the Wess-Zumino effective action}
\label{sec:WZactions}

Consider a theory that is classically invariant under a global symmetry $G$. This symmetry may be gauged by coupling the theory to a background gauge field, $A_\m$, acting as a source for the associated Noether current, $\cj^\m$. Integrating out the microscopic degrees of freedom leads to a non-local functional, $\cw[A]$, which may not be invariant under gauge transformations of the background gauge field $A_\m$, indicating the presence of a 't Hooft anomaly. 

Let $S[\Phi,A]$ be the classical action describing the dynamics of the microscopic degrees of freedom, collectively denoted by $\Phi$, in the presence of a background gauge field $A_\m$. The partition function of the theory on such a background is given by
\be\label{Z}
\cz[A]=\int \cd\Phi\,e^{iS[A,\Phi]}\,.
\ee
It is a non-local functional of $A_\m$ that corresponds to the generating function of current correlators. Analogously,  
\be\label{W}
\cw[A]\equiv -i\log \cz[A]\,,
\ee
is the generating function of connected current correlators. In what follows, $\cz[A]$ and $\cw[A]$ will refer to the renormalized versions of the quantities defined in respectively \eqref{Z} and \eqref{W}.

If the symmetry $G$ is anomalous at the quantum level, an infinitesimal gauge transformation of $\cw[A]$ with parameters $\vth^a(x)$ takes the form  
\be
\d(\vth) \cw[A]=\vth\cdot X\,\cw[A]=\vth\cdot \cg[A]\,,
\ee
where 
\be\label{Xtheta}
\vth\cdot X\,\cw[A]\equiv\int d^dx\,\vth^a(x)X_a(x)\cw[A]\,,\qquad \vth\cdot \cg[A]\equiv\int d^dx\,\vth^a(x)\cg_a[A](x)\,.
\ee
In these expressions, $\cg_a[A]$ denotes a quantum anomaly, a local functional of the gauge field $A_\m$, while $X_a$ stands for the functional differential operators 
\be\label{X-A}
X_a=-\del_\m\frac{\d}{\d A_\m^a}-f_{ab}{}^{c}A_\m^b\frac{\d}{\d A_\m^c}\,,
\ee
that satisfy the symmetry algebra 
\be\label{X-algebra}
 [X_a(x),X_b(y)]=\d^{(d)}(x-y)f_{ab}{}^c X_c(y)\,,
\ee
with $f_{ab}{}^{c}$ corresponding to the structure constants of the algebra $\frak g$ of $G$ \cite{Bertlmann:1996xk}.\footnote{In this section we take the generators $T_a$ of the algebra $\frak g$ to be anti-Hermitian, i.e. $T_a^\dag=-T_a$, so that they satisfy the algebra $[T_a,T_b]=f_{ab}{}^c T_c$ with real structure constants $f_{ab}{}^c$.}  

The fact that the operators $X_a$ obey the algebra \eqref{X-algebra} implies that the anomaly $\cg_a[A]$ satisfies the Wess-Zumino (WZ) consistency conditions \cite{Wess:1971yu}
\be\label{WZ-G}
\d(\vth')(\vth\cdot\cg[A])-\d(\vth)(\vth'\cdot\cg[A])=[\vth',\vth]\cdot\cg[A]\,,
\ee
or equivalently
\be\label{WZ}
[\d(\vth'),\d(\vth)]\cw[A]=\d([\vth',\vth])\cw[A]\,.
\ee
In the form \eqref{WZ-G} the WZ conditions comprise a set of functional differential equations for the anomaly $\cg_a[A]$. Non-trivial solutions are those that cannot be expressed as the gauge transformation of a local functional of the background gauge field $A_\m$.

Under a finite gauge transformation with parameter $\vth=\vth^a T_a$, where $T_a$ are the generators of the symmetry algebra $\frak g$, the background gauge field $A_\m\equiv A_\m^a T_a$ transforms as  
\be\label{finite-gauge}
 A_\m\to A_\m^{(\vth)}\equiv e^{-\vth}A_\m e^{\vth}+e^{-\vth}\del_\m e^{\vth}\,.
\ee
The operators \eqref{X-A} allow us to express the corresponding transformation of $\cw[A]$ as
\be
\cw[A^{(\vth)}]=e^{\vth\cdot X}\cw[A]\,.
\ee
Hence, the generalization of the WZ conditions \eqref{WZ} for finite gauge transformations is 
\be
\label{WZ-finite}
\cw[(A^{(\vth)})^{(\vth')}]-\cw[(A^{(\vth')})^{(\vth)}]=[e^{\vth'\cdot X},e^{\vth\cdot X}]\cw[A]\,.
\ee

\subsection{Coset construction and the Wess-Zumino effective action}

Given a non-trivial solution $\cg_a[A]$ of the Wess-Zumino conditions, it is often useful to construct an ``anomaly effective action'', $\cw_{\text{anom}}[A]$, whose gauge transformation reproduces the anomaly $\cg_a[A]$. The generating function $\cw[A]$ can then be written as the sum of $\cw_{\text{anom}}[A]$ and an invariant functional $\cw_{\text{inv}}[A]$, namely,
\be
\cw[A]=\cw_{\text{inv}}[A]+\cw_{\text{anom}}[A]\,.
\ee
Clearly, $\cw_{\text{anom}}[A]$ is not unique, since an invariant part can be added to it without affecting its transformation. Moreover, like the generating function $\cw[A]$, it is necessarily a non-local functional of the gauge field $A_\m$, since otherwise it could be set to zero by a local counterterm.  

A local functional, $\cw_{\rm WZ}[A,\c]$, whose variation coincides with a non-trivial solution $\cg_a[A]$ of the WZ conditions can be constructed if suitable compensator or  Stückelberg fields, $\c^a$, are introduced. Such a functional is known as a Wess-Zumino action \cite{Wess:1971yu}. The compensator fields may be identified with the would-be Goldstone fields in a phase where the global symmetry $G$ is spontaneously broken, generically to a subgroup $H$. The non-local anomaly effective action can then be obtained by integrating out the Goldstone fields, namely (see e.g. \cite{Schwimmer:2010za})     
\be\label{anomaly-EA}
e^{i\cw_{\text{anom}}[A]}=\int \cd \c\, e^{i\cw_{\rm WZ}[A,\c]}\,.
\ee
The path integral measure is invariant under the coset $G/H$, since the Goldstone fields transform additively under the coset symmetries. However, the measure may not be invariant under the subgroup $H$. It follows that $\cw_{\text{anom}}[A]$ and $\cw_{\rm WZ}[A,\c]$ have the same anomalous transformation under $G/H$, but not necessarily under $H$. The anomaly effective action can be determined systematically using the representation \eqref{anomaly-EA} once the local WZ action $\cw_{\rm WZ}[A,\c]$ is known. The main advantage of the WZ action relative to the non-local anomaly effective action is that it can be expressed directly in terms of the anomaly $\cg_a[A]$. 

The algorithm for computing $\cw_{\rm WZ}[A,\c]$ from the anomaly $\cg_a[A]$ was originally developed by Wess and Zumino in \cite{Wess:1971yu}. The effective action they obtained involves compensators for the entire symmetry group $G$, corresponding to a phase where all of $G$ is spontaneously broken. In this case, the gauge transformation of the compensators $\c^a$ is such that the quantity
\be\label{invA}
A^{(-\c)}_\m=e^{\c}A_\m e^{-\c}+e^{\c}\del_\m e^{-\c}\,,
\ee
i.e. the gauge-transformed gauge field with parameter $\vth=-\c$, is gauge invariant. Using the finite gauge transformation \eqref{finite-gauge}, this implies that the gauge transformation of $\c^a$ satisfies
\be
e^{\c^{(\vth)}}(e^{-\vth}A_\m e^{\vth}+e^{-\vth}\del_\m e^{\vth})e^{-\c^{(\vth)}}+e^{\c^{(\vth)}}\del_\m e^{-\c^{(\vth)}}=e^{\c}A_\m e^{-\c}+e^{\c}\del_\m e^{-\c}\,,
\ee
which determines that $e^\c\in G$ transforms according to the right group action, namely 
\be\label{Goldstone-trans}
e^{-\c^{(\vth)}}=e^{-\vth}e^{-\c}\,,\qquad \text{or}\qquad e^{\c^{(\vth)}}=e^{\c}e^{\vth}\,.
\ee

However, there exist situations where one would like to obtain a WZ effective action that leaves a subgroup $H$ of $G$ linearly realized, i.e. the Goldstone fields $\c^a$ correspond to the spontaneous breaking of $G$ to a non-trivial unbroken subgroup $H$. Constructing such a WZ action is more challenging, especially if the subgroup $H$ is anomalous. The algorithm of \cite{Wess:1971yu} was extended to the case of a compact and anomaly safe subgroup $H$ in \cite{Chu:1996fr}. In the following, we generalize this construction further to non-compact groups and a potentially anomalous subgroup $H$. As we will see, these generalizations are necessary for the application of the coset construction to the superconformal group. 

In the presence of a non-trivial unbroken subgroup $H$, the Goldstone fields $\c^a$ parameterize the coset $G/H$ through the so called coset construction \cite{Coleman:1969sm,Callan:1969sn,Salam:1969rq,Wu:1984pv,Chu:1996fr,Weinberg:1996kr}. This is possible provided $G$ admits a reductive decomposition, i.e. the Lie algebra $\frak g$ of $G$ can be decomposed as $\frak g=\frak h\oplus \frak m$, where $\frak h$ is the algebra of the subgroup $H$, and the commutators of the elements of $\frak h$ and $\frak m$ satisfy 
\be\label{reductive-decomp}
[\frak h, \frak h]\in \frak h\,,\qquad [\frak h, \frak m]\in \frak m\,.
\ee  
The existence of a reductive decomposition of $G$ with respect to $H$ is guaranteed if $H$ is compact. However, it is not automatic for non-compact $G$ or $H$, such as the conformal and Poincar\'e groups. 

Notice that reductivity does not impose any condition on the commutators $[\frak m, \frak m]$. However, the coset construction simplifies if, in addition to \eqref{reductive-decomp}, one has that $[\frak m, \frak m]\in \frak h$   \cite{Coleman:1969sm}. Such a decomposition of $G$ with respect to $H$ is known as a Cartan decomposition. As we will see shortly, in order for a well defined WZ action to exist when the unbroken subgroup $H$ is anomalous but not anomaly safe as in \cite{Chu:1996fr}, the commutators $[\frak m, \frak m]$ must satisfy instead
\be\label{H-anomaly-cond}
[\frak m, \frak m]\in \frak m\,.
\ee   
Together with \eqref{reductive-decomp}, this renders renders $\frak m$ an ideal of $\frak g$. 

Given the algebra splitting $\frak g=\frak h\oplus \frak m$, we may divide the generators $T_a$ of $\frak g$ into the two subsets $T_i\in \frak h$ and $T_\a\in \frak m$. The conditions \eqref{reductive-decomp} for a reductive decomposition imply that the generators satisfy the commutation relations
\be\label{reductive-decomp-T}
[T_i,T_j]=f_{ij}{}^k T_k\,,\qquad [T_i,T_\a]=f_{i\a}{}^\b T_\b\,,\qquad [T_\a,T_\b]=f_{\a\b}{}^c T_c\,,
\ee
where $f_{ab}{}^c$ are the structure constants of $\frak g$. If the further condition \eqref{H-anomaly-cond} is imposed, the last commutation relation is replaced by
\be\label{H-anomaly-cond-T}
[T_\a,T_\b]=f_{\a\b}{}^\g T_\g\,.
\ee  

Any element $g\in G$ can be decomposed as $g=he^{\c}$ for suitable $h\in H$ and $\c=\c^\a T_\a\in \frak m$. Under the right group action on $G$,
\be
g\to g^{(\vth)}=he^{\c}e^{\vth}\,.
\ee
Since $e^{\c}e^{\vth}\in G$, there exists $h_\vth\in H$ such that $e^{\c}e^{\vth}=h_\vth e^{\c^{(\vth)}}$ with $\c^{(\vth)}\in \frak m$. It follows that
\be\label{Goldstone-trans-H}
e^{\c}\to e^{\c^{(\vth)}}=h_\vth^{-1}e^{\c}e^{\vth}\,,
\ee
generalizing the Goldstone transformation \eqref{Goldstone-trans} in the presence of a non-trivial unbroken subgroup $H\subset G$ \cite{Coleman:1969sm,Callan:1969sn}. The transformation \eqref{Goldstone-trans-H} ensures that $A^{(-\c)}_\m$ in \eqref{invA}, with $\c\in \frak m$ now, is invariant up to $H$-transformations:
\bal\label{invA-H}
A^{(-\c)}_\m\to&\; e^{\c^{(\vth)}}A^{(\vth)}_\m e^{-\c^{(\vth)}}+e^{\c^{(\vth)}}\del_\m e^{-\c^{(\vth)}}\NO\\
=&\;h_\vth^{-1}e^{\c}e^{\vth}(e^{-\vth}A_\m e^{\vth}+e^{-\vth}\del_\m e^{\vth})e^{-\vth}e^{-\c}h_\vth+h_\vth^{-1}e^{\c}e^{\vth}\pa_\m(e^{-\vth}e^{-\c}h_\vth)\NO\\
=&\;h_\vth^{-1}A^{(-\c)}_\m h_\vth+h_\vth^{-1}\pa_\m h_\vth\,.
\eal

An important observation that will play a key role in the subsequent analysis is that the transformations \eqref{Goldstone-trans-H} and \eqref{invA-H} simplify if $\vth\in\frak m$ and the condition \eqref{H-anomaly-cond} is imposed as well, for in that case $h_\vth=1$. If \eqref{H-anomaly-cond} holds, $G/H$ is a subgroup of $G$ and so both $e^\c$ and $e^\th$ with $\vth\in \frak m$ are elements of $G/H$. The group property of $G/H$ in this case ensures that $e^\c e^\vth=e^{\c^{(\vth)}}\in G/H$, and hence $h_\vth=1$. It follows that if \eqref{H-anomaly-cond} holds, $A^{(-\c)}_\m$ is invariant under gauge transformations with $\vth\in \frak m$.

We are now ready to generalize the construction of the WZ action of \cite{Wess:1971yu} to the case of an anomalous but not necessarily anomaly safe unbroken subgroup. What we shall demand of the local WZ action in this case is that its infinitesimal gauge transformation reproduces the anomaly $\cg_\a[A]$ in the coset directions only, namely
\be\label{WZactionDEF}
\bar\vth\cdot X\,\cw_{\rm WZ}[A,\c]=\bar\vth\cdot X\,\cw[A]=\bar\vth\cdot \cg[A]\,,\qquad \bar\vth\in \frak m\,,
\ee
where now $X=X^{(A)}+X^{(\c)}$ with $X^{(A)}$ given in \eqref{X-A} and $X^{(\c)}$ defined through the relation
\be\label{X-chi}
e^{\c^{(\vth)}}=e^{\vth\cdot X^{(\c)}}e^{\c} =h_\vth^{-1}e^{\c}e^{\vth}\,,
\ee
so that for infinitesimal transformations
\be\label{X-chi-inf}
\d(\vth)\c(x)=\vth\cdot X^{(\c)}\c(x)=\int d^dy\,\vth^a(y) X_a^{(\c)}(y)\c(x)\,.
\ee
We will not need the explicit expression for the functional differential operators $X^{(\c)}_a$ for the subsequent analysis, but it may be obtained by making use of the Baker–Campbell–Hausdorff formula. For our present analysis it suffices to mention that $X^{(\c)}_a$ satisfy the algebra \eqref{X-algebra}.

Notice that the definition \eqref{WZactionDEF} determines the WZ action up to local functionals that are invariant under the coset $G/H$, i.e. local functionals of the invariant fields $A^{(-\c)}$. In particular, if $\cw_{\rm WZ}[A,\c]$ satisfies \eqref{WZactionDEF}, so does 
\be\label{eq:WZambiguity}
\cw'_{\rm WZ}[A,\c]\equiv \cw_{\rm WZ}[A,\c]+\cw_{\rm loc}[A^{(-\c)}]\,,
\ee
for any local functional $\cw_{\rm loc}[A^{(-\c)}]$. This freedom will help us obtain an explicit expression for the WZ action in terms of the coset anomalies.

Assuming that the condition \eqref{H-anomaly-cond-T} holds, the exponential map $e^{\bar\vth\cdot X}$ determines the gauge transformation of any functional of $A_\m$ and $\c$ along $G/H$. The relations \eqref{WZactionDEF} imply that
\bal
(e^{\bar\vth\cdot X}-1)\cw_{\rm WZ}[A,\c]=&\;\sum_{n=1}^\infty\frac{1}{n!}(\bar\vth\cdot X)^n\cw_{\rm WZ}[A,\c]\NO\\
=&\;\sum_{n=1}^\infty\frac{1}{n!}(\bar\vth\cdot X)^{n-1}\bar\vth\cdot X\,\,\cw_{\rm WZ}[A,\c]\NO\\
=&\;\sum_{n=1}^\infty\frac{1}{n!}(\bar\vth\cdot X^{(A)})^{n-1}\bar\vth\cdot X^{(A)}\,\cw[A]\NO\\
=&\;(e^{\bar\vth\cdot X^{(A)}}-1)\cw[A]\,.
\eal
Since both $\bar\vth$ and $\c$ are valued in $\frak m$, we may set $\bar\vth=-\c$ to obtain 
\be
\cw_{\rm WZ}[A,\c]=\cw_{\rm WZ}[A^{(-\c)},0]-(\cw[A^{(-\c)}]-\cw[A])\,.
\ee

As we have seen, $A^{(-\c)}$ and consequently $\cw_{\rm WZ}[A^{(-\c)},0]$ are both invariant under the coset $G/H$, but not under $H$-transformations. We can therefore utilize the freedom to add to the WZ action any local functional of the invariant fields $A^{(-\c)}$ as in \eqref{eq:WZambiguity} in order to set
\be\label{WZchoice}
\cw_{\rm WZ}[A^{(-\c)},0]=0\,.
\ee
With this choice the WZ action is given by
\be\label{WZ-W}
\cw_{\rm WZ}[A,\c]=-(\cw[A^{(-\c)}]-\cw[A])\,.
\ee
This expression determines the WZ action in terms of the generating function $\cw[A]$, but it does not help us obtain the form of the functional $\cw_{\rm WZ}[A,\c]$.

A more useful expression for the WZ action relates it directly to the anomaly. Following \cite{Wess:1971yu}, but focusing on gauge transformations along the coset $G/H$, we observe that
\bal
\cw_{\rm WZ}[A^{(\bar\vth)},\c^{(\bar\vth)}]=&\;e^{\bar\vth\cdot X}\cw_{\rm WZ}[A,\c]\NO\\
=&\;\cw_{\rm WZ}[A,\c]+(e^{\bar\vth\cdot X}-1)\cw_{\rm WZ}[A,\c]\NO\\
=&\;\cw_{\rm WZ}[A,\c]+\int_0^1dt\, e^{t\bar\vth\cdot X}\bar\vth\cdot X\cw_{\rm WZ}[A,\c]\,,\NO\\
=&\;\cw_{\rm WZ}[A,\c]+\int_0^1dt\, e^{t\bar\vth\cdot X}\bar\vth\cdot \cg[A]\,,\NO\\
=&\;\cw_{\rm WZ}[A,\c]+\int_0^1dt\, \bar\vth\cdot \cg[A^{(t\bar\vth)}]\,.
\eal
Setting again $\bar\vth=-\c$ and using \eqref{WZchoice} we obtain
\bbxd
\vskip.4cm
\be\label{WZ-transgression}
\cw_{\rm WZ}[A,\c]=\int_0^1dt\, e^{-t\c\cdot X^{(A)}}\c\cdot \cg[A]=\int_0^1dt\, \c\cdot \cg[A^{(-t\c)}]\,.
\ee
\ebxd

The result \eqref{WZ-transgression} provides a prescription for computing the local functional $\cw_{\rm WZ}[A,\c]$ once the coset anomalies $\cg_\a[A]$ are known. It is designed so that its infinitesimal variation along the coset directions reproduces the corresponding anomalies, provided the condition \eqref{H-anomaly-cond-T} holds. Crucially, no restriction has been imposed on its variation along the subgroup directions. In particular, $H$-transformations of $\cw_{\rm WZ}[A,\c]$ do not in general reproduce the subgroup anomalies $\cg_i[A]$ of the generating function $\cw[A]$. In the next subsection we determine the general form of the $H$-transformation of $\cw_{\rm WZ}[A,\c]$. In the subsequent sections we utilize \eqref{WZ-transgression} in order to compute the WZ effective action for two- and four-dimensional CFTs, and in particular, the WZ action for generic $\cn=1$ superconformal theories in four dimensions in components, including fermionic terms.

\subsection{Subgroup anomalies}

We now evaluate the infinitesimal gauge transformation of the WZ action \eqref{WZ-transgression} following \cite{Chu:1996fr,Weinberg:1996kr}, without imposing the condition \eqref{H-anomaly-cond} except at the very end of the computation. The infinitesimal form of a generic gauge transformation of the WZ action \eqref{WZ-transgression} is
\be\label{WZ-var}
\d(\vth) \cw_\text{WZ}[A,\c]=\int_0^1 dt\int d^dx \big(\d(\vth)\c^\a\cg_\a[A^{(-t\c)}]+\c^\a\d(\vth)\cg_\a[A^{(-t\c)}]\big)\,.
\ee
In order to proceed it is helpful to express the gauge transformation of the composite field $A^{(-t\c)}$ as a gauge transformation with a composite parameter, namely
\be\label{deltaAt}
\d(\vth) A_\m^{a(-t\c)}=(D_\m\vth^{(-t\c)})^a_{A\to A^{(-t\c)}}=\del_\m \vth^{a(-t\c)}+\vth^{c(-t\c)}A_\m^{b(-t\c)}f_{bc}{}^a\,,
\ee
where the effective composite parameter $\vth^{(-t\c)}$ is to be determined.

The composite field $A^{(-t\c)}$ is obtained from $A^{(-\c)}$ in \eqref{invA} by replacing $\c\to t\c$, namely
\be\label{At}
A^{(-t\c)}_\m=e^{t\c}A_\m e^{-t\c}+e^{t\c}\del_\m e^{-t\c}\,.
\ee
An infinitesimal gauge transformation of this quantity takes the form
\bal
\d(\vth) A_\m^{(-t\c)}=&\;\d(\vth)e^{t\c}A_\m e^{-t\c}+e^{t\c}\d(\vth)A_\m e^{-t\c}+e^{t\c}A_\m \d(\vth)e^{-t\c}+\d(\vth)e^{t\c}\del_\m e^{-t\c}+e^{t\c}\del_\m \d(\vth)e^{-t\c}\NO\\
=&\;e^{t\c}D_\m\vth e^{-t\c}+[A_\m^{(-t\c)},e^{t\c}\d(\vth)e^{-t\c}]+\del_\m (e^{t\c}\d(\vth)e^{-t\c})\\
=&\;\pa_\m\vth^{(-t\c)}-\pa_\m e^{t\c}\vth e^{-t\c}-e^{t\c}\vth \pa_\m e^{-t\c}+e^{t\c}[A_\m,\vth]e^{-t\c}+[A^{(-t\c)}_\m,e^{t\c}\d(\vth)e^{-t\c}]\,,\NO
\eal
where we have defined
\be\label{t-theta}
\vth^{(-t\c)}\equiv e^{t\c}(\vth+\d(\vth))e^{-t\c}\,.
\ee
Moreover,
\bal
[A^{(-t\c)}_\m,e^{t\c}\vth e^{-t\c}]=&\;[e^{t\c}A_\m e^{-t\c}+e^{t\c}\del_\m e^{-t\c},e^{t\c}\vth e^{-t\c}]\NO\\
=&\;e^{t\c}[A_\m,\vth]e^{-t\c}+e^{t\c}\del_\m e^{-t\c}e^{t\c}\vth e^{-t\c}-e^{t\c}\vth \del_\m e^{-t\c}\NO\\
=&\;e^{t\c}[A_\m,\vth]e^{-t\c}-\del_\m e^{t\c}\vth e^{-t\c}-e^{t\c}\vth \del_\m e^{-t\c}\,,
\eal
and so indeed
\be
\d(\vth) A_\m^{(-t\c)}=\del_\m \vth^{(-t\c)}+[A_\m^{(-t\c)},\vth^{(-t\c)}]\,,
\ee
with $\vth^{(-t\c)}$ given by \eqref{t-theta}.
 
The quantity $\vth^{(-t\c)}$ in \eqref{t-theta} allows us to also express $\d(\vth)\c$ in the form
\be\label{delta-chi}
\d(\vth)\c=-[\vth^{(-t\c)},\c]-\frac{d}{dt}\vth^{(-t\c)}\,,
\ee
or in components
\be\label{delta-chi-comp}
\d(\vth)\c^\a=-\vth^{c(-t\c)}\c^\b f_{c\b}{}^\a-\frac{d}{dt}\vth^{\a(-t\c)}\,,\qquad 0=-\vth^{c(-t\c)}\c^\b f_{c\b}{}^i-\frac{d}{dt}\vth^{i(-t\c)}\,.
\ee

Using these results, we can write the gauge transformation of the WZ action in \eqref{WZ-var} as
\be
\d(\vth) \cw_\text{WZ}=\int_0^1 dt\int d^dx \Big[-\Big(\vth^{c(-t\c)}\c^\b f_{c\b}{}^\a+\frac{d}{dt}\vth^{\a(-t\c)}\Big)\cg_\a+\c^\a\vth^{(-t\c)}\cdot X\cg_\a\Big]_{A\to A^{(-t\c)}}\,.
\ee
The WZ conditions \eqref{WZ-G} imply that the anomaly satisfies the relations
\be
X_a(y)\cg_b[A](x)-X_b(x)\cg_a[A](y)=f_{ab}{}^c\cg_c[A](x)\d^{(d)}(x-y)\,,
\ee
so that the gauge transformation of the WZ action takes the form
\bal
\d(\vth) \cw_\text{WZ}=&\;-\int_0^1 dt\int d^dx\Big(\vth^{c(-t\c)}\c^\b f_{c\b}{}^\a+\frac{d}{dt}\vth^{\a(-t\c)}\Big)\cg_\a[A]\Big|_{A\to A^{(-t\c)}}\NO\\
&+\int_0^1 dt\int d^dx\,\vth^{c(-t\c)}\big(\c\cdot X\cg_c[A]+\c^\b f_{c\b}{}^a\cg_a[A]\big)_{A\to A^{(-t\c)}}\\
=&\;\int_0^1 dt\int d^dx\,\Big(-\frac{d}{dt}\vth^{\a(-t\c)}\cg_\a[A]+\vth^{c(-t\c)}\big(\c\cdot X\cg_c[A]+\c^\b f_{c\b}{}^i\cg_i[A]\big)\Big)_{A\to A^{(-t\c)}}\,.\NO
\eal

Using the second equation in \eqref{delta-chi-comp}, the first and the last terms in this expression can be combined in order to obtain
\be\label{delta-WZ}
\d(\vth) \cw_\text{WZ}=\int_0^1 dt\int d^dx\,\Big(-\frac{d}{dt}\vth^{c(-t\c)}\cg_c[A]+\vth^{c(-t\c)}\c\cdot X\cg_c[A]\Big)_{A\to A^{(-t\c)}}\,.
\ee
Moreover, using the expression for $A^{(-t\c)}_\m$ in \eqref{At} gives
\be
\frac{d}{dt}A^{(-t\c)}_\m=-\del_\m\c-[A^{(-t\c)}_\m,\c]=-\d(\c)A_\m\big|_{A\to A^{(-t\c)}}\,,
\ee
and so for any local functional of $A_\m$ the following identity holds 
\be
\c\cdot X\cf[A]\big|_{A\to A^{(-t\c)}}=\d(\c)\cf[A]\big|_{A\to A^{(-t\c)}}=-\frac{d}{dt}\cf[A^{(-t\c)}]\,.
\ee
In particular, applying this to the last term in \eqref{delta-WZ} we obtain
\be\label{delta-WZ-prelim}
\d(\vth) \cw_\text{WZ}=-\int_0^1 dt\frac{d}{dt}\int d^dx\,\vth^{c(-t\c)}\cg_c[A^{(-t\c)}]
=\int d^dx\,\big(\vth^{c}\cg_c[A]-\vth^{c(-\c)}\cg_c[A^{(-\c)}]\big)\,.
\ee
This result is exactly what one expects based on the form \eqref{WZ-W} of the WZ action. 

The expression \eqref{delta-WZ-prelim} for the gauge transformation of the WZ action can be simplified by noticing that $\vth^{(-\c)}\in \frak h$, i.e. $\vth^{\a(-\c)}=0$. To show this, we can use the gauge transformation of the Goldstones in \eqref{Goldstone-trans-H} to obtain
\be
\d(\vth)e^{-\c}=-\vth e^{-\c}+e^{-\c}k_\vth\,,
\ee
where we have introduced $k_\vth\in \frak h$ through the parametrization $h_\vth=\exp(k_\vth)$. Inserting this expression in the definition of $\vth^{(-\c)}$ in \eqref{t-theta} gives
\be
\vth^{(-\c)}\equiv e^{\c}(\vth+\d(\vth))e^{-\c}=k_\vth\,,
\ee
and therefore $\vth^{(-\c)}\in \frak h$ as claimed. In fact, this result implies that
\be\label{h-theta}
h_\vth=\exp{(\vth^{(-\c)})}\,,
\ee
which also follows from a comparison of \eqref{invA-H} and \eqref{deltaAt} for $t=1$.

Since $\vth^{\a(-\c)}=0$, the gauge transformation \eqref{delta-WZ-prelim} takes the form
\bbxd
\vskip.4cm
\be\label{delta-WZ-final}
\d(\vth) \cw_\text{WZ}[A,\c] =\int d^dx\,\big(\vth^{\a}\cg_\a[A]+\vth^{i}\cg_i[A]-\vth^{i(-\c)}\cg_i[A^{(-\c)}]\big)\,.
\ee
\ebxd
This is the same result as that obtained in \cite{Chu:1996fr}. The authors of that work continue to show that if the subgroup $H$ is anomaly safe, there exists a scheme in which the subgroup anomaly $\cg_i$ vanishes so that the gauge transformation \eqref{delta-WZ-final} produces the correct coset anomaly. However, we observe that even when $\cg_i\neq 0$, the gauge transformation \eqref{delta-WZ-final} does produce the correct coset anomaly provided $\vth^{(-\c)}=0$ whenever $\vth\in \frak m$. As we showed above, this happens when we impose the condition \eqref{H-anomaly-cond} that ensures that the coset $G/H$ is a normal subgroup of $G$. This is in agreement with our analysis in the previous subsection where we showed that imposing \eqref{H-anomaly-cond} and demanding that $\cw_\text{WZ}[A,\c]$ reproduces the coset anomaly results in the expression \eqref{WZ-transgression}. However, besides confirming the transformation of the WZ action along the coset directions, the result \eqref{delta-WZ-final} tells us how $\cw_\text{WZ}[A,\c]$ transforms along the $H$-directions as well.

Notice that the last term in \eqref{delta-WZ-prelim} is the $H$-anomaly expressed in terms of the effective gauge field $A_\m^{(-\c)}$ that transforms only under $H$-transformations, and the effective gauge parameter $\vth^{(-\c)}$ that has components only along the subgroup $H$. This term satisfies the WZ conditions on its own and could itself be expressed as the $H$-transformation of a WZ action, once Goldstone fields for the subgroup $H$ were added. However, unless $H$ is anomaly safe, in which case one can set $\cg_i=0$ as in \cite{Chu:1996fr}, the last term in \eqref{delta-WZ-prelim} cannot be eliminated by a local counterterm that involves the effective subgroup gauge field $A_\m^{(-\c)}$ only.

\section{Gauging of spacetime symmetries}
\label{sec:gengauging}

The main aim of the present work is to apply the algorithm of section \ref{sec:WZactions} to the gauge theory of SU(2,2|1), the $\cn=1$ superconformal group in four dimensions. However, the gauging of spacetime symmetries presents a number of subtleties. In particular, contrary to internal symmetries, general coordinate transformations do not fit naturally within the language of gauge symmetries, while the algebra of local spacetime symmetries does not coincide with the corresponding global symmetry algebra. In this section we briefly review these subtleties in a generic context, setting the stage for the analysis of the local conformal and superconformal symmetries in the subsequent sections. We refer the interested reader to \cite{Kaku:1977pa,Kugo:1982cu, Fradkin:1985am,Freedman:2012zz} for extensive discussions on the gauging of spacetime symmetries.

\subsection{General gauge theories}

Let us consider a global symmetry with a Lie (super)algebra $\frak g$ described by
\be
[T_A,T_B\}=f_{AB}{}^C T_C\,,
\ee
with $T_A$ its anti-Hermitian generators and $f_{AB}{}^C$ the corresponding structure constants. The notation $[\cdot,\cdot\}$ stands for a commutator or an anticommutator, depending on the Grassmann parity of the generators $T_A$. For every generator we turn on a gauge field $\mathscr{A}_\m{}^A$ so that $\mathscr{A}_\m=\mathscr{A}_\m{}^A T_A$. Under an infinitesimal local gauge transformation with parameters $\vth^A(x)$, the gauge fields are required to transform as
\be\label{eq:ConnGaugeTrans}
\d(\vth)\mathscr{A}_\m{}^A=\;\del_\m\vth^A+\vth^C\mathscr{A}_\m{}^B f_{BC}{}^A\,.
\ee
Using the (graded) Jacobi identity it is straightforward to show that these transformations satisfy the local algebra
\be
[\d(\vth_1),\d(\vth_2)]\mathscr{A}_\m{}^A=(\d(\vth_3)\mathscr{A}_\m)^A=\pa_\m\vth_3^A+\vth_3^C\mathscr{A}_\m{}^B f_{BC}{}^A\,,
\ee
where $\vth_3^C=\vth_2^B\vth_1^A f_{AB}{}^C$.

A local quantity $\F(x)$, elementary or composite, transforms covariantly under the gauge algebra if it satisfies
\be\label{eq:GenCovGaugeTrans}
\d(\vth)\F=\vth^AT_A\F\,,
\ee
where the generators $T_A$ act on $\F$ according to some suitable representation of the algebra. Clearly, the derivative $\pa_\m\F$ of $\F$ does not transform covariantly. A covariant derivative, however can be defined as 
\be\label{eq:GenCovD}
\mathscr{D}_\m\F\equiv(\pa_\m-\d(\mathscr{A}_\m))\F=(\pa_\m-\mathscr{A}_\m{}^A T_A)\F\,,
\ee
where the notation $\d(\mathscr{A}_\m)\F$ stands for an infinitesimal transformation of $\F$ with parameter $\mathscr{A}_\m$, i.e. replacing $\vth^A\to\mathscr{A}_\m{}^A$ in \eqref{eq:GenCovGaugeTrans}. Using the transformation \eqref{eq:ConnGaugeTrans} of the gauge fields $\mathscr{A}_\m{}^A$, one finds that the quantity \eqref{eq:GenCovD} does transform covariantly, namely
\be
\d(\vth)(\mathscr{D}_\m\F)=\vth^A\mathscr{D}_\m(T_A\F)\,.
\ee
Moreover, the transformation \eqref{eq:ConnGaugeTrans} of the gauge fields takes the form 
\be
\d(\vth)\mathscr{A}_\m{}^A=(\mathscr{D}_\m\vth)^A\,.
\ee

The curvature or fieldstrength of the gauge fields $\mathscr{A}^A$ is defined as
\be\label{eq:GenCurv}
\mathscr{R}_{\m\n}{}^A=\pa_\m \mathscr{A}_\n{}^A-\pa_\n \mathscr{A}_\m{}^A+ f_{BC}{}^A \mathscr{A}_\m{}^B \mathscr{A}_\n{}^C\,.
\ee
The Jacobi identity ensures that this quantity transforms covariantly, namely
\be\label{eq:GenCurvTrans}
\d(\vth)\mathscr{R}_{\m\n}{}^A=\vth^C\mathscr{R}_{\m\n}{}^Bf_{BC}{}^A\,,
\ee
and satisfies the Bianchi identity
\be\label{eq:GenBianchi}
(\mathscr{D}_{[\m}\mathscr{R}_{\n\r]})^A=0\,.
\ee

\subsection{Local translations and theories of gravity}

The above results hold in general for any gauge theory based on a Lie algebra or superalgebra. However, subtleties arise when one tries to incorporate general coordinate transformations, i.e. diffeomorphisms, as is necessary in theories of gravity. A priori general coordinate transformations are not part of the gauge symmetries discussed above. However, gauge fields and other local objects possess well defined transformations under diffeomorphisms, dictated by their tensorial properties. For example, under an infinitesimal diffeomorphism with parameter $\x^\m(x)$, gauge fields transform as    
\be\label{eq:ConnGCT}
\d_D(\x)\mathscr{A}_\m{}^A=\cl_{\x}\mathscr{A}_\m{}^A=\xi^\n\del_\n \mathscr{A}_\m{}^A+\mathscr{A}_\n{}^A\del_\m\x^\n\,.
\ee
The question that arises is whether this transformation can be somehow related to gauge transformations of the Lie algebra $\frak g$.

To address this question, we observe that the transformation \eqref{eq:ConnGCT} can be expressed as
\be
\d_D(\x)\mathscr{A}_\m{}^A=\d(\x^\n\mathscr{A}_\n)\mathscr{A}_\m{}^A+\x^\n \mathscr{R}_{\n\m}{}^A=(\mathscr{D}_\m(\x^\n\mathscr{A}_\n))^A+\x^\n \mathscr{R}_{\n\m}{}^A\,,
\ee
where $\d(\x^\n\mathscr{A}_\n)$ stands for a gauge transformation with parameters $\x^\n\mathscr{A}_\n{}^A$ and $\mathscr{R}_{\n\m}{}^A$ is the curvature \eqref{eq:GenCurv}. Now suppose that the algebra $\frak g$ contains `translations' $P_a$ with the corresponding gauge field denoted by $e_\m{}^a$. The index $a$, which will be later identified with the local frame index, takes the same number of values as the spacetime dimension. From the above expression for a general coordinate transformation follows that an infinitesimal translation with parameter $\x^a\equiv \x^\m e_\m{}^a$ may be written as
\be\label{eq:TransToCGCT}
\d_P(\x)\mathscr{A}_\m{}^A=\big(\d_D(\x)-\sum_{B'}\d_{B'}(\x^\n\mathscr{A}_\n)\big)\mathscr{A}_\m{}^A-\x^\n \mathscr{R}_{\n\m}{}^A\equiv \d_{\wt P}(\x)\mathscr{A}_\m{}^A-\x^\n \mathscr{R}_{\n\m}{}^A\,,
\ee
where $B'$ runs over the Lie algebra indices except for those corresponding to translations. 

The result \eqref{eq:TransToCGCT} suggests that it may be possible to trade local translations for {\em covariant} general coordinate transformations, up to field-dependent `ordinary' gauge transformations, i.e. gauge transformations other than local translations, namely
\be\label{eq:CGCT}
\d_P(\x)\to \d_{\wt P}(\x)\equiv \d_D(\x)-\sum_{B'}\d_{B'}(\x^\n\mathscr{A}_\n)\,,
\ee
provided suitable curvature constraints are imposed. This is indeed possible for the conformal group, but not for its supersymmetric extensions \cite{Bergshoeff:1980is,deWit:1981vgr,deWit:1983qkc,deWit:1984hw}. The physical symmetry, however, is diffeomorphisms rather than local translations. For example, the action is required to be invariant under general coordinate transformations but it is not invariant under local translations, since the curvature constraints typically break local translations explicitly \cite{Kaku:1977pa}. Therefore, local translations -- together with the satisfying systematic derivation of the curvature constraints provided by demanding their formal equivalence with covariant local coordinate transformations -- are given up in the construction of physical supergravity theories. The curvature constraints are imposed instead in order to eliminate the `unphysical' gauge fields and local translations are discarded altogether in favor of local coordinate transformations. The `naive' physical local symmetry algebra therefore is    
\be
\text{diff}(\cm_d)\oplus \text{span}_{\bb R}(T_{A'})\,,\qquad T_{A'}\in \frak{g}\,,\, T_{A'}\neq P_a\,.
\ee
However, the remaining `standard' gauge transformations are in general modified if the constraints break any of the standard gauge symmetries, in addition to local translations. For example, this is the case for Q-supersymmetry in $\cn=1$ conformal supergravity. 
Moreover, the local algebra may be further reduced by suitable gauge-fixing conditions.

Isolating local translations leads to a number of further modifications in the general gauging procedure discussed above, which remains very useful despite the fact that local translations are eventually discarded. An important example is the definition of the covariant derivative in \eqref{eq:GenCovD}. Since the action of local translations involves derivatives, such transformations must be excluded from the sum over the Lie algebra generators in \eqref{eq:GenCovD} in order to obtain a meaningful covariant derivative. This leads to the modified definition
\be\label{eq:GenCovDModTrans}
\mathscr{D}'_\m\F\equiv\big(\pa_\m-\sum_{A'}\d_{A'}(\mathscr{A}_\m)\big)\F=(\pa_\m-\mathscr{A}_\m{}^{A'} T_{A'})\F\,,
\ee
where again the indices $A'$ run over all generators in $\frak g$ except for $P_a$. This derivative may be further covariantized with respect to diffeomorphisms using a suitable affine connection. 

Another modification that is required in the presence of local translations is in the general transformation of the gauge fields and the definition of the corresponding curvatures. In the presence of local translations, the gauge transformation of the standard gauge fields generically takes the modified form 
\be\label{eq:ConnGaugeTransModTrans}
\d(\vth)\mathscr{A}_\m{}^{A'}=\;\del_\m\vth^{A'}+\vth^{C'}\mathscr{A}_\m{}^{B'} f_{B'C'}{}^{A'}+\vth^{B'}\mathscr{M}_{\m B'}{}^{A'}\,,
\ee
where the new term parameterized by $\mathscr{M}_{\m B'}{}^{A'}$ may have a number of different origins. For example, restricting the original transformation \eqref{eq:ConnGaugeTrans} to standard gauge fields results in a nonzero $\mathscr{M}_{\m B'}{}^{A'}$ whenever the structure constants $f_{a B}{}^{A'}$ are nonzero, namely $\mathscr{M}_{\m B'}{}^{A'}=e_\m{}^a f_{aB'}{}^{A'}$. Another contribution to $\mathscr{M}_{\m B'}{}^{A'}$ results from imposing the curvature constraints, since the transformation laws must be adjusted in order to maintain the constraints. Finally, in supersymmetric theories $\mathscr{M}_{\m B'}{}^{A'}$ may also receive contributions from other fields in the gauge multiplet.  

The modification \eqref{eq:ConnGaugeTransModTrans} in the transformation of ordinary gauge fields requires a modification in the definition of the corresponding curvatures. In analogy with \eqref{eq:GenCurv}, we may define the following `naive' curvatures for ordinary gauge fields:
\be\label{eq:GenCurvModTransPre}
r_{\m\n}{}^{A'}=\pa_\m \mathscr{A}_\n{}^{A'}-\pa_\n \mathscr{A}_\m{}^{A'}+ f_{B'C'}{}^{A'} \mathscr{A}_\m{}^{B'} \mathscr{A}_\n{}^{C'}\,.
\ee
A simple way to determine the correct generalization of the curvatures when $\mathscr{M}_{\m B'}{}^{A'}\neq 0$ is the following. Notice that if $f_{a B}{}^{A'}=0$, the curvatures \eqref{eq:GenCurvModTransPre} coincide with the original ones in \eqref{eq:GenCurv}, once restricted to ordinary gauge fields. Moreover, the same condition, $f_{a B}{}^{A'}=0$, implies that the ordinary gauge fields are invariant under local translations, since
\be
\d_P(\x)\mathscr{A}_\m{}^{A'}=\x^a\mathscr{A}_\m{}^B f_{B a}{}^{A'}=0\,.
\ee
Hence, \eqref{eq:TransToCGCT} implies that a covariant general coordinate transformation of $\mathscr{A}_\m^{A'}$ is given by
\be
\d_{\wt P}(\x)\mathscr{A}_\m{}^{A'}=\x^\n \mathscr{R}_{\n\m}{}^{A'}=\x^\n r_{\n\m}{}^{A'}\,.
\ee

The appropriate covariant generalization of the curvature, $\widehat{\mathscr{R}}_{\m\n}{}^{A'}$, can be obtained by demanding that this property persists in the presence of a nonzero $\mathscr{M}_{\m B'}{}^{A'}$, namely
\be\label{eq:CGCTConn}
\d_{\wt P}(\x)\mathscr{A}_\m{}^{A'}=\x^\n \widehat{\mathscr{R}}_{\n\m}{}^{A'}\,.
\ee
Evaluating the covariant general coordinate transformation $\d_{\wt P}(\x)\mathscr{A}_\m^{A'}$ through \eqref{eq:TransToCGCT} and using $\mathscr{M}_{\m B'}{}^{A'}=e_\m{}^a f_{aB'}{}^{A'}$ as a special case, we find that \eqref{eq:CGCTConn} holds provided
\be\label{eq:GenCurvModTrans}
\widehat{\mathscr{R}}_{\m\n}{}^{A'}= r_{\m\n}{}^{A'}-2\mathscr{A}_{[\m}{}^{B'}\mathscr{M}_{\n]B'}{}^{A'}\,.
\ee
This expression, like \eqref{eq:GenCurvModTrans}, holds in general, irrespectively of the origin of $\mathscr{M}_{\m B'}{}^{A'}$.  

Finally, generalizing \eqref{eq:GenCurvTrans}, one can show that under standard gauge transforms \cite{Freedman:2012zz}
\be\label{eq:GenCurvTransModTrans}
\d(\vth)\widehat{\mathscr{R}}_{ab}{}^{A'}=\vth^{C'}\widehat{\mathscr{R}}_{ab}{}^{B'}f_{B'C'}{}^{A'}+2\vth^{B'}\mathscr{D}'_{[a}\mathscr{M}_{b]B'}{}^{A'}-2\vth^{C'}\mathscr{M}_{[aC'}{}^{B'}\mathscr{M}_{b]B'}{}^{A'}\,,
\ee
while the Bianchi identity \eqref{eq:GenBianchi} becomes
\be\label{eq:GenBianchiModTrans}
(\mathscr{D}'_{[a}\widehat{\mathscr{R}}_{bc]})^{A'}-2\widehat{\mathscr{R}}_{[ab}{}^{B'}\mathscr{M}_{c]B'}{}^{A'}=0\,.
\ee
In both these expressions spacetime indices are converted to local translation indices using the `vielbein' $e_\m{}^a$ and its inverse.

\section{Conformal field theories with a $U(1)$ symmetry}
\label{sec:nonSUSYCFTs}

We are now ready to discuss the gauging of the conformal and superconformal groups and to apply the algorithm of section \ref{sec:WZactions} in order to obtain the corresponding WZ effective actions. As a warm up to the construction of the WZ action for SCFTs, it is instructive to begin by considering non-supersymmetric CFTs with an anomalous U(1) symmetry that may be viewed as an avatar for R-symmetry in SCFTs. Compared to SCFTs, non-supersymmetric CFTs provide a considerably simpler setting not only for discussing the local symmetries and their anomalies, but also for the construction of the WZ effective action, since no background fermionic fields are required. 

There exist a number of different approaches to the gauging of the conformal algebra, the minimal perhaps being the so called `Weyl gauging' \cite{Iorio:1996ad,Coriano:2013nja}. In this approach, besides a vielbein, $e_\m{}^a$, or metric, $g_{\m\n}$, that gauges the Poincar\'e symmetry, one introduces a Weyl connection, $b_\m$, so that derivatives of fields with definite scaling weight can be promoted to Weyl-covariant derivatives. This enables the systematic construction of curvature couplings necessary for the local Weyl invariance of the theory. For theories that are conformally invariant in flat space, the Weyl connection can always be eliminated in the end, resulting in the desired conformal coupling of matter to the background curvature. 

Another approach involves gauging the entire conformal symmetry algebra following the general construction discussed section \ref{sec:gengauging}. This is more suitable for gauging superconformal algebras and it is the canonical approach for systematically constructing conformal theories of gravity and supergravity whose local symmetry algebra closes off-shell \cite{Kaku:1977pa,Kaku:1977rk,Kaku:1978nz,Townsend:1979ki,Bergshoeff:1980is,Kugo:1982cu} (see \cite{VanNieuwenhuizen:1981ae,deWit:1981vgr,deWit:1983qkc,deWit:1984hw,Fradkin:1985am,Freedman:2012zz} for pedagogical reviews). We begin this section by applying the general gauging procedure of section \ref{sec:gengauging} to the conformal group SO($d,2$).

\subsection{Gauging the conformal algebra}
\label{sec:GaugingConformal}

Following the gauging prescription in section \ref{sec:gengauging}, we begin with the global algebra, in this case SO($d,2$)$\times$U(1), and introduce a gauge field and a local parameter for each generator, i.e.
\be
\mathscr{A}_\m=\mathscr{A}_\m{}^A T_A=e_\m{}^a P_a+\tfrac12 \o_\m{}^{ab}M_{[ab]}+b_\m D+f_\m{}^a K_a+A_\m T\,,
\ee 
and
\be
\vth=\vth^A T_A=\x^a P_a+\tfrac12\l^{ab} M_{[ab]}-\s D+\l_K^a K_a+\th\, T\,,
\ee
where $P_a$, $M_{[ab]}$, $D$ and $K_a$ are respectively the generators of local translations, Lorentz transformations, dilatations and special conformal transformations. Moreover, $T$ generates the internal U(1) symmetry that we assume to be present in the CFTs we consider. 

Using the algebra these generators satisfy (see e.g. eq.~(15.9) in  \cite{Freedman:2012zz}), the gauge transformations \eqref{eq:ConnGaugeTrans} determine that the gauge fields introduced above transform as
\be\label{eq:GenGaugeTransConf}
\begin{split}
\d(\vth)e_\m{}^a=&\;\cd_\m\x^a-\l^a{}_b e_\m{}^b+\s e_\m{}^a\,,\\
\d(\vth)\o_\m{}^{ab}=&\;-4\x^{[a}f_\m{}^{b]}+\cd_\m\l^{ab}-4\l_K^{[a}e_\m{}^{b]}\,,\\
\d(\vth)b_\m=&\;-2\x^a f_{\m a}-\pa_\m\s+2\l_K^a e_{\m a}\,,\\
\d(\vth)f_\m{}^a=&\;\cd_\m\l_{K}^a-\l^a{}_b f_\m{}^b-\s f_\m{}^a\,,\\
\d(\vth)A_\m=&\;\pa_\m\th\,,
\end{split}
\ee
where $\cd_\m$ denotes the covariant derivative with respect to the Lorentz, Weyl and U(1) symmetries only, namely  
\be
\begin{split}
\cd_\m\x^a=&\;\pa_\m\x^a+b_\m\x^a+\o_\m{}^a{}_b\x^b\,,\\
\cd_\m\l^{ab}=&\;\pa_\m\l^{ab}+\o_\m{}^a{}_c\l^{cb}+\o_\m{}^b{}_c\l^{ac}\,,\\ \cd_\m\l_K^a=&\;\pa_\m\l_K^a-b_\m\l_K^a+\o_\m{}^a{}_b\l_K^b\,.
\end{split}
\ee

Similarly, applying the general expression \eqref{eq:GenCurv} to the conformal algebra leads to the following expressions for the curvatures:
\be
\begin{split}
\mathscr{R}_{\m\n}{}^a(P)=&\;2\cd_{[\m}e_{\n]}{}^a\,,\\
\mathscr{R}_{\m\n}{}^{ab}(M)=&\;R_{\m\n}{}^{ab}(\o)+8f_{[\m}{}^{[a}e_{\n]}{}^{b]}\,,\\
\mathscr{R}_{\m\n}(D)=&\;2\pa_{[\m}b_{\n]}-4 f_{[\m}{}^a e_{\n]a}\,,\\
\mathscr{R}_{\m\n}{}^a(K)=&\;2\cd_{[\m}f_{\n]}{}^a\,,\\
\mathscr{R}_{\m\n}{}(T)=&\;2\pa_{[\m}A_{\n]}\,,
\end{split}
\ee
where $R_{\m\n}{}^{ab}(\o)=2\pa_{[\m}\o_{\n]}{}^{ab}+2\o_{[\m}{}^a{}_c\o_{\n]}{}^{cb}$ is the standard curvature obtained from the spin connection $\o_\m{}^a{}^b$. Moreover, the Lorentz-Weyl-U(1)-covariant derivatives of the vielbein and of the gauge field of special conformal transformations are given respectively by
\be
\cd_\m e_\n{}^a=\pa_\m e_\n{}^a+b_\m e_\n{}^a+\o_\m{}^a{}_b e_\n{}^b\,, \qquad \cd_\m f_\n{}^a=\pa_\m f_\n{}^a-b_\m f_\n{}^a+\o_\m{}^a{}_b f_\n{}^b\,.
\ee

Let us now see how the requirement that local translations act as diffeomorphisms up to standard gauge transformations imply the curvature constraints. Starting from the action of local translations on the vielbein, the relation \eqref{eq:TransToCGCT} leads to the constraint
\be\label{P-constraint}
\mathscr{R}_{\m\n}{}^a(P)=2\cd_{[\m}e_{\n]}{}^a=0\,.
\ee
This constraint renders the spin connection a dependent field, namely \eqref{P-constraint} is solved provided
\be\label{eq:SpinConnConf}
\o_\m{}^{ab}(e,b)=\o_\m{}^{ab}(e)+2e_\m{}^{[a}e^{b]\n}b_\n\,, \qquad \o_\m{}^{ab}(e)=2e^{\n[a}\pa_{[\m}e_{\n]}{}^{b]}-e^{\n[a}e^{b]\s}e_{\m c}\pa_\n e_\s{}^c\,.
\ee

It remains to examine the action of local translations on the remaining gauge fields. 
Applying \eqref{eq:TransToCGCT} to the dilatation gauge field, $b_\m$, gives a curvature term of the form $\x^\n \mathscr{R}_{\m\n}(D)$. Moreover, from \eqref{eq:GenGaugeTransConf} we see that $b_\m$ transforms additively under $K$-transformations. Hence, this curvature term can be cast as a $K$-transformation with composite parameter
\be
\l_{K\m}=\tfrac12 \mathscr{R}_{\m\n}(D)\x^\n\,.
\ee
It follows that local translations act on both $e_\m{}^a$ and $b_\m$ as diffeomorphisms up to standard gauge transformations. Crucially, this automatically ensures that the same holds for the spin connection, since it is determined as a function of $e_\m{}^a$ and $b_\m$ by the constraint \eqref{P-constraint}. 

In fact, the transformation of the spin connection under local translations is not given by \eqref{eq:TransToCGCT} once the curvature constraint \eqref{P-constraint} is imposed, since the constraint is not automatically invariant under such transformations. Requiring that
\be
\d_P \mathscr{R}_{\m\n}{}^a(P)=0\,,
\ee   
determines the modified transformation of the spin connection under local translations. This transformation coincides with that obtained by applying the chain rule to the expression \eqref{eq:SpinConnConf} for the spin connection, and can be expressed as a combination of a diffeomorphism and standard gauge transformations.

Moving on to the action of local translations on $f_\m{}^a$, it suffices to impose another curvature constraint that renders $f_\m{}^a$ a composite field that is expressed in terms of $e_\m{}^a$ and $b_\m$. For $d>2$, the canonical choice that is preserved by all symmetries except local translations is    
\be\label{M-constraint}
e_b{}^\n\mathscr{R}_{\m\n}{}^{ab}(M)=0\,.
\ee
This provides an algebraic equation for $f_\m{}^a$ that can be solved to obtain\footnote{Notice that this is -1/2 the Schouten tensor in $d$ dimensions.}
\be\label{eq:fConnConf}
f_\m{}^a(e,b)=-\frac{1}{2(d-2)}\Big(R_\m{}^a(\o)-\frac{1}{2(d-1)}e_\m{}^a R(\o)\Big)\,,
\ee
where $R_\m{}^a(\o)=e_b{}^\n R_{\m\n}{}^{ab}(\o)$ and $R(\o)=e_a{}^\m R_\m{}^a(\o)$.  In two dimensions, the constraint \eqref{M-constraint} reduces to
\be\label{eq:fConnConf2D}
e_a{}^\m f_\m{}^a(e,b)=-\tfrac14 R(\o)\,,
\ee
due to the fact that the curvature $R_{\m\n}{}^{ab}(\o)$ has only one independent component and takes the form $R_{\m\n}{}^{ab}(\o)=\frac12R(\o)(e_\m{}^a e_\n{}^b-e_\n{}^a e_\m{}^b)$. Therefore, the constraint \eqref{M-constraint} determines only the trace of $f_\m{}^a(e,b)$ in two dimensions, leaving its transverse components unconstrained.

So far we have seen that the curvature constraints \eqref{P-constraint} and \eqref{M-constraint} suffice in order to trade local translations with diffeomorphisms when acting on all gauge fields of the conformal group. However, local translations act trivially on the U(1) gauge field, namely $\d_P A_\m=0$, and so trading such transformations with diffeomorphisms would clearly require that the we set the U(1) curvature to zero, i.e. $\pa_{[\m}A_{\n]}=0$. Therefore, as in the case of the superconformal group \cite{Bergshoeff:1980is,deWit:1981vgr,deWit:1983qkc,deWit:1984hw}, keeping the U(1) symmetry means that local translations cannot be made fully equivalent to general coordinate transformations. We will therefore simply discard local translations and focus exclusively on diffeomorphisms and standard gauge transformations.

Keeping the U(1) gauge field unconstrained, we are left with three independent fields, $e_\m{}^a$, $b_\m$ and $A_\m$. The corresponding local symmetries are diffeomorphisms, local Lorentz rotations, special conformal transformations, as well as Weyl and local U(1) transformations. In particular, the group transformations \eqref{eq:GenGaugeTransConf} are now replaced with
\be\label{eq:LocalTransConf}
\begin{split}
\d'(\vth)e_\m{}^a=&\;\cl_\x e_\m{}^a-\l^a{}_b e_\m{}^b+\s e_\m{}^a\,,\\
\d'(\vth)\o_\m{}^{ab}=&\;\cl_\x\o_\m{}^{ab}+\cd_\m\l^{ab}-4\l_K^{[a}e_\m{}^{b]}\,,\\
\d'(\vth)b_\m=&\;\cl_\x b_\m-\pa_\m\s+2\l_K^a e_{\m a}\,,\\
\d'(\vth)f_\m{}^a=&\;\cl_\x f_\m{}^a+\cd_\m\l_{K}^a-\l^a{}_b f_\m{}^b-\s f_\m{}^a\,,\\
\d'(\vth)A_\m=&\;\cl_\x A_\m+\pa_\m\th\,.
\end{split}
\ee

Notice that $b_\m$ transforms additively under local special conformal transformations with parameter $\l_{K\,\m}$, while the other two independent fields, the vielbein and the U(1) gauge field, remain invariant. It follows that any invariant action must be independent of the gauge field $b_\m$, and that $b_\m$ can be set to any value by gauge-fixing the local special conformal symmetry. In particular, the so called $K$-gauge corresponds to setting $b_\m=0$. This gauge is preserved provided the parameter $\l_{K\m}$ is related to the Weyl parameter, $\s$, via
\be
\l_{K\m}=\tfrac12\pa_\m\s\,.
\ee  
As a result, in the $K$-gauge, the transformations of the composite fields become
\be
\begin{split}
\d'(\vth)\o_\m{}^{ab}=&\;\cl_\x\o_\m{}^{ab}+\cd_\m\l^{ab}\underline{-2\pa^{[a}\s e_\m{}^{b]}}\,,\\
\d'(\vth)f_\m{}^a=&\;\cl_\x f_\m{}^a+\tfrac12\underline{\cd_\m(\pa^a\s)}-\l^a{}_b f_\m{}^b-\s f_\m{}^a\,,
\end{split}
\ee
where the underlined terms are determined by the $K$-gauge condition. These transformations also follow from the expressions \eqref{eq:SpinConnConf} and \eqref{eq:fConnConf} for the composite fields, using the chain rule and the transformations for the independent fields in \eqref{eq:LocalTransConf}.

The resulting local symmetry group in the $K$-gauge is
\be
G=\text{Diff}\times\text{Lorentz}\times\text{Weyl}\times U(1)\,.
\ee 
In terms of the coset construction for the WZ effective action that we discussed in section \ref{sec:WZactions}, the unbroken subgroup in this case will be $H=\text{Diff}\times\text{Lorentz}$, so that
\bbxd
\vskip.2cm
\be\label{eq:ConfCoset}
G/H\cong \text{Weyl}\times U(1)\,.
\ee
\ebxd
Importantly, in this case the coset is a normal subgroup of $G$, which means that the condition \eqref{H-anomaly-cond} indeed holds.

\subsection{Anomalies and the WZ effective action}
\label{sec:CFTanomalies}

We now consider a generic CFT with a U(1) symmetry and couple it to the independent background gauge fields $e_\m{}^a$ and $A_\m$. In fact, in the absence of external fermions, which is the case we focus on in this section, the vielbein can be replaced by the background metric, $g_{\m\n}=\h_{ab}e_\m{}^a e_{\n}{}^b$, so that all currents transform trivially under the local Lorentz symmetry. 

The generating function $\cw[g,A]$, may not in general be invariant under the local symmetries, signaling the presence of a quantum anomaly for the corresponding symmetry. We may parameterize the potential non-invariance of the effective action by writing
\be\label{CFT-anomalies}
\big(\d_D(\x)+\d_W(\s)+\d_G(\th)\big) \cw[g,A]=\int d^dx \sqrt{-g}(\x^\m \ca_{D\m}+\s \ca_W+\th \ca_{G})\,,
\ee
where $\ca_{D\m}[g,A]$, $\ca_W[g,A]$ and $\ca_{G}[g,A]$ denote respectively the potential Einstein, Weyl and U(1) anomalies, whose general form is determined by the WZ consistency conditions. Introducing the energy-momentum tensor and U(1) current operators through the relations 
\be\label{CFTcurrents}
\<\ct^{\m\n}\>=\frac{2}{\sqrt{-g}}\frac{\d \cw}{\d g_{\m\n}}\,\,,\qquad \<\cj^\m\>=\frac{1}{\sqrt{-g}}\frac{\d \cw}{\d A_\m}\,,
\ee
the transformation \eqref{CFT-anomalies} of $\cw[g,A]$ leads to the anomalous Ward identities
\be\label{CFT-WardIDs}
\nabla_\n\<\ct^\n{}_\m\>-F_{\m\n}\<\cj^\n\>+A_\m\nabla_\n\<\cj^\n\>=-\ca_{D\,\m}\,,\quad \<\ct^{\m}{}_{\m}\>=\ca_W\,,\quad
\nabla_\m\, \<\cj^\m\>=-\ca_{G}\,.
\ee
In the following we will focus on theories without a genuine gravitational anomaly, so that we can set $\ca_{D\,\m}=0$ by a choice of scheme.

Given the general form of the anomalies that solve the WZ consistency conditions, our next goal is to construct a universal effective action that reproduces the anomalous symmetry transformations \eqref{CFT-anomalies}. To this end, we follow the prescription in section \ref{sec:WZactions} and formally write the generating function as a sum of an invariant part, $\cw_{\rm inv}[g,A]$, and a non-local anomaly effective action, $\cw_{\rm anom}[g,A]$, which shares the anomalous symmetry transformations \eqref{CFT-anomalies} with $\cw[g,A]$. Subsequently, we introduce a pair of compensator fields, the dilaton $\t$, and the axion $\c$, which together parameterize the coset \eqref{eq:ConfCoset}. These compensator fields transform additively under Weyl and local U(1) transformations, i.e. 
\be\label{non-linear}
\t\to \t+\s\,,\qquad \c\to\c+\th\,.
\ee
and therefore allow us to define the invariant variables (cf. \eqref{invA})
\be\label{invariant}
\hat g_{\m\n}\equiv g^{(-\t)}_{\m\n}=e^{-2\t} g_{\m\n}\,,\qquad \hat A_\m\equiv A_\m^{(-\c)}=A_\m-\del_\m\chi\,.
\ee

The anomaly effective action $\cw_{\rm anom}[g,A]$ can be formally expressed as a path integral over the compensator fields as \cite{Schwimmer:2010za}
\be\label{eq:pathint}
e^{i\cw_{\text{anom}}[g,A]}=\int \cd\t\cd\c\, e^{i\cw_{\rm WZ}[g,A,\t,\c]}\,,
\ee
where $\cw_{\rm WZ}[g,A,\t,\c]$ is the local WZ action, given by the difference 
\be
\cw_{\rm WZ}[g,A,\t,\c]=-(\cw[\hat g,\hat A]-\cw[g,A])\,.
\ee
As we saw in section \ref{sec:WZactions}, the WZ effective action can be computed explicitly from the anomalies $\ca_W$ and $\ca_{G}$ through the identity \eqref{WZ-transgression}, which in this case takes the form 
\bbxd
\be\label{WZ-transgression-CFT}
\hskip-.2cm\cw_{\rm WZ}[g,A,\t,\c]=\int_0^1dt\int d^dx \sqrt{-g^{(-t\t)}}\big(\t\ca_{W}[g^{(-t\t)},A^{(-t\c)}]+\c\ca_{G}[g^{(-t\t)},A^{(-t\c)}]\big).\hskip-.1cm
\ee
\ebxd
In the remaining of this section we evaluate this WZ effective action explicitly for generic CFTs in two and four dimensions.

\subsection{WZ action for CFTs in two and four dimensions}
\label{sec:CFTs}

\subsubsection{2d CFTs }
\label{sec:2dCFT}

The non-local anomaly effective action and the WZ action take the simplest and most familiar form in two dimensions. The 2D conformal and U(1) anomalies are given respectively by 
\be
\label{eq:2Danomalies}
\ca_W=\frac{c}{24\p}R\,,\qquad \ca_{G}=\frac{\k_{F^2}}{8\p}\e^{\m\n}F_{\m\n}\,,
\ee
where $\e_{\m\n}$ is the Levi-Civita tensor in two dimensions. The anomaly coefficients $c$ and $\k_{F^2}$ can be expressed in terms of the left and right Virasoro ($c_{L,R}$) and Kac-Moody ($k_{L,R}$) central charges through the relations  
\be\label{anomaly_coeffs_1}
c=\tfrac12(c_L+c_R)\,,\qquad \k_{F^2}=k_L-k_R\,.
\ee
In two dimensions there is also a gravitational anomaly whose coefficient is given by $c_L-c_R$, but we will only consider non-chiral CFTs here so that $c_L=c_R$.

\subsubsection*{WZ action}

The first step in order to determine the WZ action \eqref{WZ-transgression-CFT} is to evaluate the anomalies \eqref{eq:2Danomalies} on the one-parameter family of variables 
\be\label{gtAt}
 g^{(t)}_{\m\n}\equiv g^{(-t\t)}_{\m\n}=e^{-2t\t}g_{\m\n},\qquad A^{(t)}_\m\equiv A^{(-t\c)}_\m =A_\m-t\del_\m\c\,.
\ee
Using the Weyl transformation of the Ricci scalar, $R[g^{(t)}]= e^{2t\t}(R[g]+2t\square\t)$, the fact that the combination $\sqrt{-g}\,\e^{\m\n}$ is independent of the metric, and that $F_{\m\n}$ is gauge invariant, the WZ action \eqref{WZ-transgression-CFT} becomes  
\be
\cw_{\rm WZ}[g,A,\t,\c]=\int_0^1dt\int d^2x\sqrt{-g}\Big(\frac{c}{24\p}\t(R[g]+2t\square\t)+\frac{\k_{F^2}}{8\p}\c\, \e^{\m\n}F_{\m\n}\Big)\,.
\ee
Performing the $t$-integration and integrating by parts the second term leads to the result 
\bbxd
\vskip-.1cm
\be\label{eq:2dCFTWZ}
\cw_{\rm WZ}[g,A,\t,\c]=\frac{c}{24\p}\int d^2x\sqrt{-g}\big(\t R[g]-(\nabla\t)^2\big)+\frac{\k_{F^2}}{8\p} \int d^2x\sqrt{-g}\,\c\, \e^{\m\n}F_{\m\n}\,.
\ee
\ebxd
It is straightforward to verify that this action reproduces the anomalies \eqref{eq:2Danomalies}.

\subsubsection*{Non-local anomaly action}

A non-local anomaly effective action $\cw_{\rm anom}[g,A]$ can be obtained from \eqref{eq:2dCFTWZ} by integrating out  the compensators $\t$ and $\c$ as in \eqref{eq:pathint}. It turns out that the tree-level result obtained by solving the classical equations for the compensators and plugging the solutions back in the WZ action suffices in order to obtain an effective action that correctly reproduces the anomalies. In particular, quantum effects contribute only through invariant terms \cite{Schwimmer:2010za}. 

Another important observation is that the WZ action already contains a kinetic term for the dilaton, but the axion appears algebraically. Hence, the path integral over $\c$ would impose $\e^{\m\n}F_{\m\n}=0$ as a constraint. This can be remedied by adding to the WZ action a Weyl and U(1) invariant term that is second order in derivatives of $\c$, namely   
\be\label{eq:2dCFTWZ-axion-kinetic}
-\frac{\b^2}{2}\int d^2x \sqrt{-\hat g}\, \hat A_\m \hat A^\m=-\frac{\b^2}{2}\int d^2x \sqrt{-g}\,(A_\m-\pa_\m\c)(A^\m-\pa^\m\c)\,,
\ee
where $\b$ is an arbitrary nonzero parameter. 

The equations of motion for the compensating fields obtained from the WZ action \eqref{eq:2dCFTWZ} with the additional term \eqref{eq:2dCFTWZ-axion-kinetic} read
\be\label{eq:2Deoms}
\,\square\,\tau=-\frac12R,\qquad \square\,\chi=\nabla_\n A^\n-\frac{1}{\b^2}\frac{\k_{F^2}}{8\p}\,\e^{\m\n}F_{\m\n}\,.
\ee
These decoupled equations can be solved using the Green's function of the operator $\square$. The resulting solution can be written formally in the form
\be\label{eq:2Dsols}
\tau=-\frac12\square^{-1}\,R\,, \qquad \chi=\square^{-1}\Big(\nabla_\n A^\n-\frac{1}{\b^2}\frac{\k_{F^2}}{8\p}\e^{\m\n}F_{\m\n}\Big)\,.
\ee
Inserting these expressions back in \eqref{eq:2dCFTWZ} and the invariant axion kinetic term \eqref{eq:2dCFTWZ-axion-kinetic} we obtain the nonlocal anomaly effective action
\bbxd
\vskip-.1cm
\be\label{2DWanom}
\cw_{\rm anom}[g,A]=-\frac{c}{96\p}\int d^2x\sqrt{-g}\,R\frac1\square R+\frac{\k_{F^2}}{8\p}\int d^2x\sqrt{-g}\,\e_{\m\n}F^{\m\n}\frac1\square\nabla_\m A^\m+\cdots\,,
\ee
\ebxd
where the ellipses indicate Weyl and U(1) invariant terms that can be absorbed in $\cw_{\rm inv}[g,A]$. The first term is the well-known non-local Polyakov action \cite{Polyakov:1981rd}, while the second term is known to reproduce the U(1) anomaly in two dimensions \cite{Wess:1971yu,Krasnikov:1985jp,Krasnikov:1986rq,PhysRevD.35.1268}.

\subsubsection*{WZ currents}

An important application of the WZ action is the solution of the anomalous Ward identities \eqref{CFT-WardIDs}. By construction, the currents obtained from the WZ action satisfy the inhomogeneous Ward identities \eqref{CFT-WardIDs} provided the compensators $\t$ and $\c$ satisfy their equations of motion. Invariant terms such as the axion kinetic term \eqref{eq:2dCFTWZ-axion-kinetic} correspond to the ambiguity associated with the homogeneous, i.e. non-anomalous, solution of the Ward identities. 

The WZ action \eqref{eq:2dCFTWZ} together with the invariant term \eqref{eq:2dCFTWZ-axion-kinetic} lead to the WZ currents
\bbxd
\bal\label{eq:2DWZcurrrents}
\<\ct^{\m\n}_{\rm WZ}\>=&\;\frac{c}{12\p}\big(\nabla^\m\t\nabla^\n\t-\tfrac12g^{\m\n}(\nabla\t)^2+\nabla^\m\nabla^\n\t-g^{\m\n}\square\t\big) +\b^2 \big(\hat A^{\m}\hat A^{\n}-\tfrac{1}{2} g^{\m\n} \hat A^2\big)\,,\NO\\
\<\cj^\m_{\rm WZ}\>=&\;\frac{\k_{F^2}}{4\p}\e^{\m\n}\nabla_\n\c-\b^2 \hat A^\m\,.
\eal
\ebxd
To avoid introducing further symbols, we have abused the notation and raised the index in $\hat A_\m$ with $g^{\m\n}$ here, namely $\hat A^\m=g^{\m\n}\hat A_\n$, instead of $\hat g^{\m\n}\hat A_\n$ that was used in \eqref{eq:2dCFTWZ-axion-kinetic}. 
It is straightforward to verify that these satisfy the anomalous trace and U(1) Ward identities provided the compensators satisfy their equations of motion. In particular,
\be
\<\ct^{\m}_{\rm WZ\,\m}\>=-\frac{c}{12\p}\square\t=\frac{c}{24\p}R\,,
\ee 
by virtue of the dilaton equation in \eqref{eq:2Deoms}. Similarly, using the axion equation in \eqref{eq:2Deoms},
\be
\nabla_\m\<\cj^\m_{\rm WZ}\>=\b^2(\square\c-\nabla_\m A^\m)=-\frac{\k_{F^2}}{8\p}\,\e^{\m\n}F_{\m\n}\,.
\ee

It is a little more laborious to check that the WZ currents \eqref{eq:2DWZcurrrents} satisfy the Ward identity involving the divergence of the stress tensor, but this is ensured by the diffeomorphism invariance of the WZ action. The divergence of the WZ stress tensor in \eqref{eq:2DWZcurrrents} is given by
\bal
\nabla_\m\<\ct^{\m\n}_{\rm WZ}\>=&\;\frac{c}{12\p}\Big(\square\t+\frac12 R\Big)\nabla^\n\t+\b^2 \big((\nabla_\m\hat A^{\m})\hat A^{\n}+\hat A_{\m}\nabla^\m\hat A^{\n}-\hat A_\m\nabla^\n\hat A^\m\big)\NO\\
=&\;\frac{c}{12\p}\Big(\square\t+\frac12 R\Big)\nabla^\n\t+\b^2 \Big(-\frac{1}{\b^2}\nabla_\m\<\cj^\m_{\rm WZ}\>\hat A^\n+\hat A_{\m}F^{\m\n}\Big)\,.
\eal
It follows that the l.h.s. of the diffeomorphism Ward identity in \eqref{CFT-WardIDs} takes the form
\bal
&\nabla_\m\<\ct_{\rm WZ}^\m{}_\n\>-F_{\n\m}\<\cj_{\rm WZ}^\m\>+A_\n\nabla_\m\<\cj_{\rm WZ}^\m\>\NO\\
&\;=\frac{c}{12\p}\Big(\square\t+\frac12 R\Big)\nabla_\n\t+\nabla_\n\c\nabla_\m\<\cj_{\rm WZ}^\m\>+\b^2 \hat A_{\m}F^{\m}{}_{\n}-F_{\n\m}\Big(\frac{\k_{F^2}}{4\p}\e^{\m\r}\nabla_\r\c-\b^2 \hat A^\m\Big)\NO\\
&\;=\frac{c}{12\p}\Big(\square\t+\frac12 R\Big)\nabla_\n\t+\nabla_\n\c\Big(\nabla_\m\<\cj_{\rm WZ}^\m\>+\frac{\k_{F^2}}{8\p}\,\e^{\r\s}F_{\r\s}\Big)\NO\\
&\;\hskip0.5cm-\frac{\k_{F^2}}{8\p}\,\e^{\m\r}(-2F_{\m\n}\nabla_\r\c+F_{\m\r}\nabla_\n\c)\,.
\eal
The first two terms in this expression vanish due to the dilaton and axion equations \eqref{eq:2Deoms}. The last term vanishes identically in two dimensions due to the fact that $F_{\m\n}\propto \e_{\m\n}$.

\subsubsection{4d CFTs }
\label{sec:4dCFT}

Next, let us consider generic four-dimensional CFTs. These include four-dimensional SCFTs, the main subject of this work, but crucially we do not turn on a background gravitino in this subsection, since a supercurrent does not necessarily exist in a generic CFT. The general form of the Weyl and U(1) anomalies for four-dimensional CFTs compatible with diffeomorphism invariance is respectively
\be
\label{eq:4Danomalies}
\begin{split}
\ca_W=&\;\frac{1}{16\p^2}(c\, W_{\m\n\r\s}W^{\m\n\r\s}-a E +b \square R + k_W F_{\m\n}F^{\m\n})\,,\\
\ca_{G}=&\;\frac{k_F}{27\p^2}\wt F_{\m\n}F^{\m\n}+\frac{k_R}{24\p^2}\cp\,,
\end{split}
\ee
where $a$, $b$, $c$, $k_W$, $k_F$ and $k_R$ are independent anomaly coefficients, $\wt F_{\m\n}\equiv\tfrac12 \e_{\m\n}{}^{\r\s}F_{\r\s}$, $W_{\m\n\r\s}$ is the Weyl tensor, whose square is expressible in terms of the Riemann and Ricci tensors as
\be
W_{\m\n\r\s}W^{\m\n\r\s}=R_{\m\n\r\s}R^{\m\n\r\s}-2R_{\m\n}R^{\m\n}+\tfrac13R^2\,,
\ee
while $E$ and $\cp$ are the Euler and Pontryagin densities, given respectively by
\be
E=R_{\m\n\r\s}R^{\m\n\r\s}-4R_{\m\n}R^{\m\n}+R^2\,,\qquad \cp\equiv\wt R_{\m\n\r\s}R^{\m\n\r\s}\equiv\tfrac12\e_{\m\n}{}^{\k\l}R_{\k\l\r\s}R^{\m\n\r\s}\,.
\ee

In writing the anomalies in the form \eqref{eq:4Danomalies} we have made a choice of scheme such that the mixed gravitational-U(1) anomaly is placed entirely in the U(1) symmetry, preserving diffeomorphisms. This is always possible in four dimensions due to the absence of a genuine gravitational anomaly \cite{Alvarez-Gaume:1983ihn}. In particular, the local counterterm
\be\label{eq:mixedCT}
\int A\wedge \tr\big(\G\wedge d\G+\tfrac23 \G\wedge \G\wedge \G\big)\,,
\ee
where $\G^\m{}_{\n}\equiv \G^\m{}_{\n\r}dx^\r$ is the GL(4,$\bb R$)-valued 1-form defined by the affine connection, allows us to move the mixed anomaly between diffeomorphisms and U(1) transformations, as is discussed in detail e.g. in \cite{Jensen:2012kj}. However, here we choose a scheme where the counterterm \eqref{eq:mixedCT} is absent, such that diffeomorphism invariance is preserved.

The term proportional to $\square R$ in the Weyl anomaly in \eqref{eq:4Danomalies} is also scheme-dependent, corresponding to a local counterterm proportional to $R^2$. Trivial anomalies that can be removed by local counterterms correspond to type C anomalies in the Deser-Schwimmer classification \cite{Deser:1993yx,alexakis2009}. Local Weyl invariants such as $\sqrt{-g}\,W_{\m\n\r\s}W^{\m\n\r\s}$ and  $\sqrt{-g}\,\e^{\m\n\r\s}F_{\m\n}F_{\r\s}$ correspond to type B anomalies in the same classification, while global Weyl invariants such as the Euler characteristic correspond to type A anomalies. 

Besides the local counterterm proportional to $R^2$, we may consider counterterms involving the local, weight-4, parity-even and covariant terms $R_{\m\n}R^{\m\n}$, $R_{\m\n\r\s}R^{\m\n\r\s}$ or $F_{\m\n}F^{\m\n}$ \cite{Anastasiou:2020zwc,Ntokos:2021duk}. However, these can be expressed as linear combinations of $R^2$ and the Weyl invariants $E$, $W_{\m\n\r\s}W^{\m\n\r\s}$,  $F_{\m\n}F^{\m\n}$, and hence they do not contribute any further terms to the anomalies \eqref{eq:4Danomalies}. Nevertheless, besides the Euler density, these counterterms do contribute to the stress tensor and U(1) current.

\subsubsection*{WZ action}

As in the previous example, the starting point in order to determine the WZ action \eqref{WZ-transgression-CFT} is evaluating the anomalies \eqref{eq:4Danomalies} on the one-parameter family of variables \eqref{gtAt}. Since the anomalies are U(1) invariant, we need only consider their Weyl transformation. This is straightforward for all terms except for the Euler density, $E$, and the scheme-dependent term $\square R$. In particular, the densities $\sqrt{-g}\, F_{\m\n}F^{\m\n}$ and $\sqrt{-g}\,\e^{\m\n\r\s}F_{\m\n}F_{\r\s}$ are both Weyl invariant. So is $\sqrt{-g}\,W_{\m\n\r\s}W^{\m\n\r\s}$, since the Weyl tensor transforms homogeneously according to $W_{\m\n\r\s}[e^{2\s}g]=e^{2\s}W_{\m\n\r\s}$. Finally, the Pontryagin density may expressed in terms of the Weyl tensor as $\cp=\tfrac12\e^{\k\l\m\n}W_{\k\l\r\s}W_{\m\n}{}^{\r\s}$, which implies that the density $\sqrt{-g}\,\cp$ is also Weyl invariant. It follows that for $a=b=0$, the WZ action \eqref{WZ-transgression-CFT} takes the form
\be\label{eq:4DWZab=0}
\begin{split}
\cw^{a,b=0}_{\rm WZ}[g,A,\t,\c]=\int d^4x \sqrt{-g}\Big(\frac{1}{16\p^2}\t(c\, W^2+k_W F^2)+\frac{k_F}{27\p^2}\c\wt F F+\frac{k_R}{24\p^2}\c\cp\Big)\,,
\end{split}
\ee
where we have introduced the notation $W^2\equiv W_{\m\n\r\s}W^{\m\n\r\s}$, $F^2\equiv F_{\m\n}F^{\m\n}$, $\wt F F\equiv \tfrac12\e^{\m\n\r\s}F_{\m\n}F_{\r\s}$.

In order to compute the Weyl transformation of the Euler density, it is convenient to write it in terms of the Weyl tensor as
\be\label{EWrelation}
E=W_{\m\n\r\s}W^{\m\n\r\s}-2R_{\m\n}R^{\m\n}+\frac23R^2\,.
\ee
The first term transforms homogeneously, while the transformation of the terms involving the Ricci tensor and the Ricci scalar can be determined using the identities 
\be\label{t-Ricci}
\begin{split}
R_{\m\n}[g^{(-t\t)}]=&\,R_{\m\n}[g]+2t\,\nabla_\m\nabla_\n\t+t\,g_{\m\n}\square\t+2t^2(\nabla_\m\t\nabla_\n\t-g_{\m\n}\nabla_\r\t\nabla^\r\t)\,,\\
R[g^{(-t\t)}]=&\;e^{2t\t}(R[g]+6t\,\square\t-6t^2\,\nabla_\r\t\nabla^\r\t)\,.
\end{split}
\ee
In particular, we obtain
\be
\begin{split}
[R_{\m\n} R^{\m\n}]^{(-t\t)}=&\;e^{4t\t}\big(R_{\m\n} R^{\m\n}+4t^2\nabla_\m\nabla_\n\t\nabla^\m\nabla^\n\t+8t^2(\square\t)^2+12t^4(\nabla_\r\t\nabla^\r\t)^2\\
&\qquad+4 tR^{\m\n}\nabla_\m\nabla_\n\t+2t R\square\t+4 t^2R^{\m\n}\nabla_\m\t\nabla_\n\t-4t^2R\nabla_\r\t\nabla^\r\t\\
&\qquad+8t^3\nabla_\m\nabla_\n\t\nabla^\m\t\nabla^\n\t-20t^3\square\t\nabla_\r\t\nabla^\r\t\big)\,,\\
[R^2]^{(-t\t)}=&\;e^{4t\t}\big( R^2+36t^2(\square\t)^2+36t^4(\nabla_\r\t\nabla^\r\t)^2+12tR\square\t\\
&\qquad-12t^2R\nabla_\r\t\nabla^\r\t-72t^3\nabla_\r\t\nabla^\r\t\square\t\big)\,.
\end{split}
\ee
Inserting these in \eqref{EWrelation} and simplifying the resulting expression we conclude that
\be\label{WeylE-nt}
\begin{split}
\sqrt{-g^{(-t\t)}}E^{(-t\t)}=&\;\sqrt{-g}\big(E-8\nabla_\m\big[ t^2\,\nabla_\r\t\nabla^\m\nabla^\r\t-t^2\nabla^\m\t\square\t\\
&\hspace{3.cm}+t( R^{\m\n}-\tfrac12  R g^{\m\n})\nabla_\n\t+t^3\nabla^\m\t(\nabla\t)^2\big]\big)\,.
\end{split}
\ee

Finally, the term $\sqrt{-g}\,\square R$ evaluated on the rescaled metric $g_{\m\n}^{(-t\t)}=e^{-2t\t}g_{\m\n}$ becomes 
\be\label{WeylBoxR-nt}
\begin{split}
\sqrt{-g^{(-t\t)}}\,\square^{(-t\t)}R^{(-t\t)}=&\;\pa_\m\big(\sqrt{-g^{(-t\t)}}\,g^{(-t\t)\m\n}\pa_\n R^{(-t\t)}\big)\\
=&\;\pa_\m\big(e^{-2t\t}\sqrt{-g}\,g^{\m\n}\pa_\n \big[e^{2t\t}(R+6t\,\square\t-6t^2\,\nabla_\r\t\nabla^\r\t)\big)\big]\\
=&\;\sqrt{-g}\,(\square+2t\nabla_\m\t\nabla^\m+2t\square\t)(R+6t\,\square\t-6t^2\,\nabla_\r\t\nabla^\r\t)\,.
\end{split}
\ee

Substituting the results \eqref{WeylE-nt} and \eqref{WeylBoxR-nt} in the transgression formula \eqref{WZ-transgression-CFT} for the WZ action, integrating over the parameter $t$ and performing a number of integrations by parts, leads to the WZ action
\bbxd
\be\label{eq:4DWZ}
\begin{split}
\cw_{\rm WZ}[g,A,\t,\c]=&\;\int d^4x\sqrt{-g}\,\Big[\frac{1}{16\p^2}\t(c\, W^2+k_W F^2)+\frac{k_F}{27\p^2}\c\wt F F+\frac{k_R}{24\p^2}\c\cp\\
&\;\hspace{-0.8cm}-\frac{a}{16\p^2}\big(\t E+4(R^{\m\n}-\tfrac12R g^{\m\n})\nabla_\m\t\nabla_\n\t-4(\nabla\t)^2\square\t+2(\nabla\t)^4\big)\\
&\;\hspace{-0.8cm}+\frac{b}{16\p^2}\big(\t\square R+3\t\square^2\t-R(\nabla\t)^2-6(\nabla\t)^2\square\t+3(\nabla\t)^4\big)\Big]\,.
\end{split}
\ee
\ebxd
A number of special values of the scheme parameter $b$ have been discussed in the literature. The `minimal' scheme corresponding to $b=0$ is discussed in \cite{Schwimmer:2010za,Baume:2013ika,Assel:2015nca} and arises naturally in the context of $\cn=1$ SCFTs. Adding a suitable second order invariant kinetic term for the dilaton, as we will do below, the minimal action leads to the correct analyticity properties for dilaton scattering amplitudes \cite{Schwimmer:2010za}.    

The value $b=2a/3$ results in the particularly simple dilaton action
\be\label{eq:RFT}
\begin{split}
\cw_{\rm RFT}[g,A,\t,\c]=&\;\int d^4x\sqrt{-g}\,\Big[\frac{1}{16\p^2}\t(c\, W^2+k_W F^2)+\frac{k_F}{27\p^2}\c\wt F F+\frac{k_R}{24\p^2}\c\cp\\
&\;\hspace{5.cm}-\frac{a}{16\p^2}\big(\t (E-\tfrac23\square R)-2\t\D_P\t\big)\Big]\,,
\end{split}
\ee
where $\D_P$ is the self-adjoint Weyl-covariant Paneitz operator
\be
\D_P\equiv \square^2+2\nabla_\m(R^{\m\n}-\tfrac13R g^{\m\n})\nabla_\n\,,
\ee
which is the fourth-order analogue of the conformal Laplacian $\square-\tfrac16 R$. Integrating out the dilaton in this action leads to the Riegert-Fradkin-Tseytlin (RFT) action \cite{Fradkin:1983tg,Riegert:1984kt,Deser:1999zv} (see also \cite{Mottola:2026emf}), which is the direct analogue of the Polyakov action \eqref{2DWanom} in four dimensions. However, the dilaton propagator $\square^{-2}$ that follows from \eqref{eq:RFT} violates the analyticity requirements of dilaton scattering amplitudes \cite{Deser:1996na,Deser:1999zv,Schwimmer:2010za}.

Finally, dimensional regularization naturally results in the scheme $b=2c/3$ \cite{Brown:1985ri,Tomboulis:1988gw,Coriano:2013nja}.

\subsubsection*{Non-local anomaly action}

As we have seen, a nonlocal anomaly effective action can be obtained from \eqref{eq:4DWZ} by integrating out the compensator fields. Following \cite{Schwimmer:2010za}, we work in the `minimal' scheme $b=0$ and add Weyl and U(1) invariant second order kinetic terms for the dilaton and axion, namely 
\be\label{eq:4dCFTWZ-dilaton-kinetic}
-\frac{\a^2}{12}\int d^4x \sqrt{-\hat g}\,R[\hat g]=-\frac{\a^2}{2}\int d^4x \sqrt{-g}\,e^{-2\t}(\pa_\m\t\pa^\m\t+\tfrac16R[g])\,,
\ee
and 
\be\label{eq:4dCFTWZ-axion-kinetic}
-\frac{\b^2}{2}\int d^4x \sqrt{-\hat g}\, \hat A_\m \hat A^\m=-\frac{\b^2}{2}\int d^4x \sqrt{-g}\,e^{-2\t}(A_\m-\pa_\m\c)(A^\m-\pa^\m\c)\,,
\ee
where $\a$ and $\b$ are arbitrary nonzero constants. Notice that the kinetic term for the dilaton in \eqref{eq:4dCFTWZ-dilaton-kinetic} can be expressed in terms of the conformal Laplacian as $e^{-\t}(-\square+\tfrac16 R)e^{-\t}$, while the invariant form of the axion kinetic term \eqref{eq:4dCFTWZ-axion-kinetic} coincides with the kinetic term \eqref{eq:2dCFTWZ-axion-kinetic} we used in two dimensions, except that, when written in terms of $g_{\m\n}$ and $\t$, its four-dimensional form involves a coupling to the dilaton.

The dilaton and axion equations of motion following from the WZ action \eqref{eq:4DWZ} with $b=0$, plus the invariant kinetic terms \eqref{eq:4dCFTWZ-dilaton-kinetic} and \eqref{eq:4dCFTWZ-axion-kinetic}, take respectively the form
\be\label{eq:4Deoms}
\begin{split}
&\;\a^2 e^{-\t}(\square-\tfrac16 R)e^{-\t}-\b^2 e^{-2\t}(A-\nabla\c)^2\\
&\;-\frac{a}{2\p^2}\nabla_\m\big((R^{\m\n}-\tfrac12R g^{\m\n})\nabla_\n\t-\square\t\nabla^\m\t+\tfrac12\nabla^\m(\nabla\t)^2+(\nabla\t)^2\nabla^\m\t\big)=\ca_W\,,\\
&\;\rule{0cm}{.8cm}\b^2\nabla_\m(e^{-2\t}(A^\m-\pa^\m\c))=\ca_{G}\,,
\end{split}
\ee
where the anomalies $\ca_W$ and $\ca_{G}$ are those given in \eqref{eq:4Danomalies}, but with $b=0$. This is a system of coupled nonlinear equations that cannot be solved exactly. However, if we are simply interested in obtaining a non-local effective action that reproduces the anomalies, it suffices to solve these equations perturbatively. As we will see, inserting the leading order solution for $\t$ and $\c$ in the WZ action leads to such a non-local action. Higher order corrections contribute only to invariant terms in the effective action.

Introducing the variable
\be
\vf\equiv 1-e^{-\t}\,,
\ee
the dilaton equation becomes
\be
\begin{split}
&\;(-\square+\tfrac16 R)\vf=\tfrac16 R+\co(1/\a^2)\,.
\end{split}
\ee
The perturbative solution for $\vf$, therefore, takes the form 
\be
\vf=\frac{1}{-6\square+R}R+\co(1/\a^2)\,,
\ee
while the dilaton is given on-shell by
\be\label{eq:tau-sol}
\t=-\log(1-\vf)=-\log\Big(1-\frac{1}{-6\square+R}R\Big)+\co(1/\a^2)\,.
\ee

Importantly, the quantity inside the brackets transforms homogeneously under a Weyl transformation $g_{\m\n}\to e^{2\s}g_{\m\n}$ \cite{Schwimmer:2010za}, namely 
\be
\Big(1-\frac{1}{-6\square+R}R\Big)\to e^{-\s}\Big(1-\frac{1}{-6\square+R}R\Big)\,.
\ee
Therefore, the leading order solution for $\t$ already accounts for its correct transformation, $\t\to\t+\s$. It follows that higher order corrections to the solution for $\t$ necessarily involve invariant terms.

Next, let us consider the axion equation of motion, which can be written in the form
\be
(\nabla_\m-2\nabla_\m\t)(A^\m-\pa^\m\c)=\frac{1}{\b^2}e^{2\t}\ca_{G}\,.
\ee
Solving this perturbatively we obtain
\be\label{eq:chi-sol}
\c=\frac{1}{\square-2\nabla^\n\t\nabla_\n}(\nabla_\m-2\nabla_\m\t)A^\m+\co(1/\b^2)\,,
\ee
where $\t$ is given by its on-shell value in \eqref{eq:tau-sol}. Again, this leading order solution accounts for the correct transformation of the axion under U(1) transformations, $\c\to\c+\th$, and so higher order corrections only contribute invariant terms. 

Moreover, the leading order solution \eqref{eq:chi-sol} is also invariant under Weyl transformations, as it should be. This follows from the observation that the quantities $\D\equiv \square-(d-2)\nabla^\m\t\nabla_\m$ and $\nabla_\m A^\m-(d-2)A^\m\nabla_\m\t$ transform homogeneously in $d$ dimensions, namely
\be
\begin{split}
\d_\s\D=-2\s\D\,,\qquad \d_\s(\nabla_\m A^\m-(d-2)A^\m\nabla_\m\t)=-2\s(\nabla_\m A^\m-(d-2)A^\m\nabla_\m\t)\,.
\end{split}
\ee
It follows that an infinitesimal Weyl transformation of \eqref{eq:chi-sol} is given by
\be
\begin{split}
\d_\s\Big(\frac{1}{\D}(\nabla_\m-2\nabla_\m\t)A^\m\Big)=&\;-\frac{1}{\D}\d_\s\D\frac{1}{\D}(\nabla_\m A^\m-2A^\m\nabla_\m\t)+\frac{1}{\D}\d_\s(\nabla_\m A^\m-2A^\m\nabla_\m\t)\\
=&\;\frac{2}{\D}\s(\nabla_\m A^\m-2A^\m\nabla_\m\t)
-\frac{2}{\D}\s(\nabla_\m A^\m-2A^\m\nabla_\m\t)=0\,.
\end{split}
\ee

We conclude that, up to Weyl and U(1) invariant terms that can be absorbed in $\cw_{\rm inv}[g,A]$, the anomaly effective action for generic four-dimensional CFTs takes the form 
\bbxd
\be\label{4DWanom}
\cw_{\rm anom}[g,A]=\cw_{\rm WZ}\Big[g,A,\t=-\log\big(1+\frac{1}{6\square-R}R\big),\c=\frac{1}{\D}(\nabla_\m-2\nabla_\m\t)A^\m\Big]\,,
\ee
\ebxd
where $\D\equiv \square-2\nabla^\m\t\nabla_\m$ and the WZ action is given in \eqref{eq:4DWZ}.

\subsubsection*{WZ currents}

As in two dimensions, the currents following from the WZ action \eqref{eq:4DWZ} and the invariant kinetic terms \eqref{eq:4dCFTWZ-dilaton-kinetic} and \eqref{eq:4dCFTWZ-axion-kinetic} satisfy the anomalous Ward identities \eqref{CFT-WardIDs}, provided the compensators satisfy their equations of motion \eqref{eq:4Deoms}. This is straightforward to verify for the U(1) current, which takes the form 
\bbxd
\vskip.3cm
\be\label{eq:4DWZU(1)currrent}
\<\cj_{\rm WZ}^\m\>=-\frac{k_W}{4\p^2}\nabla_\n(\t F^{\n\m})-\frac{4k_F}{27\p^2}\nabla_\n(\c\wt F^{\n\m})-\b^2e^{-2\t}\hat A^\m\,,
\ee
\ebxd
where again we have used the notation $\hat A^\m =g^{\m\n}\hat A_\n$.  Using the axion equation in \eqref{eq:4Deoms} and the fact that, in the absence of torsion, $[\nabla_\m,\nabla_\n]\ck^{\m\n}=0$ for any antisymmetric tensor $\ck^{\m\n}$, one may verify that the current \eqref{eq:4DWZU(1)currrent} satisfies the U(1) Ward identity in \eqref{CFT-WardIDs}, namely
\be
\nabla_\m\<\cj_{\rm WZ}^\m\>=-\b^2\nabla_\m(e^{-2\t}\hat A^\m)=-\ca_G\,.
\ee

The analogous calculation for the stress tensor is more involved \cite{Cappelli:1988vw}. We first compute the variation of the Euler, Pontryagin and Weyl-squared densities, which take the form
\be\label{eq:DensityVars}
\begin{split}
\d(\sqrt{-g}\,E)=&\;4\sqrt{-g}\,\nabla_\m (P^{\m\r\s\n}\nabla_\n\d g_{\r\s})\,,\\
\d(\sqrt{-g}\,\cp)=&\;4\sqrt{-g}\,\nabla_\m (\wt R^{\m\r\s\n}\nabla_\n\d g_{\r\s})\,,\\
\d(\sqrt{-g}\,W^2)=&\;-4\sqrt{-g}\,B^{\m\n}\d g_{\m\n}+4\sqrt{-g}\,\nabla_\m\big(W^{\m\r\s\n}\nabla_\n\d g_{\r\s}-(\nabla_\n W^{\m\r\s\n})\d g_{\r\s}\big)\,.
\end{split}
\ee
The tensor $P^{\m\n\r\s}$ in the variation of the Euler density is the double-dual Riemann tensor \cite{Roberts:1992ij}
\be
P^{\m\n\r\s}=R^{\m\n\r\s}-g^{\m\r}R^{\n\s}+g^{\m\s}R^{\n\r}+g^{\n\r}R^{\m\s}-g^{\n\s}R^{\m\r}+\tfrac12(g^{\m\r}g^{\n\s}-g^{\m\s}g^{\n\r})R\,.
\ee
It has the same symmetry properties as the Riemann tensor and it is covariantly conserved due to the Bianchi identity, i.e. $\nabla_\m P^{\m\n\r\s}=0$. Moreover, $B^{\m\n}$ in the variation of the Weyl-squared density is the Bach tensor 
\be
B_{\m\n}=\nabla^\r\nabla^\s W_{\m\r\n\s}+\tfrac12 R^{\r\s}W_{\m\r\n\s}\,,
\ee
which is symmetric, traceless and covariantly conserved, i.e. $B^\m_\m=0$ and $\nabla^\m B_{\m\n}=0$.

Using the variations \eqref{eq:DensityVars}, one finds that the stress tensor obtained from the WZ action \eqref{eq:4DWZ} and the kinetic terms \eqref{eq:4dCFTWZ-dilaton-kinetic} and \eqref{eq:4dCFTWZ-axion-kinetic} consists of the following terms when $b=0$
\bbxd
\be\label{eq:4DWZstresstensor}
\begin{split}
\hspace{-.35cm}\<\ct^{(\a)\m\n}_{\rm WZ}\>=&\;\a^2\big[e^{-2\t}\big(\nabla^\m\t\nabla^\n\t-\tfrac12 g^{\m\n}\nabla_\r\t\nabla^\r\t\big)+\tfrac{1}{6}e^{-2\t}\big(R^{\m\n}-\tfrac12 R g^{\m\n}\big)\\
&\;-\tfrac16(\nabla^\m\nabla^\n-g^{\m\n}\square)e^{-2\t}\big]\,,\\
\hspace{-.2cm}\<\ct^{(\b)\m\n}_{\rm WZ}\>=&\;\b^2e^{-2\t}\big(\hat A^\m\hat A^\n-\tfrac12 g^{\m\n}\hat A^\r\hat A_\r\big)\,,\\
\hspace{-.2cm}\<\ct^{(k_F)\m\n}_{\rm WZ}\>=&\;0\,,\\
\hspace{-.2cm}\<\ct^{(k_W)\m\n}_{\rm WZ}\>=&\;-\frac{k_W}{4\p^2}\t\big(F^{\m}{}_\r F^{\n\r}-\tfrac14 g^{\m\n} F^2\big)\,,\\
\hspace{-.2cm}\<\ct^{(k_R)\m\n}_{\rm WZ}\>=&\;\frac{k_R}{3\p^2}\nabla_\r(\wt R^{\s(\m\n)\r}\nabla_\s\c)\,,\\
\hspace{-.2cm}\<\ct^{(c)\m\n}_{\rm WZ}\>=&\;\frac{c}{2\p^2}\big(-\t B^{\m\n}+\nabla_\r(W^{\r(\m\n)\s}\nabla_\s\t)+\nabla_\r W^{\r(\m\n)\s}\nabla_\s\t\big)\,,\\
\hspace{-.2cm}\<\ct^{(a)\m\n}_{\rm WZ}\>=&\;-\frac{a}{2\p^2}\big(P^{\r(\m\n)\s}\nabla_\r\nabla_\s\t+\tfrac12g^{\m\n}(R^{\r\s}-\tfrac12 Rg^{\r\s})\nabla_\r\t\nabla_\s\t-R^{\r(\m}\nabla^{\n)}\t\nabla_\r\t\hspace{-1.cm}\\
&\;+\tfrac12R\nabla^\m\t\nabla^\n\t-R^{\m\r\n\s}\nabla_\r\t\nabla_\s\t+\nabla^{(\m}\nabla_\r(\nabla^{\n)}\t\nabla^\r\t)-\tfrac12\square(\nabla^\m\t\nabla^\n\t)\hspace{-1.cm}\\
&\;-\tfrac12g^{\m\n}\nabla_\r\nabla_\s(\nabla^\r\t\nabla^\s\t)+\tfrac12R^{\m\n}(\nabla\t)^2-\tfrac12\nabla^\m\nabla^\n(\nabla\t)^2+\tfrac12g^{\m\n}\square(\nabla\t)^2\hspace{-1.cm}\\
&\;+\nabla^\m\t\nabla^\n\t\square\t-\nabla^{(\m}\t\nabla^{\n)}(\nabla\t)^2+\tfrac12 g^{\m\n}\nabla^\r\t\nabla_\r(\nabla\t)^2\\
&-(\nabla\t)^2\nabla^\m\t\nabla^\n\t+\tfrac14g^{\m\n}(\nabla\t)^4\big)\,.
\end{split}
\ee
\ebxd

It is not too difficult to evaluate the trace of all the terms in \eqref{eq:4DWZstresstensor}. The result is
\bal
\<\ct^{\m}_{\rm WZ\,\m}\>=&\;\a^2e^{-\t}\big(\square-\tfrac{1}{6}R\big)e^{-\t}-\b^2e^{-2\t}\hat A^\r\hat A_\r\NO\\
&-\frac{a}{2\p^2}\nabla_\r\big((R^{\r\s}-\tfrac12g^{\r\s}R)\nabla_\s\t-\nabla^\r\t\square\t+\tfrac12\nabla^\r(\nabla\t)^2+\nabla^\r\t(\nabla\t)^2\big)\,,
\eal
where we have used the identity
\be
P^{\r\m}{}_\m{}^{\s}=R^{\r\s}-\tfrac12g^{\r\s}R\,.
\ee
It follows that, provided the dilaton satisfies its equation of motion in \eqref{eq:4Deoms}, the WZ stress tensor satisfies the trace Ward identity  
\be
\<\ct^{\m}_{\rm WZ\,\m}\>=\ca_W\,.
\ee

In order to verify the diffeomorphism Ward identity, we first evaluate the divergence of each term in \eqref{eq:4DWZstresstensor} separately. For the first two terms in \eqref{eq:4DWZstresstensor} we easily find
\be
\nabla_\m\<\ct^{(\a)\m\n}_{\rm WZ}\>=-\a^2\nabla^\n\t\, e^{-\t}\big(\square-\tfrac{1}{6}R\big)e^{-\t}\,,
\ee
and
\be
\nabla_\m\<\ct^{(\b)\m\n}_{\rm WZ}\>=\b^2e^{-2\t}\hat A^\r\hat A_\r\nabla^\n\t+\b^2\nabla_\m(e^{-2\t}\hat A^\m)\hat A^\n-F^{\n\m}\b^2e^{-2\t}\hat A_\m\,.
\ee
The divergence of the fourth term can be brought to the form
\be
\nabla_\m\<\ct^{(k_W)\m\n}_{\rm WZ}\>=-\frac{k_W}{4\p^2}\big(-\tfrac14 \nabla^\n\t F^2+F^{\n\m}\nabla_\r(\t F^{\r}{}_\m)+\t F^{\s}{}_\r \nabla_\s F^{\n\r}-\tfrac12\t F^{\r\s}\nabla^\n F_{\r\s}\big)\,.
\ee
However, the Bianchi identity implies that the last two terms add to zero, namely
\be
\begin{split}
\t F^{\s}{}_\r \nabla_\s F^{\n\r}-\tfrac12\t F^{\r\s}\nabla^\n F_{\r\s}=&\;-\t F^{\s}{}_\r (\nabla^\r F_\s{}^{\n}+\nabla^\n F^\r{}_{\s})-\tfrac12\t F^{\r\s}\nabla^\n F_{\r\s}\\
=&\;-\t F_{\s\r} \nabla^\s F^{\n\r}+\tfrac12\t F^{\r\s}\nabla^\n F_{\r\s}\,.
\end{split}
\ee
Hence,
\be
\nabla_\m\<\ct^{(k_W)\m\n}_{\rm WZ}\>=-\frac{k_W}{4\p^2}\big(-\tfrac14 \nabla^\n\t F^2+F^{\n\m}\nabla_\r(\t F^{\r}{}_\m)\big)\,.
\ee

For the divergence of $\<\ct^{(k_R)\m\n}_{\rm WZ}\>$ we obtain
\be
\begin{split}
\nabla_\m\<\ct^{(k_R)\m\n}_{\rm WZ}\>=&\;\frac{k_R}{3\p^2}\nabla_\m\nabla_\r(\wt R^{\s(\m\n)\r}\nabla_\s\c)\\
=&\;\frac{k_R}{6\p^2}\nabla_\m\nabla_\r(\wt R^{\s\m\n\r}\nabla_\s\c)+\frac{k_R}{6\p^2}\nabla_\m\nabla_\r(\wt R^{\s\n\m\r}\nabla_\s\c)\\
=&\;\frac{k_R}{6\p^2}[\nabla_\m,\nabla_\r](\wt R^{\s\m\n\r}\nabla_\s\c)+\frac{k_R}{12\p^2}[\nabla_\m,\nabla_\r](\wt R^{\s\n\m\r}\nabla_\s\c)\\
=&\;\frac{k_R}{12\p^2}(2R^\n{}_{\l\m\r}+R^\n{}_{\m\l\r})\wt R^{\s\m\l\r}\nabla_\s\c\,,
\end{split}
\ee
where we used that $\nabla_\m(\wt R^{\s\m\n\r}\nabla_\s\c)=\nabla_\m\wt R^{\s\m\n\r}\nabla_\s\c=0$ due to the second Bianchi identity. Moreover, the first Bianchi identity allows us to write
\be
\begin{split}
2R^\n{}_{\l\m\r}+R^\n{}_{\m\l\r}=&\;2(1-\a) R^\n{}_{\l\m\r}-2\a(R^\n{}_{\r\l\m}+R^\n{}_{\m\r\l})+R^\n{}_{\m\l\r}\\
=&\;2(1-2\a) R^\n{}_{\l\m\r}+(1+2\a)R^\n{}_{\m\l\r}\,,
\end{split}
\ee
where $\a$ in an arbitrary parameter. Therefore, setting $\a=1/2$ we conclude that
\be
\nabla_\m\<\ct^{(k_R)\m\n}_{\rm WZ}\>=\frac{k_R}{6\p^2}R^\n{}_{\m\l\r}\wt R^{\s\m\l\r}\nabla_\s\c=\frac{k_R}{24\p^2}R_{\s\m\l\r}\wt R^{\s\m\l\r}\nabla^\n\c\,,
\ee
where in the second equality we used the identity $\wt R^{\r\s}{}_{\m\n}R_{\l\s}{}^{\m\n}\nabla_\r\c=\tfrac14\wt R^{\r\s}{}_{\m\n}R_{\r\s}{}^{\m\n}\nabla_\l\c$ \cite{Papadimitriou:2019gel}. 

Finally, with some algebra and repeated use of the Bianchi identities we obtain
\be
\begin{split}
\nabla_\m\<\ct^{(c)\m\n}_{\rm WZ}\>=&\;\frac{c}{4\p^2}W^{\m\k\r\s}W^\n{}_{\k\r\s}\nabla_\m\t\,,\\
\nabla_\m\<\ct^{(a)\m\n}_{\rm WZ}\>=&\;\frac{a}{2\p^2}\big[-\tfrac12 P^{\m\r\s\n}R_{\m\r\s\k}\nabla^\k\t\\
&+\nabla_\r\big((R^{\r\s}-\tfrac12 R g^{\r\s})\nabla_\s\t+(\nabla\t)^2\nabla^\r\t-\square\t\nabla^\r\t+\tfrac12\nabla^\r(\nabla\t)^2\big)\nabla^\n\t\big]\,.
\end{split}
\ee
These can be brought to the desired form using the identities 
\be\label{eq:5antisymmetrization}
W^{[\m\n}{}_{\r\s}W^{\k\l}{}_{\a\b}\d^{\g]}_{\d}=0\,,\qquad P^{[\m\n}{}_{\r\s}R^{\k\l}{}_{\a\b}\d^{\g]}_{\d}=0\,,
\ee
which in turn imply that
\be
W^{\m\n\r\s}W^{\k}{}_{\n\r\s}=\tfrac14 W^2 g^{\m\k}\,,\qquad P^{\m\n\r\s}R^{\k}{}_{\n\r\s}=\tfrac14 P^{\l\n\r\s}R_{\l\n\r\s} g^{\m\k}=\tfrac14 E g^{\m\k}\,.
\ee
Hence, the divergence of the last two terms in \eqref{eq:4DWZstresstensor} becomes
\bal
\nabla_\m\<\ct^{(c)\m\n}_{\rm WZ}\>=&\;\frac{c}{16\p^2}W^2\nabla^\n\t\,,\\
\nabla_\m\<\ct^{(a)\m\n}_{\rm WZ}\>=&\;\frac{a}{2\p^2}\big[-\tfrac{1}{8}E+\nabla_\r\big((R^{\r\s}-\tfrac12 R g^{\r\s})\nabla_\s\t+(\nabla\t)^2\nabla^\r\t-\square\t\nabla^\r\t+\tfrac12\nabla^\r(\nabla\t)^2\big)\big]\nabla^\n\t\,.\NO
\eal

Collecting the above results for the divergence of all terms in \eqref{eq:4DWZstresstensor} and using the identity $\wt F^{\r\s}F_{\l\s}\nabla_\r\c=\tfrac14\wt F^{\r\s}F_{\r\s}\nabla_\l\c$ \cite{Papadimitriou:2019gel}, one may easily verify that the diffeomorphism Ward identity in \eqref{CFT-WardIDs} holds, namely
\be
\nabla_\m\<\ct_{\rm WZ}^\m{}_\n\>-F_{\n\m}\<\cj_{\rm WZ}^\m\>+A_\n\nabla_\m\<\cj_{\rm WZ}^\m\>=0\,.
\ee 

Let us conclude this section with a few remarks on the scheme dependence of the above results. The expressions \eqref{eq:4DWZU(1)currrent} and \eqref{eq:4DWZstresstensor} for the WZ currents correspond to the scheme where the coefficient of the local counterterm $R^2$, namely $b$, as well as of the potential counterterms proportional to $W_{\m\n\r\s}W^{\m\n\r\s}$ and $F_{\m\n}F^{\m\n}$, are all set to zero. However, turning on the latter two counterterms is trivially incorporated in the above results, since they simply correspond to a constant shift of the dilaton, i.e. $\t\to\t+\t_0$ for some constant $\t_0$, in the WZ action \eqref{eq:4DWZ}. This shift would also add a counterterm proportional to the Euler density $E$, but this does not contribute to the WZ currents since it is topological. Similarly, a shift of the axion by a constant does not modify the currents since it is adding topological terms to the effective action. The contribution of the counterterm $bR^2$ is less trivial, but can be obtained straightforwardly by differentiating the counterterm itself, as well as the last line of the WZ action \eqref{eq:4DWZ}, with respect to $g_{\m\n}$.  

\section{ $\cn=1$ superconformal field theories }
\label{sec:conf_sugra}

We are now ready to setup the problem we are primarily interested in, namely the construction of the WZ effective action that captures the superconformal anomalies of generic $\cn=1$ SCFTs in four dimensions. We begin in this section with a brief review of the $\cn=1$ conformal current multiplet, the corresponding local symmetry algebra, as well as the bosonic and fermionic superconformal anomalies that follow from the WZ consistency conditions. The discussion in this section follows closely \cite{Papadimitriou:2019gel}, but we extend the results of that reference and correct the form of the Q-supersymmetry anomaly in the case $a\neq c$. 

\subsection{$\cn=1$ conformal current multiplet and symmetry algebra} 

Four-dimensional $\cn=1$ SCFTs can be consistently coupled to background off-shell $\cn=1$ conformal supergravity \cite{Festuccia:2011ws}, whose field content consists of the vierbein $e_\m{}^a$, the graviphoton $A_\m$, and a Majorana gravitino $\j_{\m}$, comprising 5+3 bosonic and 8 fermionic off-shell degrees of freedom. These fields act as sources for the conformal current multiplet. In particular, if 
\be\label{sugraW}
\cw[e,A,\psi]\equiv -i\log\cz[e,A,\psi]\,,
\ee
denotes the (renormalized) partition function of the theory on an arbitrary conformal supergravity background, the conformal current multiplet operators are defined via the relations
\be\label{consistent-currents}
\<\ct^\m{}_a\>=e^{-1}\frac{\d\mathcal W}{\d e_\m{}^a}\,,\qquad \<\cj^\m\>=e^{-1}\frac{\d\mathcal W}{\d A_\m}\,,\qquad \<\cs^\m\>=e^{-1}\frac{\d\mathcal W}{\d \lbar\psi_\m}\,,
\ee
where $e\equiv\det(e_\m{}^a)$ and the notation $\<\cdot\>$ stands for a renormalized correlation function on an arbitrary supergravity background.

As is summarized in table~\ref{gauging}, off-shell $\cn=1$ conformal supergravity gauges the full $\cn=1$ superconformal group SU(2,2$|$1) \cite{Kaku:1977pa,Kaku:1977rk,Kaku:1978nz,Townsend:1979ki} (excellent reviews include \cite{VanNieuwenhuizen:1981ae,deWit:1981vgr,deWit:1983qkc,Fradkin:1985am,Freedman:2012zz}). The fields $e_\m{}^a$, $A_\m$ and $\j_{\m}$ correspond respectively to gauge connections for translations, R-symmetry and Q-supersymmetry transformations. The gauge connections for the remaining symmetries are either gauge-fixed to zero or can be expressed as composites of $e_\m{}^a$, $A_\m$ and $\j_{\m}$ through curvature constraints. In particular, the connection for local dilatations, $b_\m$, can be set to zero by gauge-fixing local special conformal transformations, while the connections for local Lorentz, special conformal and S-supersymmetry transformations, respectively $\o_\m{}^{ab}$, $f_\m{}^a$ and $\f_\m$, become composites once suitable curvature constraints are imposed. The resulting expressions for these composite connections are
\bbxd
\be\label{composites}
\begin{split}
\o_\m{}^{ab}=&\;\o_\m{}^{ab}(e)+\tfrac14(\bar\j^a\g_\m\j^b+\bar\j_\m\g^a\j^b-\bar\j_\m\g^b\j^a)\,,\\
\f_\m=&\; -\tfrac16(4\d^{[\r}_\m\d^{\s]}_\n+i\g^5 \e_{\m\n}{}^{\r\s})\g^\n \cd_\r\j_\s\,,\\
f_\m{}^a=&\;-\tfrac14\big(\Hat{\mathscr{R}}_{\m}{}^{a}-\tfrac16 \Hat{\mathscr{R}} e_\m{}^a\big)-\tfrac{1}{8}(1-k)\bar\j^\n\g_\m \mathscr{R}_{\n}{}^a(Q)-\tfrac{1}{6}\wt{\mathscr{R}}_\m{}^a(T)\,,
\end{split}
\ee
\ebxd
where $\o_\m{}^{ab}(e)$ denotes the torsion-free metric-compatible spin connection given in appendix~\ref{appendix:torsion}, $\wt{\mathscr{R}}_\m{}^a(T)\equiv \tfrac12\e_\m{}^{a\r\s} \mathscr{R}_{\r\s}(T)$, while $\Hat{\mathscr{R}}_{\m\n}{}^{ab}$, $\mathscr{R}_{\m\n}(Q)$ and $\mathscr{R}_{\m\n}(T)$ are curvatures for Lorentz, Q-supersymmetry and R-symmetry transformations respectively, given by   
\bbxd
\be\label{curvatures}
\begin{split}
\Hat {\mathscr{R}}_{\m\n}{}^{ab}=&\;R_{\m\n}{}^{ab}(\o)-\bar\j_{[\m}\g^{ab}\f_{\n]}+k\big(\bar\j_{[\m}\g^{[a}\mathscr{R}_{\n]}{}^{b]}(Q)+\tfrac12\bar\j_{[\m}\g_{\n]}\mathscr{R}^{ab}(Q)\big)\,,\\
\mathscr{R}_{\m\n}(Q)=&\;2\cd_{[\m}\j_{\n]}-2\g_{[\m}\f_{\n]}\,,\\
\mathscr{R}_{\m\n}(T)=\,& F_{\m\n}-\tfrac{3i}{2}\bar\j_{[\m}\g^5\f_{\n]}\,.
\end{split}
\ee
\ebxd
Here, $R_{\m\n}{}^{ab}(\o)=2\pa_{[\m}\o_{\n]}{}^{ab}+2\o_{[\m}{}^{ac}\o_{\n]c}{}^b$
is the Riemann curvature. Notice that we have introduced the parameter $k$ in the expressions for $f_\m{}^a$ and $\Hat{\mathscr{R}}_{\m\n}{}^{ab}$ in order to account for different conventions in the literature. In particular, $\Hat{\mathscr{R}}_{\m\n}{}^{ab}$ in eq.~(16.23) and $f_\m{}^a$ in eq.~(16.24) of \cite{Freedman:2012zz} correspond to the choice $k=1$, while the expressions given e.g. in \cite{Kaku:1978nz, Fradkin:1985am} correspond to $k=0$. In fact, once the expression for $\Hat{\mathscr{R}}_{\m\n}{}^{ab}$ in \eqref{curvatures} is substituted back in that for $f_\m{}^a$ in \eqref{composites}, the dependence on the parameter $k$ drops out, which ensures that the expression for $f_\m{}^a$ is convention independent.\footnote{The last term proportional to $\wt{\mathscr{R}}_\m{}^a(T)$ in \eqref{composites} is missing from the expression for $f_\m{}^a$ in eq.~(16.24) of \cite{Freedman:2012zz}, but this has been corrected in the Errata at \href{https://fys.kuleuven.be/itf/supergravity}{https://fys.kuleuven.be/itf/supergravity}.}  
\begin{table}
\bebx\NO
\begin{tabular}{c|ccccccc}
$T_A$ & $P_a$ & $M_{[ab]}$ & $D$ & $K_a$ &   $U(1)_R$ & $Q_\a$ &   $S_\a$\vspace{.1cm}\\
\hline\hline\rule{.0cm}{.5cm} 
$\vth^A$ & $\x^a$ & $\l^{ab}$ &  $\s$ & $\l_K^a$ &  $\th$ & $\ve^\a$ & $\h^\a$ \vspace{.1cm}\\  
\hline\rule{.0cm}{.5cm} 
$\mathscr{A}^A_\m$ & $ e_\m{}^a$ & $\o_\m{}^{ab}$ & {$b_\m$}  & {$f_\m{}^a$ }  & $A_\m$ & $\j^\a_\m$  &  {$\phi^\a_\m$} \vspace{.1cm}\\   
\hline\hline\rule{.0cm}{.5cm} 
$\ca_A$   &  $-$ & $-$ & $\ca_W$ & $-$ & $ \ca_R$ & $\ca_Q$ &$\ca_S$
\end{tabular}
\eebx
\caption{List of the $\frak su$(2,2$|$1) generators $T_A$, the corresponding gauge parameters $\vth^A$, gauge connections $\mathscr{A}^A_\m$ and superconformal anomalies $\ca_A$. The connections $\o_\m{}^{ab}$, $f_\m{}^a$ and $\f_\m^\a$ are expressed as composites of the independent fields $e_\m{}^a$, $A_\m$ and $\j_{\m}$ by imposing suitable curvature constraints, while $b_\m$ is set to zero by gauge-fixing local the special conformal symmetry.}
\label{gauging}
\end{table}

There are two distinct covariant derivatives that play a prominent role in our analysis. They are R-charge extensions respectively of the Lorentz covariant derivative, $D_\m$, and of the covariant derivative $\nabla_\m$ obtained from a torsionfull affine connection $\G^\r_{\m\n}$, namely
\be\label{covD}
\cd_\m\equiv D_\m+\text{R-charge}\,,\qquad  \dc{D}_\m\equiv \nabla_\m+\text{R-charge}\,.
\ee
For example, the fully covariant derivative of the gravitino is given by 
\be
\dc D_\m\j_\n=\big(\pa_\m+\tfrac14\o_\m{}^{ab}(e,\j)\g_{ab}+i\g^5A_\m\big)\j_\n-\G^\r_{\m\n}\j_\r=\cd_\m\j_\n-\G^\r_{\m\n}\j_\r=(\nabla_\m+i\g^5A_\m)\j_\n\,.
\ee
It should be emphasized that, even though $\cd_\m\j_\n$ is not covariant with respect to diffeomorphisms, the quantity $\cd_{[\m}\j_{\n]}$ with antisymmetrized indices that appears in the expressions in \eqref{composites} and \eqref{curvatures}, does transform tensorially because it differs from $\dc D_{[\m}\j_{\n]}$ by torsion terms. The relation between torsion and the affine connection, together with a number of further basic facts about torsionful geometry, are collected in appendix~\ref{appendix:torsion}.

Once local special conformal transformations are gauge-fixed, the remaining local symmetries of $\cn=1$ conformal supergravity are diffeomorphisms, with local parameter $\x^\m(x)$, Weyl transformations, $\s(x)$, local frame rotations, $\l^{ab}(x)$, $U(1)$ $R$-symmetry transformations, $\th(x)$, as well as Q- and S-supersymmetry transformations, parameterized respectively by the local Majorana spinors $\ve(x)$ and $\h(x)$. Under these symmetries, collectively denoted by $\vth=(\x,\l,\th,\ve,\h)$, the supergravity multiplet fields transform as
\bbxd
\be\label{sugra-transformations}
\begin{split}
\d(\vth) e_\m{}^a=&\;\x^\l\pa_\l e_\m{}^a+e_\l{}^a\pa_\m\x^\l-\l^a{}_b e_\m{}^b+\s e_\m{}^a-\tfrac12\lbar\j_\m\g^a\ve\,,\\
\d(\vth)\j_\m=&\;\x^\l\pa_\l\j_\m+\j_\l\pa_\m\x^\l-\tfrac14\l_{ab}\g^{ab}\j_\m+\tfrac12\s\j_\m - i\th\g^5\j_\m+\dc D_\m\ve-\g_\m\h\,,\\
\d(\vth)A_\m=&\;\x^\l\pa_\l A_\m+A_\l\pa_\m\x^\l+\pa_\m\th+\tfrac{3i}{4}\lbar\f_\m\g^5\ve-\tfrac{3i}{4}\lbar\j_\m\g^5\h\,,
\end{split}
\ee
\ebxd
where the covariant derivative of the spinor parameter $\ve$ is given by 
\be
\dc D_\m\ve=\big(\pa_\m+\tfrac14\o_\m{}^{ab}(e,\j)\g_{ab}+i\g^5A_\m\big)\ve=(\nabla_\m+i\g^5A_\m)\ve\,.
\ee

Before imposing any curvature constraints and gauge-fixing the special conformal transformations, the connections $\o_\m{}^{ab}$, $f_\m{}^a$ and $\f_\m$, are also independent fields and possess analogous transformations. However, imposing curvature constraints and gauge-fixing results in specific modifications of their transformations. Firstly, the constraints transform non-trivially under the local symmetries, so to maintain the constraints the transformations of the composite fields $\o_\m{}^{ab}$, $f_\m{}^a$ and $\f_\m$ must be modified. Moreover, in order to preserve the gauge-fixing condition $b_\m=0$, one must perform a compensating special conformal transformation. In contrast to $e_\m{}^a$, $A_\m$ and $\j_\m$, the connections $\o_\m{}^{ab}$, $f_\m{}^a$, $\f_\m$ transform non-trivially under special conformal transformations, and so the compensating special conformal transformation results in a further modification of their transformation under the remaining symmetries. This completely analogous to the analysis of the conformal algebra in subsection \ref{sec:GaugingConformal}. For the superconformal algebra it is explained in detail in e.g. \cite{Fradkin:1985am,Freedman:2012zz}. The resulting transformations under the remaining local symmetries coincide with those obtained by transforming the expressions \eqref{composites} for the composite connections using the transformations \eqref{sugra-transformations} of the unconstrained fields. The result is 
\bbxd
\be\label{sugra-transformations-composites}
\begin{split}
\rule{0.0cm}{.5cm}\d(\vth)\o_\m{}^{a}{}_b=&\;\x^\l\pa_\l\o_\m{}^a{}_b+\o_\l{}^a{}_b\pa_\m\x^\l+\cd_\m\l^a{}_b\underline{+2e_\m^{[a}e^\n_{b]}\pa_\n\s}\\
&+\tfrac12\bar\ve\g^a{}_b\f_\m-\tfrac12\bar\ve\g_\m \mathscr{R}^a{}_b(Q)\underline{-e_\m^{[a}e^\n_{b]}\bar\ve\f_\n}\underline{+e_\m^{[a}e^\n_{b]}\bar\h\j_\n}+\tfrac12\bar\h\g^a{}_b\j_\m\,,\\
\rule{0.0cm}{.5cm}\d(\vth)\f_\m=&\;\x^\l\pa_\l\f_\m+\f_\l\pa_\m\x^\l-\tfrac14\l_{ab}\g^{ab}\f_\m-\tfrac12\s\f_\m \underline{+\tfrac12\g^\n\j_\m\pa_\n\s}+ i\g^5\th\f_\m\\
&\hspace{-.7cm}-\big(f_\m{}^a-\tfrac{i}{3}\mathscr{R}_\m{}^a(T)\g^5+\tfrac13\wt{\mathscr{R}}_\m{}^a(T)\big)\g_a\ve\underline{-\tfrac14(\bar\f_\n\ve)\g^\n\j_\m}+\dc D_\m\h\underline{+\tfrac14(\bar\j_\n\h)\g^\n\psi_\m}\,,\\
\rule{0.0cm}{.5cm}\d(\vth)f_\m{}^a=&\;\x^\l\pa_\l f_\m{}^a+f_\l{}^a\pa_\m\x^\l-\l^a{}_b f_\m{}^b-\s f_\m{}^a\underline{+\tfrac12\cd_\m(\pa^a\s)}+\tfrac12\bar\h\g^a\phi_\m\underline{+\tfrac14\cd_\m(\bar\h\j^a)}\\
&\hspace{-.7cm}\underline{-\tfrac14\cd_\m(\bar\ve\f^a)}-\tfrac{i}{4}\bar\ve \g^5\wt{\mathscr{R}}_\m{}^a(S)-\tfrac14\bar\ve\g^{a\n} \mathscr{R}_{\m\n}(S)+\tfrac{1}{16}e_\m{}^a\bar\ve\g^{\n\r} \mathscr{R}_{\n\r}(S)\\
&\hspace{-.7cm}+\tfrac{1}{6}\bar\ve\big(\wt{\mathscr{R}}^{a\n}(T)-i\g^5\mathscr{R}^{a\n}(T)\big)\g^\m\j_\n -\tfrac{1}{24}e_\m{}^a \bar\ve\big(\wt{\mathscr{R}}^{\n\r}(T)-i\g^5\mathscr{R}^{\n\r}(T)\big)\g_\n\j_\r \,,
\end{split}
\ee
\ebxd
where the curvature of $S$-supersymmetry transformations is given by 
\be
\mathscr{R}_{\m\n}(S)=2\cd_{[\m}\f_{\n]}-2\g_af_{[\m}{}^a\j_{\n]}\,,
\ee
and the covariant derivative on the spinor parameter $\h$ takes the form
\be
\dc D_\m\h=\big(\pa_\m+\tfrac14\o_\m{}^{ab}(e,\j)\g_{ab}-i\g^5A_\m\big)\h=(\nabla_\m-i\g^5A_\m)\h\,.
\ee
The underlined terms in \eqref{sugra-transformations-composites} are generated by the compensating special conformal transformation required to preserve the gauge-fixing condition $b_\m=0$.

The transformations \eqref{sugra-transformations} and \eqref{sugra-transformations-composites} determine the local symmetry algebra. This algebra assumes a somewhat different form depending on whether the local parameters are considered inert under the local symmetry transformations, or they transform as they would do once promoted to ghost fields. Both these conventions result in a consistent algebra that closes off-shell, but we find it preferable here to treat the parameters as ghosts -- albeit without changing their Grassmann parity -- and assign them the following local transformations   
\bbxd
\be\label{parameter-transformations}
\begin{split}
&\d(\vth)\x^\m=\x'^\n\pa_\n\x^\m-\x^\n\pa_\n\x'^\m\,,\quad
\d(\vth)\l^a{}_b=\x^\m\pa_\m\l^a{}_b-[\l',\l]^a{}_b\,,\\ &\d(\vth)\s=\x^\m\pa_\m\s\,,\quad
\d(\vth)\th=\x^\m\pa_\m\th\,,\\
&\d(\vth)\ve=\x^\m\pa_\m\ve-\tfrac14\l_{ab}\g^{ab}\ve+\tfrac12\s\ve-i\th\g^5\ve\,,\\&\d(\vth)\h=\x^\m\pa_\m\h-\tfrac14\l_{ab}\g^{ab}\h-\tfrac12\s\h+i\th\g^5\h\,.
\end{split}
\ee
\ebxd
This convention was also adopted in \cite{Papadimitriou:2019gel} and it is particularly convenient for the construction of the WZ effective action, where some of the local parameters are related to compensators.  

With the parameter transformations \eqref{parameter-transformations}, the local symmetry algebra obtained from the transformations \eqref{sugra-transformations} and \eqref{sugra-transformations-composites} closes with the only nonzero commutators given by
\bbxd
\be\label{local_algebra}
\begin{split}
&[\d_D(\x_1),\d_D(\x_2)]=\d_D(\x_{DD})\,,\quad [\d_L(\l_1),\d_L(\l_2)]=\d_L(\l_{LL})\,,\\
&[\d_Q(\ve),\d_W(\s)]=\d_S(\h_{QW})\,,\\ &[\d_Q(\ve),\d_S(\h)]=\d_W(\s_{QS})+\d_R(\th_{QS})+\d_S(\h_{QS})+\d_L(\l_{QS})\,,\\
& [\d_Q(\ve_1),\d_Q(\ve_2)]=\d_D(\x_{QQ})+\d_L(\l_{QQ})+\d_Q(\ve_{QQ})+\d_R(\th_{QQ})+\d_S(\h_{QQ})\,,
\end{split}
\ee
\ebxd
where the composite parameters in these expressions are 
\bbxd
\be\label{composite-parameters}
\begin{split}
&\x_{DD}^\m=\x_1^\n\pa_\n\x^\m_2-\x_2^\n\pa_\n\x^\m_1\,,\quad \l_{LL}{}^a{}_b=[\l_2,\l_1]^a{}_b\,, \quad \h_{QW}=\tfrac12\g^\m\ve\,\pa_\m\s\,,\\
&\s_{QS}=\tfrac12\lbar\ve\h\,,\quad\l_{QS}{}^a{}_b=-\tfrac12\lbar\ve\g^a{}_b\h\,,\quad \th_{QS}=-\tfrac{3i}{4}\lbar\ve\g^5\h\,,\quad \h_{QS}=\tfrac14(\lbar\h\j_\m)\g^\m\ve\,,\\
&\x_{QQ}^\m=\tfrac12\lbar\ve_2\g^\m\ve_1\,,\quad \l_{QQ}^a{}_b=-\x_{QQ}^\m\,\o_\m{}^a{}_b\,,\quad \th_{QQ}=-\x_{QQ}^\m\,A_\m\,,\\
& \ve_{QQ}=-\x_{QQ}^\m\,\j_\m\,,\quad \h_{QQ}=-\x_{QQ}^\m\f_\m+\tfrac14((\lbar\f_\m\ve_1)\g^\m\ve_2-(\lbar\f_\m\ve_2)\g^\m\ve_1)\,.
\end{split}
\ee
\ebxd
The local algebra \eqref{local_algebra} is exact in the gravitino and coincides with that obtained in \cite{Papadimitriou:2019gel} to leading order in the gravitino, except for the $[\d_Q(\ve),\d_W(\s)]$ commutator that was incorrectly set to zero there.

\subsection{Superconformal anomalies}

Let us now consider the transformation of the generating function \eqref{sugraW} under the symmetries of conformal supergravity. As in our earlier discussion of 4d CFTs in subsection \ref{sec:4dCFT}, we work in a scheme where the mixed gravitational-U(1)$_R$ anomaly can be placed entirely in the R-symmetry transformations. With this choice of scheme, the general symmetry transformation of the generating function takes the form   
\be\label{W-trans}
\d(\vth)\cw[e,A,\psi]=\int d^4x\, e\,\big(\s\mathcal A_W-\theta\mathcal A_R-\bar\ve\mathcal A_Q+\bar\h\mathcal A_S\big)\,,
\ee
where the anomalies $\ca_W$, $\ca_R$, $\ca_Q$ and $\ca_S$ are local functionals of the independent supergravity fields $e_\m{}^a$, $A_\m$ and $\j_\m$. The choice of signs in \eqref{W-trans} is a convention for later convenience.

The symmetry transformation of the generating function can be evaluated alternatively using the chain rule, namely
\be\label{W-trans-chain}
\d(\vth)\cw[e,A,\psi]=\int d^4x\, e\,\Big(\d(\vth)e_\m{}^a\frac{\d\cw}{\d e_\m{}^a}+\d(\vth)A_\m\frac{\d\cw}{\d A_\m}+\d(\vth)\bar\j_\m\frac{\d\cw}{\d \bar\j_\m}\Big)\,.
\ee
To derive the anomalous superconformal Ward identities we simply need to substitute the transformations \eqref{sugra-transformations} of the supergravity fields in this expression, use the definition of the conformal multiplet currents in \eqref{consistent-currents}, and equate the result with \eqref{W-trans}. However, in order to obtain the Ward identities in a manifestly diffeomorphism covariant form, we write the general coordinate transformations of the supergravity fields in \eqref{sugra-transformations} in the covariant form
\be
\begin{split}
\d_D(\x) e_\m{}^a=&\;\nabla_\m(\x^\n e_\n{}^a)-\x^\n\o_\n{}^a{}_b e_\m{}^b+\x^\n T_{\n\m}{}^\r e_\r{}^a\,,\\
\d_D(\x)A_\m=&\;\x^\n F_{\n\m}+\nabla_\m(\x^\n A_\n)\,,\\
\d_D(\x) \j_\m=&\;2\x^\n\dc D_{[\n}\j_{\m]}+\dc D_\m(\x^\n\j_\n)-\tfrac14\x^\n\o_\n{}^{ab}\g_{ab}\j_\m-i\x^\n A_\n\g^5\j_\m+\x^\n T_{\n\m}{}^\r\j_\r\,,
\end{split}
\ee
where the torsion is given by
\be\label{eq:torsion}
T_{\m\n}{}^\r=\frac12\lbar\j_\m\g^\r\j_\n\,.
\ee
Using these, and taking into account the torsion terms when discarding total derivative terms (see appendix~\ref{appendix:torsion}), we obtain the Ward identities \cite{Papadimitriou:2019gel}\footnote{The torsion terms were dropped in the Ward identities derived in \cite{Papadimitriou:2019gel}. In contrast, the Ward identities in \eqref{WardIDs} are valid to all orders in the gravitino.}
\bbxd
\be\label{WardIDs}
\begin{split}
&\nabla_\n(e_\m{}^a\<\ct^\n{}_a\>+\lbar\j_\m\<\cs^\n\>)-(T_{\r\n}{}^\r \d^\s_\m+T_{\m\n}{}^\r \d_\r^\s)(e_\s{}^a\<\ct^\n{}_a\>+\lbar\j_\s\<\cs^\n\>)\\
&\hspace{1.cm}-\lbar\j_\n\overleftarrow{\dc D}_\m \<\cs^\n\>-F_{\m\n}\<\cj^\n\>+A_\m\big(\nabla_\n\<\cj^\n\>-T_{\r\n}{}^\r\<\cj^\n\>+i\lbar\j_\n \g^5\<\cs^\n\>\big)\\
&\hspace{1.cm}-\o_\m{}^{ab}\big(e_{\n[a}\<\ct^\n{}_{b]}\>+\tfrac14\lbar\j_\n\g_{ab}\<\cs^\n\>\big)=0\,,\\
\rule{0.0cm}{.5cm}&e_{\m[a} \<\ct^\m{}_{b]}\>+\tfrac14\lbar\j_\m\g_{ab} \<\cs^\m\>=0\,,\\
\rule{0.0cm}{.5cm}&e_\m{}^a\<\ct^\m{}_a\>+\tfrac12\lbar\j_\m \<\cs^\m\>=\ca_W\,,\\
\rule{0.0cm}{.5cm}&\nabla_\m \<\cj^\m\>-T_{\n\m}{}^\n\<\cj^\m\>+i\lbar\j_\m\g^5 \<\cs^\m\>=\ca_R\,,\\
\rule{0.0cm}{.5cm}&\dc D_\m \<\cs^\m\>-T_{\n\m}{}^\n\<\cs^\m\>-\tfrac12\g^a\j_\m \<\ct^\m{}_a\>-\tfrac{3i}{4}\g^5\f_\m\<\cj^\m\>=\ca_Q\,,\\
\rule{0.0cm}{.5cm}&\g_\m \<\cs^\m\>-\tfrac{3i}{4}\g^5\j_\m\<\cj^\m\>=\ca_{S}\,.
\end{split}
\ee
\ebxd

The anomalies $\ca_W$, $\ca_R$, $\ca_Q$, $\ca_S$ satisfy the WZ consistency conditions for the local symmetry algebra \eqref{local_algebra}. We solved these conditions to first subleading order in the gravitino using the Mathematica tensor algebra package FieldsX \cite{Frob:2020gdh}. We find two independent nontrivial anomaly cocycles, controlled respectively by the anomaly coefficients $a$ and $c$, corresponding to supersymmetric completions of the two conformal anomaly cocycles of non-supersymmetric CFTs. The anomaly coefficients $a$ and $c$ depend on the specific SCFT and are normalized such that for $N_\c$ free chiral and $N_v$ free vector multiplets \cite{Anselmi:1997am}
\be
\label{acNchiNv}  
a=\frac{1}{48}(N_\c+9N_v)\,,\qquad c=\frac{1}{24}(N_\c+3N_v)\,.
\ee
The general solution we obtain is parameterized in addition by two arbitrary constants, $s_1$ and $s_2$, corresponding to the fermion bilinear local counterterm
\bbxd
\be\label{counterterm}
\begin{split}
\cw^{f}_{\rm ct}[e,A,\j|\substack{\blue{s_1}\\ \red{s_2}}]=&\;-\tfrac{(a-c)}{72\p^2}\int d^4x\,e\Big(\blue{s_1}\big(12\wt F^{\m\n}\bar\j_\m\f_\n+\bar\j^\r(\wt F^{\m\n}+iF^{\m\n}\g^5)\g_\r\cd_\m\j_\n\\
&\hspace{-2.3cm}+\bar\j^\r(\wt F^{\m\n}+iF^{\m\n}\g^5)\g_\m\cd_{[\r}\j_{\n]}+\bar\j_\m(\wt F^{\m\n}+iF^{\m\n}\g^5)\g^\r\cd_{[\n}\j_{\r]}+\co(\j^4)\big)\\
&\hspace{-2.3cm}+\red{s_2}\,\e^{\m\n\r\s}A_\m\big(12\bar\f_\n(\g_\r\f_\s-2\cd_\r\j_\s)+i F_{\r\s}\bar\j^\k\g^5(2\g_\k\j_\n+\g_\n\j_\k)+\co(\j^4)\big)\Big)\,.
\end{split}
\ee
\ebxd

The complete expressions for the anomalies $\ca_W$, $\ca_R$, $\ca_Q$ and $\ca_S$ up to first subleading order in the gravitino are given respectively in eqs.~\eqref{W-anomaly}, \eqref{R-anomaly}, \eqref{S-anomaly} and \eqref{Q-anomaly}.
\bbxd
\begin{subequations}
\bal
\ca_W=&\;-\tfrac{a}{16\p^2}\Big[E+\nabla_\m\big(8i\e^{\m\n\r\s}\bar\f_\n\g^5\cd_\r\psi_\s-8\bar\f^\m\g^\n\f_\n\NO\\
&\hskip1.9cm+\tfrac{4\blue{s_1}}{3}\wt F^{\r\s}\bar\j_\r\g^\m\j_\s+\tfrac{8(1-\red{s_2})}{3}\e^{\k\l\r\s}\bar\j_\k\g^\m\cd_\l\j_\r A_\s\big)\Big]\NO\\
&+\tfrac{c}{16\p^2}\Big[W^2-\tfrac83F^2+8i\e^{\m\n\r\s}\bar\f_\m\g^5\g_\n\cd_\r\f_\s+8i\e^{\m\n\r\s}\nabla_\m(\bar\f_\n\g^5\cd_\r\j_\s)\NO\\
&-2R^{\m\n}\big(\bar\j^\r\g_{\r\n}\f_\m-\bar\j_\m\g_{\r\n}\f^\r-\tfrac23g_{\m\n}\bar\j^\r\g_{\r\s}\f^\s+2\bar\j^\r\g_\n(\cd_{[\m}\j_{\r]}-\g_{[\m}\f_{\r]})\big)\NO\\
&+8\wt F^{\m\n}\bar\j_\m\f_\n+\tfrac{16i}{3}F^{\m\n}\bar\j_\m\g^\r\g^5\cd_{[\n}\j_{\r]}-\tfrac43\wt F^{\m\n}\bar\j^\r(\g_\r\cd_\m\j_\n+2\g_\m\cd_{[\n}\j_{\r]})\NO\\
&+\tfrac43\nabla_\m\big(\blue{s_1}\wt F^{\r\s}\bar\j_\r\g^\m\j_\s+2(1-\red{s_2})\e^{\k\l\r\s}\bar\j_\k\g^\m\cd_\l\j_\r A_\s\big)\Big]+\co(\j^4)\label{W-anomaly}\,,\\
\rule{0.0cm}{.5cm}\ca_R=&\;\tfrac{5a-3c}{27\p^2}\;\wt F_{\m\n} F^{\m\n}+\tfrac{(c-a)}{24\p^2}\Big[\cp-\tfrac{\red{s_2}}{3}\e^{\m\n\r\s}\nabla_\m\big(12\bar\f_\n(\g_\r\f_\s-2\cd_{\r}\j_\s)\NO\\
&\hskip4.3cm+i F_{\r\s}\bar\j_\k\g^5(2\g^\k\j_\n+\g_\n\j^\k)\big)\Big]+\co(\j^4)\label{R-anomaly}\,,\\
\rule{0.0cm}{.5cm}\ca_{S}=&\;\tfrac{(5+2(\blue{s_1}-\red{s_2}))a-(3+2(\blue{s_1}-\red{s_2}))c}{6\p^2}\wt F^{\m\n}\cd_\m\j_{\n}-\tfrac{i(5a-3c)}{9\p^2}\wt F^{\m\n}A_\m\g^5\j_{\n}\NO\\
&-\tfrac{ic}{6\p^2} F^{\m\n}\g^{\s}\g_{\m}\g^5\cd_{[\n}\j_{\s]}+\tfrac{3a}{4\p^2}P_{\m\n}g^{\m[\n}\g^{\r\s]}\cd_\r\j_\s\NO\\
&+\tfrac{(a-c)}{8\p^2}\big(W^{\m\n\r\s}\g_{\m\n}\cd_\r\j_\s-\tfrac{2(1-\red{s_2})}{3}\wt R^{\m\n}{}_{\r\s}A_\m\g^{\r\s}\j_\n\big)+\co(\j^3)\label{S-anomaly}\,,\\
\rule{0.0cm}{.5cm}\ca_Q=&\;-\tfrac{(5a-3c)i}{9\p^2}\wt F^{\m\n}A_\m\g^5\f_\n-\tfrac{(a-c)}{12\p^2}\Big[\wt R^{\m\n\r\s}A_\m(2\g_\r\cd_{[\n}\j_{\s]}+\g_\n\cd_\r\j_\s)\NO\\
&+\red{s_2}\wt W^{\m\n\r\s}A_\m\g_{\r\s}\f_\n-2\red{s_2}\e^{\m\n\r\s}P_{\m\k}A_\n\g^\k\cd_\r\j_\s\NO\\
&+\tfrac{4\red{s_2}}{3}\g^{\m\s}{}_{\k}\cd_{[\r}\j_{\s]}\big(\d^\r_{(\m}F_{\n)}{}^\k-\tfrac52g_{\m\n}F^{\r\k}\big)A^\n+\tfrac{4\red{s_2}}{3}\g^{\m}\cd_{[\r}\j_{\s]}(\d^\r_{(\m}F_{\n)}{}^\s-g_{\m\n}F^{\r\s})A^\n\NO\\
&-\tfrac{\red{s_2}}{3}\dc D^\m\big((3F_{[\r\s}A^{\k]}g_{\m\n}-6A_{[\n} \d^\k_\m F_{\r]\s}-2F_{\r\s}\d^\k_{(\m}A_{\n)})\g^{\n\r\s}\j_\k\big)\NO\\
&-\tfrac{\blue{s_1}}{6}\dc D_\m\big(12\wt F^{\m\n}\f_\n-(\wt F^{\m\n}+iF^{\m\n}\g^5)\g^\r\cd_{[\r}\j_{\n]}-(\wt F_{\n\r}+iF_{\n\r}\g^5)\g^\r\cd^{[\m}\j^{\n]}\NO\\
&-(\wt F^{\n\r}+iF^{\n\r}\g^5)\g^\m\cd_{[\r}\j_{\n]}\big)-\tfrac{\blue{s_1}}{6}(\wt F^{\m\n}-i F^{\m\n}\g^5)(\wt F_{\m\r}+i F_{\m\r}\g^5)\g_\n\j^\r\NO\\
&-\tfrac{i\blue{s_1}}{4}(W_{\m\n\r\s}-g_{\m\s}\wt F_{\n\r})F^{\n\r}\g^\m\g^5\j^\s+\tfrac{\blue{s_1}}{48}\g^\m\j^\n(\wt R_{\m\n\r\s}F^{\r\s}+5\wt F_{\r\s}R_{\m\n}{}^{\r\s}\NO\\
&-3\wt F_{\m\n}R+8\wt F_{[\m}{}^{\r}R_{\n]\r}-12\wt F_{(\m}{}^{\r}R_{\n)\r})\Big]+\co(\j^3)\label{Q-anomaly}\,.
\eal
\end{subequations}
\ebxd
Besides quantities that were already introduced in subsection \ref{sec:4dCFT}, these expressions contain also the dual Weyl tensor 
\be
\wt W_{\m\n\r\s}=\tfrac12\e_{\m\n}{}^{\k\l}W_{\k\l\r\s}\,,
\ee
and the Schouten tensor in four dimensions 
\be\label{Schouten}
P_{\m\n}=\tfrac12\big(R_{\m\n}-\tfrac16Rg_{\m\n}\big)\,.
\ee

From the expressions \eqref{W-anomaly}-\eqref{Q-anomaly} it is clear that the form of the superconformal anomalies simplifies considerably in the scheme corresponding to $s_1=s_2=0$, i.e. in the absence of the local counterterm \eqref{counterterm}. However, for $s_1=0$ and $s_2=1$ the $c$-cocycle of the Weyl anomaly, as well as the supersymmetric completion of the Pontryagin density in the R-symmetry anomaly, become local superconformal invariants, i.e. pointwise invariant under Weyl, R-symmetry and S-supersymmetry transformations. We will see later on that this property is particularly convenient for determining the WZ action. 

Let us now compare our result with those in the earlier literature. For $a=c$, in which case the values of the scheme parameters $s_1$ and $s_2$ drop out, the anomalies \eqref{W-anomaly}, \eqref{R-anomaly}, \eqref{S-anomaly} and \eqref{Q-anomaly} coincide with those obtained holographically in \cite{Papadimitriou:2017kzw}, as well as those in \cite{Papadimitriou:2019gel}, which, as in the present work, were determined by solving the WZ consistency conditions. For $a\neq c$, the results we obtain here extend and partially correct those found in \cite{Papadimitriou:2019gel}. In particular, for $s_1=s_2=1$, the $\ca_S$ anomaly in \eqref{S-anomaly} coincides with the corresponding anomaly found in \cite{Papadimitriou:2019gel}, while the bosonic anomalies $\ca_W$ and $\ca_R$ in respectively \eqref{W-anomaly} and \eqref{R-anomaly} are determined to quadratic order in the gravitino, generalizing the corresponding results in \cite{Papadimitriou:2019gel}, which were given only to zeroth order in the gravitino. However, the part of the $\ca_Q$ anomaly proportional to $a-c$ given in \eqref{Q-anomaly} differs from and corrects the corresponding expression in \cite{Papadimitriou:2019gel}. The $\cn=1$ superconformal anomalies have also been computed recently in \cite{Imbimbo:2023sph} and \cite{Imbimbo:2025ffw}. Only the $a$-cocycle is discussed in \cite{Imbimbo:2023sph}, but their result agrees with \eqref{W-anomaly}, \eqref{R-anomaly}, \eqref{S-anomaly} and \eqref{Q-anomaly} in the scheme corresponding to $s_1=s_2=0$. In \cite{Imbimbo:2025ffw}, both the $a$ and $c$ anomaly cocycles are determined in a number of different schemes. However, the family of schemes presented there differs from the family parameterized by the local counterterm \eqref{counterterm}. Nevertheless, our results for $s_1=s_2=0$ do agree with a specific scheme in \cite{Imbimbo:2025ffw}.

\subsection{Bardeen-Zumino currents and covariant anomalies}

The structure of the superconformal anomalies \eqref{W-anomaly}-\eqref{Q-anomaly} becomes more transparent once the Ward identities are expressed in terms of the covariant currents instead of the consistent ones \cite{Papadimitriou:2017kzw,Minasian:2021png,Panopoulos:2023cdg}. As the name suggests, the defining property of the covariant currents is that they transform covariantly, i.e. without derivatives on the local transformation parameters. They are obtained from the consistent ones by adding the Bardeen-Zumino (BZ) currents \cite{Bardeen:1984pm}, which are local expressions of the supergravity fields, but cannot be expressed as functional derivatives of a four-dimensional functional.

The covariant transformation of the covariant currents provides a constructive way to determine the BZ currents.\footnote{Alternatively, one should be able to obtain the superconformal BZ currents using the recently developed superconformal descent  \cite{Imbimbo:2023sph,Imbimbo:2025ffw,Aminov:2026zrv}.} The local symmetry transformations of the consistent currents \eqref{consistent-currents} are dictated by the corresponding transformations of the supergravity fields \eqref{sugra-transformations} and the superconformal anomalies \eqref{W-anomaly}-\eqref{Q-anomaly}. For the purpose of determining the BZ currents we may work in the specific scheme $s_1=s_2=0$ since the currents for any other value of the scheme parameters $s_1$ and $s_2$ differ by terms that correspond to derivatives of the local counterterm \eqref{counterterm}.

Consider first the R-symmetry transformation of the R-symmetry current. In the scheme $s_1=s_2=0$ it is given by 
\be
\begin{split}
\d(\th)\<\cj^\m\>=&\;e^{-1}\frac{\d}{\d A_\m}\d(\th)\mathcal W=-e^{-1}\frac{\d}{\d A_\m}\int d^4x\, e\,\theta\mathcal A_R\\
=&\;-\frac{5a-3c}{27\p^2}e^{-1}\frac{\d}{\d A_\m}\int d^4x\, e\,\theta\wt F^{\m\n}F_{\m\n}=-\frac{4(5a-3c)}{27\p^2} \wt F^{\m\n}\pa_\n\th\,.
\end{split}
\ee
It follows that the covariant R-symmetry current $\<\cj^\m_{\text{cov}}\>\equiv\<\cj^\m\>+\cj_{\rm BZ}^\m$, where
\bbxd
\vskip.4cm
\be\label{eq:JBZ}
\cj_{\rm BZ}^\m=\frac{4(5a-3c)}{27\p^2}\wt F^{\m\n}A_\n\,,
\ee 
\ebxd
is the local BZ current, is invariant under R-symmetry transformations.

The BZ currents for the stress tensor and the supercurrent can be obtained analogously. Starting with the latter, in the scheme $s_1=s_2=0$, the R-symmetry transformation of the consistent supercurrent is given by
\be
\begin{split}
\d(\th)\<\cs^\m\>=&\;e^{-1}\d(\th)\Big(\frac{\d}{\d \lbar\psi_\m}\Big)\mathcal W+e^{-1}\frac{\d}{\d \lbar\psi_\m}\d(\th)\mathcal W\\
=&\;i\g^5\<\cs^\m\>\th-e^{-1}\frac{\d}{\d \lbar\psi_\m}\int d^4x\, e\,\theta\mathcal A_R\\
=&\;i\g^5\<\cs^\m\>\th+\frac{a-c}{24\p^2}e^{-1}\frac{\d}{\d \lbar\psi_\m}\int d^4x\, e\,\theta \wt R_{\m\n\r\s}R^{\m\n\r\s}\,.
\end{split}
\ee
Since the Pontryagin density is topological, it is a functional of the spin connection only and so its generic variation takes the form
\be\label{eq:varPontryagin}
\d\int d^4x\, e\,\theta \wt R_{\m\n\r\s}R^{\m\n\r\s}=4\int d^4x\,\theta\,\pa_\m(e\,\wt R^{\m\n ab}\d\o_{\n ab})=-4\int d^4x\,e\,\pa_\m\theta\wt R^{\m\n ab}\d\o_{\n ab}\,.
\ee
Inserting the variation of the spin connection \eqref{composites} with respect to the gravitino,
\be
\d_\j\o_\m{}^{ ab}=\frac12e_\r{}^{[a}e_\s^{b]}(\bar\d\j^\r\g^\s\j_\m+\bar\d\j^\r\g_\m\j^\s+\bar\d\j_\m\g^\r\j^\s)\,,
\ee
in the variation \eqref{eq:varPontryagin}, we may read off the R-symmetry transformation of the consistent supercurrent and construct the corresponding BZ current by demanding that the covariant supercurrent $\<\cs^\m_{\text{cov}}\>\equiv\<\cs^\m\>+\cs_{\rm BZ}^\m$ transforms covariantly under R-symmetry. We find
\bbxd
\vskip.4cm
\be\label{eq:SBZ}
\cs_{\rm BZ}^\m=\frac12Z^{\m\r\s}\g_\r\j_\s\,,
\ee
\ebxd
where we have introduced the shorthand notation 
\be\label{eq:YZ}
Z^{\r\m\n}\equiv Y^{\r\m\n}+Y^{\m\r\n}+Y^{\n\r\m}\,,\qquad Y^{\m\n\r}\equiv \frac{a-c}{6\p^2}A_\s\wt R^{\s\m\n\r}\,.
\ee

Finally, the R-symmetry transformation of the consistent stress tensor in the scheme $s_1=s_2=0$ is given by
\be
\begin{split}
\d(\th)\<\ct^\m{}_a\>=&\;e^{-1}\frac{\d}{\d e_\m{}^a}\d(\th)\mathcal W=-e^{-1}\frac{\d}{\d e_\m{}^a}\int d^4x\, e\,\theta\mathcal A_R\\
=&\;\frac{a-c}{24\p^2}e^{-1}\frac{\d}{\d e_\m{}^a}\int d^4x\, e\,\theta \wt R_{\m\n\r\s}R^{\m\n\r\s}\,.
\end{split}
\ee
As above, inserting the variation of the spin connection \eqref{composites} with respect to the vielbein,
\be
\d_e\o_\m{}^{ab}=e^{\r [a}\cd_\m e_\r{}^{b]}-e^{\r [a}\cd_\r e_\m{}^{b]}-e^{\r[ a}e_\s{}^{b]}e_\m{}^c\cd_\r e_c{}^\s\,,
\ee
in the variation \eqref{eq:varPontryagin}, we may read off the R-symmetry transformation of the consistent stress tensor and construct the corresponding BZ current by demanding that the covariant stress tensor $\<\ct^\m_{\text{cov}\,a}\>\equiv\<\ct^\m{}_a\>+\ct_{\text{BZ}\, a}^\m$ does not transform under R-symmetry. We find that the BZ current for the stress tensor is given by\footnote{This agrees with the BZ stress tensor given in \cite{Panopoulos:2023cdg} up to torsion terms, which were not included there since they vanish on supersymmetric backgrounds.}
\bbxd
\vskip.4cm
\be\label{eq:TBZ}
\ct_{\rm BZ}^{\m\n}=\nabla_{\r}Z^{\r\m\n}-Z^{\r\m\n}T_{\s\r}{}^\s
-\frac12Z^{\r\n\s}T_{\r\s}{}^\m-L^{\m\n}\,,
\ee
\ebxd
where $Z^{\r\m\n}$ is given in \eqref{eq:YZ} and we have defined the covariant antisymmetric tensor
\be\label{eq:L}
L^{\m\n}\equiv-\frac{a-c}{12\p^2}F_{\r\s}\wt R^{\r\s\m\n}\,.
\ee

Notice that the BZ stress tensor is defined as $\ct_{\rm BZ}^{\m\n}\equiv\ct^{\m}_{\text{BZ}\,a}e^{a\n}$ and it is {\em not} symmetric. Moreover, as expected, the BZ stress tensor \eqref{eq:TBZ} agrees with the known expression for non-supersymmetric theories (see e.g. \cite{Jensen:2012kj}) only to leading order in the gravitino. In particular, 
\be
\ct_{\rm BZ}^{\m\n}=-\frac12\nabla_{\r}(X^{\r\m\n}+X^{\r\n\m}-X^{\m\n\r})+\co(\j^2)\,,
\ee
with
\be
X^{\m\l\n}\equiv\frac{a-c}{6\p^2}(\wt R^{\m\r\l\n}+\wt R^{\l\r\m\n})A_\r\,.
\ee

We should stress that the BZ currents \eqref{eq:JBZ}, \eqref{eq:SBZ} and \eqref{eq:TBZ} are valid to all orders in the gravitino -- they are exact. The corresponding covariant currents transform covariantly under R-symmetry, as well as general coordinate transformations and frame rotations. It may be possible to obtain additional contributions to the BZ currents which will covariantize the currents with respect to Weyl, Q- and S-supersymmetry transformations, while preserving covariance under R-symmetry, general coordinate transformations and frame rotations. However, we will not pursue this question here. The BZ currents \eqref{eq:JBZ}, \eqref{eq:SBZ} and \eqref{eq:TBZ} already allow us to write the \eqref{WardIDs} and the superconformal anomalies \eqref{W-anomaly}-\eqref{Q-anomaly} in a more compact form. 

Namely, in terms of the covariant currents, the Ward identities \eqref{WardIDs} become   
\bbxd
\be\label{covWardIDs}
\begin{split}
&\nabla_\n(e_\m{}^a\<\ct^\n_{\text{cov}\,a}\>+\lbar\j_\m\<\cs^\n_{\text{cov}}\>)-(T_{\r\n}{}^\r \d^\s_\m+T_{\m\n}{}^\r \d_\r^\s)(e_\s{}^a\<\ct^\n_{\text{cov}\,a}\>+\lbar\j_\s\<\cs^\n_{\text{cov}}\>)\\
&\hspace{1.cm}-\lbar\j_\n\overleftarrow{\dc D}_\m \<\cs^\n_{\text{cov}}\>-F_{\m\n}\<\cj^\n_{\text{cov}}\>=\nabla_\n L^{\n}{}_\m\,,\\
\rule{0.0cm}{.5cm}&e_{\m[a} \<\ct^\m_{\text{cov}\,b]}\>+\tfrac14\lbar\j_\m\g_{ab} \<\cs^\m_{\text{cov}}\>=0\,,\\
\rule{0.0cm}{.5cm}&e_\m{}^a\<\ct^\m_{\text{cov}\,a}\>+\tfrac12\lbar\j_\m \<\cs^\m_{\text{cov}}\>=\ca^{\text{cov}}_W\,,\\
\rule{0.0cm}{.5cm}&\nabla_\m \<\cj^\m_{\text{cov}}\>-T_{\n\m}{}^\n\<\cj^\m_{\text{cov}}\>+i\lbar\j_\m\g^5 \<\cs^\m_{\text{cov}}\>=\ca^{\text{cov}}_R\,,\\
\rule{0.0cm}{.5cm}&\dc D_\m \<\cs^\m_{\text{cov}}\>-T_{\n\m}{}^\n\<\cs^\m_{\text{cov}}\>-\tfrac12\g^a\j_\m \<\ct^\m_{\text{cov}\,a}\>-\tfrac{3i}{4}\g^5\f_\m\<\cj^\m_{\text{cov}}\>=\ca^{\text{cov}}_Q\,,\\
\rule{0.0cm}{.5cm}&\g_\m \<\cs^\m_{\text{cov}}\>-\tfrac{3i}{4}\g^5\j_\m\<\cj^\m_{\text{cov}}\>=\ca^{\text{cov}}_{S}\,.
\end{split}
\ee
\ebxd
where in the scheme $s_1=s_2=0$ the covariant superconformal anomalies take the form 
\bbxd
\begin{subequations}
\bal
\rule{0.0cm}{.5cm}\ca_W^{\text{cov}}=&\;-\tfrac{a}{16\p^2}\big(E+\nabla_\m(8i\e^{\m\n\r\s}\bar\f_\n\g^5\cd_\r\psi_\s-8\bar\f^\m\g^\n\f_\n)\big)\NO\\
&+\tfrac{c}{16\p^2}\big[W^2-\tfrac83F^2+8i\e^{\m\n\r\s}\bar\f_\m\g^5\g_\n\cd_\r\f_\s+8i\e^{\m\n\r\s}\nabla_\m(\bar\f_\n\g^5\cd_\r\j_\s)+8\wt F^{\m\n}\bar\j_\m\f_\n\NO\\
&-2R^{\m\n}\big(\bar\j^\r\g_{\r\n}\f_\m-\bar\j_\m\g_{\r\n}\f^\r-\tfrac23g_{\m\n}\bar\j^\r\g_{\r\s}\f^\s+2\bar\j^\r\g_\n(\cd_{[\m}\j_{\r]}-\g_{[\m}\f_{\r]})\big)\NO\\
&+\tfrac{16i}{3}F^{\m\n}\bar\j_\m\g^\r\g^5\cd_{[\n}\j_{\r]}-\tfrac43\wt F^{\m\n}\bar\j^\r(\g_\r\cd_\m\j_\n+2\g_\m\cd_{[\n}\j_{\r]})\big]+\co(\j^4)\,,\label{cov-W-anomaly}\\
\rule{0.0cm}{.5cm}\ca_R^{\text{cov}}=&\;\tfrac{5a-3c}{9\p^2}\;\wt F_{\m\n} F^{\m\n}+\tfrac{(c-a)}{24\p^2}\cp\label{cov-R-anomaly}\,,\\
\rule{0.0cm}{.5cm}\ca_{S}^{\text{cov}}=&\;\tfrac{5a-3c}{6\p^2}\wt F^{\m\n}\cd_\m\j_{\n}-\tfrac{ic}{6\p^2} F^{\m\n}\g^{\s}\g_{\m}\g^5\cd_{[\n}\j_{\s]}\NO\\
&+\tfrac{3a}{4\p^2}P_{\m\n}g^{\m[\n}\g^{\r\s]}\cd_\r\j_\s+\tfrac{(a-c)}{8\p^2}W^{\m\n\r\s}\g_{\m\n}\cd_\r\j_\s+\co(\j^3)\label{cov-S-anomaly}\,,\\
\rule{0.0cm}{.5cm}\ca_Q^{\text{cov}}=&\;-\tfrac12\g_\r\j_\s L^{\s\r}\label{cov-Q-anomaly}\,.
\eal
\end{subequations}
\ebxd

\section{Compensator multiplet, super-Poincar\'e subalgebra and super-Weyl symmetry}
\label{sec:compensator_multiplet}

Our next goal is to construct a local WZ action that reproduces the superconformal anomalies \eqref{W-anomaly}-\eqref{Q-anomaly}. As we saw, this requires the introduction of suitable compensator fields, which should now come in the form of a superconformal multiplet. However, contrary to the earlier examples, the role of the compensator fields in conformal supergravity serves {\em three} distinct purposes. Firstly, it allows us to {\em twist} the local supergravity algebra \eqref{local_algebra} so that it admits a closed subalgebra. The corresponding subgroup $H$ permits the implementation of the coset construction described in section \ref{sec:WZactions}, with the compensator fields parameterizing the coset $G/H$ as before. The twist of the algebra facilitated by the compensator multiplet is essential for the coset construction, since there exists no subalgebra that contains Q-supersymmetry in the untwisted local symmetry algebra \eqref{local_algebra}.

The coset construction for conformal supergravity can be viewed as a generalization of the conformal calculus used in the construction of off-shell Poincar\'e supergravities \cite{Kugo:1982cu}. The crucial difference is that in the coset construction we pursue here we {\em do not} impose any gauge fixing conditions. We keep all the fields of conformal supergravity, as well as all fields in the compensator multiplet. This ensures that the symmetries in the coset $G/H$ are still present in the theory and are not eliminated by any gauge-fixing conditions. Indeed, as we have seen, it is precisely the anomalies under the coset symmetries that the WZ action captures.

Finally, the compensator multiplet facilitates an enhancement of the local symmetries of conformal supergravity, mimicking its superspace formulation \cite{Butter:2009cp,Kugo:2016zzf}. This symmetry is ``hidden'' in the sense that it acts trivially on the conformal multiplet, but non-trivially on the compensator multiplet. It will enable us not only to obtain a generalized WZ action that reproduces the Q-supersymmetry anomaly as well, but also to trivialize the Poincar\'e supersymmetry anomaly in the extended space of fields, through a local counterterm that moves the anomaly to the hidden symmetry.

\subsection{Compensator multiplet}

The compensator multiplet must contain a dilaton and an axion, which act as compensators respectively for Weyl and R-symmetry transformations. It should also contain a fermion -- a Goldstino -- which acts as a compensator for S-supersymmetry. Even though we have not yet implemented the coset construction described in section \ref{sec:WZactions}, we will refer to these three symmetries collectively as `coset symmetries'. They are the symmetries for which the compensators act as Goldstone fields. The minimal multiplet that contains these fields is the compensator multiplet of old-minimal supergravity \cite{Stelle:1978ye,Ferrara:1978em}, a chiral multiplet of Weyl weight $w=1$ in the conventions of \cite{Freedman:2012zz}. Its field content consists of a complex scalar, $Z$, a chiral fermion, $\z_L$, and an auxiliary complex scalar field, $F$.  

In our conventions, a chiral multiplet of Weyl weight $w=1$ transforms under the symmetries of conformal supergravity according to
\bal\label{eq:ChiralMultipletTransformations}
\d(\vth) Z=&\;\xi^\m\del_\m Z -\s Z-\tfrac{2i}{3}\theta Z + \bar\ve_L \z_L\,,\\
\d(\vth) \z_L=&\;\xi^\m\del_\m \z_L-\tfrac{3}{2}\s \z_L+\frac{i}{3}\theta \z_L-\frac{1}{4}\l_{ab}\g^{ab}\z_L+\tfrac{1}{2}\big(\g^\m(\cd_\m Z-\bar\psi_{\m L}\z_L)\ve_R+F\ve_L+2Z\h_L\big)\,,\NO\\
\d(\vth) F=&\;\xi^\m\del_\m F-2\s F+\tfrac{4i}{3}\theta F+\tfrac12\bar\ve_R\g^\m\big(2\,\cd_\m\z_L-\g^\n(\cd_\n Z-\bar\psi_{\n L}\z_L)\psi_{\m R}-F\psi_{\m L}-2 Z \phi_{\m L}\big)\,,\NO
\eal
where the covariant derivatives of the $Z$ and $\z$ are given respectively by 
\be
\cd_\m Z=\big(\del_\m +\tfrac{2i}{3}A_\m \big)Z\,,\qquad \cd_\m\z_L=\big(D_\m-\tfrac i3 A_\m\big)\z_L\,.
\ee
Notice that for a chiral multiplet of Weyl weight $w=1$ the top component, $F$, is invariant under S-supersymmetry. 

In the Majorana formulation of conformal supergravity we are using, the chiral multiplet consists of a chiral and an antichiral multiplet, $X=(Z, \z_L, F)$ and $\bar X=(\bar Z, \z_R, \bar F)$, where $Z$ and $\bar Z$, as well as $F$ and $\bar F$ are related by complex conjugation rather than being independent fields, while $\z_{L,R}$ are chiral projections of a single Majorana spinor, $\z$.  

The compensator fields do not map directly to the components of the chiral multiplet. They are embedded non-trivially in the lowest two components of the chiral multiplet, while the top component does not act as a compensator for any symmetry of conformal supergravity. In order to identify the compensators within the chiral multiplet we utilize the fact that they must transform according to the right action \eqref{Goldstone-trans} under the corresponding symmetry. In particular, for Abelian symmetries as is the case here, the compensators must transform by a shift, namely $\c^A\to\c^A+\vth^A$. In combination with the transformations \eqref{eq:ChiralMultipletTransformations}, this requirement determines that the compensators $\t$, $\c$ and $\vf$, for respectively Weyl, R-symmetry and S-supersymmetry are embedded in the components of the chiral multiplet as 
\bbxd
\vskip.3cm
\be\label{eq:Goldstone-embedding}
Z=e^{-\t-\tfrac{2i}{3}\c}\,,\qquad \z_{L}=Z\vf_{L}\,,\qquad \z_{R}=\bar Z\vf_{R}\,,\qquad F=Z\cf\,.
\ee
\ebxd
Although $F$ does not correspond to a compensator for any conformal supergravity symmetry, what appears in the symmetry transformations of the compensators $\t$, $\c$ and $\vf$ is the rescaled complex scalar $\cf$, rather than $F$.

The symmetry transformations of the Goldstone multiplet $(\t,\c,\vf,\cf)$ take the form\footnote{\label{GoldstoneMultiplet} The canonical construction of the Goldstone multiplet proceeds via multiplet calculus. The transformation laws \eqref{eq:ChiralMultipletTransformations} determine that, given a chiral multiplet $X=(Z, \z_L, F)$, one may construct a composite chiral multiplet  $W(X)=\big(W(Z),W'(Z)\z_L,W'(Z)F-W''(Z)\bar\z_L\z_L\big)$, for any function $W(Z)$ of the lowest component. Taking $W(X)=\log X$ gives $Z_{\log X}=-\t-\tfrac{2i}{3}\c$, $\z_{\log X}=\vf$ and $F_{\log X}=\cf+\bar\vf_L\vf_L$. The top component of the canonical Goldstone multiplet, therefore, differs from $\cf$ by the dilatino bilinear $\bar\vf_L\vf_L$, as is also clear from the dilatino transformation in \eqref{eq:CompensatorMultipletTransformations}. However, we find it convenient to parameterize the Goldstone multiplet in terms of $\cf$ rather than $F_{\log X}=\cf+\bar\vf_L\vf_L$, due to the fact that $\cf$ is S-supersymmetry invariant, in contrast to $\cf+\bar\vf_L\vf_L$.}  
\bbxd
\be\label{eq:CompensatorMultipletTransformations}
\begin{split}
\d(\vth)\t=&\;\x^\m\pa_\m\t+\s-\tfrac12\bar\ve\,\vf\,,\\
\d(\vth)\c=&\;\x^\m\pa_\m\c+\th+\tfrac{3i}{4}\bar\ve\g^5\vf\,,\\
\d(\vth)\vf_L=&\;\x^\m\pa_\m\vf_L-\tfrac{1}{4}\l_{ab}\g^{ab}\vf_L-\tfrac12 \s\vf_L+i\th\vf_L\\
&-\tfrac{1}{2}(\cd_\n\t+\tfrac{2i}{3}\cd_\n\c+\bar\psi_{\n L}\vf_L)\g^\n\ve_R+\tfrac12(\cf+\bar\vf_L\vf_L)\ve_L+\h_L\,,\\
\d(\vth)\cf=&\;\x^\m\pa_\m\cf-\s \cf+2i\th \cf+(2\bar\ve_R\vf_R-\bar\ve_L\vf_L)\cf\\
&-\bar\ve_R\g^\m\big(\phi_{\m}-\cd_\m\vf-(\bar\vf_L\j_{\m L})\vf\\
&-\tfrac12(\cd_\n \t+\tfrac{2i}{3}\cd_\n\c+\bar\j_{\n L}\vf_L)\g^\n(\j_{\m}+\g_\m\vf)+\tfrac12\cf(\j_{\m}+\g_\m\vf)\big)\,,
\end{split}
\ee
\ebxd
where
\be
\cd_\m\t=\pa_\m\t\,,\qquad \cd_\m\c=\pa_\m\c-A_\m\,,\qquad \cd_\m\vf=D_\m\vf-i \g^5 A_\m\vf\,.
\ee
The transformation of $\cf$ in \eqref{eq:CompensatorMultipletTransformations} is purposefully not written in the most compact form possible, but rather in the form that will be most convenient for the subsequent analysis.

\subsection{Coset-invariant fields}

Using the compensator fields $\t$, $\c$ and $\vf$, we can construct a supergravity multiplet that is invariant under the coset symmetries, namely under Weyl, R-symmetry and S-supersymmetry transformations. This multiplet defines the invariant gauge fields \eqref{invA} and generalizes the expressions in \eqref{invariant} that we obtained for non-supersymmetric CFTs. The invariant gravity multiplet obtained from the compensating chiral multiplet $(Z,\z,F)$ was originally constructed in \cite{Kaku:1978ea} and takes the form
\be\label{eq:ChiralMultipletInvariantFields}
\begin{split}
\hat e_\m{}^a =&\;|Z|e_\m{}^a\,\,,\\
\hat A_\m =&\;A_\m-\frac{3i}{4} \frac{1}{|Z|^2}\big(\bar Z\del_\m Z-Z\del_\m\bar Z+\tfrac{1}{2}\bar\z\g_\m\g^5\z-\bar\psi_\m(\bar Z\z_L-Z\z_R)\big)\,,\\
\hat\psi_\m =&\;\frac{1}{|Z|}(Z\,P_R+\bar Z \,P_L)^{3/2}\psi_\m+\frac{1}{|Z|}\g_\m(ZP_L+\bar Z\,P_R)^{1/2}\z\,,\\
\hat\cf=&\;\frac{Z^2}{|Z|^4}F\,.
\end{split}
\ee
Recall that $F$ is already invariant under S-supersymmetry. $\hat\cf $ is invariant in addition under Weyl and R-symmetry transformations.

These expressions simplify considerably once written in terms of the actual compensator fields. Inserting the parametrization \eqref{eq:Goldstone-embedding} of the chiral multiplet in \eqref{eq:ChiralMultipletInvariantFields} we obtain 
\be\label{hatpsi1}
\begin{split}
\hat e_\m{}^a =&\;e^{-\t}e_\m{}^a\,\,,\\
\hat A_\m =&\;A_\m-\del_\m\c-\tfrac{3i}{4} e^{2\t}\big(\tfrac{1}{2}\bar\z\g_\m\g^5\z-\, e^{-\t}\bar\psi_\m\big(\g_5\cos(\tfrac{2\c}{3})\z+i\sin(\tfrac{2\c}{3})\z\big)\big)\,,\\
\hat \psi_\m =&\;e^{-\t/2}\big(\cos(\tfrac{2\c}{3})+i\g_5\sin(\tfrac{2\c}{3})\big)^{3/2}\psi_\m+ e^{\t/2}\g_\m\big(\cos(\tfrac{2\c}{3})-i\g_5\sin(\tfrac{2\c}{3})\big)^{1/2}\z\,,\\
\hat\cf =&\;e^{\t-2i\c}\cf\,,
\end{split}
\ee
which can be further simplified using the identity 
\be
(\cos a +i\g_5\sin a)^{b}=\cos (ab)+i\g_5\sin (ab)\,.
\ee
In particular, introducing the matrix
\be\label{eq:U}
U\equiv e^{-\tfrac{\t}{2}+ i\g^5\c}=e^{-\tfrac{\t}{2}}(\cos \c+ i\g^5\sin \c)\,,
\ee
the set of coset-invariant fields can be written compactly in the form
\bbxd
\be\label{eq:InvariantFields}
\begin{split}
\hat e_\m{}^a =&\;e^{-\t}\,e_\m{}^a\,,\\
\hat A_\m =&\; A_\m-\del_\m\c+\tfrac{3i}{4}\, \bar\psi_{\m}\g^5\,\vf-\tfrac{3i}{8}\,\bar\vf\g_\m\g^5\vf\,,\\
\hat\psi_{\m } =&\;U\,(\psi_{\m }+\,\g_\m \vf)\,,\\
\hat\cf =&\;e^{\t-2i\c}\cf\,.
\end{split}
\ee
\ebxd

A key property of the fields \eqref{eq:InvariantFields} is that their Q-supersymmetry transformations, when accompanied by a Lorentz transformation with a specific field-dependent parameter, close to invariant fields only. In particular, under the modified Q-supersymmetry transformation 
\be\label{eq:HattedQ}
\hat \d_Q(\hat\ve)\equiv\d_Q(\ve)+\d_L(\l(\ve))\,,\qquad \l^{ab}(\ve)=-\tfrac12\bar\ve\g^{ab}\vf\,,
\ee
the fields \eqref{eq:InvariantFields}  transform as
\bbxd
\be\label{susy-trans-inv}
\begin{split}
\hat\d_Q(\hve)\hat e_\m{}^a=&\;\tfrac12\bar\hve \g^a\hat\j_\m\,,\\
\hat\d_Q(\hve)\hat A_\m=&\;\tfrac{3i}{4}\bar\hve \g^5\hat\f_\m+\tfrac14\hat A_\n\,\bar\hve \hat\g^\n\hat\j_\m+\tfrac{3i}{8}\bar\hve (\hat \cf P_L-\bar{\hat \cf}P_R)\hat\j_\m\,,\\
\hat\d_Q(\hve)\hat\j_\m=&\;\hat D_\m\hve-\tfrac{i}{3}\hat A_\n(\hat\g_\m{}^\n-2\d_\m^\n)\g^5\hve+\frac12\hat\g_\m(\hat \cf P_L+\bar{\hat \cf} P_R)\hat\ve\,,\\
\hat\d_Q(\hve)\hat \cf=&\;\tfrac{i}{3}\hat A_\m\,\bar\hve P_R(\hat\g^{\m\n}-\hat g^{\m\n})\hat\j_\n-\bar\hve  P_R\hat\g^\m(\hat\f_\m+\tfrac12\hat \cf\hat\j_\m)\,.
\end{split}
\ee
\ebxd
where $\hat D_\m\hve\equiv\pa_\m\hat\ve+\tfrac14\hat\o_{\m}{}^{ab}\g_{ab}\hat\ve$ with  $\hat \ve$ the coset-invariant composite parameter
\be\label{inv-epsilon}
\hat \ve\equiv U\ve\,,
\ee
and $\hat\o_\m{}^{ab}\equiv\o_\m{}^{ab}(\hat e,\hat\j)$, $\hat\f_\m\equiv \f_\m(\hat e,\hat A,\hat\j)$. In terms of the original fields they take the form
\be\label{eq:hatomega-hatphi}
\begin{split}
\hat\o_\m{}^a{}_b=&\;\o_\m{}^a{}_b-2e_\m{}^{[a}e_{b]}{}^\n(\pa_\n\t+\tfrac12\bar\vf\j_\n)-\tfrac12\bar\vf\g^a{}_b(\j_\m+\tfrac12\g_\m\vf)\,,\\
\hat\f_\m=&\;U^{-1}\big[\f_\m-\cd_\m\vf+\tfrac18(\bar\vf\g_\n\g^5\vf)\g^\n\g^5(\j_\m-\g_\m\vf)\\
&\;\hspace{.8cm}+\big(-\tfrac12(\pa_\n\t+\tfrac12\bar\vf\j_\n)\g^\n+\tfrac14(\bar\vf\vf)+\tfrac14(\bar\vf\g^5\vf)\g^5\big)(\j_\m+\g_\m\vf)\big]\,.
\end{split}
\ee 

An observation that might seem serendipitous at this point is that the transformations \eqref{susy-trans-inv} can be written compactly in terms of the quantity
\be\label{eq:hateta0}
\hat\h_0(\hve)\equiv\tfrac{i}{3}\hat A_\n\hat\g^\n\g^5\hve-\tfrac12(\hat\cf P_L+\bar{\hat\cf}P_R)\hve\,.
\ee
In particular, defining $\hat A_a=\hat e_a{}^\m\hat A_\m$, and analogously $\hat\f_a$ and $\hat\j_a$, the transformations \eqref{susy-trans-inv} can be written in the form
\bbxd
\be\label{susy-trans-inv-compact}
\begin{split}
\hat\d_Q(\hve)\hat e_\m{}^a=&\;\tfrac12\bar\hve \g^a\hat\j_\m\,,\\
\hat\d_Q(\hve)\hat A_a=&\;\tfrac{3i}{4}\bar\hve \g^5(\hat\f_a-\hat\h_0(\hat\j_a))\,,\\
\hat\d_Q(\hve)\hat\j_\m=&\;\hat\cd_\m\hve-\hat\g_\m\hat\h_0(\hat\ve)\,,\\
\hat\d_Q(\hve)\hat \cf=&\;-\bar\hve P_R\hat\g^\m(\hat\f_\m-\hat\h_0(\hat\j_\m))\,,
\end{split}
\ee
\ebxd
where $\hat\cd_\m\hve=\hat D_\m\hve+i\hat A_\m\g^5\hat\ve$. We will see in the following subsection that the fact that the transformations of the coset-invariant fields admit this compact form is a direct consequence of how local super-Poincar\'e algebra is embedded within the $\cn=1$ conformal supergravity algebra in the presence of the compensator multiplet.  

Indeed, the transformations \eqref{susy-trans-inv} or \eqref{susy-trans-inv-compact} of the invariant fields, together with diffeomorphisms and Lorentz transformations, close into the local super-Poincar\'e algebra
\bbxd
\vskip.3cm
\be\label{eq:super-Poincare-algebra}
[\hat \d_Q(\hve_1),\hat \d_Q(\hve_2)]=\d_D(\hat\x_{QQ})+\d_L(\hat\l_{QQ})+\hat\d_Q(\hve_{QQ})\,,
\ee
\ebxd
with field-dependent parameters
\bbxd
\be\label{eq:super-Poincare-parameters}
\begin{split}
\hat\x_{QQ}^\m=&\;\tfrac12\lbar\hve_2\hat\g^\m\hve_1\,,\qquad \hve_{QQ}=-\hat\x_{QQ}^\m\,\hat\j_\m\,,\\ \hat\l_{QQ}^a{}_b=&\;-\hat\x_{QQ}^\m\,\hat\o_\m{}^a{}_b-\tfrac12\lbar\hve_1\g^a{}_b\hat\h_0(\hve_2)+\tfrac12\lbar\hve_2\g^a{}_b\hat\h_0(\hve_1)\,.
\end{split}
\ee
\ebxd
This is precisely the algebra of old-minimal supergravity \cite{Stelle:1978ye,Ferrara:1978em}, obtained here without gauge fixing the coset symmetries.

The fact that the Q-supersymmetry transformations (as well as diffeomorphisms and local Lorentz transformations) of the invariant fields \eqref{eq:InvariantFields} close to invariant fields means that these fields furnish a representation -- a new multiplet -- of the local algebra of conformal supergravity. Since these transformations satisfy the local super-Poincar\'e algebra \eqref{eq:super-Poincare-algebra} and the coset symmetries act trivially on the invariant fields, the representation they transform under is a faithful representation of a local super-Poincar\'e subalgebra of the conformal supergravity algebra. Our next goal is to identify the local super-Poincar\'e subalgebra itself on the {\em full} field space  -- not just on the invariant fields -- and {\em without} imposing any gauge fixing conditions. Both of these requirements are essential in our analysis. We need to identify the local super-Poincar\'e subalgebra in order to implement the coset construction, while keeping the coset symmetries intact is a necessary condition for being able to construct the WZ action that reproduces the superconformal anomalies.

\subsection{Local super-Poincar\'e subalgebra}
\label{sec:PoincareSubalgebra}

In order to implement the coset construction of section \ref{sec:WZactions} we need to identify the local subalgebra of the conformal supergravity algebra \eqref{local_algebra} that remains unbroken in the coset construction. This subalgebra should comprise all symmetries for which we have not introduced a compensator field, that is diffeomorphisms, local Lorentz, as well as Q-supersymmetry transformations. The algebra of old-minimal supergravity incorporates exactly these symmetries, but it is realized either after gauge-fixing Weyl, R-symmetry and S-supersymmetry by setting the compensator fields to zero, or only on the invariant fields \eqref{eq:InvariantFields}. Our goal here is to realize a local super-Poincar\'e algebra \eqref{eq:super-Poincare-algebra} as a subalgebra of the conformal supergravity algebra on the {\em full} field space and {\em without} any gauge fixing. 

The symmetries of old-minimal supergravity are obtained by fixing the compensator fields to zero and identifying the subset of conformal supergravity symmetries that preserve these values of the compensator fields \cite{Kugo:1982cu}. We are going to replicate this procedure here, but without fixing the compensators to specific values. From the conformal supergravity algebra \eqref{local_algebra} we see that diffeomorphisms and local Lorentz transformations already close among themselves. However, the commutator of Q-supersymmetry transformations generates R-symmetry and S-supersymmetry transformations, violating the closure of the putative subalgebra. To remedy this, we will define a new field-dependent Q-supersymmetry transformation under which the compensator fields are invariant. Closing the super-Poincar\'e algebra then requires that diffeomorphisms and local Lorentz transformations must also be modified by field-dependent coset transformations. 

Let us therefore consider the transformations
\bbxd
\be\label{eq:RingedTransformations}
\begin{split}
\mathring{\d}_D(\x)\equiv&\; \d_D(\x)+\d_W(\s(\x))+\d_R(\th(\x))+\d_S(\h(\x))\,,\\
\mathring{\d}_L(\l)\equiv&\; \d_L(\l)+\d_W(\s(\l))+\d_R(\th(\l))+\d_S(\h(\l))\,,\\
\mathring{\d}_Q(\ve)\equiv&\;\d_Q(\ve)+\d_W(\s(\ve))+\d_R(\th(\ve))+\d_S(\h(\ve))\,,
\end{split}
\ee
\ebxd
and demand that they annihilate all the Goldstone fields, namely 
\be\label{eq:deformation-conditions}
\begin{split}
&\mathring{\d}_D(\x)\t=0\,,\qquad \mathring{\d}_D(\x)\c=0\,, \qquad \mathring{\d}_D(\x)\vf=0\,,\\
&\mathring{\d}_L(\l)\t=0\,,\qquad \mathring{\d}_L(\l)\c=0\,, \qquad \mathring{\d}_L(\l)\vf=0\,,\\
&\mathring{\d}_Q(\ve)\t=0\,,\qquad \mathring{\d}_Q(\ve)\c=0\,, \qquad \mathring{\d}_Q(\ve)\vf=0\,.
\end{split}
\ee
The existence of a unique solution to these conditions is ensured by the fact that the coset symmetries act transitively on the compensator fields. Each row of \eqref{eq:deformation-conditions} provides a non-degenerate system of three equations for the three corresponding composite coset parameters. The transformations \eqref{eq:RingedTransformations} are unique in that they annihilate the compensators, but also reduce to the corresponding conformal supergravity transformations when they act on the invariant fields \eqref{eq:InvariantFields}. Clearly, any linear combination of \eqref{eq:RingedTransformations} also annihilates the compensators, but reduces to the corresponding linear combinations of conformal supergravity transformations when acting on the invariant fields. In the following, we will use this observation in order to identify the correct extension of the transformations that satisfy the local super-Poincar\'e algebra \eqref{eq:super-Poincare-algebra} to the entire field space. 

Applying the conditions \eqref{eq:deformation-conditions} to the compensator transformations \eqref{eq:CompensatorMultipletTransformations} determines that the composite parameters that define the transformations \eqref{eq:RingedTransformations} are given by
\bbxd
\be\label{eq:RingedParameters}
\begin{split}
\s(\x)=&\;-\x^\m\pa_\m\t\,,\quad \th(\x)=-\x^\m\pa_\m\c\,,\quad \h_L(\x)=-\x^\m(\pa_\m+\tfrac12\pa_\m\t-i\pa_\m\c)\vf_L\,,\\
\rule{0cm}{0.5cm}\s(\l)=&\;0\,,\quad \th(\l)=0\,,\quad \h_L(\l)=\tfrac14\l_{ab}\g^{ab}\vf_L\,,\\
\rule{0cm}{0.5cm}\s(\ve)=&\;\tfrac12\bar\ve\vf\,,\quad \th(\ve)=-\tfrac{3i}{4}\bar\ve\g^5\vf\,,\\
\h_L(\ve)=&\;\tfrac12 \s(\ve)\vf_L-i\th(\ve)\vf_L+\tfrac{1}{2}(\cd_\n \t+\tfrac{2i}{3}\cd_\n\c+\bar\j_{\n L}\vf_L)\g^\n\ve_R+(\bar\vf\ve_L)\vf_L-\tfrac12\cf\ve_L\,.\hspace{-.5cm}
\end{split}
\ee
\ebxd  
However, as we saw earlier, the Poincar\'e Q-supersymmetry transformation on the invariant fields differs from that of conformal supergravity by a Lorentz transformation according to \eqref{eq:HattedQ}. It follows that the correct extension of the Poincar\'e Q-supersymmetry transformation to the complete space of fields is $\mathring{\d}_Q(\ve)$ in \eqref{eq:RingedTransformations}, accompanied by a field dependent Lorentz transformation $\mathring{\d}_L(\l)$ with parameter $\l^{ab}(\ve)=-\tfrac12\bar\ve\g^{ab}\vf$, while the extension of diffeomorphisms and local Lorentz transformations that satisfy the super-Poincar\'e algebra is provided respectively by the transformations $\mathring{\d}_D(\x)$ and $\mathring{\d}_L(\l)$ in \eqref{eq:RingedTransformations}. In total, the extension of the local symmetry generators that satisfy the super-Poincar\'e algebra on the entire field space is provided by the identifications
\bbxd
\vskip.2cm
\be\label{eq:HattedPoincare}
\hat\d_D(\x)\equiv\mathring{\d}_D(\x)\,,\qquad \hat\d_L(\l)\equiv\mathring{\d}_L(\l)\,,\qquad \hat\d_Q(\hve)\equiv\mathring{\d}_Q(\ve)+\mathring{\d}_L(\l(\ve))\,.
\ee
\ebxd

By construction, the transformations \eqref{eq:HattedPoincare} annihilate the compensators and, therefore, trivially close on the local super-Poincar\'e algebra \eqref{eq:super-Poincare-algebra} when acting on these fields. When acting on the invariant fields they reduce to the transformations \eqref{susy-trans-inv}, which also close to the same super-Poincar\'e algebra. Therefore, we have identified a local super-Poincar\'e subalgebra of the $\cn=1$ conformal supergravity algebra, on the full field space spanned by the conformal gravity multiplet and the compensating chiral multiplet, without gauge fixing.  

The closure of the transformations \eqref{eq:HattedPoincare} to the local super-Poincar\'e subalgebra \eqref{eq:super-Poincare-algebra} is facilitated by the fact that the composite parameters \eqref{eq:RingedParameters} -- crucially including the Lorentz shift in the case of $\hat\d_Q(\hve)$ -- possess well defined transformations under the coset symmetries. In particular, they are functions of adapted coordinates along the coset fibers. To see this, let us introduce the modified S-supersymmetry parameter
\be\label{inv-eta}
\hat \h\equiv U^{-1}\h\,,
\ee
in analogy with the modified parameter $\hat\ve$ introduced earlier in \eqref{inv-epsilon}. Recall that the dilaton and axion dependent matrix $U$ was defined in \eqref{eq:U}. Like $\hve$, the S-supersymmetry parameter \eqref{inv-eta} does not transform under Weyl or R-symmetry transformations. 

Accordingly, we introduce the modified dilatino  
\be\label{inv-varphi}
\hat\vf\equiv U^{-1}\vf\,,
\ee
which is also invariant under both Weyl and R-symmetry transformations, and shifts by $\hat\eta$ under S-supersymmetry. The variables $(\t,\c,\hat\vf)$, therefore, transform only under the coset symmetry for which they act as a compensator. They are the correct Goldstone fields, comprising a set of adapted coordinates on the coset fibers. Moreover, $(\s,\th,\hat \h)$ are the adapted parameters for the coset under which the Goldstones $(\t,\c,\hat\vf)$ shift, while $(\x,\l,\hat \ve)$ comprise a set of coset-invariant parameters for the base symmetries. 

The fact that the transformations \eqref{eq:HattedPoincare} close to the local super-Poincar\'e subalgebra \eqref{eq:super-Poincare-algebra} on the entire field space suggests that the composite parameters \eqref{eq:RingedParameters}, including the Lorentz shift in the case of $\hat\d_Q(\hve)$, must correspond to composite adapted coset parameters $(\s,\th,\hat \h)$. This is guaranteed if they are expressible exclusively in terms of the coset-invariant base parameters $(\x,\l,\hat \ve)$, the adapted Goldstones $(\t,\c,\hat\vf)$, and possibly the invariant fields \eqref{eq:InvariantFields}. We will now demonstrate that this is indeed the case, with the caveat that we need to include an additional Goldstone field, $\hat b_\m$, corresponding to the broken special conformal transformations. Since we have gauge-fixed special conformal transformations, this will not be an independent Goldstone field, but rather a composite one, expressed in terms of the independent fields in the gravity and compensator multiplets.

Starting with the composite coset parameters that define the Poincar\'e  diffeomorphisms, $\hat\d_D(\x)$, and Lorentz transformations, $\hat\d_L(\l)$, we can easily express them in terms of the coset-invariant base parameters $(\x,\l,\hat \ve)$ and the adapted Goldstones $(\t,\c,\hat\vf)$, namely 
\be
\begin{split}
&\s(\x)=-\x^\m\pa_\m\t\,,\qquad \th(\x)=-\x^\m\pa_\m\c\,,\qquad \hat\h(\x)=-\x^\m\pa_\m\hat\vf\,,\\
&\s(\l)=0\,,\qquad \th(\l)=0\,,\qquad \hat\h(\l)=\tfrac14\l_{ab}\g^{ab}\hat\vf\,.
\end{split}
\ee
The same applies to the composite Weyl, R-symmetry and Lorentz parameters that define the Poincar\'e Q-supersymmetry transformation, $\hat\d_Q(\hve)$, which take the form
\be
\s(\hve)=\tfrac12\bar{\hat\ve}\hat\vf\,,\qquad \th(\hve)=-\tfrac{3i}{4}\bar{\hat\ve}\g^5\hat\vf\,,\qquad \l^{ab}(\hat\ve)=-\tfrac12\bar{\hve}\g^{ab}\hat\vf\,,
\ee
where, by abuse of notation, we have changed the argument of these parameters to the coset-invariant spinor parameter $\hve$. A little more work is needed for $\h(\ve)$, however.

The accompanying Lorentz transformation $\mathring{\d}_L(\l(\ve))$ in the definition of the Poincar\'e Q-supersymmetry transformation, $\hat\d_Q(\hve)$, contributes an additional term in the total expression for $\h(\ve)$, namely $\h_{\text{tot}}(\ve)\equiv\h(\ve)+\h(\l(\ve))$. Multiplying this by $U^{-1}$, we obtain
\be
\begin{split}
\hat\h_{\text{tot}}(\ve)=&\;\tfrac14 (\bar\ve\vf) U^{-1}\vf-\tfrac{3}{4}(\bar\ve\g^5\vf) U^{-1}\g^5\vf+\tfrac{1}{2}(\pa_\n\t+\tfrac{2i}{3}(\pa_\n\c-A_\n)+\bar\psi_{\n}\vf_L)U^{-1}\g^\n\ve_R\\
&+\tfrac{1}{2}(\pa_\n\t-\tfrac{2i}{3}(\pa_\n\c-A_\n)+\bar\psi_{\n}\vf_R)U^{-1}\g^\n\ve_L+(\bar\vf\ve_L)U^{-1}\vf_L+(\bar\vf\ve_R)U^{-1}\vf_R\,\\
&-\tfrac12U^{-1}(\cf P_L+\bar\cf P_R)\ve-\tfrac18(\bar\ve\g^{ab}\vf)U^{-1}\g_{ab}\vf\,,\\
\rule{0.0cm}{0.6cm}=&\;\tfrac{1}{2}(\pa_\n\t+\tfrac12\bar\j_\n\vf)U^{-1}\g^\n\ve+\tfrac{i}{3}(A_\n-\pa_\n\c+\tfrac{3i}{4} \bar\j_\n\g^5\vf)U^{-1}\g^\n\g^5\ve+\tfrac34(\bar\vf\ve)U^{-1}\vf\\
&-\tfrac14(\bar\vf\g^5\ve)U^{-1}\g^5\vf-\tfrac18(\bar\ve\g^{ab}\vf)U^{-1}\g_{ab}\vf-\tfrac12U^{-1}(\cf P_L+\bar\cf P_R)\ve\,.
\end{split}
\ee
All terms in this expression can be straightforwardly expressed in terms of the proper coset variables. For example, 
\be
U^{-1}\g^\n\ve=e^{\tfrac{\t}{2}-i\g^5\c}\g^\n\ve=\g^\n e^{\tfrac{\t}{2}+i\g^5\c}\ve=e^{\t}\g^\n U\ve=\hat\g^\n\hat\ve\,.
\ee
Similarly,
\be
\begin{split}
U^{-1}(\cf P_L+\bar\cf P_R)\ve=&\;U^{-2}(\cf P_L+\bar\cf P_R)U\ve\\
=&\,(e^{\t-2i\c}\cf P_L+e^{\t+2i\c}\bar\cf P_R)\hve=(\hat\cf P_L+\bar{\hat\cf} P_R)\hve\,.
\end{split}
\ee

Proceeding analogously for all terms in $\hat\h_{\text{tot}}(\ve)$, recalling the expression for the invariant vector field $\hat A_\m$ in \eqref{eq:InvariantFields} and abusing again the notation in order to change the argument of $\hat\h_{\text{tot}}$ from $\ve$ to $\hve$, we obtain the following expression for the total S-supersymmetry parameter:
\be
\begin{split}
\hat\h_{\text{tot}}(\hve)=&\;\tfrac{1}{2}(\pa_\n\t+\tfrac12\bar{\hat\j}_\n\hat\vf)\hat\g^\n\hve+\tfrac{i}{3}(\hat A_\n+\tfrac{3i}{8}\,\bar{\hat\vf}\g_\n\g^5\hat\vf)\hat\g^\n\g^5\hve\\
&\;+\tfrac34(\bar{\hat\vf}\hve)\hat\vf-\tfrac14(\bar{\hat\vf}\g^5\hve)\g^5\hat\vf-\tfrac18(\bar\hve\g^{ab}\hat\vf)\g_{ab}\hat\vf-\tfrac12(\hat\cf P_L+\bar{\hat\cf} P_R)\hve\,.
\end{split}
\ee
Finally, using Fierz rearrangement, the dilatino bilinear terms can be brought to the form
\be
\begin{split}
&-\tfrac{1}{8}(\bar{\hat\vf}\g_\n\g^5\hat\vf)\hat\g^\n\g^5\hve
+\tfrac34(\bar{\hat\vf}\hve)\hat\vf-\tfrac14(\bar{\hat\vf}\g^5\hve)\g^5\hat\vf-\tfrac18(\bar\hve\g^{ab}\hat\vf)\g_{ab}\hat\vf\\
=&\;\tfrac14(\bar{\hat\vf}\hat\vf)\hve+\tfrac14(\bar{\hat\vf}\g^5\hat\vf)\g^5\hve+\tfrac{1}{8}(\bar{\hat\vf}\g_\n\g^5\hat\vf)\hat\g^\n\g^5\hve\,,
\end{split}
\ee
so that 
\be\label{eq:etahattotal}
\hat\h_{\text{tot}}(\hve)=\tfrac{1}{2}(\pa_\n\t+\tfrac12\bar{\hat\j}_\n\hat\vf)\hat\g^\n\hve+\tfrac{i}{3}(\hat A_\n-\tfrac{3i}{8}\,\bar{\hat\vf}\g_\n\g^5\hat\vf)\hat\g^\n\g^5\hve-\tfrac12(\hat\cf-\bar{\hat\vf}\hat\vf_L)\hve_L-\tfrac12(\bar{\hat\cf}-\bar{\hat\vf}\hat\vf_R)\hve_R\,.
\ee
This expression can be simplified further by noticing that the first two terms correspond to the composite Goldstone field of the broken special conformal transformations that we anticipated above, namely
\be\label{eq:SCGoldstone}
\hat b_\m\equiv -\pa_\m\t-\tfrac12\bar{\j}_\m\vf=-\pa_\m\t-\tfrac12\bar{\hat\j}_\m\hat\vf\,,
\ee
where in the second equality we have utilized the fact that $\bar\vf\g_\m\vf=0$.
Had we not gauge-fixed special conformal transformations by imposing $b_\m=0$, this would be $\hat b_\m\equiv b_\m -\pa_\m\t-\tfrac12\bar{\j}_\m\vf$, which is invariant under Weyl, R-symmetry and S-supersymmetry.

Collecting the above results, we can summarize the composite parameters of the local Poincar\'e transformations that act on the {\em full} field space, consisting of the conformal supergravity fields and the chiral compensator multiplet. They are given by
\bbxd
\be\label{eq:PoincareFullSpace}
\begin{split}
\hat\d_D(\x):&\;\hspace{.3cm} \s(\x)=-\x^\m\pa_\m\t\,,\quad \th(\x)=-\x^\m\pa_\m\c\,,\quad \hat\h(\x)=-\x^\m\pa_\m\hat\vf\,,\\
\hat\d_L(\l):&\;\hspace{.3cm} \s(\l)=0\,,\quad \th(\l)=0\,,\quad \hat\h(\l)=\tfrac14\l_{ab}\g^{ab}\hat\vf\,,\\
\hat\d_Q(\hve):&\;\hspace{.3cm} \s(\hve)=\tfrac12\bar{\hat\ve}\hat\vf\,,\hspace{.3cm} \th(\hve)=-\tfrac{3i}{4}\bar{\hat\ve}\g^5\hat\vf\,,\hspace{.3cm} \l^{ab}(\hat\ve)=-\tfrac12\bar{\hve}\g^{ab}\hat\vf\\
&\hspace{.1cm}\hat\h_{\text{tot}}(\hve)=-\tfrac{1}{2}\hat b_\n\hat\g^\n\hve+\tfrac{i}{3}(\hat A_\n-\tfrac{3i}{8}\,\bar{\hat\vf}\g_\n\g^5\hat\vf)\hat\g^\n\g^5\hve\\
&\hspace{1.7cm}-\tfrac12(\hat\cf-\bar{\hat\vf}\hat\vf_L)\hve_L-\tfrac12(\bar{\hat\cf}-\bar{\hat\vf}\hat\vf_R)\hve_R\,.\hspace{.0cm}
\end{split}
\ee
\ebxd

Although it is straightforward to verify that the action of the local Poincar\'e transformations \eqref{eq:PoincareFullSpace} on the coset-invariant fields reduces to the expressions \eqref{susy-trans-inv-compact} we found previously, the reason behind the specific form of the composite Poincar\'e parameters is still somewhat unclear at this point. This applies especially to the relation between the S-supersymmetry parameter $\hat\h_{\text{tot}}(\hve)$ in \eqref{eq:PoincareFullSpace} and $\hat\eta_0(\hat\ve)$ in \eqref{susy-trans-inv-compact}. As we will now demonstrate, this relation is revealed by evaluating the action of the local Poincar\'e generators, $\hat\d_D(\x)$, $\hat\d_L(\l)$ and $\hat\d_Q(\hve)$, on the invariant fields in two different ways.  

Let us for concreteness focus on the local Poincar\'e Q-supersymmetry generator $\hat\d_Q(\hve)$, which is relevant for the comparison with \eqref{susy-trans-inv-compact}. As we have seen, on the full field space, this generator is defined as the composite transformation  
\be\label{eq:HattedPoincare-Q-explicit}
\hat{\d}_Q(\hve)\equiv \d_Q(\ve)+\d_W(\s(\ve))+\d_R(\th(\ve))+\d_L(\l(\ve))+\d_S(\h_{\text{tot}}(\ve))\,,
\ee
with the local parameters shown in \eqref{eq:PoincareFullSpace}. When acting on the coset-invariant fields $\hat\F\equiv\{\hat e_\m{}^a,\hat A_\m,\hat\j_\m,\hat\cf\}$ this reduces to 
\be
\hat{\d}_Q(\hve)\hat\F=(\d_Q(\ve)+\d_L(\l(\ve)))\hat\F\,,
\ee
which coincides with \eqref{eq:HattedQ} and, therefore, leads immediately to the transformations \eqref{susy-trans-inv-compact}.

However, there is an alternative, indirect but particularly instructive way to evaluate the action of \eqref{eq:HattedPoincare-Q-explicit} on the invariant fields, $\hat\F$. Treating the invariant fields as functionals of the bare fields, $\F$, and the compensators, $\t$, $\c$ and $\vf$, and using the fact that the local Poincar\'e generators -- by construction -- annihilate the compensators, we can use the chain rule to obtain, schematically,
\be\label{eq:ChainRule}
\hat{\d}_Q(\hve)\hat\F=\frac{\pa\hat\F}{\pa\F}\hat{\d}_Q(\hve)\F=\frac{\pa\hat\F}{\pa\F}\big(\d_Q(\ve)+\d_W(\s(\ve))+\d_R(\th(\ve))+\d_L(\l(\ve))+\d_S(\h_{\text{tot}}(\ve))\big)\F\,.
\ee  
Clearly, the action of $\d_W$, $\d_R$ and $\d_S$ now is non-trivial, but of course, the final result must again reproduce \eqref{susy-trans-inv-compact}. 

Let us see how this works out explicitly in the simplest possible example, the vielbein. Using the chain rule \eqref{eq:ChainRule} and the conformal supergravity transformations \eqref{sugra-transformations} with the composite parameters given in \eqref{eq:PoincareFullSpace}, we get 
\be\label{eq:ChainRuleVielbein}
\begin{split}
\hat{\d}_Q(\hve)\hat e_\m{}^a=&\;e^{-\t}\big(\d_Q(\ve)+\d_W(\s(\ve))+\d_L(\l(\ve))\big)e_\m{}^a\\
=&\;e^{-\t}\big(\tfrac12\bar\ve\g^a\j_\m+\tfrac12(\bar\ve\vf)e_\m{}^a+\tfrac12(\bar\ve\g^a{}_b\vf)e_\m{}^b\big)\\
=&\;\tfrac12 e^{-\t}\bar\ve\g^a(\j_\m+\g_\m\vf)=\tfrac12\bar\hve \g^a\hat\j_\m\,,
\end{split}
\ee
in agreement with \eqref{susy-trans-inv-compact}.

A considerably more involved, but also more illustrative example, is the action of $\hat{\d}_Q(\hve)$ on the invariant gravitino. Recalling the form of the invariant gravitino, $\hat\j_\m$, given in \eqref{eq:InvariantFields}, applying again the chain rule \eqref{eq:ChainRule} and using the conformal supergravity transformations \eqref{sugra-transformations} with the composite parameters given in \eqref{eq:PoincareFullSpace}, we obtain
\be\label{eq:ChainRuleGravitino}
\begin{split}
\hat{\d}_Q(\hve)\hat \j_\m{}=&\;U\big(\d_Q(\ve)+\d_W(\s(\ve))+\d_R(\th(\ve))+\d_L(\l(\ve))+\d_S(\h_{\text{tot}}(\ve))\big)\j_\m\\
&\;+U\big[\big(\d_Q(\ve)+\d_W(\s(\ve))+\d_L(\l(\ve))\big)e_\m{}^a\big]\g_a\vf\\
=&\;U\big(\tfrac12\s(\ve)\j_\m - i\th(\ve)\g^5\j_\m-\tfrac14\l_{ab}(\ve)\g^{ab}\j_\m+\cd_\m\ve-\g_\m\h_{\text{tot}}(\ve)\big)\\
&\;+\tfrac12(\bar\hve \g^a\hat\j_\m) e^\t U\g_a\vf\,,
\end{split}
\ee
where in writing the last line we used the result \eqref{eq:ChainRuleVielbein}.
In order to verify that this coincides with the expression given in  \eqref{susy-trans-inv-compact}, let us examine the various terms in turn. 

Starting from the last term in \eqref{eq:ChainRuleGravitino}, it follows immediately from the form of $U$ given in \eqref{eq:U} that $e^\t U\g_a\vf=\g_a U^{-1}\vf=\g_a\hat\vf$. Hence, using Fierz rearrangement we obtain
\be
\begin{split}
\tfrac12(\bar\hve \g^a\hat\j_\m) \g_a\hat\vf=&\;\tfrac12(\bar{\hat\vf}\hat\j_\m)\hve+\tfrac14(\bar{\hat\vf}\hat\g_\n\hat\j_\m)\hat\g^\n\hve-\tfrac12(\bar{\hat\vf}\g^5\hat\j_\m)\g^5\hve-\tfrac14(\bar{\hat\vf}\hat\g_\n\g^5\hat\j_\m)\hat\g^\n\g^5\hve\,.
\end{split}
\ee

Turning next to the covariant derivative of the spinor parameter $\ve$, we need to recall the relation between the spin connection $\o_\m{}^{ab}(e,\j)$ and $\hat\o_\m{}^{ab}\equiv\o_\m{}^{ab}(\hat e,\hat\j)$, given in \eqref{eq:hatomega-hatphi}. In terms of the composite Goldstone field $\hat b_\m$ given in \eqref{eq:SCGoldstone}, this becomes 
\be\label{eq:hatomega}
\begin{split}
\hat\o_\m{}^{a}{}_b=&\;\o_\m{}^{a}{}_b+2\hat e_\m{}^{[a}\hat e_{b]}{}^\n\hat b_\n-\tfrac12\bar\vf\g^a{}_b(\j_\m+\tfrac12\g_\m\vf)\,.
\end{split}
\ee 
Notice that although the $b_\m$ dependence of the spin connection $\o_\m{}^{ab}(e,\j)$ has been eliminated by gauge-fixing, $\hat\o_\m{}^{a}{}_b$ retains this information through $\hat b_\m$. Using \eqref{eq:hatomega}, we can write
\be
\begin{split}
U\cd_\m\ve=&\;U(\pa_\m\ve+\tfrac14\o_{\m}{}^{ab}\g_{ab}\ve+iA_\m\g^5\ve)\\
=&\;\pa_\m\hve+\tfrac14\hat\o_{\m}{}^{ab}\g_{ab}\hve+\tfrac12\pa_\m\t\hve-i\pa_\m\c\g^5\hve-\tfrac12\hat b_\n\hat\g_{\m}{}^\n\hve\\
&\;+\tfrac18[\bar\vf\g^{ab}(\j_\m+\tfrac12\g_\m\vf)]\g_{ab}\hve+iA_\m\g^5\hve\\
=&\;\hat\cd_\m\hve-\tfrac12\hat b_\n\hat\g_{\m}\hat\g{}^\n\hve-\tfrac14(\bar{\j}_\m\vf)\hve+\tfrac{3}{4} (\bar\psi_{\m}\g^5\,\vf)\g^5\hve\\
&\;-\tfrac{3}{8}(\bar\vf\g_\m\g^5\vf)\g^5\hve+\tfrac18[\bar\vf\g^{ab}(\j_\m+\tfrac12\g_\m\vf)]\g_{ab}\hve\,.
\end{split}
\ee

Finally, using Fierz rearrangement, the Weyl, R-symmetry and Lorentz transformations in \eqref{eq:ChainRuleGravitino} can be brought to the form
\be
\begin{split}
\tfrac12\s(\ve)\j_\m - i\th(\ve)\g^5\j_\m-\tfrac14\l_{ab}(\ve)\g^{ab}\j_\m=&\;\tfrac14(\bar\ve\vf)\j_\m - \tfrac34(\bar\ve\g^5\vf)\g^5\j_\m+\tfrac18(\bar\ve\g^{ab}\vf)\g_{ab}\j_\m\\
=&\;-\tfrac14(\bar\vf\j_\m)\ve-\tfrac14(\bar\vf\g_\n\j_\m)\g^\n\ve-\tfrac18(\bar\vf\g_{ab}\j_\m)\g^{ab}\ve\\
&\;-\tfrac14(\bar\vf\g^5\j_\m)\g^5\ve+\tfrac14(\bar\vf\g_\n\g^5\j_\m)\g^\n\g^5\ve\,.
\end{split}
\ee

Putting together the above results and using the expression for $\hat\j_\m$ in \eqref{eq:InvariantFields} and for $\hat\h_{\text{tot}}(\hve)$ in \eqref{eq:PoincareFullSpace}, one may verify with a bit more algebra that \eqref{eq:ChainRuleGravitino} indeed reproduces the transformation of the invariant gravitino given in \eqref{susy-trans-inv-compact}. Besides demonstrating the mechanism that relates the composite parameters $\hat\h_{\text{tot}}(\hve)$ and $\hat\h_0(\hve)$, this amounts to a nontrivial check of the expression for $\hat\h_{\text{tot}}(\hve)$. The transformation of the remaining two invariant fields, namely $\hat{\d}_Q(\hve)\hat A_a$ and $\hat{\d}_Q(\hve)\hat\cf$, may be evaluated similarly via the chain rule \eqref{eq:ChainRule}. Recalling the relation between the quantities $\f_\m{}(e,A,\j)$ and $\hat\f_\m{}\equiv\f_\m{}(\hat e,\hat A,\hat\j)$ given in \eqref{eq:hatomega-hatphi}, one can confirm that these transformations reproduce the corresponding ones in \eqref{susy-trans-inv-compact}.

\subsection{Super-Weyl symmetry}
\label{sec:SuperWeyl}

In the previous subsection we identified a local super-Poincar\'e subalgebra of the conformal supergravity algebra acting on the full field space and without any gauge fixing. This is the subalgebra that remains unbroken in the coset construction we discussed in section \ref{sec:WZactions} and enables us to determine the WZ effective action for the conformal supergravity coset anomalies utilizing the transgression formula \eqref{WZ-transgression}. Before we proceed with this construction, however, let us make a brief digression to discuss an additional set of symmetries that exist in the combined field space comprising the conformal and chiral multiplets. As we will see later on, these extra symmetries are the key to understanding the Q-supersymmetry transformation of the WZ functional and to constructing a suitable local counterterm that renders the Poincar\'e supersymmetry non-anomalous.       

\subsubsection{Super-Weyl symmetry as an Abelian ideal}

A somewhat peculiar feature of the compensator multiplet we introduced earlier is that all its components correspond to compensators for symmetries of conformal supergravity, except for the top component, $\cf$. There is no remaining conformal supergravity symmetry that can be compensated by a complex scalar parameter. To remedy this asymmetry, we will {\em introduce} an extra symmetry for which $\cf$ can act as a compensator. In fact, we will introduce a chiral multiplet worth of extra symmetries with local parameters $\varpi=(\wt\s,\wt\th,\wt\h,\L)$, where $\wt\s$ and $\wt\th$ are real scalars, $\wt\h$ is a Majorana spinor and $\L$ a complex scalar. The defining properties of these extra symmetries are:
\begin{enumerate}
\item They act trivially on the conformal supergravity multiplet, i.e.
\be
\d(\varpi)e_\m{}^a=0\,,\qquad \d(\varpi)\j_\m=0\,,\qquad \d(\varpi)A_\m{}=0\,.
\ee

\item The action of $\wt\s$, $\wt\th$ and $\wt\h$ on the compensator multiplet coincides with the action of respectively Weyl symmetry, R-symmetry and S-supersymmetry, given in \eqref{eq:CompensatorMultipletTransformations}.

\item The top component, $\L$, acts only on the top component of the compensator multiplet, $\cf$, which transforms by a linear shift, namely
\be
\d_\L\t=\d_L\c=\d_\L\vf_L=0\,,\qquad \d_\L\cf=-\L\,.
\ee
In particular, $\cf$ acts as a compensator for $\L$-symmetry.

\end{enumerate}

In summary, the action of the extra symmetries on the conformal and compensator multiplets takes the form
\bbxd
\be\label{eq:superWeyl}
\begin{split}
&\d(\varpi)e_\m{}^a=0\,,\quad \d(\varpi)\j_\m=0\,,\quad \d(\varpi)A_\m{}=0\,,\\
&\d(\varpi)\t=\wt\s\,,\quad
\d(\varpi)\c=\;\wt\th\,,\quad
\d(\varpi)\vf_L=-\tfrac12 \wt\s\vf_L+i\wt\th\vf_L+\wt\h_L\,,\\
&\d(\varpi)\cf=-\wt\s \cf+2i\wt\th \cf-\L\,.
\end{split}
\ee
\ebxd
Using these transformations, we find that $\d(\varpi)$ acts on the coset-invariant fields \eqref{eq:InvariantFields} as
\be\label{eq:superWeyl-InvariantFields}
\begin{split}
\d(\varpi) \hat e_\m{}^a=&\;-\wt\s \hat e_\m{}^a\,,\\
\d(\varpi)\hat A_\m=&\;-\tfrac{3i}{8}\wt\s\bar{\hat\vf}\g^5\hat\j_\m-\pa_\m\wt\th-\tfrac{3}{4}\wt\th\bar{\hat\vf}\hat\j_\m+\tfrac{3i}{4}\bar{\hat{\wt\h}}\g^5\hat\j_\m\,,\\
\d(\varpi)\hat\j_\m=&\;-\tfrac12\wt\s(\hat\j_\m+\hat\g_\m\hat\vf) + i\wt\th\g^5(\hat\j_\m-\hat\g_\m\hat\vf)+\hat\g_\m\hat{\wt\h}\,,\\
\d(\varpi) \hat \cf=&\;-\hat{\L}\,,
\end{split}
\ee
where $\hat{\wt\h}\equiv U^{-1}\wt\h$ and $\hat{\L}\equiv e^{\t-2i\c}\L$ are invariant under $\s$ and $\th$, or equivalently $\wt\s$ and $\wt\th$ transformations.

We will dub the chiral multiplet of symmetries $\varpi=(\wt\s,\wt\th,\wt\h,\L)$  {\em ``super-Weyl symmetry''} because it is a supersymmetric {\em analogue} of the local Weyl symmetry. It is also analogous to super-Weyl symmetry in superspace \cite{Howe:1978km,Buchbinder:1998twe,Kuzenko:2019vvi}, whose parameter also corresponds to an unconstrained chiral superfield. It should also be emphasized that the Weyl symmetry, R-symmetry and S-supersymmetry of conformal supergravity {\em are not} part of super-Weyl symmetry, even though they act on the compensators exactly as the lower components of the super-Weyl symmetry. This distinction, together with the super-Weyl $\L$-symmetry, will play pivotal role in the subsequent analysis.

Combining the conformal supergravity and super-Weyl symmetries forms a local semi-direct product algebra, with the conformal supergravity algebra as a subalgebra and super-Weyl symmetry as an Abelian ideal, that is all mixed commutators are linear combinations of field-dependent super-Weyl transformations. Assigning the usual tensorial, Lorentz, Weyl, R-symmetry and super-Weyl transformations to the parameters $\wt\s,\wt\th,\wt\h,\L$, namely  
\be\label{eq:SW-parameter-transformations}
\begin{split}
\d(\vth,\varpi)\wt\s=&\;\x^\m\pa_\m\wt\s\,,\quad
\d(\vth,\varpi)\wt\th=\x^\m\pa_\m\wt\th\,,\\
\d(\vth,\varpi)\wt\h=&\;\x^\m\pa_\m\wt\h-\tfrac14\l_{ab}\g^{ab}\wt\h-\tfrac12(\s+\wt\s)\wt\h+i(\th+\wt\th)\g^5\h\,,\\
\d(\vth,\varpi)\L=&\;\x^\m\pa_\m\L-(\s+\wt\s)\L+2i(\th+\wt\th)\L\,,
\end{split}
\ee
we find that the nonzero mixed commutators take the form
\be\label{eq:SWideal}
\begin{split}
[\d_W(\s),\d_{\wt S}(\wt\h)]=&\;\d_{\wt S}(\tfrac12\s\wt\h)\,,\hspace{1.45cm}
[\d_W(\s),\d_{\wt F}(\L)]=\d_{\wt F}(\s\L)\,,\\
[\d_R(\th),\d_{\wt S}(\wt\h)]=&\;\d_{\wt S}(-i\th\g^5\wt\h)\,,\hspace{1.0cm} [\d_R(\th),\d_{\wt F}(\L)]=\d_{\wt F}(-2i\th\L)\,,\\
[\d_S(\h),\d_{\wt W}(\wt\s)]=&\;\d_{\wt S}(-\tfrac12\wt\s \h)\,,\hspace{1.35cm}
[\d_S(\h),\d_{\wt R}(\wt\th)]=\d_{\wt S}(i\wt\th\g^5\h)\,,\\
\rule{0cm}{0.7cm}[\d_Q(\ve),\d_{\wt W}(\wt\s)]=&\;\d_{\wt W}(-\tfrac14\wt\s\bar\ve\vf)+\d_{\wt R}(\tfrac{3i}{8}\wt\s\bar\ve\g^5\vf)+\d_{\wt S}(\wt\h_{Q\wt W})+\d_{\wt F}(\L_{Q\wt W})\,,\\
[\d_Q(\ve),\d_{\wt R}(\wt\th)]=&\;\d_{\wt W}(\tfrac{i}{2}\wt\th\bar\ve\g^5\vf)+\d_{\wt R}(\tfrac{3}{4}\wt\th\bar\ve\vf)+\d_{\wt S}(\wt\h_{Q\wt R})+\d_{\wt F}(\L_{Q\wt R})\,,\\ 
[\d_Q(\ve),\d_{\wt S}(\wt\h)]=&\;\d_{\wt W}(\tfrac12\bar\ve\wt\h)+\d_{\wt R}(-\tfrac{3i}{4}\bar\ve\g^5\wt\h)+\d_{\wt S}(\wt\h_{Q\wt S})+\d_{\wt F}(\L_{Q\wt S})\,,\\
[\d_Q(\ve),\d_{\wt F}(\L)]=&\;\d_{\wt S}(\wt\h_{Q\wt F})+\d_{\wt F}(\L_{Q\wt F})\,,
\end{split}
\ee
with the following composite super-Weyl parameters:
\be\label{eq:SWideal-composite-parameters}
\begin{split}
\wt\h_{Q\wt W}=&\;\tfrac12(\pa_\n\wt\s+\tfrac12\wt\s\pa_\n\t)\g^\n\ve+\tfrac{i}{6}\wt\s(A_\n-\pa_\n\c+\tfrac{3i}{4}\bar\vf\g_\n\g^5\vf)\g^\n\g^5\ve\\
&+\tfrac14\wt\s(\cf P_L+\bar\cf P_R)\ve+\tfrac{1}{16}\wt\s (\bar\vf\vf)\ve+\tfrac{1}{16}\wt\s(\bar\vf\g^5\vf)\g^5\ve\,,\\
\L_{Q\wt W}=&\;-\tfrac12\bar\ve_R\g^\m\big(2\f_\m-\cd_\m\vf-(\cd_\n\t+\tfrac{2i}{3}\cd_\n\c+\tfrac12\bar\j_{\n  L}\vf_L)\g^\n\j_\m\\
&+(\cd_\m\t+\tfrac{2i}{3}\cd_\m\c)\vf-\tfrac12\cf\g_\m\vf\big)\wt\s+\tfrac12\bar\ve_R\g^\m\g^\n(\j_\m+\tfrac32\g_\m\vf)\pa_\n\wt\s\,,\\
\wt\h_{Q\wt R}=&\;-\tfrac{i}{3}(\pa_\n\wt\th-\tfrac32\wt\th\pa_\n\t)\g^\n\g^5\ve-\tfrac13\wt\th(A_\n-\pa_\n\c-\tfrac{3i}{4}\bar\vf\g_\n\g^5\vf)\g^\n\ve\\
&-\tfrac{i}{2}\wt\th(\cf P_L-\bar\cf P_R)\ve-\tfrac{i}{8}\wt\th(\bar\vf\vf)\g^5\ve-\tfrac{i}{8}\wt\th(\bar\vf\g^5\vf)\ve\,,\\
\L_{Q\wt R}=&\;i\bar\ve_R\g^\m\big(2\f_\m-\cd_\m\vf-(\cd_\n\t+\tfrac{2i}{3}\cd_\n\c+\tfrac12\bar\j_{\n L}\vf_L)\g^\n\j_\m\\
&+(\cd_\m\t+\tfrac{2i}{3}\cd_\m\c)\vf+\tfrac12\cf\g_\m\vf\big)\wt\th+\tfrac{i}{3}\bar\ve_R\g^\m\g^\n(\j_\m-\tfrac12\g_\m\vf)\pa_\n\wt\th\,,\\
\wt\h_{Q\wt S}=&\;\tfrac14\bar{\wt \h}(\j_\n-\g_\n\vf)\g^\n\ve-\tfrac14\bar{\wt\h}\g^5(\j_\n+\g_\n\vf)\g^\n\g^5\ve-\tfrac18(\bar{\wt\h}\g^{ab}\ve)\g_{ab}\vf\,,\\
\L_{Q\wt S}=&\;\bar\ve_R\g^\m\big(\cd_\m\wt\h-(\cd_\m\t+\tfrac{2i}{3}\cd_\m\c)\wt\h+\tfrac12(\bar\j_{\n L}\wt\h)\g^\n\j_\m-\tfrac12\cf\g_\m\wt\h\big)\,,\\
\wt\h_{Q\wt F}=&\;\tfrac12(\L P_L+\bar{\L} P_R)\ve\,,\\
\L_{Q\wt F}=&\;\tfrac12\bar\ve\g^\n(\j_{\n L}+\tfrac12\g_\n\vf_L)\L\,.
\end{split}
\ee
This semi-direct product of local symmetries is a component version of the superspace setup in section 3 of \cite{Kuzenko:2019vvi}.

\subsubsection{Old minimal supergravity symmetries}

The super-Weyl symmetry can also be combined with the super-Poincar\'e subalgebra of the conformal supergravity algebra. This realization, however, requires a suitable redefinition of the super-Weyl generators, analogous to the construction of the local super-Poincar\'e generators. As in \eqref{eq:RingedTransformations}, we begin by defining a version of super-Weyl transformations that annihilate the Goldstone fields $\t$, $\c$ and $\vf$, namely
\bbxd
\be\label{eq:SWRingedTransformations}
\begin{split}
\mathring\d_{W}(\s)\equiv&\; \d_{W}(\s)-\d_{\wt W}(\s)\,,\qquad \mathring \d_{R}(\th)\equiv \d_{R}(\th)-\d_{\wt R}(\th)\,,\\
\mathring \d_{S}(\h)\equiv&\; \d_{S}(\h)-\d_{\wt S}(\h)\,,\qquad\hspace{.28cm} \mathring\d_{F}(\L)\equiv -\d_{\wt F}(\L)\,.
\end{split}
\ee
\ebxd
Notice that the same local parameter is used for both the coset and super-Weyl transformations, while the overall sign is a convention chosen for reasons that will become clear momentarily. Also, $\mathring\d_{F}(\L)$ is identical (up to a sign) to the super-Weyl generator $\d_{\wt F}(\L)$, since there is no corresponding symmetry in conformal supergravity and it annihilates $\t$, $\c$ and $\vf$ on its own. With the exception of $\mathring\d_{F}(\L)$, these transformations annihilate $\cf$ as well. 

Next, as in \eqref{eq:HattedPoincare} in the case of super-Poincar\'e generators, we dress the generators \eqref{eq:SWRingedTransformations} so that they still annihilate the Goldstone fields $\t$, $\c$ and $\vf$, but also act on the invariant fields \eqref{eq:InvariantFields} in the same way that the coset symmetries of conformal supergravity act on the corresponding conformal supergravity fields. This fixes the overall sign convention in the definition of the generators \eqref{eq:SWRingedTransformations} and leads to the new super-Weyl generators  
\bbxd
\be\label{eq:HattedSW}
\begin{split}
\hat\d_{W}(\s)\equiv&\; \mathring\d_{W}(\s)+\mathring\d_{S}(\tfrac12\s\vf)+\mathring\d_{F}(-\s)\,,\\
\hat \d_{R}(\th)\equiv&\; \mathring \d_{R}(\th)+\mathring\d_{S}(-i\th\g^5\vf)+\mathring\d_{F}(2i\th)\,,\\
\hat \d_{S}(\h)\equiv&\; \mathring\d_{S}(\h)\,,\quad \hat\d_{F}(\L)\equiv \mathring\d_{F}(\L)\,.
\end{split}
\ee
\ebxd

Recalling the Poincar\'e supersymmetry transformations \eqref{susy-trans-inv-compact} and reinstating diffeomorphisms and Lorentz transformations, the combined super-Poincar\'e and super-Weyl symmetries act on the invariant fields as  
\bbxd
\be\label{eq:OM-transformations-extended}
\begin{split}
\hskip-.2cm\hat\d(\hat\vth) \hat e_\m{}^a=&\;\x^\l\pa_\l \hat e_\m{}^a+\hat e_\l{}^a\pa_\m\x^\l-\l^a{}_b \hat e_\m{}^b+\s \hat e_\m{}^a+\tfrac12\bar\hve \g^a\hat\j_\m\,,\\
\hskip-.2cm\hat\d(\hat\vth)\hat A_\m=&\;\x^\l\pa_\l \hat A_\m+\hat A_\l\pa_\m\x^\l+\pa_\m\th-\tfrac{3i}{4}\bar{\hat{\h}}\g^5\hat\j_\m+\tfrac{3i}{4}\bar\hve \g^5(\hat\f_\m-\hat\h_0(\hat\j_\m))+\tfrac12\bar\hve \hat\g^\n\hat\j_\m \hat A_\n\,,\\
\hskip-.2cm\hat\d(\hat\vth)\hat\j_\m=&\;\x^\l\pa_\l\hat \j_\m+\hat\j_\l\pa_\m\x^\l-\tfrac14\l_{ab}\g^{ab}\hat\j_\m+\tfrac12\s\hat\j_\m - i\th\g^5\hat\j_\m+\hat\cd_\m\hve-\hat\g_\m(\hat\h+\hat\h_0(\hat\ve))\,,\hspace{-.3cm}\\
\hskip-.2cm\hat\d(\hat\vth) \hat \cf=&\;\x^\l\pa_\l\hat\cf-\s \hat\cf+2i\th \hat\cf+\hat{\L}-\bar\hve_R\hat\g^\m(\hat\f_\m-\hat\h_0(\hat\j_\m))\,.
\end{split}
\ee
\ebxd
These transformations comprise the full set of local symmetries of old-minimal supergravity. They are exactly the symmetries of the ``minimal Q-supersymmetric scheme'' of \cite{Kuzenko:2019vvi}.\footnote{There is no analogue of the transverse components of the parameter $s^m$ in our formulation. Only the longitudinal parameter $\pa_m s^m$ appears in our formulation, corresponding to the complex scalar parameter $\L$.} However, a key difference of our analysis in comparison to \cite{Kuzenko:2019vvi} is that we have constructed the local super-Poincar\'e and super-Weyl generators without gauge-fixing the compensators. In particular, we have obtained an exact nonlinear embedding of the invariant fields in the full field space of conformal supergravity with a compensating chiral multiplet, and identified the old-minimal supergravity symmetries as a subalgebra of the full semi-direct product of the conformal supergravity algebra with the super-Weyl symmetries.    

Notice that the old-minimal local symmetries are closely related with -- but distinct from -- the local symmetries of conformal supergravity. In particular, the coset symmetries of conformal supergravity act in exactly the same way as the corresponding super-Weyl symmetries in \eqref{eq:HattedSW}. The same holds for diffeomorphisms and local Lorentz transformations. The only differences are in the action of Q-supersymmetry and the fact that there is no analogue of $\hat\d_F(\hat\L)$ in conformal supergravity. However, as we have seen, Poincar\'e supersymmetry can be expressed as a linear combination of Q- and S-supersymmetry in conformal supergravity.

The local symmetry algebras are also closely related. Together with the super-Poincar\'e algebra \eqref{eq:super-Poincare-algebra}, the old-minimal supergravity algebra comprises the mixed commutators   
\bbxd
\be\label{eq:OM-algebra-mixed}
\begin{split}
[\hat\d_Q(\hve),\hat\d_{W}(\s)]=&\;\hat\d_{S}(\hat{\h}_{QW})+\hat\d_{F}(\hat{\L}_{QW})\,,\\
[\hat\d_Q(\hve),\hat\d_{R}(\th)]=&\;\hat\d_{S}(\hat{\h}_{QR})+\hat\d_{F}(\hat{\L}_{QR})\,,\\ 
[\hat\d_Q(\hve),\hat\d_{S}(\hat{\h})]=&\;\hat\d_{W}(\s_{QS})+\hat\d_{R}(\th_{QS})+\hat\d_L(\l_{QS})+\hat\d_{S}(\hat{\h}_{QS})+\hat\d_{F}(\hat{\L}_{QS})\,,\\
[\hat\d_Q(\hve),\hat\d_{F}(\hat{\L})]=&\;\hat\d_{S}(\hat{\h}_{QF})+\hat\d_{F}(\hat{\L}_{QF})\,,
\end{split}
\ee
\ebxd
with composite parameters
\bbxd
\be
\begin{split}
\hat{\h}_{QW}=&\; \tfrac12\hat\g^\n\hve\,\pa_\n\s\,,\quad \hat{\L}_{QW}=\tfrac12\bar\hve_R\hat\g^\m\hat\g^\n\hat\j_\m\pa_\n\s\,,\\
\hat{\h}_{QR}=&\;\tfrac{i}{3}\g^5\g^\n\hve\,\pa_\n\th\,,\quad \hat{\L}_{QR}=\tfrac{i}{3}\bar{\hve}_R\hat\g^\m\hat\g^\n\hat\j_\m\pa_\n\th\,,\\
\s_{QS}=&\;\tfrac12\bar\hve\hat{\h}\,,\quad \th_{QS}=-\tfrac{3i}{4}\bar\hve\g^5\hat{\h}\,,\quad \l^{ab}_{QS}=-\tfrac12\bar\hve\g^{ab}\hat{\h}\,,\\
\hat{\h}_{QS}=&\;\tfrac14(\bar{\hat\h}\hat\j_\n)\hat\g^\n\hve-\tfrac14(\bar{\hat\h}\g^5\hat\j_\n)\hat\g^\n\g^5\hve\,,\\
\hat{\L}_{QS}=&\;\bar\hve_R\hat\g^\m(\hat\cd_\m\hat\h+\tfrac{2i}{3}\hat A_\m\hat\h+\tfrac12(\bar{\hat\j}_{\n L}\hat\h)\hat\g^\n\hat\j_\m)-\hat\cf\bar\hve_L\hat\eta\,,\\
\hat{\h}_{QF}=&\;\tfrac12(\hat{\L} P_L+\bar{\hat{\L}} P_R)\hve\,,\quad \hat{\L}_{QF}=\tfrac12\hat{\L}\,\bar{\hve}\hat\g^\m\hat\j_{\m\,L}\,.
\end{split}
\ee
\ebxd
These commutators imply that the local super-Poincar\'e and super-Weyl algebras do not form a semi-direct product due to the composite Lorentz transformation produced by the  commutator $[\hat\d_Q(\hve),\hat\d_{S}(\hat{\h})]$. However, the structure of these commutators and composite parameters closely resembles that of conformal supergravity in \eqref{local_algebra} and \eqref{composite-parameters}. Any discrepancies can be traced to the different action of Q-supersymmetry and the absence of $\hat\d_F(\hat\L)$ in conformal supergravity. The close relation between these algebras, however, will play a key role in the analysis of the WZ action in the next section.

\section{Wess-Zumino action for $\cn=1$ SCFTs}
\label{sec:WZaction}

At this point we have assembled all the ingredients necessary to compute the WZ effective action for generic $\cn=1$ SCFTs using the transgression formula \eqref{WZ-transgression}. Having identified a local super-Poincar\'e subalgebra of the conformal supergravity algebra acting on the full field space, and without any gauge fixing, we can apply the coset construction discussed in section \ref{sec:WZactions} in order to write the WZ action in the form
\bbxd
\vspace{.4cm}
\be\label{super-transgression}
\cw_\text{WZ}[e,A,\j,\t,\c,\vf]=\int_0^1 dt\int d^4x\, e^{(t)}\big(\t \ca_W^{(t)}-\c \ca_R^{(t)}+\bvf^{(t)} \ca_S^{(t)}\big)\,,
\ee
\ebxd
where the coset anomalies $\ca_W$, $\ca_R$ and $\ca_S$ are given respectively in eqs.~\eqref{W-anomaly}, \eqref{R-anomaly} and \eqref{Q-anomaly}. This WZ action is an extension of \eqref{WZ-transgression-CFT} to $\cn=1$ supersymmetric CFTs. 

We begin this section by evaluating the WZ transgression formula \eqref{super-transgression}, pointing out and clarifying several technical subtleties. Subsequently, we verify that the resulting WZ action reproduces the coset anomalies $\ca_W$, $\ca_R$ and $\ca_S$. Finally, we discuss the transformation of the WZ action under the Q-supersymmetry of the unbroken base symmetry algebra, namely Poincar\'e supersymmetry. In particular, we argue that $\cw[e,A,\j]-\cw_\text{WZ}[e,A,\j,\t,\c,\vf]$ is a functional of the coset-invariant old-minimal supergravity fields, $\hat e_\m{}^a$, $\hat A_\m$ and $\hat \j_\m$ only, and we construct a local counterterm that removes the Poincar\'e supersymmetry anomaly in old-minimal supergravity. The construction of the counterterm is facilitated by essential use of the super-Weyl symmetry discussed at the end of the previous section.

\subsection{Evaluation of the WZ transgression formula }

In order to evaluate the WZ transgression formula \eqref{super-transgression} we must determine the homotopies of the elementary and composite fields it depends on. Moreover, the computation can be simplified drastically by a judicial choice of the scheme parameters $s_1$ and $s_2$ on which the coset anomalies in eqs.~\eqref{W-anomaly}, \eqref{R-anomaly} and \eqref{Q-anomaly} depend. As we will see, a certain choice of the scheme parameters ensures that several anomaly terms acquire well defined transformations under the coset symmetries, which significantly simplifies the evaluation of the transgression formula. After determining the homotopies of the elementary and composite fields, therefore, we will discuss the optimal choice of the scheme parameters and will decompose the anomalies into terms that possess well defined transformations under the coset symmetries. We will then proceed with the actual evaluation of the WZ transgression formula \eqref{super-transgression}.

\subsubsection{Elementary and composite homotopies}
\label{sec:homotopies}

The elementary gauge field homotopies $e_\m{}^{a(-t\t,-t\c,-t\vf)}$, $A^{(-t\t,-t\c,-t\vf)}_\m$ and $\psi^{(-t\t,-t\c,-t\vf)}_{\m }$ can be read off from the invariant fields \eqref{eq:InvariantFields}, namely
\be\label{t-fields}
\begin{split}
 e^{(t)a}_\m\equiv&\; e_\m{}^{a(-t\t,-t\c,-t\vf)}=e^{-t\t}\,e_\m{}^a\,,\\
 \psi^{(t)}_{\m } \equiv &\; \psi^{(-t\t,-t\c,-t\vf)}_{\m }= U^t(\psi_{\m }+\,t\g_\m \vf)\,,\\
  A^{(t)}_\m \equiv &\;A^{(-t\t,-t\c,-t\vf)}_\m = A_\m-t\,\del_\m\c+\tfrac{3i}{4}\,t \bar\psi_{\m}\g^5\vf-\tfrac{3i}{8}t^2 \,\bar\vf\g_\m\g^5\vf\,.
\end{split}
\ee
These quantities enter as arguments of the coset anomalies in the transgression formula \eqref{super-transgression}. Namely, in writing \eqref{super-transgression} we have used the shorthand notation
$\ca^{(t)}_W\equiv \ca_W[e^{(t)},A^{(t)}, \j^{(t)}]$, $\ca_R^{(t)}\equiv \ca_R[e^{(t)},A^{(t)}, \j^{(t)}]$ and $\ca_S^{(t)}\equiv \ca_S[e^{(t)},A^{(t)}, \j^{(t)}]$. The transgression formula depends also on the dilatino homotopy $\vf^{(t)}$, which can be read off \eqref{inv-varphi}, namely
\be\label{t-varphi}
\vf^{(t)}\equiv \vf^{(-t\t,-t\c,-t\vf)}= U^{-t}\vf\,.
\ee

Next, let us consider the homotopies of composite fields that enter in the expressions for the coset anomalies. From \eqref{eq:hatomega-hatphi} we read off the spin connection homotopy
\be\label{t-omega}
\o^{(t)}_\m{}^{ab}\equiv \o_\m{}^{ab}(e^{(t)},\j^{(t)}) =\o_\m{}^{ab}(e,\j)-2te_\m{}^{[a}e^{b]\n}(\pa_\n\t+\tfrac12\bar\vf\j_\n)-\tfrac{1}{2}t\bar\vf\g^{ab}(\j_\m+\tfrac{1}{2}t\g_\m\vf)\,.
\ee
In the following it will sometimes be useful to isolate the dilatino-dependent fermion bilinear terms by writing
\be\label{t-omega-deltaK}
\o_\m{}^{(t)ab}=\o_\m{}^{ab}(e,\j)-2te_\m{}^{[a}e^{b]\n}\pa_\n\t+\D K^{(t)}_{\m}{}^{ab}\,,
\ee
where 
\bbxd
\vskip.3cm
\be\label{deltaK}
\D K^{(t)}_{\m}{}^{ab}=-te_\m{}^{[a}e^{b]\n}\bar\vf\j_\n-\tfrac{1}{2}t\bar\vf\g^{ab}(\j_\m+\tfrac{1}{2}t\g_\m\vf)\,,
\ee
\ebxd
is the $t$-dependent shift of the contorsion\footnote{Recall that the spin connection can be expressed in terms of the contorsion tensor $K_\m{}^{ab}$ as  $\o_\m{}^{ab}(e,\j)=\o_\m{}^{ab}(e)+K_\m{}^{ab}(e,\j)$ (see appendix \ref{appendix:torsion}), where in this case the contorsion tensor is given by \eqref{contorsion} (see \eqref{composites}). It follows that the spin connection homotopy may be written as $\o_\m{}^{(t)ab}=[\o_\m{}^{ab}(e)]^{(-t\t)}+[K_\m{}^{ab}(e,\j)]^{(-t\t,-t\vf)}$, with $[K_{\m ab}(e,\j)]^{(-t\t,-t\vf)}=K_{\m ab}(e,\j)+\D K^{(t)}_{\m ab}$, and $\D K^{(t)}_{\m ab}$ given by \eqref{deltaK}.} 
\be\label{contorsion}
K_\m{}_{ab}(e,\j)=\tfrac14(\lbar\j_a\g_\m\j_b+\lbar\j_\m\g_a\j_b-\lbar\j_\m\g_b\j_a)\,.
\ee

From \eqref{eq:hatomega-hatphi} follows also the homotopy of the composite gauge field $\f_\m$, namely
\be\label{t-phi}
\begin{split}
\f_\m^{(t)}=&\;U^{-t}\big[\f_\m-t\cd_\m\vf+\tfrac18t^2(\bar\vf\g_\n\g^5\vf)\g^\n\g^5(\j_\m-t\g_\m\vf)\\
&\;\hspace{.8cm}+\big(-\tfrac12t(\pa_\n\t+\tfrac12\bar\vf\j_\n)\g^\n+\tfrac14t^2(\bar\vf\vf)+\tfrac14t^2(\bar\vf\g^5\vf)\g^5\big)(\j_\m+t\g_\m\vf)\big]\,.
\end{split}
\ee 
However, we only need the terms linear in the fermions for the subsequent computation, i.e. 
\be\label{t-phi-linear}
\f_\m^{(t)}=U^{-t}\big(\f_\m-t\cd_\m\vf-\tfrac12t\pa_\n\t\g^\n(\j_\m+t\g_\m\vf)+\co(\text{fermions}^3)\big)\,.
\ee 
This can be easily verified using the expression for $\f_\m$ in \eqref{composites} and the linear homotopy 
\be\label{t-Dpsi-linear}
[\cd_\m\j_{\n}]^{(t)}=U^t\big(\cd_\m\j_{\n }+\, t \,\g_\n \cd_\m\vf+\tfrac 12t\,\pa^\r\t\,(\g_{\r}{}_\m-g_{\r\m})(\j_\n+t\g_\n\vf)+\co(\text{fermions}^3)\big)\,.
\ee

The final homotopy we will need is that of the Riemann curvature. Since we have already determined the spin connection homotopy \eqref{t-omega}, it is convenient to express the Riemann tensor homotopy in terms of the spin connection one through the identity 
\be\label{R-loc-coord}
[R_{\m\n\r\s}]^{(t)}=e^{(t)a}_\r e^{(t)b}_\s \,[R_{\m\n ab}(\o)]^{(t)}=e^{-2t\t}e^a_\r e^b_\s [R_{\m\n ab}(\o)]^{(t)}\,.
\ee
Isolating the dilatino-dependent terms as before by writing
\be\label{Riemann-split}
[R_{\m\n\r\s}]^{(t)}=[R_{\m\n\r\s}]^{(-t\t,-t\vf)}=[R_{\m\n\r\s}]^{(-t\t)}+\D R^{(t)}_{\m\n\r\s}\,,
\ee
we find
\begin{subequations}
\bal
[R_{\m\n\r\s}]^{(-t\t)}=&\; e^{-2t\t}\big(R_{\m\n\r\s}+2tg_{\m[\r}\nabla_{\s]}\nabla_\n\t-2tg_{\n[\r}\nabla_{\s]}\nabla_\m\t\NO\\
&\hspace{1.5cm}+2t^2g_{\m[\r}\nabla_{\s]}\t\nabla_\n\t-2t^2g_{\n[\r}\nabla_{\s]}\t\nabla_\m\t-2t^2g_{\m[\r}g_{\s]\n}(\nabla\t)^2\NO\\
&\hspace{1.5cm}-2t\, T_{\m\n[\r}\nabla_{\s]}\t+2t\,g_{\m[\r}T_{\s]\n}{}^\k\nabla_\k\t-2t\,g_{\n[\r}T_{\s]\m}{}^\k\nabla_\k\t\big)\,,\label{Riemann-tau}\\
\D R^{(t)}_{\m\n\r\s}=&\; e^{-2t\t}\big(\nabla_{[\m}\D K^{(t)}_{\n]\r\s}+2t\,\D T^{(t)}_{\m\n[\s}\nabla_{\r]}\t+2t\,g_{\r[\m}\D K^{(t)}_{\n]\s}{}^{\k}\nabla_\k\t\NO\\
&\hspace{3.5cm}- 2t\,g_{\s[\m}\D K^{(t)}_{\n]\r}{}^{\k}\nabla_\k\t+2\D K^{(t)}_{[\m|\r|}{}^{\k}\D K^{(t)}_{\n]\k\s}\big)\,,\label{Riemann-tauvf}
\eal
\end{subequations}
where $T_{\m\n}{}^\r$ is the torsion tensor \eqref{eq:torsion}, $\D K^{(t)}_{\m}{}^{ab}$ is given in \eqref{deltaK}, and $\D T^{(t)}_{\m\n\r}\equiv-\D K^{(t)}_{\m\n\r}+\D K^{(t)}_{\n\m\r}$, as follows from the relation \eqref{torsion-contorsion} between the torsion and the contorsion. Note also that the notation $|\r|$ in the last term in \eqref{Riemann-tauvf} means that this index is not antisymmetrized.
 
From \eqref{Riemann-tau}-\eqref{Riemann-tauvf} follow the corresponding expressions for the Ricci tensor and scalar:
\begin{subequations}
\bal
[R_{\m\n}]^{(-t\t)}=&\; R_{\m\n}+t(2\nabla_\m\nabla_\n\t+g_{\m\n}\square\t)+2t^2\big(\nabla_\m\t\nabla_\n\t-g_{\m\n}(\nabla\t)^2\big)\NO\\
&-2t T_{\m}{}^\r{}_{[\n}\nabla_{\r]}\t\,,\label{Ricci-tau}\\
\D R^{(t)}_{\m\n}= &\; \nabla_{[\m}\D K^{(t)}_{\r]\n}{}^{\r}+t\, \D T^{(t)}_{\m\r}{}^\r\nabla_\n\t-t\,\D T^{(t)}_{\m\r\,\n}\nabla^\r\t\NO\\
&\qquad+t \,g_{\m\n}\D K^{(t)}_{\r}{}^{\r\s}\nabla_\s\t+2t\,\D K^{(t)}_{\m\n}{}^\r\nabla_\r\t+2g_{\n\s}\D K^{(t)}_{[\m}{}^{\s\k}\D K^{(t)}_{\r]\k}{}^\r\,,\label{Ricci-tauvf}\\
\rule{0.0cm}{0.6cm}
[R]^{(-t\t)}= &\; e^{2t\t}\big(R+6t\square\t-6t^2(\nabla\t)^2-2t\,T_{\m\r}{}^\m\nabla^\r\t\big)\,,\label{R-tau}\\
\D R^{(t)}=&\;e^{2t\t}(\nabla_{\m}\D K^t_{\r}{}^{\m\r}+2t\,\D T^{(t)}_{\r\m}{}^\m\nabla^\r\t+6t\D K^{(t)}_\r{}^{\r\k}\nabla_\k\t)\NO\\
&+\D K^{(t)}_\m{}^{\m\r}\D K^{(t)}_{\n\r}{}^\n-\D K^{(t)}_\m{}^{\n\r}\D K^{(t)}_{\n\r}{}^\m\,.\label{R-tauvf}
\eal
\end{subequations}
Notice that setting the torsion and dilatino to zero in \eqref{Ricci-tau} and \eqref{R-tau} we recover the expressions \eqref{t-Ricci}, while \eqref{Ricci-tauvf} and \eqref{R-tauvf} vanish.
In the following analysis we will need all curvatures to quadratic order in the fermions, so the quartic terms $(\D K)^2$ may be dropped.

\subsubsection{Symmetry decomposition of coset anomalies}

Before proceeding with the evaluation of the transgression formula \eqref{super-transgression}, let us take a closer look at the structure of the coset anomalies \eqref{W-anomaly}, \eqref{R-anomaly} and \eqref{Q-anomaly}. A key observation that helps considerably simplify the evaluation of the transgression formula is that setting the scheme parameter $s_2=1$ renders several terms in the coset anomalies local superconformal invariants. No significant simplification follows from any specific choice of the scheme parameter $s_1$, which we therefore keep generic.

The simplifications the scheme $s_2=1$ affords us can be identified by decomposing the anomalies as 
$\ca_W=\ca_W^{(A)}+\ca_W^{(B_1)}+\ca_W^{(B_2)}$, $\ca_R=\ca_R^{(F)}+\ca_R^{(\cp)}$ and $\ca_S=\ca^{(FP)}_S+\ca_S^{(W)}$, where 
\bal\label{eq:coset-anomaly-split-s1}
\ca_W^{(A)}=&\;-\tfrac{a}{16\p^2}\big[E+\nabla_\m\big(8i\e^{\m\n\r\s}\bar\f_\n\g^5\cd_\r\psi_\s-8\bar\f^\m\g^\n\f_\n+\tfrac{4\blue{s_1}}{3}\wt F^{\r\s}\bar\j_\r\g^\m\j_\s\big)\big]+\co(\j^4)\,,\NO\\
\ca_W^{(B_1)}=&\;\tfrac{c}{16\p^2}\big[W^2-\tfrac83F^2+8i\e^{\m\n\r\s}\bar\f_\m\g^5\g_\n\cd_\r\f_\s+8i\e^{\m\n\r\s}\nabla_\m(\bar\f_\n\g^5\cd_\r\j_\s)\NO\\
&\;-2R^{\m\n}\big(\bar\j^\r\g_{\r\n}\f_\m-\bar\j_\m\g_{\r\n}\f^\r-\tfrac23g_{\m\n}\bar\j^\r\g_{\r\s}\f^\s+2\bar\j^\r\g_\n(\cd_{[\m}\j_{\r]}-\g_{[\m}\f_{\r]})\big)\NO\\
&\;+8\wt F^{\m\n}\bar\j_\m\f_\n+\tfrac{16i}{3}F^{\m\n}\bar\j_\m\g^\r\g^5\cd_{[\n}\j_{\r]}-\tfrac43\wt F^{\m\n}\bar\j^\r(\g_\r\cd_\m\j_\n+2\g_\m\cd_{[\n}\j_{\r]})\big]+\co(\j^4)\,,\NO\\
\ca_W^{(B_2)}=&\;\tfrac{c}{16\p^2}\nabla_\m(\tfrac{4\blue{s_1}}{3}\wt F^{\r\s}\bar\j_\r\g^\m\j_\s)+\co(\j^4)\,,\\
\rule{0cm}{.8cm}\ca_R^{(F)}=&\;\tfrac{5a-3c}{27\p^2}\;\wt F_{\m\n} F^{\m\n}\NO\\
\ca_R^{(\cp)}=&\;\tfrac{(c-a)}{24\p^2}\big[\cp-\tfrac{1}{3}\e^{\m\n\r\s}\nabla_\m\big(12\bar\f_\n(\g_\r\f_\s-2\cd_{\r}\j_\s)+i F_{\r\s}\bar\j_\k\g^5(2\g^\k\j_\n+\g_\n\j^\k)\big)\big]+\co(\j^4)\,,\NO\\
\rule{0cm}{.8cm}\ca_{S}^{(FP)}=&\;\tfrac{(3+2\blue{s_1})a-(1+2\blue{s_1})c}{6\p^2}\wt F^{\m\n}\cd_\m\j_{\n}-\tfrac{i(5a-3c)}{9\p^2}\wt F^{\m\n}A_\m\g^5\j_{\n}\NO\\
&-\tfrac{ic}{6\p^2} F^{\m\n}\g^{\s}\g_{\m}\g^5\cd_{[\n}\j_{\s]}+\tfrac{3a}{4\p^2}P_{\m\n}g^{\m[\n}\g^{\r\s]}\cd_\r\j_\s+\co(\j^3)\,,\NO\\
\ca_{S}^{(W)}=&\;\tfrac{(a-c)}{8\p^2}W^{\m\n\r\s}\g_{\m\n}\cd_\r\j_\s+\co(\j^3)\,.\NO
\eal
In table \ref{anomaly-table} we summarize the transformation properties of each of these terms under the coset symmetries. 
\begin{table}
\bebx\NO
\begin{tabular}{c|ccccccc}
$e \ca_A$ & $e \ca_W^{(A)}$ & $e \ca_W^{(B_1)}$ &  $e\ca_W^{(B_2)}$ & $e \ca_R^{(F)}$ &   $e \ca_R^{(\cp)}$ & $e U\hspace{-2pt}\ca_S^{(FP )}$ & $e U\hspace{-2pt}\ca_S^{(W)}$\vspace{.1cm}\\
\hline\hline\rule{.0cm}{.5cm} 
$\d_W$ & $\neq0$ & $0$ &  $0$ & $0$ & $0$ & $\neq0$ & $0$ \vspace{.1cm}\\  
\hline\rule{.0cm}{.5cm} 
$\d_R$ & $0$ & $0$ &  $0$ & $0$ & $0$ & $\neq0$ & $0$ \vspace{.1cm}\\   
\hline\hline\rule{.0cm}{.5cm} 
$\d_S$ & $\neq0$ & $0$ &  $\neq0$ & $\neq0$ & $0$ & $\neq0$  & $\neq0$
\end{tabular}
\eebx
\caption{Transformation properties of the various terms in the coset anomalies under the coset symmetries in the one-parameter family of schemes corresponding to $\red{s_2}=1$ and arbitrary $\blue{s_1}$. Notice that $e \ca_W^{(B_1)}$ and $e \ca_R^{(\cp)}$ become local superconformal invariants in this family of schemes. Note that the S-supersymmetry anomaly terms have been multiplied by the matrix $U$ in \eqref{eq:U} in order to compensate for the transformations of the corresponding spinor parameter.}
\label{anomaly-table}
\end{table}

The transformation properties in table \ref{anomaly-table} lead to a significant simplification of the evaluation of the transgression formula \eqref{super-transgression} due to the following key observation. For any local superconformal invariant term $e\,\ca[e,A,\j]$, such as $e \ca_W^{(B_1)}$ and $e \ca_R^{(\cp)}$, the transgression becomes trivial, namely  
\be
e^{(t)}\ca[e^{(t)},A^{(t)},\j^{(t)}]=e\,\ca[e,A,\j]\,.
\ee
Hence, the integral over $t$ simply gives 1. However, the transgression simplifies even if a term is invariant under only a subset of the coset symmetries. For example, for a term $e\,\ca[e,A,\j]$ that is W- and R-invariant but not S-invariant, such as $e\ca_W^{(B_2)}$, $e \ca_R^{(F)}$ or $e \ca_S^{(W)}$, we have
\be
e^{(t)}\ca[e^{(t)},A^{(t)},\j^{(t)}]=e\,\ca[e^{(0,0,-t\vf)},A^{(0,0,-t\vf)},\j^{(0,0,-t\vf)}]\,.
\ee
Namely, all $\t$ and $\c$ dependence drops out.

In order to utilize these simplifications, we decompose the WZ action according to the decomposition of the coset anomalies in \eqref{eq:coset-anomaly-split-s1}, i.e. 
\bbxd
\vskip.3cm
\be\label{eq:wz-decomposition}
\cw_\text{WZ}=\cw_W^{(A)}+\cw_W^{(B_1)}+\cw_W^{(B_2)}+\cw_R^{(F)}+\cw_R^{(\cp)}+\cw_S^{(FP )}+\cw_S^{(W)}\,.
\ee
\ebxd
The transformation properties of each term under the coset symmetries in table \ref{anomaly-table} lead to the simplified transgression formulas
\bbxd
\be\label{eq:wz-decomposed-transgressions}
\begin{split}
\cw_W^{(A)}=&\;\int_0^1dt\int d^4x\,e^{(t)}\,\t \ca_W^{(A)(-t\t,-t\vf)}\,,\\
\cw_W^{(B_1)}=&\;\int d^4x\,e\,\t\ca_W^{(B_1)}\,,\\
\cw_W^{(B_2)}=&\;\int_0^1dt\int d^4x\,e\,\t\ca_W^{(B_2)(-t\vf)}\,,\\
\cw_R^{(F)}=&\;-\int_0^1dt\int d^4x\,e\,\c\ca_R^{(F)(-t\vf)}\,,\\
\cw_R^{(\cp)}=&\;-\int d^4x\,e\,\c\ca_R^{(\cp)}\,,\\
\cw_S^{(FP)}=&\;\int_0^1dt\int d^4x\,e^{(t)}\,\bvf^{(t)} \ca_S^{(FP)(-t\t,-t\c,-t\vf)}\,,\\
\cw_S^{(W)}=&\;\int_0^1dt\int d^4x\,e\,\bvf\,\ca_S^{(W)(-t\vf)}\,.
\end{split}
\ee
\ebxd
Notice that the transgression integral is completely eliminated from $\cw_W^{(B_1)}$ and $\cw_R^{(\cp)}$, which give directly the corresponding contribution to the WZ action. Our next task is to explicitly evaluate the remaining reduced transgressions in \eqref{eq:wz-decomposed-transgressions}.

\subsubsection{Evaluation of the transgression formula}

\subsubsection*{Evaluation of $\cw_W^{(A)}$}

Using the expression for $\ca_W^{(A)}$ in \eqref{eq:coset-anomaly-split-s1} and integrating by parts the fermion bilinear terms, the first reduced transgression in \eqref{eq:wz-decomposed-transgressions} takes the form
\bal
\cw_W^{(A)}=&\;-\frac{a}{16\p^2}\int_0^1dt\int d^4x\,e^{(t)}\,\t E^{(t)}\\
&\;+\frac{a}{16\p^2}\int_0^1dt\int d^4x\,\pa_\m\t \big[e\big(8i\e^{\m\n\r\s}\bar\f_\n\g^5\cd_\r\psi_\s-8\bar\f^\m\g^\n\f_\n+\tfrac{4\blue{s_1}}{3}\wt F^{\r\s}\bar\j_\r\g^\m\j_\s\big)+\co(\j^4)\big]^{(t)}.\NO
\eal
Notice that the integration by parts generates additionally a torsion term (see \eqref{tor-der}), which in this case contributes only quartic in the fermion terms that we ignore. 

Let us first consider the Euler density transgression. In contrast to the non-supersymmetric case we discussed in section \ref{sec:CFTs}, the square of the Weyl tensor is not Weyl invariant on its own -- only its supersymmetric completion, $\ca_W^{(B_1)}$, is Weyl invariant -- and so expressing the Euler density in terms of the square of the Weyl tensor ceases to be a useful approach in the supersymmetric case. Hence, we work directly with the Euler density and decompose it as   
\be
E^{(t)}=E^{(-t\t)}+\D E^{(t)}\,,
\ee
with
\be
E^{(-t\t)}=[R_{\m\n\r\s}R^{\m\n\r\s}-4R_{\m\n}R^{\m\n}+R^2)]^{(-t\t)}\,,
\ee
and 
\be
\D E^{(t)}=2 [R_{\m\n\r\s}]^{(-t\t)}\D R^{(t)\m\n\r\s}-8[R_{\m\n}]^{(-t\t)}\D R^{(t)\m\n}+2[R]^{(-t\t)}\D R^{(t)}+\co(\text{fermions}^4)\,,
\ee
where $[R_{\m\n\r\s}]^{(-t\t)}$, $[R_{\m\n}]^{(-t\t)}$ and $[R]^{(-t\t)}$ are given respectively in
\eqref{Riemann-tau}, \eqref{Ricci-tau} and \eqref{R-tau}, while $\D R^{(t)\m\n\r\s}$, $\D R^{(t)\m\n}$ and $\D R^{(t)}$ are given in \eqref{Riemann-tauvf}, \eqref{Ricci-tauvf} and \eqref{R-tauvf}. Notice that we have dropped all $(\Delta R)^2$ terms in $\D E^{(t)}$ since they are quartic in the fermions. 

Performing the $t$ integration and simplifying the result we arrive at the following expressions for the Euler term transgression:
\bbxd
\be\label{WA-E}
\begin{split}
&-\tfrac{a}{16\p^2}\int_0^1dt\int d^4x\,e^{(t)}\,\t E^{(-t\t)}=\\
&\hspace{1.cm}-\tfrac{a}{16\p^2} \int d^4x\,e\,\big[\t E+4( R^{\m\n}-\tfrac12  R g^{\m\n})\nabla_\m\t\nabla_\n\t-4(\nabla\t)^2\square\t+2(\nabla\t)^4\\
&\hspace{10.0cm}-\tfrac43\t \,T_{\m\n}{}^\n\t(\nabla\t)^2\nabla^\m\t\big]\,,
\end{split}
\ee
\ebxd
and
\bbxd
\be\label{WA-deltaE}
\begin{split}
&-\tfrac{a}{16\p^2}\int_0^1dt\int d^4x\,e^{(t)}\,\t \D E^{(t)}=\\
&\hspace{1.cm}\tfrac{a}{16\p^2}\int d^4 x\,e\,\t\big[2 R^{\m\n}\big(\nabla_{\r}(\bar\vf\g_\m\g^\r\,\j_{\n})-\nabla_\m( \bar\vf\g_{\n}{}^\r\j_{\r})\big)-R\,\nabla_\m(\bvf\,\g^\n\g^\m\j_\n)\\
&\hspace{1.cm}-2 R^{\m\n\r\s}\nabla_\m(\bvf\g_{\n\r}\j_\s)+\tfrac43 R^{\m\n}\bvf(2\d_\n^\r+ \g^\r{}_\n)\j_\r \,\nabla_\m\t -\tfrac43R^{\m\n\r\s}\,\nabla_\m\t\,\bvf\g_{\n\r}\j_\s\\
&\hspace{1.cm}-\tfrac43\nabla^{\n}(\bvf(2\d^\r_\n+\g^{\r}{}_{\n})\j_{\r})\nabla^2\t+\tfrac83\nabla_{[\m}(\bar\vf\g^{\r\n}\j_{\r]})\nabla_\n\nabla^\m\t+\tfrac83\nabla_\m(\bar\vf\j^\n)\nabla_\n\nabla^\m\t\\
&\hspace{1.cm}+\nabla_{\m}\big(\bar\vf\j^\m(\nabla\t)^2+\bar\vf\g^\n\g^\r\j_\n\nabla^\m\t\nabla_\r\t+\bar\vf\g^\r\g^\m\j_\n\nabla^\n\t\nabla_\r\t\big)\big]\,.
\end{split}
\ee
\ebxd
These results are correct to quadratic order in the fermions. Setting the torsion to zero in \eqref{WA-E} reproduces the corresponding terms in the non-supersymmetric WZ action in \eqref{eq:4DWZ}.

The fermion transgression can be evaluated similarly using the homotopies we computed earlier in section \ref{sec:homotopies}. The result is
\bbxd
\be\label{WA-fermions}
\begin{split}
&\tfrac{a}{16\p^2}\int_0^1dt\int d^4x\,\nabla\t\,[\text{fermions}]^{(t)}=\\
&\hspace{1.cm}\tfrac{a}{16\p^2}\int d^4x\,e\,\nabla_\m\t \Big( i\e^{\m\n\r\s}\big[2\bar\f_\n\g^5\big(4\cd_\r\j_{\s}+2\g_\s \cd_\r\vf-\nabla^\k\t\,\g_\r\g_\k(\j_\s+\tfrac23\g_\s\vf)\big)\\
&\hspace{1.cm}-\bvf \overset{\leftarrow}{\cd_\n}\g^5\big(4\cd_\r\j_{\s}+\tfrac83\g_\s \cd_\r\vf-\nabla^\k\t\,\g_\r\g_\k(\tfrac43\j_\s+\g_\s\vf)\big)\\
&\hspace{1.cm}+\nabla_\k\t\,\bar\j_\n\g^\k\g^5\big(2\cd_\r\j_{\s}+ \tfrac43\g_\s \cd_\r\vf\big)-\nabla_\k\t\,\bvf\g_\n\g^\k\g^5\big(\tfrac43\cd_\r\j_{\s}+\g_\s \cd_\r\vf\big)\big]\\
&\hspace{1.cm}-2\bphi^\m\g^\n\big(4\f_\n- 2\cd_\n\vf-\g^\r(\j_\n+\tfrac23\g_\n\vf)\nb_\r\t\big)\\
&\hspace{1.cm}+(\bvf\overset{\leftarrow}{\cd{\,}^\m}-\tfrac12\nabla_\a\t\,\bar\j^\m\g^\a)\g^\n \big(4\f_\n- \tfrac83\cd_\n\vf-\g^\r(\tfrac43\j_\n+\g_\n\vf)\nb_\r\t\big)\\
&\hspace{1.cm}+\nabla_\r\t\,\bar\vf\g^\m\g^\r\g^\n\big(\tfrac43\f_\n-\cd_\n\vf-\tfrac12\nabla_\s\t\,\g^\s\j_\n\big)\\
&\hspace{1.cm}+\tfrac{4\blue{s_1}}{3}\widetilde F^{\r\s}(\bar\j_\r\g^\m\j_\s-\bvf\g_\r\g^\m\j_\s-\tfrac13\bvf\g_\r\g^\m\g_\s\vf)\Big)\,.
\end{split}
\ee
\ebxd
Again, this result is correct to quadratic order in the fermions.

\subsubsection*{Evaluation of $\cw_W^{(B_2)}$}

Moving on to $\cw_W^{(B_2)}$, from \eqref{eq:coset-anomaly-split-s1} and  \eqref{eq:wz-decomposed-transgressions} follows that 
\be
\cw^{(B_2)}_W = \frac{c}{16\p^2}\int_0^1dt\int d^4x\,e\,\t\nabla_\m\big(\tfrac{4\blue{s_1}}{3}\wt F^{\r\s}\bar\j_\r\g^\m\j_\s+\co(\j^4)\big)^{(-t\vf)}\,,
\ee
which coincides with the last line in \eqref{WA-fermions}. Namely, up to terms quartic in the fermions,
\bbxd
\vspace{0.3cm}
\be
\label{WB2}
\cw^{(B_2)}_W=-\frac{\blue{s_1}c}{12\p^2}\int d^4x\,e\,\nabla_\m\t\,\widetilde F^{\r\s}(\bar\j_\r\g^\m\j_\s-\bvf\g_\r\g^\m\j_\s-\tfrac13\bvf\g_\r\g^\m\g_\s\vf)\,.
\ee
\ebxd

\subsubsection*{Evaluation of $\cw_R^{(F)}$}

From \eqref{eq:coset-anomaly-split-s1} and \eqref{eq:wz-decomposed-transgressions} we get
\be
\cw_R^{(F)}=-\frac {5a-3c}{27\p^2}\int_0^1dt \int d^4x\,e\,\c\, [\wt F F]^{(-t\vf)}\,,
\ee
while the homotopies \eqref{t-fields} determine that
\be\label{FF}
[\wt F F]^{(-t\vf)}=
\wt FF+\wt F^{\m\n}\del_\m(-\tfrac{3i}{2}t^2 \,\bar\vf\g_\n\g^5\vf+3it\,\bar\psi_{\n }\g^5\,\vf)\,.
\ee
Inserting this in the transgression formula, integrating by parts the fermion terms and using the Bianchi identity (see appendix \ref{appendix:torsion}) gives 
\bbxd
\vspace{-.1cm}
\be\label{cwRfinal}
\begin{split}
\cw_R^{(F)}=&- \int d^4x\,e\,\c\,\ca_R^{(F)}+\frac {5a-3c}{27\p^2} \int d^4x\,e\,\nabla_\m\c\,\wt F^{\m\n}(-\tfrac{i}{2} \,\bar\vf\g_\n\g^5\vf+\,\tfrac{3i}{2}\,\bar\psi_{\n}\g^5\,\vf)\,.
\end{split}
\ee
\ebxd
Both the integration by parts and the Bianchi identity contribute torsion terms, so this result holds again up to quartic in the fermions terms.

\subsubsection*{Evaluation of $\cw_S^{(FP)}$}

Since the term $\ca_S^{(FP)}$ of the S-supersymmetry anomaly in \eqref{eq:coset-anomaly-split-s1} is not invariant under any coset symmetry, the corresponding transgression in \eqref{eq:wz-decomposed-transgressions} cannot be reduced -- it necessarily involves the homotopies of all fields. However, there is a useful decomposition of $\cw_S^{(FP)}$ when working to quadratic order in the fermions. Evaluating the transgression integral at zero dilaton and axion results in a compact expression in terms of the original anomaly $\ca_S^{(FP)}$. In particular, since $\ca_S^{(FP)}$ depends linearly on the gravitino in our approximation, the transgression $\j_\m^{(t)}$ in \eqref{t-fields}, evaluated at zero $\t$ and $\c$, amounts to the linear shift $\j_\m\to\j_\m+t\g_\m\vf$. Integration over the parameter $t$ then simply gives   
\be\label{WS-FP-A}
\int_0^1dt\int d^4x\,e\,\bvf \ca_S^{(FP)(-t\vf)}=\int d^4x \,e\,\bar\vf \ca^{(FP)}_S[e,A,\j+\tfrac12\g\vf]\,.
\ee
This relies only on the linear dependence of $\ca^{(FP)}_S[e,\j,A]$ on $\j_\m$ and not on its specific form. The complete transgression integral $\cw_S^{(FP)}$ can therefore be decomposed as 
\bbxd
\vskip.4cm
\be\label{WS-FP}
\cw_S^{(FP)}=\int d^4x \,e\,\bar\vf \ca^{(FP)}_S[e,A,\j+\tfrac12\g\vf]+\D \cw_S^{(FP)}\,,
\ee
\ebxd
where $\D \cw_S^{(FP)}$ accounts for all the dilaton and axion dependence.

All terms in $\D \cw_S^{(FP)}$ can be evaluated using the elementary and composite homotopies in section \eqref{sec:homotopies}. After extensive spinor and gamma matrix manipulations we obtain
\bbxd
\be\label{WS-deltaFP}
\begin{split}
\D \cw_S^{(FP)}=&\; -\tfrac{(3+2\blue{s_1})a-(1+2\blue{s_1})c}{24\p^2}\int d^4x\,\,e\,\big(\widetilde F^{\m\n}\,\nabla^\r\t\,\bar\vf\g_\m\g_\r\j_{\n}+\tfrac{4i}{3}F^{\m\n}\nabla_\m\t\,\bar\vf\g_\n\g^5\vf\big)\\
&+\tfrac{i(5a-3c)}{18\p^2} \int d^4x\,e\,\wt F^{\m\n}\nb_\m\c\,\bar\vf\g^5(\psi_\n+\tfrac23\g_\n\vf)\\
&+\tfrac{ic}{24\p^2} \int d^4x \,e\, F^{\m\n}\big(\nb_\r\t\,\bvf \g^\s\g_\m\g_{[\n}\g^\r\g^5\j_{\s]}+\tfrac43\nb_\m\t\,\bvf\g^5\g_\n\vf\big)\,,\\
&+\tfrac{3a}{4\p^2} \int d^4x\,e\,\bvf\, g^{\m[\n}\g^{\r\s]}\big[\tfrac12\big(\nabla_\m\nabla_\n\t+\tfrac23\nabla_\m\t\nabla_\n\t-\tfrac13 g_{\m\n}(\nabla\t)^2\big)\cd_\r\j_\s\\
&+\tfrac{1}{3}\big(\nabla_\m\nabla_\n\t+\tfrac34\nabla_\m\t\nabla_\n\t-\tfrac38g_{\m\n}(\nabla\t)^2\big)\g_\s \cd_\r\vf\\
&-\tfrac{1}{4}\big(P_{\m\n}+\tfrac23\nabla_\m\nabla_\n\t+\tfrac12\nabla_\m\t\nabla_\n\t-\tfrac14g_{\m\n}(\nabla\t)^2\big)\nb_\k\t\,\g_\r\g^\k\j_\s\big]\,.
\end{split}
\ee
\ebxd
Once again this result is correct to quadratic order in the fermions.

\subsubsection*{Evaluation of $\cw_S^{(W)}$}

Finally, the reduced transgression for $\cw_S^{(W)}$ in \eqref{eq:wz-decomposed-transgressions} involves only the gravitino homotopy. In analogy to \eqref{WS-FP}, therefore, to quadratic order in the fermions the result of the transgression integral can be expressed in terms of the corresponding anomaly in \eqref{eq:coset-anomaly-split-s1} with a linear shift in the gravitino, namely 
\bbxd
\vspace{.4cm}
\be\label{WS-W}
\cw_S^{(W)}=\int d^4x \,e\,\bar\vf \ca^{(W)}_S[e,A,\j+\tfrac12\g\vf]+\co(\text{fermions}^4)\,.
\ee
\ebxd

\subsection{Coset anomalies from the WZ action}

We have now computed the WZ action to quadratic order in the fermions by evaluating the transgression formula \eqref{super-transgression}. Before proceeding further, however, we should verify that it reproduces the coset anomalies, $\ca_W$, $\ca_R$ and $\ca_S$ that it is designed to encode. To this end, we observe that the WZ action takes the form
\be\label{WZ-split}
\begin{split}
\cw_\text{WZ}[e,A,\j,\t,\c,\vf]=&\;\int d^4x\, e\big(\t \ca_W[e,A,\j]-\c \ca_R[e,A,\j]+\bvf \ca_S[e,A,\j]\big)\\
&\hspace{0.0cm}+\cw^{\text{(NL)}}_\text{WZ}[e,A,\j,\t,\c,\vf]\,,
\end{split}
\ee
where $\cw^{\text{(NL)}}_\text{WZ}$ contains all terms nonlinear in the Goldstone fields, $\t$, $\c$ and $\vf$. Crucially, the only terms linear in the Goldstone fields are those involving their pairing with the anomaly of the corresponding coset symmetry. 

Demanding that the WZ action in the form \eqref{WZ-split} reproduces the coset anomalies implies that the coset transformations of the nonlinear term $\cw^{\text{(NL)}}_\text{WZ}$ satisfy the three conditions
\be\label{NL-transformations}
\begin{split}
&\d_W(\s)\cw^{\text{(NL)}}_\text{WZ}+\int d^4x\,\t\,\d_W(\s)(e\,\ca_W)+\d_W(\s)\int d^4x\,e(-\c \ca_R+\bvf \ca_S)=0\,,\\
&\d_R(\th)\cw^{\text{(NL)}}_\text{WZ}-\int d^4x\,e\, \c\,\d_R(\th)\ca_R+\d_R(\th)\int d^4x\,e(\t\ca_W+\bvf\ca_S)=0\,,\\
&\d_S(\h)\cw^{\text{(NL)}}_\text{WZ}+\int d^4x\,e\,\bvf\, \d_S(\h)\ca_S+\d_S(\h)\int d^4x\,e(\t\ca_W-\c \ca_R)=0\,.
\end{split}
\ee
Using the decomposition \eqref{eq:coset-anomaly-split-s1} of the anomalies and the symmetry properties of each term listed in table \eqref{anomaly-table}, the conditions \eqref{NL-transformations} become
\be\label{NL-transformations-simple}
\begin{split}
&\d_W(\s)\cw^{\text{(NL)}}_\text{WZ}+\int d^4x\,\t\,\d_W(\s)(e\,\ca_W^{(A)})+\d_W(\s)\int d^4x\,e\,\bvf \ca_S^{(FP)}=0\,,\\
&\d_R(\th)\cw^{\text{(NL)}}_\text{WZ}+\d_R(\th)\int d^4x\,e\,\bvf\ca_S^{(FP)}=0\,,\\
&\d_S(\h)\cw^{\text{(NL)}}_\text{WZ}+\int d^4x\,e\,\bvf\, \d_S(\h)\ca_S+\d_S(\h)\int d^4x\,e(\t\ca_W^{(A)}+\t\ca_W^{(B_2)}-\c \ca_R^{(F)})=0\,.
\end{split}
\ee

Verifying these transformations using the explicit form of $\cw^{\text{(NL)}}_\text{WZ}$ we determined earlier confirms that the WZ action reproduces the coset anomalies, $\ca_W$, $\ca_R$ and $\ca_S$. We have checked these transformations with FieldsX \cite{Frob:2020gdh}, but it may be instructive to see explicitly how the conditions \eqref{NL-transformations-simple} are satisfied for some specific parts of the WZ action. 

\subsubsection*{$\ca_R$ anomaly}

The simplest example is the second equation in \eqref{NL-transformations-simple}. The terms that contribute to the R-symmetry transformation of the nonlinear part $\cw^{\text{(NL)}}_\text{WZ}$ of the WZ action are the terms in $\cw_R^{(F)}$ in \eqref{cwRfinal} and $\cw_S^{(FP)}$ in \eqref{WS-FP}-\eqref{WS-deltaFP} that involve $A_\m$ or $\nabla_\m\c$. 
\be\label{eq:NLWZ-Achi}
\begin{split}
\cw^{\text{(NL)}}_\text{WZ}|_{A,\c}=&\;\frac {5a-3c}{27\p^2} \int d^4x\,e\,\wt F^{\m\n}\nb_\m\c\,(-\tfrac{i}{2} \,\bar\vf\g_\n\g^5\vf+\,\tfrac{3i}{2}\,\bar\psi_{\n}\g^5\,\vf)\\
&\;-\frac{i(5a-3c)}{18\p^2}\int d^4x \,e\,\wt F^{\m\n}A_\m\bar\vf\g^5\g_\n\vf\\
&\;+\frac{i(5a-3c)}{18\p^2} \int d^4x\,e\,\wt F^{\m\n}\nb_\m\c\,\bar\vf\g^5(\psi_\n+\tfrac23\g_\n\vf)\\
=&\;\frac{i(5a-3c)}{9\p^2} \int d^4x\,e\,\wt F^{\m\n}\big(\nb_\m\c\,\bar\vf\g^5\psi_\n+\tfrac12(A_\m-\nabla_\m\c)\bar\vf\g_\n\g^5\vf\big)\,.
\end{split}
\ee

Their combined transformation under R-symmetry takes the form
\be
\d_R(\th)\cw^{\text{(NL)}}_\text{WZ}=\frac{i(5a-3c)}{9\p^2}\int d^4x \,e\,\wt F^{\m\n}\pa_\m\th\,\bar\vf\g^5\j_{\n}\,.
\ee
Moreover, 
\be
\begin{split}
\d_R(\th)\int d^4x\,e\,\bvf\ca_S^{(FP)}[e,A,\j]=&\;-\frac{i(5a-3c)}{9\p^2}\int d^4x \,e\,\wt F^{\m\n}\pa_\m\th\,\bar\vf\g^5\j_{\n}\,,
\end{split}
\ee
and therefore the second equation in \eqref{NL-transformations-simple} indeed holds.

\subsubsection*{$\ca_W$ anomaly}

Confirming the full equation for the Weyl transformation of $\cw^{\text{(NL)}}_\text{WZ}$ in \eqref{NL-transformations-simple} is a somewhat involved task that we have carried out using FieldsX. However, the transformation of the terms proportional to the scheme parameter $s_1$ can be easily checked analytically. Collecting all terms of the WZ action proportional to $s_1$ that contribute $\cw^{\text{(NL)}}_\text{WZ}$ gives
\be
\begin{split}
\cw_{\text{WZ}}^{(\text{NL})}\big|_{\blue{s_1}}=&\;-\frac{\blue{s_1}(a-c)}{12\p^2}\int d^4x\,e\,\nabla^\m\t\,\wt F^{\r\s}(2\,\bvf\g_\r\g_\m\j_\s+\tfrac{1}{3}\bvf\g_{\r\m\s}\vf)\\
&\;+\frac{\blue{s_1}(a-c)}{6\p^2}\int d^4x\,e\,\wt F^{\r\s}\bvf\g_\s\cd_\r\vf-\frac{ i\blue{s_1}(a-c)}{9\p^2}\int d^4x\,e\,\nb_\r\t\,F^{\r\s}\bvf\g_\s\g^5\vf\,.
\end{split}
\ee
Using the relation $F^{\r\s}=-\tfrac12\e^{\r\s\m\n}\wt F_{\m\n}$ and gamma matrix identities, this simplifies further to
\be\label{eq:NLWZ-s1}
\cw_{\text{WZ}}^{(\text{NL})}\big|_{\blue{s_1}}=-\frac{\blue{s_1}(a-c)}{6\p^2}\int d^4x\,e\,\wt F^{\r\s}\big(\nabla^\m\t\,(\bvf\g_\r\g_\m\j_\s+\tfrac12\bvf\g_{\r\m\s}\vf)+\bvf\g_\r\cd_\s\vf\big)\,.
\ee

The Weyl transformation of \eqref{eq:NLWZ-s1} is given by
\bal
\d_W(\s)\cw_{\text{WZ}}^{(\text{NL})}\big|_{\blue{s_1}}=&\;-\frac{\blue{s_1}(a-c)}{6\p^2}\int d^4x\,e\,\wt F^{\r\s}\big(\nabla^\m\s\,(\bvf\g_\r\g_\m\j_\s+\tfrac12\bvf\g_{\r\m\s}\vf)+\tfrac12\bvf\g_\r\g_{\s\m}\vf\nabla^\m\s\big)\NO\\
=&\;-\frac{\blue{s_1}(a-c)}{6\p^2}\int d^4x\,e\,\wt F^{\r\s}\nabla^\m\s\,\bvf\g_\r\g_\m\j_\s\,.
\eal
Moreover, from \eqref{eq:coset-anomaly-split-s1} follows that
\be
\begin{split}
\d_W(\s)\int d^4x\,e\,\bvf \ca_S^{(FP)}\big|_{\blue{s_1}}=&\;\frac{2\blue{s_1}(a-c)}{6\p^2}\d_W(\s)\int d^4x\,e\,\bvf\wt F^{\m\n}\cd_\m\j_{\n}\\
=&\;\frac{\blue{s_1}(a-c)}{6\p^2}\int d^4x\,e\,\wt F^{\m\n}\bvf\g_\m\g_\r\j_{\n}\nabla^\r\s\,,
\end{split}
\ee
and hence,
\be
\d_W(\s)\Big(\cw^{\text{(NL)}}_\text{WZ}+\int d^4x\,e\,\bvf \ca_S^{(FP)}\Big)\Big|_{\blue{s_1}}=0\,,
\ee
which confirms the Weyl transformation equation in \eqref{NL-transformations-simple} for the $s_1$ sector of the WZ action.

\subsubsection*{$\ca_S$ anomaly}

Verifying the full S-supersymmetry equation for the WZ action in  \eqref{NL-transformations-simple} is also an elaborate task that we have carried out using FieldsX. However, it can be easily checked for the two sectors we discussed above. Starting with the $A_\m$- and $\c$-dependent part of $\cw^{\text{(NL)}}_\text{WZ}$, given explicitly in \eqref{eq:NLWZ-Achi}, the S-supersymmetry equation in  \eqref{NL-transformations-simple} reduces to
\be
\d_S(\h)\cw^{\text{(NL)}}_\text{WZ}\big|_{A,\c}+\int d^4x\,e\,\bvf\, \d_S(\h)\ca_S^{(FP)}\big|_{A}-\d_S(\h)\int d^4x\,e\,\c \ca_R^{(F)}=0\,.
\ee
To quadratic order in the fermions,
\be
\begin{split}
\d_S(\h)\cw_\text{WZ}^{(\text{NL})}\big|_{A,\c}=&\;\frac {i(5a-3c)}{9\p^2} \int d^4x\,e\,\wt F^{\m\n}
(\nb_\m\c\,\bar\h\g^5\, \psi_{\n}+A_\m\bar\vf\g_\n\g^5\h)\,,\\
\int d^4x e\, \bvf\,\d_S(\h)\ca_S^{(FP)}\big|_{A}=&\;\frac{i(5a-3c)}{9\p^2} \int d^4x\,e\,\wt F^{\m\n}A_\m\,\bvf\g^5\g_\n\vf\,,\\
-\int d^4x \,e\,\c\d_S(\h)\ca_R^{(F)}=&\;\frac {i(5a-3c)}{9\p^2} \int d^4x\,e\, \c\wt F^{\m\n}\,\nb_\m(\bar\h\g^5\j_\n)\,.
\end{split}
\ee
Integrating by parts and using the Bianchi identity these terms sum to zero, as required. 

Similarly, we can check the S-supersymmetry equation in  \eqref{NL-transformations-simple} for the $s_1$ sector of $\cw^{\text{(NL)}}_\text{WZ}$, given explicitly in \eqref{eq:NLWZ-s1}. In this sector, the S-supersymmetry equation in \eqref{NL-transformations-simple} becomes
\be
\Big(\d_S(\h)\cw^{\text{(NL)}}_\text{WZ}+\int d^4x\,e\,\bvf\, \d_S(\h)\ca_S+\d_S(\h)\int d^4x\,e(\t\ca_W^{(A)}+\t\ca_W^{(B_2)})\Big)\Big|_{\blue{s_1}}=0\,.
\ee
Evaluating each term to quadratic order in the fermions we obtain 
\bal
\d_S(\h)\cw_{\text{WZ}}^{(\text{NL})}\big|_{\blue{s_1}}=&\;-\frac{\blue{s_1}(a-c)}{6\p^2}\int d^4x\,e\,\wt F^{\r\s}(\nabla^\m\t\,\bar\h\g_\r\g_\m\j_\s+2\bar\h\g_\r\cd_\s\vf)\,,\NO\\
\int d^4x\,e\,\bvf\, \d_S(\h)\ca_S^{(FP)}\big|_{\blue{s_1}}=&\;-\frac{\blue{s_1}(a-c)}{3\p^2}\int d^4x\,e\,\wt F^{\m\n}\,\bvf\g_\n\cd_\m\h\,,\\
\d_S(\h)\int d^4x\,e(\t\ca_W^{(A)}+\t\ca_W^{(B_2)})\big|_{\blue{s_1}}=&\;-\frac{\blue{s_1}(a-c)}{6\p^2}\int d^4x\,e\,\nabla_\m\t\wt F^{\r\s}\bar\j_\r\g^\m\g_\s\h\,.\NO
\eal
Integrating by parts and using the Bianchi identity these again sum to zero, as required.

\subsection{Poincar\'e supersymmetry and local counterterms}

Having computed the WZ action and verified that it reproduces the coset anomalies, we next look at its transformation under Q-supersymmetry. We will demonstrate that there exists a local counterterm in the space of {\em old-minimal fields} which allows us to define a generalized WZ action that, besides the coset anomalies of $\cw[e,A,\j]$, reproduces also its {\em Poincar\'e} Q-supersymmetry anomaly, i.e. the anomaly corresponding to the Q-supersymmetry of the unbroken Poincar\'e subalgebra we identified in section \ref{sec:PoincareSubalgebra}. Moreover, we will see that the same local counterterm, together with the WZ action we obtained above, can be used to eliminate the Poincar\'e supersymmetry anomaly of the {\em old-minimal} generating functional $\cw[\hat e,\hat A,\hat\j]$, at the expense of introducing an anomaly in the $\L$-symmetry we discussed in section \ref{sec:SuperWeyl}. Throughout this section we comment on the relation of our analysis with the superspace construction of \cite{Kuzenko:2019vvi}.

\subsubsection{Invariant Goldstone kinetic term}

Before discussing the Q-supersymmetry transformation of the WZ action, let us briefly comment on the possible Goldstone kinetic terms that could be added to the WZ action in order to ensure that it has the correct analyticity properties and the Goldstone fields can be consistently integrated out to obtain a non-local anomaly effective action. Recall that for generic four-dimensional CFTs the coset invariant dilaton and axion kinetic terms were given respectively in \eqref{eq:4dCFTWZ-dilaton-kinetic} and \eqref{eq:4dCFTWZ-axion-kinetic}. For $\cn=1$ SCFTs we need in addition an invariant kinetic term for the dilatino, $\vf$. Such a kinetic term is provided by the Rarita-Schwinger term for the invariant gravitino, $\hat \j_\m$.

Adding all three kinetic terms with independent coefficients would preserve the coset symmetries and ensure the correct analyticity properties of the WZ action when viewed as a Goldstone effective action. However, these kinetic terms are not independently invariant under Q-supersymmetry and, therefore, adding them to the WZ action would change its transformation under Q-supersymmetry. However, there exists a unique linear combination of these invariant kinetic terms, augmented by a term involving the auxiliary field $\hat\cf$, that does preserve all symmetries, including Q-supersymmetry. This term is the well known Poincar\'e supergravity action \cite{Ferrara:1978em,Stelle:1978ye,Fradkin:1978jq,Kaku:1978ea}  
\bbxd
\vskip.3cm
\be\label{eq:PoincareAction}
\cw_{\text{kin}}[\hat e,\hat A,\hat\j,\hat\cf]=-\frac{\a^2}{12}\int d^4x\,\hat e\,\big(\hat R-\bar{\hat\j}_\m\hat\g^{\m\n\r}\hat D_\n\hat\j_\r+\frac{8}{3}\hat A^a\hat A_a-6|\hat\cf|^2\big)\,, 
\ee
\ebxd
written here in terms of the {\em coset-invariant} fields. In particular, in contrast to the standard textbook derivation of the Poincar\'e supergravity action by gauge-fixing the coset symmetries (see e.g. \cite{Freedman:2012zz}), there is {\em no gauge-fixing} involved in writing the local action \eqref{eq:PoincareAction}.

Notice that the covariant derivative in the Rarita-Schwinger term in \eqref{eq:PoincareAction} is the simple Lorentz spinor covariant derivative without the R-symmetry connection (see \eqref{covD}). Moreover, comparing with the generic kinetic terms \eqref{eq:4dCFTWZ-dilaton-kinetic} and \eqref{eq:4dCFTWZ-axion-kinetic}, we see that the supersymmetric kinetic term \eqref{eq:PoincareAction} sets $\b=2\a/3$, with $\a$ an arbitrary real and nonzero parameter. Finally, the action \eqref{eq:PoincareAction} is invariant under Poincar\'e Q-supersymmetry transformations $\hat\d_Q(\hve)$. As we have shown, however, when acting on coset-invariant fields Poincar\'e supersymmetry reduces to the conformal supergravity Q-supersymmetry, up to a field-dependent Lorentz transformation -- see eq.~\eqref{eq:HattedQ}. Since the action \eqref{eq:PoincareAction} depends only on coset-invariant fields and is invariant under local Lorentz transformations, it is also invariant under the conformal supergravity Q-supersymmetry.

\subsubsection{Local Poincar\'e transformations and anomalies}

Let us now consider the Q-supersymmetry transformation of the WZ action. In order to apply the coset construction of section \ref{sec:WZactions} it is necessary to change the basis of local symmetries from the conformal supergravity ones to the coset symmetries $\d_W$, $\d_R$, $\d_S$, and the symmetries of the unbroken Poincar\'e subalgebra identified in section \ref{sec:PoincareSubalgebra}. In terms of this symmetry parameterization, the anomalies $\ca_W$, $\ca_R$ and $\ca_S$ remain unchanged, but the Q-supersymmetry anomaly $\ca_Q$ is traded for the anomaly under the Poincar\'e supersymmetry $\hat\d_Q$. Its form follows directly from the Poincar\'e supersymmetry generator \eqref{eq:HattedPoincare-Q-explicit} when acting on the full field space -- not just on the coset invariant fields. In particular,
\be\label{eq:PoincareQAnomaly0}
\begin{split}
\hat\d_Q(\hve)\cw[e,A,\j]=&\;\int d^4x\,e\,\big(\s(\ve)\ca_W-\th(\ve)\ca_R-\bar\ve\ca_Q+\bar\h_{\text{tot}}(\ve)\ca_S\big)\\
\equiv&\; - \int d^4x\,e\,\bar\ve\hat\ca^{(0)}_Q[e,A,\j,\t,\c,\vf]\,,
\end{split}
\ee
where the composite parameters are given in \eqref{eq:PoincareFullSpace} and the anomalies $\ca_W$, $\ca_R$ and $\ca_S$ in the first line are evaluated on the conformal supergravity fields $e_\m{}^a$, $A_\m$ and $\j_\m$. $\hat\ca^{(0)}_Q[e,A,\j,\t,\c,\vf]$ denotes the Poincar\'e supersymmetry anomaly of the conformal supergravity generating functional $\cw[e,A,\j]$. Although the latter does not depend on the Goldstone fields, the Poincar\'e supersymmetry anomaly $\hat\ca^{(0)}_Q[e,A,\j,\t,\c,\vf]$ does depend on the Goldstone fields through the composite parameters $\s(\ve)$, $\th(\ve)$ and $\h_{\text{tot}}(\ve)$.

Besides modifying the Q-supersymmetry anomaly, the regrouping of the local symmetries produces effective diffeomorphism and Lorentz anomalies, even though the corresponding conformal supergravity symmetries are non-anomalous. The Poincar\'e diffeomorphism and local Lorentz generators were defined in \eqref{eq:HattedPoincare} with composite parameters given in \eqref{eq:RingedParameters}. These act on the conformal supergravity generating functional $\cw[e,A,\j]$ as
\be\label{eq:PoincareDLAnomalies0}
\begin{split}
\hat\d_D(\x)\cw[e,A,\j]=&\;\int d^4x\,e\,\big(\s(\x)\ca_W-\th(\x)\ca_R+\bar\h(\x)\ca_S\big)\\
\equiv&\;\int d^4x\,e\,\x^\m\hat\ca^{(0)}_{D\m}[e,A,\j,\t,\c,\vf]\,,\\
\hat\d_L(\l)\cw[e,A,\j]=&\;\int d^4x\,e\,\bar\h(\l)\ca_S\\
\equiv&\;\int d^4x\,e\,\l^{ab}\hat\ca^{(0)}_{L\,ab}[e,A,\j,\t,\c,\vf]\,.
\end{split}
\ee
Notice that these anomalies vanish when the Goldstone fields are set to zero.

We therefore conclude that in terms of the coset parameterization of the conformal supergravity symmetries the conformal supergravity generating functional $\cw[e,A,\j]$ transforms anomalously under all symmetries with corresponding anomalies
\bebx\NO
\begin{tabular}{ll}
\raisebox{.1cm}{$\cw[e,A,\j]$}&\\ 
\hline
\rule{0.cm}{.6cm}Coset anomalies: & $\ca_W[e,A,\j]$, $\ca_R[e,A,\j]$, $\ca_S[e,A,\j]$,\\
Poincar\'e anomalies: & $\hat\ca^{(0)}_Q[e,A,\j,\t,\c,\vf]$, $\hat\ca^{(0)}_{D\m}[e,A,\j,\t,\c,\vf]$, $\hat\ca^{(0)}_{L\,ab}[e,A,\j,\t,\c,\vf]$.
\end{tabular}  
\eebx

Recall that the WZ action can be expressed as the difference \eqref{WZ-W}, which in the present case takes the form 
\be\label{WZdiff}
\cw_{\text{WZ}}[e,A,\j,\t,\c,\vf]=-(\cw[\hat e,\hat A,\hat\j]-\cw[e,A,\j])\,.
\ee
Hence, in order to determine the transformations of the WZ action under the local Poincar\'e symmetries, it suffices to know those of the functionals $\cw[e,A,\j]$ and $\cw[\hat e,\hat A, \hat\j]$. We already discussed the transformations of the conformal supergravity generating functional $\cw[e,A,\j]$, so let us consider those of the coset-invariant functional $\cw[\hat e,\hat A, \hat\j]$.

Since $\cw[\hat e,\hat A, \hat\j]$ depends only on the coset-invariant fields, its Poincar\'e supersymmetry transformation follows from that of the invariant fields in \eqref{susy-trans-inv-compact}. These amount to a conformal supergravity Q-supersymmetry transformation accompanied by an S-supersymmetry transformation with composite parameter $\eta_0(\ve)$ given in \eqref{eq:hateta0}, all evaluated on the hatted fields and parameters. It follows that 
\be\label{eq:PoincareQAnomaly1}
\begin{split}
\hat\d_Q(\hve)\cw[\hat e,\hat A,\hat\j]=&\;\int d^4x\,\hat e\,(-\bar\hve\ca_Q[\hat e,\hat A,\hat\j]+\bar{\hat\h}_0(\hve)\ca_S[\hat e,\hat A,\hat\j])\\
\equiv &\;-\int d^4x\,\hat e\,\bar\hve\hat\ca_Q^{(1)}[\hat e,\hat A,\hat\j,\hat\cf]\,,
\end{split}
\ee
where $\ca_Q[\hat e,\hat A,\hat\j]$ and $\ca_S[\hat e,\hat A,\hat\j]$ are the corresponding conformal supergravity anomalies evaluated on the coset-invariant fields. Moreover, since the Poincar\'e diffeomorphism and local Lorentz transformations \eqref{eq:HattedPoincare} consist of the corresponding conformal supergravity transformations, accompanied by effective coset symmetries, the functional $\cw[\hat e,\hat A, \hat\j]$ is invariant under these transformations. Hence, the corresponding anomalies vanish identically:
\be\label{eq:DL-anomalies-1}
\hat\ca^{(1)}_{D\m}[\hat e,\hat A,\hat\j,\hat\cf]=\hat\ca^{(1)}_{L\,ab}[\hat e,\hat A,\hat\j,\hat\cf]=0\,.
\ee
Summarizing, the transformations of the functional $\cw[\hat e,\hat A, \hat\j]$ under the reorganized local symmetries are characterized by the following anomalies 
\bebx\NO
\begin{tabular}{ll}
\raisebox{.1cm}{$\cw[\hat e,\hat A,\hat\j]$}&\\ 
\hline
\rule{0.cm}{.6cm}Coset anomalies: & $0$, $0$, $0$,\\
Poincar\'e anomalies: & $\hat\ca^{(1)}_Q[\hat e,\hat A,\hat\j,\hat\cf]$, $0$, $0$.
\end{tabular}  
\eebx

Using the expression \eqref{WZdiff} for the WZ action and the transformations of the functionals $\cw[e,A,\j]$ and $\cw[\hat e,\hat A,\hat\j]$ we just determined, we conclude that Poincar\'e supersymmetry transformation of the WZ action is given by
\bbxd
\be\label{eq:PoincareQAnomalyWZ}
\hspace{-.2cm}\hat\d_Q(\hve)\cw_{\text{WZ}}[e,A,\j,\t,\c,\vf]=\int d^4x\,\big(\hspace{-1pt}-e\,\bar\ve\hat\ca^{(0)}_Q[e,A,\j,\t,\c,\vf]+\hat e\,\bar\hve\hat\ca_Q^{(1)}[\hat e,\hat A,\hat\j,\hat\cf]\big)\,,
\ee
\ebxd
in agreement with the general result \eqref{delta-WZ-final}, except that in this case the anomaly $\hat\ca^{(0)}_Q$ at the lower end of the transgression depends on the Goldstone fields due to the dependence of the corresponding subalgebra transformation. Analogous expressions hold for the Poincar\'e diffeomorphism and local Lorentz transformations of the WZ action, but since the corresponding anomalies of $\cw[\hat e,\hat A,\hat\j]$ vanish, the transformations of the WZ action coincides with those of the conformal supergravity generating functional $\cw[e,A,\j]$.

\subsubsection{Old minimal counterterm and restoration of Poincar\'e supersymmetry}

We just showed that the WZ action reproduces all transformations of the conformal supergravity generating functional, $\cw[e,A,\j]$, except for that under Poincar\'e supersymmetry. However, the WZ action would reproduce the Poincar\'e supersymmetry transformation of $\cw[e,A,\j]$ as well if the term proportional to $\ca_Q^{(1)}[\hat e,\hat A,\hat\j,\hat\cf]$ in \eqref{eq:PoincareQAnomalyWZ} could be removed. As we now show, this is indeed possible due to the existence of a local functional of all the invariant fields, $\cw_{\text{ct}}[\hat e,\hat A,\hat\j,\hat\cf]$, that possesses the key property  
\bbxd
\vskip.4cm
\be\label{eq:ctdef}
\begin{split}
\hat\d_Q(\hve)\cw_{\text{ct}}[\hat e,\hat A,\hat\j,\hat\cf]=&\;\int d^4x\,\hat e\,\bar\hve\hat\ca_Q^{(1)}[\hat e,\hat A,\hat\j,\hat\cf]\\
=&\;\int d^4x\,\hat e\,(\bar\hve\ca_Q[\hat e,\hat A,\hat\j]-\bar{\hat\h}_0(\hve)\ca_S[\hat e,\hat A,\hat\j])\,,
\end{split}
\ee
\ebxd
where $\ca_Q[\hat e,\hat A,\hat\j]$, $\ca_S[\hat e,\hat A,\hat\j]$ are respectively the conformal supergravity anomalies \eqref{Q-anomaly} and \eqref{S-anomaly}, evaluated on the invariant fields, and $\hat\h_0(\hve)$ is given in \eqref{eq:hateta0}.

There are two important observations that help us to systematically solve the condition \eqref{eq:ctdef} as a functional differential equation for the local functional $\cw_{\text{ct}}[\hat e,\hat A,\hat\j,\hat\cf]$. Firstly, any local functional that satisfies \eqref{eq:ctdef} admits a natural grading in powers of the auxiliary field $\hat\cf$ and its conjugate. Moreover, if we make the reasonable assumption that the expansion in powers of $\hat\cf$ and $\bar{\hat\cf}$ terminates, the top term in the expansion must be S-supersymmetry invariant. This is because $\hat\d_Q(\hve)$ acts on $\cw_{\text{ct}}[\hat e,\hat A,\hat\j,\hat\cf]$ as
\be
\hat\d_Q(\hve)\cw_{\text{ct}}[\hat e,\hat A,\hat\j,\hat\cf]= (\d_Q(\hat\ve)+\d_S(\hat\h_0(\hat\ve)))\cw_{\text{ct}}[\hat e,\hat A,\hat\j,\hat\cf]\,,
\ee
and $\hat\h_0(\hat\ve)$ contains terms linear in $\hat\cf$ and $\bar{\hat\cf}$. Hence, unless the top term in the expansion in $\hat\cf$ and $\bar{\hat\cf}$ is S-supersymmetry invariant, the action of $\hat\d_Q(\hve)$ on it would produce terms of higher order in $\hat\cf$ and $\bar{\hat\cf}$, contradicting the assumption that it is the top term. The highest order terms that satisfy this property and have Weyl weight four are $\hat e\,\hat\cf^2\bar{\hat\cf}^2$, as well as $\hat e\,\hat\cf^4$, $\hat e\,\hat\cf^3\bar{\hat\cf}$, $\hat e\,\hat\cf\bar{\hat\cf}^3$, and $\hat e\,\bar{\hat\cf}^4$. Among these, however, only the first has definite R-charge zero.
  
These observations lead us to seek a solution in the form 
\be\label{eq:CountertermAnsatz}
\cw_{\text{ct}}[\hat e,\hat A,\hat\j,\hat\cf]=\sum_{\substack{m,\,\bar m=0\\m+\bar m \leq 4}}^{4}\sum_{p=0}^{m-1}\sum_{\bar p=0}^{\bar m-1}\int d^4x\,\hat e\, \cl_{\text{ct}\,\{p,\bar p\}}^{(m,\bar m)}(\hat e,\hat A,\hat\j)\,\pa^p\hat\cf^{m-p}\pa^{\bar p}\bar{\hat\cf}^{\bar m-\bar p}\,,
\ee
with $\cl_{\text{ct}\,\{0,\bar 0\}}^{(4,0)}=\cl_{\text{ct}\,\{0,\bar 0\}}^{(3,1)}=\cl_{\text{ct}\,\{0,\bar 0\}}^{(1,3)}=\cl_{\text{ct}\,\{0,\bar 0\}}^{(0,4)}=0$ and $\cl_{\text{ct}\,\{0,\bar 0\}}^{(2,2)}$ a nonzero constant. Inserting this ansatz in the condition \eqref{eq:ctdef} results in a tower of functional differential equations for the coefficients $\cl_{\text{ct}\,\{p,\bar p\}}^{(m,\bar m)}(\hat e,\hat A,\hat\j)$. We have solved these equations recursively using the Mathematica package FieldsX \cite{Frob:2020gdh}, starting from the highest term with $m=\bar m=2$ and $p=\bar p=0$, and progressing downward. Of course, the solution is not unique because the kernel of $\hat\d_Q(\hve)$ is non-empty. Any supersymmetric action of weight four is a trivial solution of the condition \eqref{eq:ctdef}. However, this ambiguity can be eliminated by making further assumptions on the ansatz. For example, demanding that no term of the form $\pa_\m \hat\cf\pa^\m\bar{\hat\cf}$ appears excludes the supersymmetric $R^2$ density \cite{Ferrara:1978rk,Ferrara:1988qx}. In this way we obtain a specific solution of the inhomogeneous condition \eqref{eq:ctdef}. Adding to this solution any supersymmetric invariant of weight four is an equally acceptable solution of the condition \eqref{eq:ctdef}.  

For reasons that will become clear below, it is instructive to present our solution of the condition \eqref{eq:ctdef} as a sum of two distinct terms: a local functional $\cw_{\text{ct}}^{(\hat\cf)}[\hat e,\hat A,\hat\j,\hat\cf]$ that is at least linear in  the auxiliary field $\hat\cf$ and its conjugate, and a local functional $\cw_{\text{ct}}^{(0,0)}[\hat e,\hat A,\hat\j]$ that is independent of $\hat\cf$ and $\bar{\hat\cf}$, i.e. contains only the terms corresponding to $m=\bar m=0$ in the ansatz \eqref{eq:CountertermAnsatz}. In particular, the solution takes the form
\bbxd
\vspace{.2cm}
\be\label{eq:OM-counterterm-sum}
\cw_{\text{ct}}[\hat e,\hat A,\hat\j,\hat\cf]=\cw_{\text{ct}}^{(\hat\cf)}[\hat e,\hat A,\hat\j,\hat\cf]+\cw_{\text{ct}}^{(0,0)}[\hat e,\hat A,\hat\j]\,,
\ee
\ebxd
where
\bbxd
\be\label{eq:OM-counterterm-F}
\begin{split}
\cw_{\text{ct}}^{(\hat\cf)}[\hat e,\hat A,\hat\j,\hat\cf]=&\;-\tfrac{a}{8\p^2}\int d^4x\,\hat e\,\Big(\hat\cf^2\bar{\hat\cf}^2-2(\bar{\hat\cf}\lbar{\hat\F}^\m\hat\F_{\m L}+\hat\cf\lbar{\hat\F}^\m\hat\F_{\m R})+\tfrac{1}{3}\hat\cf\bar{\hat\cf}\hat R\\
&+\tfrac{40}{9}\hat\cf\bar{\hat\cf}\hat A^\m\hat A_\m+\tfrac{4i}{3}\hat A^\m(\bar{\hat\cf}\pa_\m\hat\cf-\hat\cf\pa_\m\bar{\hat\cf})-\tfrac{1}{3}\hat\cf\bar{\hat\cf}\bar{\hat\j}_\m\hat\g^{\m\n}\hat\f_\n\\
&+\tfrac{2i}{3}\hat\cf\bar{\hat\cf}\hat A^\m\bar{\hat\j}_\m\g^5\hat\g^\n\hat\j_\n+\tfrac{8i}{3}\hat A^\m(\bar{\hat\cf}\bar{\hat\j}_{\m R}\hat\g^\n\hat\F_\n-\hat\cf\bar{\hat\j}_{\m L}\hat\g^\n\hat\F_\n)\Big)\\
&+\tfrac{(a-c)}{16\p^2}\int d^4x\,\hat e\,\big(\hat\cf\lbar{\hat{\mathscr{R}}}_{\m\n}(Q)\hat{\mathscr{R}}(Q)^{\m\n}_L+\bar{\hat\cf}\lbar{\hat{\mathscr{R}}}_{\m\n}(Q)\hat{\mathscr{R}}(Q)^{\m\n}_R\big)\,,
\end{split}
\ee
\ebxd
and
\bbxd
\be\label{eq:OM-counterterm-0}
\begin{split}
\cw^{(0,0)}_{\rm ct}[\hat e,\hat A,\hat\j]=&\;-\tfrac{a}{\p^2}\int d^4x\,\hat e\,\big(-\tfrac49\big(\hat f^{\m\n}-\tfrac1{18}\hat A^\m\hat A^\n+\tfrac{1}{8}\bar{\hat\j}_\r\hat\g^\m\hat\g^{\r\s}\hat\g^\n\hat\f_{\s}\big)\hat A_\m \hat A_\n\\
&\hspace{.5cm}+\tfrac{i}{6}\bar{\hat\F}_{0 \m}(2\hat A^\m\hat\g^\n-\hat A_\r\hat\g^\r\hat\g^{\m\n})\g^5\hat\F_{0 \n}\big)\\
&\;+\tfrac{(a-c)}{72\p^2}\int d^4x\,\hat e\,\hat\e^{\m\n\r\s}\hat A_\m\big(i \hat F_{\r\s}\bar{\hat\j}^\k\g^5(2\hat\g_\k\hat\j_\n+\hat\g_\n\hat\j_\k)\\
&\hspace{.5cm}-\tfrac12\hat W_{\r\s\k\l}\bar{\hat\j}^\k(2\hat\g^\l\hat\j_\n-\hat\g_\n\hat\j^\l)+\co(\j^4)\big)-\cw^{f}_{\rm ct}[\hat e,\hat A,\hat\j|\substack{\blue{s_1}\\ \red{s_2}-1}]\,.
\end{split}
\ee
\ebxd

In these expressions, $\hat{\mathscr{R}}(Q)_{\m\n}$ is the gravitino curvature in \eqref{curvatures} evaluated on the invariant fields, $\hat f^{\m\n}$ is the composite connection in \eqref{composites} again evaluated on the invariant fields, $\hat\F_\m\equiv\hat\f_\m-\hat\eta_0(\hat\j_\m)$, $\hat\F_{0 \m}\equiv\hat\f_\m-\hat\eta_0(\hat\j_\m)\big|_{\hat\cf=\bar{\hat\cf}=0}$, and $\cw^{f}_{\rm ct}[\hat e,\hat A,\hat\j|\substack{\blue{s_1}\\ \red{s_2}-1}]$ is the local counterterm \eqref{counterterm}, evaluated on the invariant fields and with the replacement $s_2\to s_2-1$. In particular, setting $s_1=0$ and $s_2=1$, this term drops from \eqref{eq:OM-counterterm-0} and, in this case, $\cw_{\text{ct}}[\hat e,\hat A,\hat\j,\hat\cf]$ solves the condition \eqref{eq:ctdef} with the anomalies $\ca_Q[\hat e,\hat A,\hat\j]$ and $\ca_S[\hat e,\hat A,\hat\j]$, given respectively in \eqref{Q-anomaly} and \eqref{S-anomaly}, evaluated at $s_1=0$ and $s_2=1$. Keeping the values of $s_1$ and $s_2$ generic in \eqref{eq:OM-counterterm-0}, the resulting local functional $\cw_{\text{ct}}[\hat e,\hat A,\hat\j,\hat\cf]$ solves the condition \eqref{eq:ctdef} with the anomalies $\ca_Q[\hat e,\hat A,\hat\j]$ and $\ca_S[\hat e,\hat A,\hat\j]$ evaluated with generic $s_1$ and $s_2$.

One of the reasons for separating the terms $\cw_{\text{ct}}^{(\hat\cf)}[\hat e,\hat A,\hat\j,\hat\cf]$ and $\cw^{(0,0)}_{\rm ct}[\hat e,\hat A,\hat\j]$ is that the expression \eqref{eq:OM-counterterm-F} for the former is {\em exact} in the fermions. Due to its dependence on $\hat\cf$ and $\bar{\hat\cf}$, the condition \eqref{eq:ctdef} allows us to determine the form of $\cw_{\text{ct}}^{(\hat\cf)}[\hat e,\hat A,\hat\j,\hat\cf]$ exactly in the fermions, even without having determined the all orders in the gravitino expressions for the anomalies $\ca_Q[\hat e,\hat A,\hat\j]$ and $\ca_S[\hat e,\hat A,\hat\j]$. This is not the case for $\cw^{(0,0)}_{\rm ct}[\hat e,\hat A,\hat\j]$, whose form is directly tied to the form of the anomalies $\ca_Q[\hat e,\hat A,\hat\j]$ and $\ca_S[\hat e,\hat A,\hat\j]$. The expression \eqref{eq:OM-counterterm-0} for $\cw^{(0,0)}_{\rm ct}[\hat e,\hat A,\hat\j]$ is therefore correct to quadratic order in the fermions.\footnote{In fact, even though we have verified it only to quadratic order in the fermions, we believe the part of \eqref{eq:OM-counterterm-0} proportional to the anomaly coefficient $a$ to be exact in the fermions.}

The existence of a local functional $\cw_{\text{ct}}[\hat e,\hat A,\hat\j,\hat\cf]$ of the old-minimal fields that satisfies the condition \eqref{eq:ctdef} has two important and immediate consequences. Firstly, it enables us to construct a generalized WZ action that reproduces the transformations of the conformal supergravity generating functional, $\cw[e,A,\j]$, under {\em all} local symmetries -- both the coset and base symmetries. Indeed, from \eqref{eq:PoincareQAnomalyWZ} and \eqref{eq:ctdef} follows that the functional  
\bbxd
\vspace{.2cm}
\be\label{eq:genWZ}
\cw^{\text{gen}}_{\text{WZ}}[e,A,\j,\t,\c,\vf,\cf]\equiv \cw_{\text{WZ}}[e,A,\j,\t,\c,\vf]-\cw_{\text{ct}}[\hat e,\hat A,\hat\j,\hat\cf]\,,
\ee
\ebxd
satisfies
\be
\hat\d_Q(\hve)\cw^{\text{gen}}_{\text{WZ}}[e,A,\j,\t,\c,\vf,\cf]=\hat\d_Q(\hve)\cw[e,A,\j]\,.
\ee
The same holds for Poincar\'e diffeomorphisms, local Lorentz transformations and all the coset symmetries, since both $\cw[\hat e,\hat A,\hat\j]$ and $\cw_{\text{ct}}[\hat e,\hat A,\hat\j,\hat\cf]$ are invariant under these symmetries. The functional \eqref{eq:genWZ}, therefore, is the result we sought from the onset -- the complete local WZ functional that reproduces all anomalies of generic four-dimensional $\cn=1$ SCFTs coupled to {\em conformal supergravity}.

A second, distinct consequence of the existence of a local functional $\cw_{\text{ct}}[\hat e,\hat A,\hat\j,\hat\cf]$ satisfying \eqref{eq:ctdef} is that it enables us to construct a functional of the {\em old-minimal fields} that is free of a Poincar\'e supersymmetry anomaly. From \eqref{eq:PoincareQAnomaly1} and \eqref{eq:ctdef} follows that the functional 
\bbxd
\vspace{.2cm}
\be\label{eq:OMW}
\cw_{\text{om}}[\hat e,\hat A,\hat\j,\hat\cf]\equiv\cw[\hat e,\hat A,\hat\j]+\cw_{\text{ct}}[\hat e,\hat A,\hat\j,\hat\cf]\,,
\ee
\ebxd
is invariant under all local symmetries, including local Poincar\'e supersymmetry, i.e.
\be
\hat\d_Q(\hve)\cw_{\text{om}}[\hat e,\hat A,\hat\j,\hat\cf]=0\,.
\ee
It cannot be overemphasized that the functional $\cw[\hat e,\hat A,\hat\j]$ is obtained by coupling a generic four-dimensional $\cn=1$ SCFT to {\em old-minimal supergravity} -- not conformal supergravity. In this -- and only this -- context the functional $\cw_{\text{ct}}[\hat e,\hat A,\hat\j,\hat\cf]$ can be interpreted as a local counterterm that eliminates the Poincar\'e supersymmetry anomaly of the functional $\cw[\hat e,\hat A,\hat\j]$ (but {\em not} of $\cw[e,A,\j]$), albeit at the expense of introducing another anomaly, as we will discuss momentarily. This leads to the conclusion that Poincar\'e supersymmetry is non-anomalous {\em in old-minimal supergravity} \cite{Kuzenko:2019vvi,Katsianis:2020hzd,Bzowski:2020tue}.\footnote{However, we believe that crucial parts of the argument given for this statement in \cite{Bzowski:2020tue}, including the claim that the compensators satisfy a different algebra, are incorrect.}       

The old-minimal functional \eqref{eq:OMW} corresponds to the ``minimal supersymmetric scheme'' in the superspace analysis of \cite{Kuzenko:2019vvi}. However, as we will explain below, their ``WZ gauge'' and local counterterm are related to, but do not coincide with, the functional $\cw[\hat e,\hat A,\hat\j]$ and our local counterterm $\cw_{\text{ct}}[\hat e,\hat A,\hat\j,\hat\cf]$. This difference is important in certain applications of these results. In fact, the defining relation \eqref{eq:ctdef} of our counterterm coincides -- up to an overall sign -- with that of the local counterterm introduced in \cite{Katsianis:2020hzd}. However, only specific terms of the counterterm that could be deduced from a limited set of flat space correlation functions were obtained in \cite{Katsianis:2020hzd}, and only for the specific values of the $a$ and $c$ anomaly coefficients, corresponding to the Wess-Zumino model. Moreover, the expression for the local counterterm in \cite{Katsianis:2020hzd} is affected by the partially incorrect expression for the $a-c$ part of the $\ca_Q$ anomaly quoted in \cite{Papadimitriou:2019gel} and was discussed in section \ref{sec:conf_sugra}. Finally, a different choice for the supersymmetric invariants is made in \cite{Katsianis:2020hzd}, as is evident from the presence of a term of the form $\pa_\m \hat\cf\pa^\m\bar{\hat\cf}$, which, as we discussed, arises from the $R^2$ supersymmetric invariant.

\subsubsection{Super-Weyl anomalies of old-minimal supergravity}

The old-minimal functional \eqref{eq:OMW} is invariant under all symmetries of conformal supergravity -- it is trivially invariant under the coset symmetries since it is a functional of the coset-invariant old-minimal fields, and less trivially under the local Poincar\'e subalgebra. However, it transforms non-trivially under the super-Weyl symmetries \eqref{eq:HattedSW}. Recall that, with the exception of $\hat\L$-transformations that act only on $\hat\cf$ and $\bar{\hat\cf}$, super-Weyl symmetries act on the old-minimal fields exactly as the corresponding coset symmetries act on the bare conformal supergravity fields. From the definition \eqref{eq:OMW}, therefore, follows that the super-Weyl transformations of $\cw_{\text{om}}[\hat e,\hat A,\hat\j,\hat\cf]$ take the form 
\bbxd
\be\label{eq:sWanomalies}
\begin{split}
\hat\d_W(\s)\cw_{\text{om}}\equiv&\;\int d^4x\,\hat e\,\s\ca_W^{\text{om}}[\hat e,\hat A,\hat\j,\hat\cf]=\int d^4x\,\hat e\,\s\ca_W[\hat e,\hat A,\hat\j]+\hat\d_W(\s)\cw_{\text{ct}}\,,\\
\hat\d_R(\th)\cw_{\text{om}}\equiv&\;-\int d^4x\,\hat e\,\th\ca_R^{\text{om}}[\hat e,\hat A,\hat\j,\hat\cf]=-\int d^4x\,\hat e\,\th\ca_R[\hat e,\hat A,\hat\j]+\hat\d_R(\th)\cw_{\text{ct}}\,,\\
\hat\d_S(\hat\h)\cw_{\text{om}}\equiv&\;\int d^4x\,\hat e\,\bar{\hat\h}\ca_S^{\text{om}}[\hat e,\hat A,\hat\j,\hat\cf]=\int d^4x\,\hat e\,\bar{\hat\h}\ca_S[\hat e,\hat A,\hat\j]+\hat\d_S(\hat\h)\cw_{\text{ct}}\,,\\
\hat\d_F(\hat\L)\cw_{\text{om}}\equiv&\;\int d^4x\,\hat e\,\big(\hat\L\ca_F^{\text{om}}[\hat e,\hat A,\hat\j,\hat\cf]+\bar{\hat\L}\ca_{\bar F}^{\text{om}}[\hat e,\hat A,\hat\j,\hat\cf]\big)=\hat\d_F(\hat\L)\cw_{\text{ct}}\,,
\end{split}
\ee 
\ebxd
where the anomalies $\ca_W[\hat e,\hat A,\hat\j]$, $\ca_R[\hat e,\hat A,\hat\j]$ and $\ca_S[\hat e,\hat A,\hat\j]$ are the corresponding conformal supergravity anomalies in respectively \eqref{W-anomaly}, \eqref{R-anomaly} and \eqref{S-anomaly}, evaluated on the old-minimal fields, and the transformations of the counterterm $\cw_{\text{ct}}[\hat e,\hat A,\hat\j,\hat\cf]$ can be determined from its explicit form in \eqref{eq:OM-counterterm-F} and \eqref{eq:OM-counterterm-0}. 

For $s_1=0$, $s_2=1$, and to quadratic order in the gravitino, the super-Weyl transformations of the local counterterm $\cw_{\text{ct}}[\hat e,\hat A,\hat\j,\hat\cf]$  are given by 
\be\label{eq:sWctW}
\begin{split}
\hat\d_W(\s) \cw_{\text{ct}}=
&\;-\tfrac{a}{8\p^2}\int d^4x\,\hat e\,\Big(-\big(\tfrac{16}{9}\hat A^\m \hat A^\n+2\hat\cf\bar{\hat\cf}\hat g^{\m\n}\big)\hat\nabla_\m\hat\nabla_\n\s\\
&+\hat\nabla_\k\s\,\hat{\bar\j}_\m\hat\g^\k\big(2(\hat\cf\hat\F_R^\m+\bar{\hat\cf}\hat\F_L^\m)-\tfrac{4i}{3}(2\hat A^\m\hat\g^\n-\hat A_\r\hat\g^\r\hat\g^{\m\n})\g^5\hat\F_{0\n}\\
&+\tfrac29\hat A_\r\hat A_\s\hat\g^\r\hat\g^{\m\n}\hat\g^\s\hat\j_\n-\tfrac{4i}{3}\hat A^\n\hat\g^\m(\hat\cf\hat\j_{\n L}-\bar{\hat\cf}\hat\j_{\n R})\big)\Big)+\co(\hat{\j}^4)\,,
\end{split}
\ee
\be\label{eq:sWctR}
\begin{split}
\hat\d_R(\th) \cw_{\text{ct}}=
&\;-\tfrac{a}{8\p^2}\int d^4x\,\hat e\,\hat\nabla_\m\th\Big(-\tfrac{64}{9}\hat f^{(\m\n)}\hat A_\n+\big(\tfrac{64}{81}\hat A^2+\tfrac{32}{9}\hat\cf\bar{\hat\cf}\big)\hat A^\m\\
&+\tfrac{4i}{3}(\bar{\hat\cf}\hat\nabla^\m\hat\cf-\hat\cf\hat\nabla^\m\bar{\hat\cf})-\tfrac49\hat A_\k\bar{\hat\j}_\r(\hat\g^\m\hat\g^{\r\s}\hat\g^\k+\hat\g^\k\hat\g^{\r\s}\hat\g^\m)\hat\f_\s\\
&+\tfrac{4i}{3}\bar{\hat\F}_{0\r}(2\hat g^{\r\m}\hat\g^{\s}-\hat\g^\m\hat\g^{\r\s})\g^5\hat\F_{0\s}+\tfrac{2i}{3}\hat\cf\bar{\hat\cf}\bar{\hat\j}^\m\g^5\hat\g^\n\hat\j_\n\\
&+\tfrac{4i}{3}(2\bar{\hat{\psi}}^\m\hat\g^\n+\bar{\hat{\psi}}^\n\hat\g^\m)(\bar{\hat{\cf}}\hat{\F}_{\n L}-\hat{\cf}\hat{\F}_{\n R})-\tfrac{8}{9}\bar{\hat{\j}}_\r\hat\g^\m (2\hat A^{\r}\hat\g^{\s}-\hat{\slashed A}\hat\g^{\r\s})\hat{\F}_{0\s}\\
&-\tfrac{8}{9}\hat A^{\s}\bar{\hat{\j}}_\r\hat\g^\m\hat\g^{\r}(\hat{\cf}\hat{\j}_{\s L}+\bar{\hat{\cf}}\hat{\j}_{\s R})\Big)\\
&+\tfrac{a-c}{72\p^2}\int d^4x\,\hat e\,
 \hat\nabla_\m\th\,\hat\e^{\m\n\r\s}
 \big(i\hat F_{\r\s}\bar{\hat{\j}}^\k\g^5
 (2\hat\g_\k\hat{\j}_\n+\hat\g_\n\hat{\j}_\k)\\
&-\tfrac12\hat W_{\r\s\k\l}\bar{\hat{\j}}^\k
 (2\hat\g^\l\hat{\j}_\n-\hat\g_\nu\hat{\j}^\l)
 \big)+\co(\hat{\j}^4)\,,
\end{split}
\ee
\be\label{eq:sWctS}
\begin{split}
\hat\d_S(\hat\h) \cw_{\text{ct}}=
&\;-\tfrac{a}{8\p^2}\int d^4x\,\hat e\Big(-(\tfrac{8}{9} \hat A^\m \hat A^\n + \hat\cf \bar{\hat\cf} \hat g^{\m\n})\hat\nabla_\m (\bar{\hat\h}\hat\j_\n)\\
&+(\bar{\hat\h}\stackrel{\leftarrow}{\hat\cd}_\m) \big(\tfrac{8i}{3}(2\hat A^\m \hat\g^\n - \hat{\slashed{A}}\hat\g^{\m\n})\g^5\hat \F_{0 \n} - 4(\bar{\hat\cf} \hat\F^\m_{L} +\hat\cf \hat\F^\m_{R})\\
&+\tfrac{8i}{3} \hat A^\n \hat\g^\m(\hat\cf \j_{\n L} - \bar{\hat\cf} \j_{\n R}) - \tfrac{4}{9}\hat{\slashed{A}}\hat\g^{\m\n}\hat{\slashed{A}}\hat\j_\n\big)\\
&+\bar{\hat\h}\big(\tfrac{4}{9}\hat A^2 \hat\g^\m \hat\F_{0 \m} - 2 \hat\cf\bar{\hat\cf} \g^\m (\hat\F_\m+\hat\F_{0 \m})+\tfrac{16}{9}\hat A^\m \hat{\slashed A}(\hat\cf\hat\j_{\m L}+\bar{\hat\cf}\hat\j_{\m R})\\
& +\tfrac{4i}{3}(\hat{\slashed A}\hat\g^\m + 2\hat A^\m)(\bar{\hat\cf}\hat\F_{\m L}-\hat\cf\hat\F_{\m R})+ (\bar{\hat\cf}\hat\nabla^\m\hat\cf - \hat\cf \hat\nabla^\m\bar{\hat\cf})\g^5\hat\j_\m\\ 
&+\tfrac{16i}{3}\hat A_\n(\hat f^{(\m\n)} - (\hat\cf\bar{\hat\cf} + \tfrac{1}{6}\hat A^2)\hat g^{\m\n})\g^5\hat\j_\m + \tfrac{4i}{27} \hat A^2\hat{\slashed A} \hat\g^\m\g^5\hat\j_\m\big)\Big) \\
&+\tfrac{a-c}{72\p^2}\int d^4x\,\hat e\,\hat\e^{\m\n\r\s}\hat A_\m \bar{\hat\h}\big(W_{\r\s\k\l}(2 \hat\g_{\n}{}^\l\hat\j^\k - \hat\g^{\k\l}\hat\j_\n) - 12i \hat F_{\r\s}\g^5\hat\j_\n\big)\\
&+\co(\hat\j^3)\,,
\end{split}
\ee
\be\label{eq:sWctF}
\begin{split}
\d_{F}(\hat\L)\cw_{\text{ct}}=&\;\tfrac{(a-c)}{16\p^2}\int d^4x\,\hat e\,\big(\hat\L\lbar{\hat{\mathscr{R}}}_{\m\n}(Q)\hat{\mathscr{R}}(Q)^{\m\n}_L+\bar{\hat\L}\lbar{\hat{\mathscr{R}}}_{\m\n}(Q)\hat{\mathscr{R}}(Q)^{\m\n}_R\big)\\
&-\tfrac{a}{8\p^2}\int d^4x\,\hat e\,\Big(\big(2\hat\cf\bar{\hat\cf}+\tfrac{1}{3}\hat R+\tfrac{40}{9}\hat A^2-\tfrac{1}{3}\bar{\hat\j}_\m\hat\g^{\m\n}\hat\f_\n\big)(\hat\L\bar{\hat\cf}+\hat\cf\bar{\hat\L})\\
&-2\lbar{\hat\F}^\m(\bar{\hat\L}\hat\F_{\m L}+\hat\L\hat\F_{\m R})+\tfrac{4i}{3}\hat A^\m(\bar{\hat\L}\pa_\m\hat\cf-\hat\L\pa_\m\bar{\hat\cf}+\bar{\hat\cf}\pa_\m\hat\L-\hat\cf\pa_\m\bar{\hat\L})\\
&+\tfrac{2i}{3}(\hat\L\bar{\hat\cf}-\hat\cf\bar{\hat\L})\hat A^\m\bar{\hat\j}_\m\hat\g^\n\hat\j_\n-2\bar{\hat\j}^\m(\hat\L\bar{\hat\cf}\hat\F_{\m L}+\bar{\hat\L}\hat\cf \hat\F_{\m R})\\
&+\tfrac{8i}{3}\hat A^\m(\bar{\hat\L}\bar{\hat\j}_{\m R}-\hat\L\bar{\hat\j}_{\m L})\hat\g^\n\hat\F_\n\Big)\,.
\end{split}
\ee

Notice that the $\hat\L$-transformation \eqref{eq:sWctF} of the counterterm is exact in the gravitino, since it is entirely due to the $\hat\cf$-dependent part of the counterterm in \eqref{eq:OM-counterterm-F}, which is exact in the gravitino. Moreover, although these counterterm transformations are valid specifically for $s_1=0$ and $s_2=1$, the super-Weyl anomalies, $\ca_W^{\text{om}}[\hat e,\hat A,\hat\j,\hat\cf]$, $\ca_R^{\text{om}}[\hat e,\hat A,\hat\j,\hat\cf]$, $\ca_S^{\text{om}}[\hat e,\hat A,\hat\j,\hat\cf]$ and $\ca_F^{\text{om}}[\hat e,\hat A,\hat\j,\hat\cf]$ defined in \eqref{eq:sWanomalies} are in fact independent of the values of the scheme parameters $s_1$ and $s_2$. This is because the counterterm 
$\cw^{f}_{\rm ct}[\hat e,\hat A,\hat\j|\substack{\blue{s_1}\\ \red{s_2}-1}]$ appears with a minus sign in the expression for $\cw^{(0,0)}_{\rm ct}[\hat e,\hat A,\hat\j]$ in \eqref{eq:OM-counterterm-0} and cancels the $s_1$ and $s_2$ dependence of the anomalies $\ca_W[\hat e,\hat A,\hat\j]$, $\ca_R[\hat e,\hat A,\hat\j]$ and $\ca_S[\hat e,\hat A,\hat\j]$, given respectively in \eqref{W-anomaly}, \eqref{R-anomaly} and \eqref{S-anomaly}.

\subsection{Old-minimal WZ action and relation to superspace}

The old-minimal super-Weyl anomalies $\ca_W^{\text{om}}[\hat e,\hat A,\hat\j,\hat\cf]$, $\ca_R^{\text{om}}[\hat e,\hat A,\hat\j,\hat\cf]$, $\ca_S^{\text{om}}[\hat e,\hat A,\hat\j,\hat\cf]$ and $\ca_F^{\text{om}}[\hat e,\hat A,\hat\j,\hat\cf]$ defined in \eqref{eq:sWanomalies} comprise the standard chiral multiplet of superspace anomalies \cite{Buchbinder:1998twe,Schwimmer:2010za,Kuzenko:2019vvi}. As we have demonstrated, they are closely related -- but distinct -- from the corresponding anomalies of conformal supergravity. In the context of old-minimal supergravity, the counterterm $\cw_{\text{ct}}[\hat e,\hat A,\hat\j,\hat\cf]$ enables one to interpolate between two different schemes: a scheme where $\hat\L$-transformations are non-anomalous but Poincar\'e supersymmetry is anomalous, corresponding to the functional $\cw[\hat e,\hat A, \hat\j]$, and the standard supersymmetric scheme where Poincar\'e supersymmetry is non-anomalous but $\hat\L$-transformations are anomalous, corresponding to the functional $\cw_{\text{om}}[\hat e,\hat A,\hat\j,\hat\cf]$. This provides a clean understanding of the role of the counterterm $\cw_{\text{ct}}[\hat e,\hat A,\hat\j,\hat\cf]$, interpolating between two different schemes {\em within old-minimal supergravity}.

The framework we have developed also allows us to describe very concretely the results of \cite{Schwimmer:2010za} and \cite{Kuzenko:2019vvi}, and their relation to ours. Recall that in the seminal work \cite{Schwimmer:2010za}, Schwimmer and Theisen obtain the superspace WZ effective action for super-Weyl anomalies using the superspace version of the WZ transgression formula. In the present framework, this is the WZ action for the super-Weyl symmetries of old-minimal supergravity obtained via the transgression formula from the anomalies 
$\ca_W^{\text{om}}[\hat e,\hat A,\hat\j,\hat\cf]$, $\ca_R^{\text{om}}[\hat e,\hat A,\hat\j,\hat\cf]$, $\ca_S^{\text{om}}[\hat e,\hat A,\hat\j,\hat\cf]$ and $\ca_F^{\text{om}}[\hat e,\hat A,\hat\j,\hat\cf]$. This WZ action is distinct from -- but related to -- the {\em conformal supergravity} WZ action \eqref{eq:genWZ} we have constructed here. 

A key difference is that the superspace WZ action of \cite{Schwimmer:2010za} involves a transgression for the $\hat\L$-symmetry, for which $\hat\cf$ is the corresponding Goldstone field. There is no $\hat\L$-symmetry in conformal supergravity and, hence, no corresponding transgression in our WZ action \eqref{super-transgression}. Moreover, the old-minimal Weyl, R-symmetry and S-supersymmetry anomalies defined in \eqref{eq:sWanomalies} also depend on $\hat\cf$, while the corresponding conformal supergravity anomalies do not. Nevertheless, the generalized WZ action \eqref{eq:genWZ} does depend on $\hat\cf$. Could the superspace/old-minimal WZ action of \cite{Schwimmer:2010za} coincide with our generalized WZ action \eqref{eq:genWZ}?  

We can answer this question very concretely. In analogy with the conformal supergravity WZ action \eqref{WZdiff}, the old-minimal WZ action is given by
\be\label{omWZ}
\cw_{\text{WZ}}^{\text{om}}[\hat e,\hat A,\hat\j,\hat\cf,-\t,-\c,-\vf,-\cf]\equiv-(\cw_{\text{om}}[e,A,\j,0]-\cw_{\text{om}}[\hat e,\hat A,\hat\j,\hat\cf])\,.
\ee
Notice that the negative sign in front of the Goldstones implements an inverse super-Weyl transformation so that we can connect this WZ action to objects we have already discussed. Moreover, the auxiliary field $\cf$ plays the role of both an old-minimal field and a compensator. As a result, at the upper point of the transgression only the bare conformal supergravity fields, $e_\m{}^a$, $A_\m$ and $\j_\m$ appear; these are the invariant fields under super-Weyl transformations acting as in \eqref{eq:superWeyl}. The old-minimal WZ action \eqref{omWZ} is precisely the component version of the superspace WZ action obtained in \cite{Schwimmer:2010za}. 

However, we can now determine the explicit form of this WZ action in components, by relating it to objects we have already constructed in our analysis. Using the definition of the old-minimal functional in \eqref{eq:OMW}, we can write \eqref{omWZ} as 
\be
\begin{split}
&\cw_{\text{WZ}}^{\text{om}}[\hat e,\hat A,\hat\j,\hat\cf,-\t,-\c,-\vf,-\cf]\\
&\hspace{3.0cm}=-\cw_{\text{WZ}}[e,A,\j,\t,\c,\vf]+\cw_{\text{ct}}[\hat e,\hat A,\hat\j,\hat\cf]-\cw_{\text{ct}}[e,A,\j,0]\,.
\end{split}
\ee
The first two terms combine into the generalized WZ action \eqref{eq:genWZ}, while from the decomposition \eqref{eq:OM-counterterm-sum} of the counterterm follows that only $\cw^{(0,0)}_{\rm ct}[e,A,\j]$ survives at $\cf=0$. Hence,\hspace{-.2cm}
\bbxd
\vspace{-.2cm}
\be\label{omWZ-explicit}
\begin{split}
\cw_{\text{WZ}}^{\text{om}}[\hat e,\hat A,\hat\j,\hat\cf,-\t,-\c,-\vf,-\cf]=-\cw_{\text{WZ}}^{\text{gen}}[e,A,\j,\t,\c,\vf,\cf]-\cw^{(0,0)}_{\rm ct}[e,A,\j]\,.
\end{split}
\ee
\ebxd
Therefore, the old-minimal WZ action differs from the generalized WZ action \eqref{eq:genWZ} by the counterterm $\cw^{(0,0)}_{\rm ct}[e,A,\j]$ in \eqref{eq:OM-counterterm-0}. Since we have explicitly computed all quantities on the r.h.s. of \eqref{omWZ-explicit}, this relation provides an explicit expression for the old-minimal/superspace WZ action in components.

The old-minimal WZ action \eqref{omWZ-explicit} can be written in a form that can be directly compared with the superspace result of \cite{Schwimmer:2010za}. Using again \eqref{eq:genWZ} and \eqref{eq:OM-counterterm-sum}, \eqref{omWZ-explicit} can be expressed as  
\be\label{omWZ-explicit-ST}
\begin{split}
&\cw_{\text{WZ}}^{\text{om}}[\hat e,\hat A,\hat\j,\hat\cf,-\t,-\c,-\vf,-\cf]\\
&\hspace{1.0cm}=-\cw_{\text{WZ}}[e,A,\j,\t,\c,\vf]+\cw^{(\hat\cf)}_{\text{ct}}[\hat e,\hat A,\hat\j,\hat\cf]+\cw^{(0,0)}_{\text{ct}}[\hat e,\hat A,\hat\j]-\cw^{(0,0)}_{\text{ct}}[e,A,\j]\,.
\end{split}
\ee
The first term is the WZ action we obtained through the transgression 
\eqref{super-transgression} along the coset symmetries of conformal supergravity. Since the super-Weyl Weyl, R-symmetry and S-supersymmetry transformations in \eqref{eq:HattedSW} act on the old-minimal fields exactly as the coset transformations act on the conformal supergravity fields, the first term is equivalent to a transgression for the super-Weyl Weyl, R-symmetry and S-supersymmetry transformations of old-minimal supergravity. Remarkably, the second term can also be written as a transgression along $\hat\L$-transformations of the super-Weyl symmetry, namely 
\bbxd
\be\label{Ftransgression}
\cw^{(\hat\cf)}_{\text{ct}}[\hat e,\hat A,\hat\j,\hat\cf]=\int_0^1dt\int d^4x\,\hat e\,\big(\hat\cf\ca_F^{\text{om}}[\hat e,\hat A,\hat\j,t\hat\cf]+\bar{\hat\cf}\ca_{\bar F}^{\text{om}}[\hat e,\hat A,\hat\j,t\hat\cf]\big)\,,
\ee
\ebxd
where the anomalies $\ca_F^{\text{om}}[\hat e,\hat A,\hat\j,\hat\cf]$ and $\ca_{\bar F}^{\text{om}}[\hat e,\hat A,\hat\j,\hat\cf]$ are given in \eqref{eq:sWctF}. Together, the first two terms in \eqref{omWZ-explicit-ST} correspond to a transgression along all super-Weyl symmetries, paralleling the superspace transgression of \cite{Schwimmer:2010za}. Moreover, the bosonic part of the last two terms in \eqref{omWZ-explicit-ST}, $\cw^{(0,0)}_{\text{ct}}[\hat e,\hat A,\hat\j]-\cw^{(0,0)}_{\text{ct}}[e,A,\j]$, coincides with the fourth line of the superspace WZ action given in eq.~(3.23) of \cite{Schwimmer:2010za}. In particular, the bosonic part of the counterterm $\cw^{(0,0)}_{\text{ct}}[\hat e,\hat A,\hat\j]$ in \eqref{eq:OM-counterterm-0} coincides with the local term in eq.~(3.26) of \cite{Schwimmer:2010za}.

Even though the old-minimal WZ action \eqref{omWZ-explicit-ST} coincides with the superspace result of \cite{Schwimmer:2010za}, the above comparison of the calculations that produced these results is only qualitative. The superspace WZ action is obtained from a transgression for the super-Weyl symmetries of old-minimal supergravity using the anomalies 
$\ca_W^{\text{om}}[\hat e,\hat A,\hat\j,\hat\cf]$, $\ca_R^{\text{om}}[\hat e,\hat A,\hat\j,\hat\cf]$, $\ca_S^{\text{om}}[\hat e,\hat A,\hat\j,\hat\cf]$ and $\ca_F^{\text{om}}[\hat e,\hat A,\hat\j,\hat\cf]$. We could have evaluated this transgression in components here as well, but our main goal was instead the WZ action for conformal supergravity. This still does allow us to determine the old-minimal WZ action as well by relating it to the WZ action of conformal supergravity and suitable local counterterms, as is explicitly done in \eqref{omWZ-explicit} or \eqref{omWZ-explicit-ST}. However, the transgression along the Weyl, R-symmetry and S-supersymmetry directions is implemented using the conformal supergravity anomalies rather than the old-minimal ones, while the transgression along $\hat\L$-transformations is done separately. Moreover, the transgression of the counterterm $\cw^{(0,0)}_{\text{ct}}[\hat e,\hat A,\hat\j]$ appears separately in \eqref{omWZ-explicit-ST}, while in the superspace/old-minimal transgression it is an integral part of the old-minimal anomalies. Indeed, even though \cite{Schwimmer:2010za} identifies part of the superspace WZ action as the transformation of a local term, the role of this local term as an essential local counterterm that interpolates between the old-minimal and conformal supergravity WZ actions was not pointed out.   

The counterterm $\cw^{(0,0)}_{\text{ct}}[e,A,\j]$ plays a key role also in the comparison of our results with those of \cite{Kuzenko:2019vvi}. That work uses the superspace WZ action constructed in \cite{Schwimmer:2010za} in order to obtain a local functional that interpolates between the old-minimal functional $\cw_{\text{om}}[\hat e,\hat A,\hat\j,\hat\cf]$ and a functional that depends only on the conformal supergravity fields, corresponding to what \cite{Kuzenko:2019vvi} refer to the ``WZ gauge''. However, we now show that this functional, which we will call $\cw_{\text{KST}}[e,A,\j]$ here, differs from the conformal supergravity functional, $\cw[e,A,\j]$, precisely by the local counterterm $\cw^{(0,0)}_{\text{ct}}[e,A,\j]$.  

Let us begin by noticing that combining \eqref{WZdiff}, \eqref{eq:genWZ} and \eqref{eq:OMW} we can write
\be
\cw[e,A,\j]=\cw_{\text{om}}[\hat e,\hat A,\hat\j,\hat\cf]+W_{\text{WZ}}^{\text{gen}}[e,A,\j,\t,\c,\vf,\cf]\,.
\ee
We can trade the generalized WZ action in this expression for the old-minimal/superspace
WZ functional using \eqref{omWZ-explicit}. Namely,
\be
\cw[e,A,\j]=\cw_{\text{om}}[\hat e,\hat A,\hat\j,\hat\cf]-\cw_{\text{WZ}}^{\text{om}}[\hat e,\hat A,\hat\j,\hat\cf,-\t,-\c,-\vf,-\cf]-\cw^{(0,0)}_{\rm ct}[e,A,\j]\,,
\ee
or equivalently
\bbxd
\vspace{-.2cm}
\be
\hspace{-.2cm}\cw[e,A,\j]+\cw^{(0,0)}_{\rm ct}[e,A,\j]=\cw_{\text{om}}[\hat e,\hat A,\hat\j,\hat\cf]-\cw_{\text{WZ}}^{\text{om}}[\hat e,\hat A,\hat\j,\hat\cf,-\t,-\c,-\vf,-\cf]\,.
\ee
\ebxd
Therefore, the old-minimal/superspace WZ action interpolates between the old minimal functional $\cw_{\text{om}}[\hat e,\hat A,\hat\j,\hat\cf]$ and the functional
\bbxd
\vskip.3cm
\be\label{KST-WZgauge}
\cw_{\text{KST}}[e,A,\j]=\cw[e,A,\j]+\cw^{(0,0)}_{\rm ct}[e,A,\j]\,,
\ee
\ebxd
which depends only on the conformal supergravity fields. This is precisely the construction of \cite{Kuzenko:2019vvi}, stated here in components rather than superspace and without gauge-fixing the super-Weyl symmetry. As anticipated, the functional $\cw_{\text{KST}}[e,A,\j]$ corresponding to the ``WZ gauge'' of \cite{Kuzenko:2019vvi} {\em is not} the generating function of conformal supergravity -- the two differ by the local counterterm $\cw^{(0,0)}_{\text{ct}}[e,A,\j]$ in \eqref{eq:OM-counterterm-0}. This can also be seen by gauge-fixing the super-Weyl symmetry, in which case \eqref{omWZ-explicit-ST} determines that $\cw_{\text{WZ}}^{\text{om}}[\hat e,\hat A,\hat\j,\hat\cf,0,0,0,-\cf]=\cw^{(\hat\cf)}_{\text{ct}}[\hat e,\hat A,\hat\j,\hat\cf]$. The counterterm of \cite{Kuzenko:2019vvi}, therefore, gives only the $\hat\cf$-dependent part of the counterterm \eqref{eq:OM-counterterm-sum} that interpolates between old-minimal and conformal supergravity.


\section*{Acknowledgments}

We would like to thank Piljin Yi and Matthew M. Roberts for valuable discussions. The work of PP is supported by KIAS Grant PG 105601. Part of this work has been supported by the National Research Foundation of Korea (NRF) grant funded by the Korea government (MSIT) (No.~2023R1A2C1006975) and from the JRG Program at the Asia Pacific Center for Theoretical Physics (APCTP) through the Science and Technology Promotion Fund and Lottery Fund of the Korean Government. IP was supported in part by the Brain Pool program funded by the Ministry of Science and ICT through the National Research Foundation of Korea (RS-2025-25457100). PP thanks the National and Kapodistrian University of Athens for  warm hospitality at several stages of this work and the APCTP, where this project was formally initiated. IP thanks the APCTP and Kyung Hee University for the hospitality during the completion of this work. We would also like to acknowledge extensive use of the symbolic computation package FieldsX \cite{Frob:2020gdh}.

\appendix

\renewcommand{\theequation}{\Alph{section}.\arabic{equation}}

\section*{Appendix}
\setcounter{section}{0}

\section{Spinor conventions and identities}
\label{appendix:spinors}

In this appendix we summarize our spinor conventions, which generally follow those of \cite{Freedman:2012zz}, and collect several identities that we use extensively in our analysis. The tangent space metric is given by the diagonal matrix $\h=\diag(-1,1,1,1)$ and the Levi-Civita symbol $\ve_{\m\n\r\s}=\pm 1$ satisfies $\ve_{0123}=1$. The Levi-Civita tensor is defined by  $\e_{\m\n\r\s}=\sqrt{-g}\;\ve_{\m\n\r\s}=e\;\ve_{\m\n\r\s}$. The chirality matrix in four dimensions  and the projection operators on a  spinor $\l$ are given by 
\be
\g^5=i\g_0\g_1\g_2\g_3\,,\qquad \l_L=P_L\l=\tfrac12(1+\g^5)\l\,,\qquad \l_R=P_R\l=\tfrac12(1-\g^5)\l\,.
\ee
Antisymmetrized products of gamma matrices are defined as
\be
\g^{\m_1\m_2\ldots\m_n}\equiv\g^{[\m_1}\g^{\m_2}\cdots\g^{\m_n]},
\ee
where antisymmetrization is done with weight one.

The {\em Majorana} conjugate of a four-component spinor is defined by $\bar\l=\l^T C$ where $C$ is the conjugation matrix satisfying $(C\g^{\m_1\dots\m_r})^T=-t_r  C \g^{\m_1\dots\m_r}$. In $d=4$, $t_0=t_3=t_4=1$ and $t_1=t_2=-1$, which translates to 
$C^T=-C$, together with
\be
(\g^\m)^T=-C\g^\m C^{-1}\,,\qquad (\g^{\m\n})^T=-C\g^{\m\n} C^{-1}\,, \qquad(\g^5)^T=C\g^5C^{-1}\,.
\ee

The {\em charge} conjugate of a four-component spinor $\l$ is defined by $\l^C=B^{-1}\l$ where $B=iC\g^0$. A conventional realization of the conjugation matrix is $C=i\g^3\g^1$, and thus $B=\g^0\g^1\g^3$. For any realization of $C$, charge conjugation on acts on the gamma matrices as
\be\label{C-conjugation}
(\g^\m)^C=B^{-1}\g^\m{}^*B=\g^\m\,,\qquad (\g^5)^C=B^{-1}\g^5{}^*B=-\g^5\,.
\ee

Majorana spinors are defined as four-component spinors satisfying the condition $\l^C=\l$. In $d=4$ they obey the following Majorana flip and complex conjugation relations
\be
\lbar\l\g^{\m_1}\g^{\m_2}\cdots\g^{\m_p}\c=(-1)^p\lbar\c\g^{\m_p}\cdots\g^{\m_2}\g^{\m_1}\l\,,\qquad (\lbar\c\g_{\m_1\ldots\m_r}\l)^*=\lbar\c\g_{\m_1\ldots\m_r}\l\,.
\ee

Spinor products are re-arranged by employing the Fierz identities, with the basic one 
\be\label{Fierz}
\bar\l_1\l_2\,\bar\l_3\l_4=-\tfrac14\sum_A \bar\l_1\G^A\l_4\,\bar\l_3\G_A\l_2\,,
\ee
where $\G^A=\{\mathbb1,\g^\m,\g^{\m_1\m_2},\g^{\m_1\m_2\m_3},\g^{\m_1\m_2\m_3\m_4}\}$, with $\m_1<\m_2<\dots$, is the Clifford basis in four dimensions. The chirally projected forms of \eqref{Fierz} are use extensively throughout our analysis. Their explicit form is
\be\label{PL-Fierz}
\begin{split}
P_L\l_1\bar\l_2 P_L=&\;-\tfrac12P_L(\bar\l_2 P_L\l_1)+\tfrac18 P_L\g^{\m\n}(\bar\l_2\g_{\m\n} P_L\l_1)\,,\\
P_L\c\bar\l P_R=&\;-\tfrac12P_L\g^\m(\bar\l\g_\m P_L\c)\,.
 \end{split}
\ee
Opposite chirality identities are obtained by applying the charge conjugation \eqref{C-conjugation} on \eqref{PL-Fierz}.

We also make extensive use of several identities involving antisymmetrized products of gamma matrices, which we collect here for general dimension $d$:
\be\label{d-gamma-ids-1}
\begin{split} 
\g^{\m\n\r}&=\tfrac12\{\g^\m,\g^{\n\r}\}\,,\\
\g^{\m\n\r\s}&=\tfrac12[\g^\m,\g^{\n\r\s}]\,,\\
\g^{\m\n}\g_{\r\s}&=\g^{\m\n}{}_{\r\s}+4\g^{[\m}{}_{[\s}\d^{\n]}{}_{\r]}+2\d^{[\m}{}_{[\s}\d^{\n]}{}_{\r]}\,,\\
\g_{\m}\g^{\n_1\ldots\n_p}&=\g_\m{}^{\n_1\ldots\n_p}+p\d ^{[\n_1}_{\m}\g^{\n_2\ldots\n_p]}\,,\\
\g^{\n_1\ldots\n_p}\g_{\m}&=\g^{\n_1\ldots\n_p}{}_{\m}+p\g^{[\n_1\ldots\n_{p-1}}\d ^{\n_p]}_{\m}\,,\\
\g^{\m\n\r}\g_{\s\t}&=\g^{\m\n\r}{}_{\s\t}+6\g^{[\m\n}{}_{[\t}\d ^{\r]}{}_{\s]}+6\g^{[\m}\d^\n{}_{[\t}\d^{\r]}{}_{\s]}\,,\\
\g^{\m\n\r\s}\g_{\t\l}&=\g^{\m\n\r\s}{}_{\t\l}+8\g^{[\m\n\r}{}_{[\l}\d^{\s]}{}_{\t]}+12\g^{[\m\n}\d^\r{}_{[\l}\d^{\s]}{}_{\t]}\,,\\
\g^{\m\n\r}\g_{\s\t\l}&=\g^{\m\n\r}{}_{\s\t\l}+9\g^{[\m\n}{}_{[\t\l}\d^{\r]}{}_{\s]}+18\g^{[\m}{}_{[\l}\d^\n{}_{\t}\d^{\r]}{}_{\s]}+6\d^{[\m}{}_{[\l}\d^\n{}_{\t}\d^{\r]}{}_{\s]}\,,
\end{split}
\ee
\be\label{d-gamma-ids-2}
\begin{split}
\g^{\m_1\ldots\m_r\n_1\ldots\n_s}\g_{\n_s\ldots\n_1}&=\tfrac{(d-r)!}{(d-r-s)!}\g^{\m_1\ldots\m_r}\,,\\
\g^{\m\r}\g_{\r\n}&=(d-2)\g^{\m}{}_\n+(d-1)\d^\m_\n\,,\\
\g^{\m\n\r}\g_{\r\s}&=(d-3)\g^{\m\n}{}_\s+2(d-2)\g^{[\m}\d^{\n]}{}_\s\,,\\
\g_{\m\n}\g^{\n\r\s}&=(d-3)\g_{\m}{}^{\r\s}+2(d-2)\d^{[\r}_\m\g^{\s]}\,,\\
\g^{\m\n\l}\g_{\l\r\s}&=(d-4)\g^{\m\n}{}_{\r\s}+4(d-3)\g^{[\m}{}_{[\s}\d^{\n]}{}_{\r]}+2(d-2)\d^{[\m}{}_{[\s}\d^{\n]}{}_{\r]}\,,\\
\g_{\m\r}\g^{\r\s\t}\g_{\t\n}&=(d-4)^2\g_\m{}^\s{}_\n+(d-4)(d-3)(\g_\m\d_\n^\s-\g^\s g_{\m\n})\\
&\hskip0.5cm+(d-3)(d-2)\d_\m^\s\g_\n-(d-3)\g^\s\g_{\m\n}\,,\\
\g_\r \g^{\m_1\m_2\ldots \m_p}\g^\r&=(-1)^p(d-2p)\g^{\m_1\m_2\ldots \m_p}\,.
\end{split}
\ee

Specifically in $d=4$ dimensions, we use extensively the identities
\be\label{d=4-gamma-ids}
\begin{split}
&\g^\r\g^\m\g^\s+\g^\s\g^\m\g^\r=2(g^{\m\r}\g^\s+g^{\m\s}\g^\r-g^{\r\s}\g^\m)\,,\\
&\g^\r\g^\m\g^\s-\g^\s\g^\m\g^\r=2\g^{\r\m\s}\,,\\
&\g^\r\g^\m\g^\s=g^{\m\r}\g^\s+g^{\m\s}\g^\r-g^{\r\s}\g^\m+\g^{\r\m\s}\,,\\
&\g^{\m\r}\g^\s=\g^{\m\r\s}+\g^\m g^{\r\s}-\g^\r g^{\m\s}\,,\\
&\g^{\m\r\s}=i\e^{\m\r\s\n}\g_\n\g^5,\quad\g^{\m\n}=\tfrac{i}{2}\e^{\m\n\r\s}\g_{\r\s}\g^5,\quad\g^{\m\n\r\s}=-i\e^{\m\n\r\s}\g^5\,.
\end{split}
\ee

\section{Geometry with torsion}
\label{appendix:torsion}

In this appendix we summarize a number of useful properties and identities of torsionful geometry, again in the conventions of \cite{Freedman:2012zz}. In the presence of torsion, the curvature constraint $R_{\m\n}(P^a)=0$ expresses the spin connection in terms of the vielbein and the torsion tensor, $T_{\m\n}{}^\r$, through the relation
\be
\o_\m{}^{ab}=\o_\m{}^{ab}(e)+K_\m{}^{ab}\,,
\ee
where the torsion-free spin connection is given by 
\be
\o_\m{}^{ab}(e)=2e^{\n[a}\del_{[\m}e_{\n]}{}^{b]}-e^{\n[a}e^{b]\s} e_{\m c}\del_\n e_\s{}^c\,,
\ee
and the contorsion tensor, $K_\m{}^{ab}$, is antisymmetric in the last two indices and is defined as
\be
K_\m{}^{ab}=-\tfrac12(T_\m{}^{ab}-T^{ab}{}_\m+T^b{}_\m{}^a)\,.
\ee
Conversely, the torsion tensor, $T_{\m\n}{}^\r$, is antisymmetric in the first two indices and can be expressed in terms of the contorsion as 
\be\label{torsion-contorsion}
T_{\m\n\r}=-K_{\m\n\r}+K_{\n\m\r}\,.
\ee

Similarly, the affine connection $\G_{\m\n}^\r$ can be expressed in terms of the Christoffel connection $\G_{\m\n}^\r(g)$ and the contorsion as
\be 
\G^\r_{\m\n}=\G^\r_{\m\n}(g)-K_{\m\n}{}^\r\,,\qquad \G_{\m\n}^\r(g)=\tfrac12 g^{\r\s}(\pa_\m g_{\s\n}+\pa_\n g_{\m\s}-\pa_\s g_{\m\n})\,.
\ee
When using this expression for the affine connection it is important to remember that the contorsion is antisymmetric in the last two indices, i.e. $K_{\m\n}{}^\r=-K_\m{}^\r{}_\n$.

The presence of torsion modifies some identities that are used repeatedly throughout our analysis. Firstly, integration by parts picks a term proportional to the torsion: 
\be\label{tor-der}
\int d^dx\sqrt{-g}\,\nabla_\m V^\m=\int d^dx\,\del_\m(\sqrt{-g}V^\m)+\int d^dx\sqrt{-g}\,T_{\n\m}{}^\n V^\m\,.
\ee
Moreover, the presence of torsion modifies the Bianchi identities as follows:\footnote{The second Bianchi identity in eq.~(7.16) of \cite{Freedman:2012zz} contains a typo -- all signs on the r.h.s. must be reversed.}
\be\label{eq:genBianchi}
\begin{split}
R_{\m\n\r}{}^\s+R_{\n\r\m}{}^\s+R_{\r\m\n}{}^\s=&\;-\nabla_\m T_{\n\r}{}^\s-\nabla_\n T_{\r\m}{}^\s-\nabla_\r T_{\m\n}{}^\s\\
&\;+T_{\m\n}{}^\k T_{\r\k}{}^\s+T_{\r\m}{}^\k T_{\n\k}{}^\s+T_{\n\r}{}^\k T_{\m\k}{}^\s\,,\\
\nabla_\m R_{\n\r}{}^{\s\t}+\nabla_\r R_{\m\n}{}^{\s\t}+\nabla_\n R_{\r\m}{}^{\s\t}=&\;-T_{\m\n}{}^\k R_{\k\r}{}^{\s\t}-T_{\n\r}{}^\k R_{\k\m}{}^{\s\t}-T_{\r\m}{}^\k R_{\k\n}{}^{\s\t}\,,\\
\nabla_\m F_{\n\r}+\nabla_\r F_{\m\n}+\nabla_\n F_{\r\m}=&\;F_{\m\s}T_{\n\r}{}^\s+F_{\n\s}T_{\r\m}{}^\s+F_{\r\s}T_{\m\n}{}^\s\,.
\end{split}
\ee


\bibliographystyle{JHEP}
\bibliography{refs}

\providecommand{\href}[2]{#2}\begingroup\raggedright\begin{thebibliography}{10}

\bibitem{tHooft:1979rat}
G.~'t~Hooft, \emph{{Naturalness, chiral symmetry, and spontaneous chiral
  symmetry breaking}},
  \href{https://doi.org/10.1007/978-1-4684-7571-5_9}{\emph{NATO Sci. Ser. B}
  {\bfseries 59} (1980) 135}.

\bibitem{Komargodski:2011vj}
Z.~Komargodski and A.~Schwimmer, \emph{{On Renormalization Group Flows in Four
  Dimensions}}, \href{https://doi.org/10.1007/JHEP12(2011)099}{\emph{JHEP}
  {\bfseries 12} (2011) 099} [\href{https://arxiv.org/abs/1107.3987}{{\ttfamily
  1107.3987}}].

\bibitem{Wess:1971yu}
J.~Wess and B.~Zumino, \emph{{Consequences of anomalous Ward identities}},
  \href{https://doi.org/10.1016/0370-2693(71)90582-X}{\emph{Phys. Lett. B}
  {\bfseries 37} (1971) 95}.

\bibitem{Komargodski:2010rb}
Z.~Komargodski and N.~Seiberg, \emph{{Comments on Supercurrent Multiplets,
  Supersymmetric Field Theories and Supergravity}},
  \href{https://doi.org/10.1007/JHEP07(2010)017}{\emph{JHEP} {\bfseries 07}
  (2010) 017} [\href{https://arxiv.org/abs/1002.2228}{{\ttfamily 1002.2228}}].

\bibitem{omct}
P.~Panopoulos and I.~Papadimitriou, \emph{{Interpolating between 4d $\cn=1$
  SCFT supercurrent multiplet anomalies}},
  \href{https://arxiv.org/abs/2611.04967}{{\ttfamily 2611.04967}}.

\bibitem{Kaku:1977pa}
M.~Kaku, P.~K. Townsend and P.~van Nieuwenhuizen, \emph{{Gauge Theory of the
  Conformal and Superconformal Group}},
  \href{https://doi.org/10.1016/0370-2693(77)90552-4}{\emph{Phys. Lett.}
  {\bfseries 69B} (1977) 304}.

\bibitem{Kaku:1977rk}
M.~Kaku, P.~K. Townsend and P.~van Nieuwenhuizen, \emph{{Superconformal Unified
  Field Theory}},
  \href{https://doi.org/10.1103/PhysRevLett.39.1109}{\emph{Phys. Rev. Lett.}
  {\bfseries 39} (1977) 1109}.

\bibitem{Kaku:1978nz}
M.~Kaku, P.~K. Townsend and P.~van Nieuwenhuizen, \emph{{Properties of
  Conformal Supergravity}},
  \href{https://doi.org/10.1103/PhysRevD.17.3179}{\emph{Phys. Rev.} {\bfseries
  D17} (1978) 3179}.

\bibitem{Townsend:1979ki}
P.~K. Townsend and P.~van Nieuwenhuizen, \emph{{Simplifications of Conformal
  Supergravity}}, \href{https://doi.org/10.1103/PhysRevD.19.3166}{\emph{Phys.
  Rev.} {\bfseries D19} (1979) 3166}.

\bibitem{Stelle:1978ye}
K.~Stelle and P.~C. West, \emph{{Minimal Auxiliary Fields for Supergravity}},
  \href{https://doi.org/10.1016/0370-2693(78)90669-X}{\emph{Phys. Lett. B}
  {\bfseries 74} (1978) 330}.

\bibitem{Ferrara:1978em}
S.~Ferrara and P.~van Nieuwenhuizen, \emph{{The Auxiliary Fields of
  Supergravity}},
  \href{https://doi.org/10.1016/0370-2693(78)90670-6}{\emph{Phys. Lett. B}
  {\bfseries 74} (1978) 333}.

\bibitem{Papadimitriou:2019gel}
I.~Papadimitriou, \emph{{Supersymmetry anomalies in $\mathcal{N}=1$ conformal
  supergravity}}, \href{https://doi.org/10.1007/JHEP04(2019)040}{\emph{JHEP}
  {\bfseries 04} (2019) 040}
  [\href{https://arxiv.org/abs/1902.06717}{{\ttfamily 1902.06717}}].

\bibitem{Imbimbo:2023sph}
C.~Imbimbo, D.~Rovere and A.~Warman, \emph{{Superconformal anomalies from
  superconformal Chern-Simons polynomials}},
  \href{https://doi.org/10.1007/JHEP05(2024)277}{\emph{JHEP} {\bfseries 05}
  (2024) 277} [\href{https://arxiv.org/abs/2311.05684}{{\ttfamily
  2311.05684}}].

\bibitem{Imbimbo:2025ffw}
C.~Imbimbo and L.~Porro, \emph{{One Ring to Rule Them All: A Unified
  Topological Framework for 4D Superconformal Anomalies}},
  \href{https://arxiv.org/abs/2507.16505}{{\ttfamily 2507.16505}}.

\bibitem{Kuzenko:2019vvi}
S.~M. Kuzenko, A.~Schwimmer and S.~Theisen, \emph{{Comments on Anomalies in
  Supersymmetric Theories}},
  \href{https://doi.org/10.1088/1751-8121/ab64a8}{\emph{J. Phys. A} {\bfseries
  53} (2020) 064003} [\href{https://arxiv.org/abs/1909.07084}{{\ttfamily
  1909.07084}}].

\bibitem{Schwimmer:2010za}
A.~Schwimmer and S.~Theisen, \emph{{Spontaneous Breaking of Conformal
  Invariance and Trace Anomaly Matching}},
  \href{https://doi.org/10.1016/j.nuclphysb.2011.02.003}{\emph{Nucl. Phys. B}
  {\bfseries 847} (2011) 590}
  [\href{https://arxiv.org/abs/1011.0696}{{\ttfamily 1011.0696}}].

\bibitem{Bertlmann:1996xk}
R.~Bertlmann, \emph{{Anomalies in quantum field theory}}. Oxford University
  Press, 1996.

\bibitem{Chu:1996fr}
C.-S. Chu, P.-M. Ho and B.~Zumino, \emph{{NonAbelian anomalies and effective
  actions for a homogeneous space G/H}},
  \href{https://doi.org/10.1016/0550-3213(96)00322-7}{\emph{Nucl. Phys. B}
  {\bfseries 475} (1996) 484}
  [\href{https://arxiv.org/abs/hep-th/9602093}{{\ttfamily hep-th/9602093}}].

\bibitem{Coleman:1969sm}
S.~R. Coleman, J.~Wess and B.~Zumino, \emph{{Structure of phenomenological
  Lagrangians. 1.}},
  \href{https://doi.org/10.1103/PhysRev.177.2239}{\emph{Phys. Rev.} {\bfseries
  177} (1969) 2239}.

\bibitem{Callan:1969sn}
C.~G. Callan, Jr., S.~R. Coleman, J.~Wess and B.~Zumino, \emph{{Structure of
  phenomenological Lagrangians. 2.}},
  \href{https://doi.org/10.1103/PhysRev.177.2247}{\emph{Phys. Rev.} {\bfseries
  177} (1969) 2247}.

\bibitem{Salam:1969rq}
A.~Salam and J.~A. Strathdee, \emph{{Nonlinear realizations. 1: The Role of
  Goldstone bosons}},
  \href{https://doi.org/10.1103/PhysRev.184.1750}{\emph{Phys. Rev.} {\bfseries
  184} (1969) 1750}.

\bibitem{Wu:1984pv}
Y.-S. Wu, \emph{{Local Cohomology of Gauge Fields and Explicit Construction of
  {Wess-Zumino} Lagrangians in Nonlinear $\sigma$ Models Over $G/H$}},
  \href{https://doi.org/10.1016/0370-2693(85)91444-3}{\emph{Phys. Lett. B}
  {\bfseries 153} (1985) 70}.

\bibitem{Weinberg:1996kr}
S.~Weinberg, \emph{{The quantum theory of fields. Vol. 2: Modern
  applications}}. Cambridge University Press, 8, 2013,
  \href{https://doi.org/10.1017/CBO9781139644174}{10.1017/CBO9781139644174}.

\bibitem{Kugo:1982cu}
T.~Kugo and S.~Uehara, \emph{{Conformal and Poincare Tensor Calculi in $N=1$
  Supergravity}},
  \href{https://doi.org/10.1016/0550-3213(83)90463-7}{\emph{Nucl. Phys. B}
  {\bfseries 226} (1983) 49}.

\bibitem{Fradkin:1985am}
E.~S. Fradkin and A.~A. Tseytlin, \emph{{CONFORMAL SUPERGRAVITY}},
  \href{https://doi.org/10.1016/0370-1573(85)90138-3}{\emph{Phys. Rept.}
  {\bfseries 119} (1985) 233}.

\bibitem{Freedman:2012zz}
D.~Z. Freedman and A.~Van~Proeyen, \emph{{Supergravity}}. Cambridge Univ.
  Press, Cambridge, UK, 2012.

\bibitem{Bergshoeff:1980is}
E.~Bergshoeff, M.~de~Roo and B.~de~Wit, \emph{{Extended Conformal
  Supergravity}},
  \href{https://doi.org/10.1016/0550-3213(81)90465-X}{\emph{Nucl. Phys. B}
  {\bfseries 182} (1981) 173}.

\bibitem{deWit:1981vgr}
B.~de~Wit, \emph{{CONFORMAL INVARIANCE IN EXTENDED SUPERGRAVITY}},  in
  \emph{{First School on Supergravity Trieste, Italy, April 22-May 6, 1981}},
  p.~0267, 1981.

\bibitem{deWit:1983qkc}
B.~de~Wit, \emph{{MULTIPLET CALCULUS}},  in \emph{{September School on
  Supergravity and Supersymmetry}}, 3, 1983.

\bibitem{deWit:1984hw}
B.~de~Wit, \emph{{INTRODUCTION TO SUPERGRAVITY}},  in \emph{{Spring School on
  Supergravity and Supersymmetry}}, 6, 1984.

\bibitem{Iorio:1996ad}
A.~Iorio, L.~O'Raifeartaigh, I.~Sachs and C.~Wiesendanger, \emph{{Weyl gauging
  and conformal invariance}},
  \href{https://doi.org/10.1016/S0550-3213(97)00190-9}{\emph{Nucl. Phys. B}
  {\bfseries 495} (1997) 433}
  [\href{https://arxiv.org/abs/hep-th/9607110}{{\ttfamily hep-th/9607110}}].

\bibitem{Coriano:2013nja}
C.~Coriano, L.~Delle~Rose, C.~Marzo and M.~Serino, \emph{{The dilaton
  Wess\textendash{}Zumino action in six dimensions from Weyl gauging: local
  anomalies and trace relations}},
  \href{https://doi.org/10.1088/0264-9381/31/10/105009}{\emph{Class. Quant.
  Grav.} {\bfseries 31} (2014) 105009}
  [\href{https://arxiv.org/abs/1311.1804}{{\ttfamily 1311.1804}}].

\bibitem{VanNieuwenhuizen:1981ae}
P.~Van~Nieuwenhuizen, \emph{{Supergravity}},
  \href{https://doi.org/10.1016/0370-1573(81)90157-5}{\emph{Phys. Rept.}
  {\bfseries 68} (1981) 189}.

\bibitem{Polyakov:1981rd}
A.~M. Polyakov, \emph{{Quantum Geometry of Bosonic Strings}},
  \href{https://doi.org/10.1016/0370-2693(81)90743-7}{\emph{Phys. Lett. B}
  {\bfseries 103} (1981) 207}.

\bibitem{Krasnikov:1985jp}
N.~V. Krasnikov, \emph{{RENORMALIZATION IN GAUGE THEORIES WITH GAMMA(5)
  ANOMALIES}}, {\emph{JETP Lett.} {\bfseries 40} (1984) 1170}.

\bibitem{Krasnikov:1986rq}
N.~V. Krasnikov, \emph{{$\gamma$ (5) Anomaly as the Source of the
  Nonlocality}}, \href{https://doi.org/10.1007/BF02906448}{\emph{Nuovo Cim. A}
  {\bfseries 95} (1986) 325}.

\bibitem{PhysRevD.35.1268}
D.~S. Hwang, \emph{Gauge and gravitational anomalies in two dimensions},
  \href{https://doi.org/10.1103/PhysRevD.35.1268}{\emph{Phys. Rev. D}
  {\bfseries 35} (1987) 1268}.

\bibitem{Alvarez-Gaume:1983ihn}
L.~Alvarez-Gaume and E.~Witten, \emph{{Gravitational Anomalies}},
  \href{https://doi.org/10.1016/0550-3213(84)90066-X}{\emph{Nucl. Phys. B}
  {\bfseries 234} (1984) 269}.

\bibitem{Jensen:2012kj}
K.~Jensen, R.~Loganayagam and A.~Yarom, \emph{{Thermodynamics, gravitational
  anomalies and cones}},
  \href{https://doi.org/10.1007/JHEP02(2013)088}{\emph{JHEP} {\bfseries 02}
  (2013) 088} [\href{https://arxiv.org/abs/1207.5824}{{\ttfamily 1207.5824}}].

\bibitem{Deser:1993yx}
S.~Deser and A.~Schwimmer, \emph{{Geometric classification of conformal
  anomalies in arbitrary dimensions}},
  \href{https://doi.org/10.1016/0370-2693(93)90934-A}{\emph{Phys. Lett. B}
  {\bfseries 309} (1993) 279}
  [\href{https://arxiv.org/abs/hep-th/9302047}{{\ttfamily hep-th/9302047}}].

\bibitem{alexakis2009}
S.~Alexakis, \emph{The decomposition of global conformal invariants i: On a
  conjecture of deser and schwimmer},
  \href{https://arxiv.org/abs/0711.1685}{{\ttfamily 0711.1685}}.

\bibitem{Anastasiou:2020zwc}
G.~Anastasiou, O.~Miskovic, R.~Olea and I.~Papadimitriou, \emph{{Counterterms,
  Kounterterms, and the variational problem in AdS gravity}},
  \href{https://doi.org/10.1007/JHEP08(2020)061}{\emph{JHEP} {\bfseries 08}
  (2020) 061} [\href{https://arxiv.org/abs/2003.06425}{{\ttfamily
  2003.06425}}].

\bibitem{Ntokos:2021duk}
P.~Ntokos and I.~Papadimitriou, \emph{{Black hole superpotential as a unifying
  entropy function and BPS thermodynamics}},
  \href{https://doi.org/10.1007/JHEP03(2022)058}{\emph{JHEP} {\bfseries 03}
  (2022) 058} [\href{https://arxiv.org/abs/2112.05954}{{\ttfamily
  2112.05954}}].

\bibitem{Baume:2013ika}
F.~Baume and B.~Keren-Zur, \emph{{The dilaton Wess-Zumino action in higher
  dimensions}}, \href{https://doi.org/10.1007/JHEP11(2013)102}{\emph{JHEP}
  {\bfseries 11} (2013) 102} [\href{https://arxiv.org/abs/1307.0484}{{\ttfamily
  1307.0484}}].

\bibitem{Assel:2015nca}
B.~Assel, D.~Cassani, L.~Di~Pietro, Z.~Komargodski, J.~Lorenzen and
  D.~Martelli, \emph{{The Casimir Energy in Curved Space and its Supersymmetric
  Counterpart}}, \href{https://doi.org/10.1007/JHEP07(2015)043}{\emph{JHEP}
  {\bfseries 07} (2015) 043}
  [\href{https://arxiv.org/abs/1503.05537}{{\ttfamily 1503.05537}}].

\bibitem{Fradkin:1983tg}
E.~S. Fradkin and A.~A. Tseytlin, \emph{{Conformal Anomaly in Weyl Theory and
  Anomaly Free Superconformal Theories}},
  \href{https://doi.org/10.1016/0370-2693(84)90668-3}{\emph{Phys. Lett. B}
  {\bfseries 134} (1984) 187}.

\bibitem{Riegert:1984kt}
R.~J. Riegert, \emph{{A Nonlocal Action for the Trace Anomaly}},
  \href{https://doi.org/10.1016/0370-2693(84)90983-3}{\emph{Phys. Lett. B}
  {\bfseries 134} (1984) 56}.

\bibitem{Deser:1999zv}
S.~Deser, \emph{{Closed form effective conformal anomaly actions in D $\geq$
  4}}, \href{https://doi.org/10.1016/S0370-2693(00)00315-4}{\emph{Phys. Lett.}
  {\bfseries B479} (2000) 315}
  [\href{https://arxiv.org/abs/hep-th/9911129}{{\ttfamily hep-th/9911129}}].

\bibitem{Mottola:2026emf}
E.~Mottola, \emph{{The Wess-Zumino Action of the Conformal Anomaly}},  10,
  2026, \href{https://arxiv.org/abs/2610.36735}{{\ttfamily 2610.36735}}.

\bibitem{Deser:1996na}
S.~Deser, \emph{{Conformal anomalies: Recent progress}}, {\emph{Helv. Phys.
  Acta} {\bfseries 69} (1996) 570}
  [\href{https://arxiv.org/abs/hep-th/9609138}{{\ttfamily hep-th/9609138}}].

\bibitem{Brown:1985ri}
M.~R. Brown and A.~C. Ottewill, \emph{{EFFECTIVE ACTIONS AND CONFORMAL
  TRANSFORMATIONS}},
  \href{https://doi.org/10.1103/PhysRevD.31.2514}{\emph{Phys. Rev. D}
  {\bfseries 31} (1985) 2514}.

\bibitem{Tomboulis:1988gw}
E.~T. Tomboulis, \emph{{Dynamically Adjusted Cosmological Constant and
  Conformal Anomalies}},
  \href{https://doi.org/10.1016/0550-3213(90)90150-C}{\emph{Nucl. Phys. B}
  {\bfseries 329} (1990) 410}.

\bibitem{Cappelli:1988vw}
A.~Cappelli and A.~Coste, \emph{{On the Stress Tensor of Conformal Field
  Theories in Higher Dimensions}},
  \href{https://doi.org/10.1016/0550-3213(89)90414-8}{\emph{Nucl. Phys. B}
  {\bfseries 314} (1989) 707}.

\bibitem{Roberts:1992ij}
M.~D. Roberts, \emph{{The Double duals and gauges of the Lanczos tensor}},
  \href{https://doi.org/10.1007/BF01606365}{\emph{Czech. J. Phys.} {\bfseries
  42} (1992) 151}.

\bibitem{Festuccia:2011ws}
G.~Festuccia and N.~Seiberg, \emph{{Rigid Supersymmetric Theories in Curved
  Superspace}}, \href{https://doi.org/10.1007/JHEP06(2011)114}{\emph{JHEP}
  {\bfseries 06} (2011) 114} [\href{https://arxiv.org/abs/1105.0689}{{\ttfamily
  1105.0689}}].

\bibitem{Frob:2020gdh}
M.~B. Fr\"ob, \emph{{FieldsX -- An extension package for the xAct tensor
  computer algebra suite to include fermions, gauge fields and BRST
  cohomology}},  \href{https://arxiv.org/abs/2008.12422}{{\ttfamily
  2008.12422}}.

\bibitem{Anselmi:1997am}
D.~Anselmi, D.~Z. Freedman, M.~T. Grisaru and A.~A. Johansen,
  \emph{{Nonperturbative formulas for central functions of supersymmetric gauge
  theories}}, \href{https://doi.org/10.1016/S0550-3213(98)00278-8}{\emph{Nucl.
  Phys. B} {\bfseries 526} (1998) 543}
  [\href{https://arxiv.org/abs/hep-th/9708042}{{\ttfamily hep-th/9708042}}].

\bibitem{Papadimitriou:2017kzw}
I.~Papadimitriou, \emph{{Supercurrent anomalies in 4d SCFTs}},
  \href{https://doi.org/10.1007/JHEP07(2017)038}{\emph{JHEP} {\bfseries 07}
  (2017) 038} [\href{https://arxiv.org/abs/1703.04299}{{\ttfamily
  1703.04299}}].

\bibitem{Minasian:2021png}
R.~Minasian, I.~Papadimitriou and P.~Yi, \emph{{Anomalies and supersymmetry}},
  \href{https://doi.org/10.1103/PhysRevD.105.065005}{\emph{Phys. Rev. D}
  {\bfseries 105} (2022) 065005}
  [\href{https://arxiv.org/abs/2104.13391}{{\ttfamily 2104.13391}}].

\bibitem{Panopoulos:2023cdg}
P.~Panopoulos and I.~Papadimitriou, \emph{{Supersymmetric Casimir energy on
  $\mathcal{N}=1$ conformal supergravity backgrounds}},
  \href{https://arxiv.org/abs/2312.17740}{{\ttfamily 2312.17740}}.

\bibitem{Bardeen:1984pm}
W.~A. Bardeen and B.~Zumino, \emph{{Consistent and Covariant Anomalies in Gauge
  and Gravitational Theories}},
  \href{https://doi.org/10.1016/0550-3213(84)90322-5}{\emph{Nucl. Phys. B}
  {\bfseries 244} (1984) 421}.

\bibitem{Aminov:2026zrv}
G.~Aminov, C.~Cs{\'a}ki, O.~Telem and S.~Yankielowicz, \emph{{Non-Abelian and
  Type-A Conformal Anomalies from Euler Descent}},
  \href{https://arxiv.org/abs/2601.18892}{{\ttfamily 2601.18892}}.

\bibitem{Butter:2009cp}
D.~Butter, \emph{{N=1 Conformal Superspace in Four Dimensions}},
  \href{https://doi.org/10.1016/j.aop.2009.09.010}{\emph{Annals Phys.}
  {\bfseries 325} (2010) 1026}
  [\href{https://arxiv.org/abs/0906.4399}{{\ttfamily 0906.4399}}].

\bibitem{Kugo:2016zzf}
T.~Kugo, R.~Yokokura and K.~Yoshioka, \emph{{Component versus superspace
  approaches to D = 4, N = 1 conformal supergravity}},
  \href{https://doi.org/10.1093/ptep/ptw090}{\emph{PTEP} {\bfseries 2016}
  (2016) 073} [\href{https://arxiv.org/abs/1602.04441}{{\ttfamily
  1602.04441}}].

\bibitem{Kaku:1978ea}
M.~Kaku and P.~K. Townsend, \emph{{POINCARE SUPERGRAVITY AS BROKEN
  SUPERCONFORMAL GRAVITY}},
  \href{https://doi.org/10.1016/0370-2693(78)90098-9}{\emph{Phys. Lett.}
  {\bfseries 76B} (1978) 54}.

\bibitem{Howe:1978km}
P.~S. Howe and R.~Tucker, \emph{{Scale Invariance in Superspace}},
  \href{https://doi.org/10.1016/0370-2693(78)90327-1}{\emph{Phys. Lett. B}
  {\bfseries 80} (1978) 138}.

\bibitem{Buchbinder:1998twe}
I.~L. Buchbinder and S.~M. Kuzenko, \emph{{Ideas and Methods of Supersymmetry
  and Supergravity or A Walk Through Superspace: A Walk Through Superspace}}.
  Taylor {\&} Francis, 1998,
  \href{https://doi.org/10.1201/9780367802530}{10.1201/9780367802530}.

\bibitem{Fradkin:1978jq}
E.~Fradkin and M.~A. Vasiliev, \emph{{S Matrix for Theories That Admit Closure
  of the Algebra With the Aid of Auxiliary Fields: The Auxiliary Fields in
  Supergravity}}, \href{https://doi.org/10.1007/BF02783437}{\emph{Lett. Nuovo
  Cim.} {\bfseries 22} (1978) 651}.

\bibitem{Ferrara:1978rk}
S.~Ferrara, M.~T. Grisaru and P.~van Nieuwenhuizen, \emph{{Poincare and
  Conformal Supergravity Models With Closed Algebras}},
  \href{https://doi.org/10.1016/0550-3213(78)90389-9}{\emph{Nucl. Phys. B}
  {\bfseries 138} (1978) 430}.

\bibitem{Ferrara:1988qx}
S.~Ferrara and M.~Villasante, \emph{{Curvatures, {Gauss-Bonnet} and
  Chern-simons Multiplets in Old Minimal $N=1$ Supergravity}},
  \href{https://doi.org/10.1063/1.528576}{\emph{J. Math. Phys.} {\bfseries 30}
  (1989) 104}.

\bibitem{Katsianis:2020hzd}
G.~Katsianis, I.~Papadimitriou, K.~Skenderis and M.~Taylor,
  \emph{{Supersymmetry anomaly in the superconformal Wess-Zumino model}},
  \href{https://doi.org/10.1007/JHEP02(2021)209}{\emph{JHEP} {\bfseries 02}
  (2021) 209} [\href{https://arxiv.org/abs/2011.09506}{{\ttfamily
  2011.09506}}].

\bibitem{Bzowski:2020tue}
A.~Bzowski, G.~Festuccia and V.~Proch{\'a}zka, \emph{{Consistency of
  supersymmetric {\textquoteright}t Hooft anomalies}},
  \href{https://doi.org/10.1007/JHEP02(2021)225}{\emph{JHEP} {\bfseries 02}
  (2021) 225} [\href{https://arxiv.org/abs/2011.09978}{{\ttfamily
  2011.09978}}].

\end{thebibliography}\endgroup

\end{document}